%% file: main.tex
\documentclass[aps,preprint]{revtex4}
\usepackage{psfig}
\usepackage{amsfonts,amssymb}
\usepackage{graphics}

\begin{document}

\title{Propagation of Surface Plasmons in Ordered and Disordered Chains of
Metal Nanospheres}
\author{Vadim A. Markel}
\affiliation{Departments of Radiology and Bioengineering, University
  of Pennsylvania, Philadelphia, PA 19104}

\author{Andrey K. Sarychev}
\affiliation{Ethertronics Inc., San Diego, CA 92121 }

\begin{abstract}
  We report a numerical investigation of surface plasmon (SP)
  propagation in ordered and disordered linear chains of metal
  nanospheres.  In our simulations, SPs are excited at one end of a
  chain by a near-field tip. We then find numerically the SP amplitude
  as a function of propagation distance. Two types of SPs are
  discovered. The first SP, which we call the ordinary or quasistatic,
  is mediated by short-range, near-field electromagnetic interaction
  in the chain. This excitation is strongly affected by Ohmic losses
  in the metal and by disorder in the chain.  These two effects result
  in spatial decay of the quasistatic SP by means of absorptive and
  radiative losses, respectively. The second SP is mediated by longer
  range, far-field interaction of nanospheres. We refer to this SP as
  the extraordinary or non-quasistatic. The non-quasistatic SP can not
  be effectively excited by a near-field probe due to the small
  integral weight of the associated spectral line. Because of that, at
  small propagation distances, this SP is dominated by the quasistatic
  SP.  However, the non-quasistatic SP is affected by Ohmic and
  radiative losses to a much smaller extent than the quasistatic one.
  Because of that, the non-quasistatic SP becomes dominant
  sufficiently far from the exciting tip and can propagate with little
  further losses of energy to remarkable distances. The unique
  physical properties of the non-quasistatic SP can be utilized in
  all-optical integrated photonic systems.
\end{abstract}

\date{\today}
\maketitle

\section{Introduction}
\label{sec:intro}

Surface plasmons (SPs) are states of polarization that can propagate
along metal-dielectric interfaces or along other structures without
radiative losses. Polarization in an SP excitation can be spatially
confined on scales that are much smaller than the free-space
wavelength. This property proved to be extremely valuable for
manipulation of light energy on subwavelength
scales~\cite{sarychev_00_1,stockman_04_2}, miniaturization of optical
elements~\cite{engheta_05_1}, and achieving coherent temporal control
at remarkably short times~\cite{podolskiy_02_2,stockman_04_1}. SP
excitations in ordered one-dimensional arrays of nanoparticles have
attracted significant attention in recent years due to numerous
potential application in
nanoplasmonics~\cite{quinten_98_1,brongersma_00_1,burin_04_1,quidant_04_1,simovski_05_1}.
A periodic chain of high conductivity metal nanospheres can be used as
an SP wave guide - an analog of an optical
waveguide~\cite{maier_03_1}. High-quality SP modes in ordered and
disordered chains may be utilized in random lasers~\cite{burin_04_1}.
Electromagnetic forces acting on linear chains of nanoparticles can
produce the effect of optical trapping~\cite{guillon_06_1}. Various
spectroscopic and sensing applications have also been
discussed~\cite{markel_93_1,zou_04_1,zou_06_1}.

In this paper we study theoretically and numerically propagation of SP
excitations in long ordered and disordered chains of nanospheres.
Although, under ideal conditions, SP excitations can propagate without
loss of energy, in practical situations this is not so.  There are two
physical effects that can result in decay of SP excitations as they
propagate along the chain. The first effect is Ohmic losses due to the
finite conductivity of the metal.  The second effect is radiative
losses due to disorder in the chain (scattering from imperfections).
This effect is more subtle and is closely related to the phenomenon of
localization. In this paper we discuss both effects and illustrate
them with numerical examples.

We first focus on decay due to Ohmic losses and show that it can be
suppressed at sufficiently large propagation distances. The main idea
is based on exploiting an exotic non-Lorentzian resonance in the chain
which originates due to radiation-zone interaction of
nanoparticles~\cite{markel_93_1,markel_05_2} and can not be understood
within the quasistatics, even when both the nanoparticles in the chain
and the inter-particle spacing are much smaller than the wavelength.
From the spectroscopic point of view, the non-Lorentzian resonances
are manifested by very narrow lines in extinction
spectra~\cite{markel_93_1,zou_04_1}. One of the authors (V.A.M.) has
argued previously that the small integral weight of the spectral lines
associated with these resonances precludes them from being excited by
a near-field probe~\cite{markel_05_2}. This property would make the
non-Lorentzian resonance a curiosity which is rather useless for
nanoplasmonics.  However, numerical simulations shown below reveal
that the corresponding SP has relatively small yet nonzero amplitude
and is also characterized by very slow spatial decay. Therefore, in
sufficiently long chains, this SP becomes dominant and can propagate,
without significant further losses, to remarkable distance.  We stress
that the non-Lorentzian SP is an excitation specific to discrete
systems; it does not exist, for example, in metal nanowires.

But the above consideration applies only to ordered chains. Therefore,
we consider next the effects of disorder. To isolate radiative losses
due to scattering on imperfections from Ohmic losses, we consider
nanoparticles with infinite conductivity (equivalently, zero Drude
relaxation constant).  Although such metals do not exist in nature, an
equivalent system can be constructed experimentally by embedding
metallic particles into a dielectric medium with positive gain. We
show that while the ordinary (defined more precisely in the text
below) SP excitations are very sensitive to off-diagonal (position)
disorder, the SP due to the non-Lorentzian resonance is not. Diagonal
disorder (disorder in nanoparticle properties) is also considered.

The paper is organized as follows. In section~\ref{sec:theory}, we
describe the theoretical model and introduce basic equations.
Conditions for resonance excitation of SPs in a chain are considered
ins section~\ref{sec:disp}. Numerical results for propagation in
ordered and disordered chains are reported in
sections~\ref{sec:prop_ord} and~\ref{sec:prop_dord}, respectively.
Finally, section~\ref{sec:summary} contains a summary of obtained
results.

\section{Theoretical Model}
\label{sec:theory}

Consider a linear chain of $N$ nanospheres with radiuses $a_n$
centered at points $x_n$. We work in the dipole approximation which is
valid if $x_{n+1} - x_n \gtrsim (a_{n+1} + a_n) / 2$ and has been
widely used in the
literature~\cite{brongersma_00_1,burin_04_1,simovski_05_1,weber_04_1,citrin_05_1,citrin_06_1}.
The $n$-th nanosphere is then characterized by a dipole moment with
amplitude $d_n$ oscillating at the electromagnetic frequency $\omega$.
The dipole moments are coupled to each other and to external field by
the coupled-dipole equation~\cite{markel_93_1}

\begin{equation}  
\label{CDE}
d_n = \alpha_n \left[ E_n + \sum_{n^{\prime}\neq n} G_k(x_n,x_{n^{\prime}})
d_{n^{\prime}} \right] \ ,
\end{equation}

\noindent 
where $\alpha_n$ is the polarizability of the $n$-th nanosphere, $E_n$
is the external electric field at the point $x_n$, $k=\omega/c$ is the
free space wave number and $G_k(x,x^{\prime})$ is the appropriate
element of the free space, frequency-domain Green's tensor for the
electric field. The latter is translationally invariant with respect
to spatial variables, namely, $G_k(x,x^{\prime}) = G_k(x-x^{\prime},0)
= G_k(x^{\prime}-x,0)$. For an SP polarized perpendicular and parallel
to the chain, the respective functions $G_k^{\perp}$ and
$G_k^{\parallel}$ are given by

\begin{eqnarray}  
\label{G_perp}
G_k^{\perp}(x,0) = \left(\frac{k^2}{\vert x \vert} + \frac{ik}{\vert x
\vert^2} - \frac{1}{\vert x \vert^3} \right)\exp( i k \vert x \vert) \ , \\
\label{G_par}
G_k^{\parallel}(x,0) = \left( - \frac{2ik}{\vert x \vert^2} + \frac{2}{\vert
x \vert^3} \right)\exp(i k \vert x \vert) \ .
\end{eqnarray}

\noindent 
Polarizability of the $n$-th sphere is taken in the form

\begin{equation}  
\label{alpha_def}
\alpha_n = \frac{1}{1/\alpha_n^{\mathrm{(LL)}} - 2ik^3/3} \ ,
\end{equation}

\noindent 
where $\alpha_n^{\mathrm{(LL)}}$ is the Lorenz-Lorentz quasistatic
polarizability of a sphere of radius $a_n$ and $2ik^3/3$ is the first
non-vanishing radiative correction to the inverse polarizability;
account of this correction is important to ensure that the system
conserves energy~\cite{draine_88_1}. The Lorenz-Lorentz polarizability
is given, in terms of the complex permeability of the $n$-th
nanosphere $\epsilon_n$, by

\begin{equation}  
\label{alpha_LL}
\alpha_n^{\mathrm{(LL)}} = a_n^3 \frac{\epsilon_n - 1}{\epsilon_n + 2} \ .
\end{equation}

\noindent 
We further adopt, for simplicity, the Drude model for $\epsilon_n$:

\begin{equation}  
\label{Drude}
\epsilon_n = 1 - \frac{\omega_{\mathrm{p}n}^2}{\omega(\omega +
  i\gamma_n)} \ ,
\end{equation}

\noindent 
where $\omega$ is the electromagnetic frequency,
$\omega_{\mathrm{p}n}$ is the plasma frequency, and $\gamma_n$ is the
Drude relaxation constant in the $n$-th nanosphere. The inverse
polarizability of the $n$-th nanosphere is then given by

\begin{equation}  
\label{Re_Im_alpha}
\mathrm{Re}\left(\frac{1}{\alpha_n}\right) = \frac{1}{a_n^3} \left[ 1 -
\left( \frac{\omega}{\omega_{\mathrm{F}n}} \right)^2 \right] \ , \ \
\mathrm{Im} \left(\frac{1}{\alpha_n}\right) = -i \left(\frac{2k^3}{3}
  + \frac{1}{a_n^3} \frac{\omega\gamma_n}{\omega_{\mathrm{F}n}^2} \right) \ ,
\end{equation}

\noindent 
where $\omega_{\mathrm{F}n}=\omega_{\mathrm{p}n}/\sqrt{3}$ is the
Frohlich frequency. The SP resonance of an isolated $n$-th nanosphere
takes place when $\omega=\omega_{\mathrm{F}n}$. Polarizability
$\alpha_n$ at the Frohlich resonance is purely imaginary; if, in
addition, there are no Ohmic losses in the material ($\gamma_n=0$),
the resonance value of the polarizability becomes $\alpha_n =
\alpha_{\mathrm{res}}= -i3/2k^3$, irrespectively of the particle
radius.

Suppose that SP is excited at a given site (say, $n=n_0$) by a
near-field probe. Then the external field can be set to
$E_n=E_0\delta_{n,n_0}$. Of course, this is an idealization: the field
produced even by a very small near-field tip is, strictly speaking,
non-zero at all sites. However, this approximation is physically
reasonable because of the fast (cubic) spatial decay of the dipole
field in the near-field zone. The solution with $E_n = \delta_{n,n_0}$
is, essentially, the Green's function for polarization. We denote this
Green's function by ${\mathcal D}_{k}(x_{n},x_{n_{0}})$. It satisfies

\begin{equation}  
\label{GFD}
{\mathcal D}_k(x_n,x_{n_0}) = \alpha_n \left[ \delta_{n,n_0} + \sum_{n^{\prime}\neq
n} G_k(x_n,x_{n^{\prime}}) {\mathcal D}_k(x_{n^{\prime}},x_{n_0})
\right] \ , 
\end{equation}

\noindent 
where either (\ref{G_perp}) or (\ref{G_par}) should be used for $G_k$,
depending on polarization of the SP. In the case of a finite or
disordered chain, one can find ${\mathcal D}_k(x_n,x_{n_0})$ by
solving Eq.~(\ref{CDE}) numerically. However, in infinite ordered
chains such that $\alpha_n=\alpha=\mathrm{const}$ and
$x_{n+1}-x_n=h=\mathrm{const}$, the following analytic solution is
obtained by Fourier transform~\cite{markel_05_2}:

\begin{equation}  
\label{d_n_Fur}
{\mathcal D}_k(x_n,0) = \int_{-\pi/h}^{\pi/h} {\frac{{\exp(iq x_n)} }{{1/\alpha -
S(k,q)}}} {\frac{h dq }{2\pi}} \ , 
\end{equation}

\noindent 
where $S(k,q)$ is the ``dipole sum'' given by

\begin{equation}  
\label{S_def}
S(k,q) = 2\sum_{n>0} G_{k}(0,x_n)\cos(q x_n) \ .
\end{equation}

\noindent 
Obviously, in infinite chains ${\mathcal D}_k(x_n,x_{n^{\prime}}) =
{\mathcal D}_k(x_n-x_{n^{\prime}},0)={\mathcal D}_k(x_{n^{\prime}} -
x_n,0)$. Note that the dipole sum (\ref{S_def}) is independent of
material properties. It can be shown~\cite{markel_93_1} that, for all
values of parameters, $\mathrm{Im} S(k,q) \geq -2k^3/3$. The equality
holds when $q>k$. This is a manifestation of the fact that SPs with
$q>k$ are non-radiating due to the light-cone
constraint~\cite{burin_04_1}.  Non-radiating modes exist if $\pi/h>k$
or, equivalently, if $\lambda>2h$. Obviously, these SPs do not couple
to running waves but can be excited by a near-field probe. The
dimensionless radiative relaxation parameter can be defined as $Q(k,q)
= [\mathrm{Im}S(k,q) + 2k^3/3]/(2k^3/3)$; this factor is identically
zero for $q>k$.

\section{Dispersion Relations and Resonant Excitation of SP}
\label{sec:disp}

In this section, we consider periodic chains with $a_n=a$,
$\omega_{\mathrm{p}n}=\omega_{\mathrm{p}}$,
$\omega_{\mathrm{F}n}=\omega_{\mathrm{F}}$, $\gamma_n=\gamma$ and,
consequently, with the constant polarizability $\alpha_n=\alpha$.

It follows from formula (\ref{d_n_Fur}) that the wave numbers $q$ of
SP excitations that can propagate effectively in an infinite periodic
chain are such that $1/\alpha \approx S(k,q)$. This condition must be
satisfied for some range of $q$ which is, in some sense, small.
Indeed, there is no effective interaction in the chain if $1/\alpha -
S(k,q) \approx \mathrm{const}$, in which case integration according to
(\ref{d_n_Fur}) yields~\cite{fn2} $d_n \propto \delta_{n,0}$. We now
examine the conditions under which the denominator of (\ref{d_n_Fur})
can become small. The dispersion relation in the usual sense, e.g.,
the dependence of the resonant SP frequency on its wavenumber is
obtained by solving the equation $1/\alpha(\omega) - S(\omega/c,q) =
0$. This approach was adopted, for example, in
Refs.~\onlinecite{brongersma_00_1,weber_04_1,burin_04_1,simovski_05_1}.
Solution to the above equation depends on the model for
$\epsilon(\omega)$ and may result in several branches of the complex
function $\omega(q)$. Here we adopt a slightly different point of
view. Namely, we note that for real frequencies $ \omega$ and wave
numbers $q$, the real part of the denominator can change sign while
the imaginary part is always non-negative. Physically,
$\mathrm{Re}[1/\alpha(\omega) - S(\omega/c,q)]$ can be interpreted as
the generalized \emph{detuning} from a resonance while
$\mathrm{Im}[1/\alpha(\omega) - S(\omega/c,q)]$ gives total (radiative
and absorptive) losses. We thus define the resonance condition to be
$\mathrm{Re}[1/\alpha - S(k,q)]=0$ and view $\omega$ and $q$ as
independent purely real variables.

\begin{figure}[tbp]
\centerline{\input{disp_re_ort.tex} \input{disp_re_par.tex}}
\centerline{\input{disp_im_ort.tex} \input{disp_im_par.tex}}
\caption{(a,b): Real parts of the dipole sum $S(k,q)$ as a function of $q$,
  for different values of $kh$, as indicated. Sharp peaks
  corresponding to divergence of $\mathrm{Re}S(k,q)$ at the point
  $q=k$ are not completely resolved. Quasistatic results obtained in
  the limit $k=0$ are shown by curves labeled ``QS''. Only half of the
  Brillouin zone is shown since $ S(k,-q)=S(k,q)$. Polarization of SP
  is perpendicular (a) and parallel (b) to the chain. (c,d):
  Dimensionless radiative relaxation parameter
  $Q(k,q)=[\mathrm{Im}S(k,q) + 2k^3/3]/(2k^3/3)$ as a function of $q$
  for the same sets of parameters as above. Polarization of SP is
  perpendicular (c) and parallel (d) to the chain. Quasistatic result
  is not shown since $Q$ is not defined in the quasistatic limit.}
\label{fig:disp}
\end{figure}
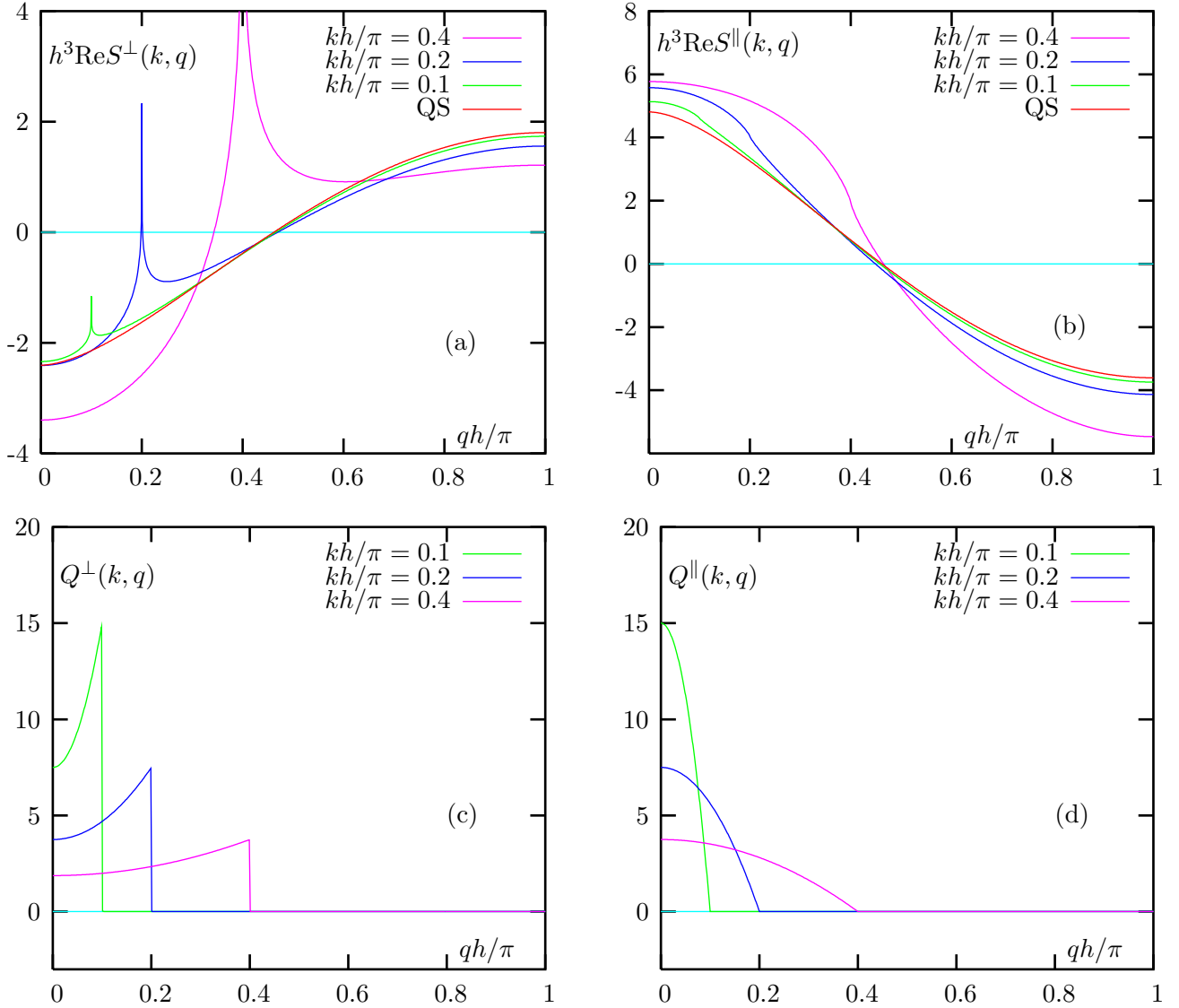

Plots of the dimensionless function $h^3 S(k,q)$ are shown in
Fig.~\ref{fig:disp} for some typical sets of parameters and for two
orthogonal polarizations of the SP. Note that the imaginary part of
$h^3 S(k,q)$ is related to radiative relaxation parameter $Q(k,q)$
[shown in Fig.~\ref{fig:disp}(c,d)] by $h^3 \mathrm{Im}S(k,q) =
[2(kh)^3/3][Q(k,q)-1]$. Apart from the very narrow peaks appearing in
the case of perpendicular polarization and centered at
$q=k$,~\cite{fn3} the numerical values of $h^3 \vert \mathrm{Re}S
\vert$ do not exceed, approximately, $6$. On the other hand, according
to (\ref{Re_Im_alpha}), we have $h^3\mathrm{Re}(1/\alpha) = (h/a)^3[1
-(\omega/\omega_{\mathrm{F}})^2]$. The dipole approximation is valid
when $(h/a)^3 \gtrsim 64$. Therefore, if we stay within the range of
parameters in which the dipole approximation is valid, the real part
of the denominator in (\ref{d_n_Fur}) is relatively small only if
$\omega\approx \omega_{\mathrm{F}}$, i.e., if $\omega$ is near the
Frohlich frequency of an isolated sphere. This is the case that will
be considered below.

Consider first oscillations polarized orthogonally to the chain and
let $\omega=\omega_{\mathrm{F}}$. The condition of resonant excitation
of SP is then $\mathrm{Re}S(k,q) = 0$. It can be seen from
Fig.~\ref{fig:disp}(a) that, for sufficiently small values of the
dimensionless parameter $kh$, there are two different values of $q$
that satisfy the above resonance condition.

The first solution is $q=q_1\approx 0.47\pi/h$ (for $kh=0.2\pi$). The
value of $q_1$ depends only weakly on $kh$, as long as $kh\lesssim
0.2\pi$, and is approximately the same as in the quasistatic limit
$kh=0$, in which case $q_1\approx 0.46\pi/h$. We will refer to the SP
with this wave number as the ordinary, or the quasistatic SP. This is
because propagation of this SP is mediated by near-field interaction,
while the far-field interaction is suppressed by destructive
interference of waves radiated by different nanoparticles in the
chain. Since the wave number of the ordinary SP is greater than $k$,
it propagates without radiative losses.  However, it may experience
spatial decay due to absorption in metal.  The characteristic
exponential length of decay can be easily inferred from
(\ref{d_n_Fur}) by making the quasi-particle pole approximation.
Namely, we approximate $S(k,q)$ as

\begin{equation}  
\label{QPP_approx}
S(k,q) \approx \mathrm{Re} S(k,q_1) + (q-q_1) \left\vert \frac{\partial 
\mathrm{Re} S(k,q)}{\partial q} \right \vert_{q=q_1} - i\frac{2k^3}{3} \ .
\end{equation}

\noindent 
We then extend integration in (\ref{d_n_Fur}) to the real axis and
obtain the following characteristic exponential decay length $\ell$:

\begin{equation}  
\label{ell_def}
\ell = \frac{1}{\delta} \left \vert \frac{\partial \mathrm{Re}S(k,q)}
{\partial q} \right \vert_{q=q_1} \ ,
\end{equation}

\noindent 
where

\begin{equation}  
\label{delta_def}
\delta = -\mathrm{Im}(1/\alpha) - 2k^3/3
\end{equation}

\noindent is a positive parameter characterizing the absorption strength of
the nanosphere. In general, it can be shown that $\delta=0$ in non-absorbing
particles whose dielectric function $\epsilon(\omega)$ is purely real at the
given frequency $\omega$. For the Drude model adopted in this paper, we have 
$\delta=\omega\gamma/a^3\omega_{\mathrm{F}}^2$. Thus, the ordinary SP can
decay exponentially due to absorption in metal with the characteristic scale
given by (\ref{ell_def}).

The second solution is obtained at $q = q_2 \approx k$, when
$\mathrm{Re} S(k,q)$ has a narrow sharp peak as a function of $q$ (for
fixed $k$)~\cite{fn1}. This peak is explained by far-field interaction
in an infinite chain~\cite{markel_93_1,markel_05_2}.  It does not
disappear in the limit $kh\rightarrow 0$, but becomes increasingly
(super-exponentially) narrow~\cite{markel_05_2}. Note that in the
above limit, this peak appears as a singularity of zero integral
weight which can not be obtained from the formal quasistatic
approximation. We will refer to this SP as extraordinary, or
non-quasistatic. It is mediated by the far-field interaction. The
latter is important in the case of extraordinary SP because of the
constructive interference of far field contributions from all
nanoparticles arriving at a given one. Obviously, the extraordinary SP
can be excited only in chains with $h<\lambda/2$, where $\lambda$ is
the wavelength of light in free space.

It is interesting to consider radiative losses of the extraordinary SP. As
can be seen from Fig.~\ref{fig:disp}(c), the factor $Q(k,q)$ is
discontinuous at $q=k$: it is zero for $q>k$ but positive for $q<k$. Since
the extraordinary SP has $q\approx k$, it can experience some radiative
losses, although the exact law of its decay depends on parameters of the
problem in a complicated manner. This is, in part, related to
inapplicability of the quasi-particle pole approximation for evaluating the
integral (\ref{d_n_Fur}) in the vicinity of $q\approx q_2$.

In the case of oscillations polarized along the chain, the resonance
condition can be satisfied only at $q=q_1\approx 0.45\pi/h$. This is
the wave number of an ordinary (quasistatic) SP which depends on $k$
only weakly, as long as $kh \lesssim 0.2\pi$. However, the
extraordinary (non-quasistatic) SP can be excited even for
longitudinal oscillations.  Mathematically, this can be explained by
observing that $\partial \mathrm{Re} S(k,q)/\partial q$ diverges at
$q=k$ while $\partial \mathrm{Im} S(k,q)/\partial q$ is discontinuous
at $q=k$ and performing integration (\ref{d_n_Fur}) by parts. However,
the amplitude of the extraordinary SP is much smaller for longitudinal
oscillations than for transverse oscillations; this will be
illustrated numerically in the next section.

Finally, we note, that when inter-sphere separations become smaller
than the radiuses, higher-order multipole resonances can be
excited~\cite{sansonetti_80_1,park_04_1,markel_04_3}. In this case,
resonant excitation of SP can become possible even at frequencies
which are far from the Frohlich resonance of an isolated sphere, e.g.,
in the IR part of the spectrum.

\section{Propagation in Finite Ordered Chains}
\label{sec:prop_ord}

We now turn to propagation of SP in ordered chains of finite length $N$. We,
however, emphasize that the finite size effects play a very minor role in
the computations shown below. Citrin~\cite{citrin_05_1} has studied
dispersion relations in finite chains and has found that the infinite-chain
limit is reached at $N\approx 10$ (although we anticipate that longer chains
are needed for accurate description of the extraordinary SP). In this and
following sections, we work with chains of $N\ge 1000$. In this limit,
propagation of both ordinary and extraordinary SP is not much different from
the case of infinite chains. In particular, we have verified numerically
that the Green's function ${\mathcal D}_k(x_n,x_{501})$ (for $kh=0.2\pi$) in a
chain of $N=1001$ particles does not differ in any significant way from that
in an infinite chain, except for values of $n$ very close to either end of
the finite chain. The Green's function $G_k(x_n,x_1)$ (here $n_0=1$ is the
end-point of the finite chain) differed by a trivial factor in finite and
infinite chains (results not shown). However, proper numerical evaluation of
integral (\ref{d_n_Fur}) required very fine discretization of $q$ and was a
more demanding and less stable procedure than direct numerical solution of
the system of equations (\ref{GFD}).

In this section, we take $a_n=a=\mathrm{const}$,
$\alpha_n=\alpha=\mathrm{const}$ and $x_n = nh$, $n=1,\ldots,N$. We
also assume that $kh=0.2\pi$ and $ h=4a$. Practically, this can be
realized for silver particles in a transparent host matrix with
refractive index of approximately $n=1.4$ so that the wavelength at
the Frohlich frequency is $\lambda_{\mathrm{F}}=2\pi c /
\omega_{\mathrm{F}} \approx 400\mathrm{nm}$, the chain spacing is
$h=40 \mathrm{nm}=0.1\lambda_{\mathrm{F}}$ and the sphere radius is
$a=10\mathrm{nm}=h/4$. The dipole approximation is very accurate for
this set of parameters. We then obtain the SP Green's function
${\mathcal D}_k(x_n,x_1)$ by solving Eq.~(\ref{GFD}) numerically.

\begin{figure}[tbp]
\centerline{\psfig{file=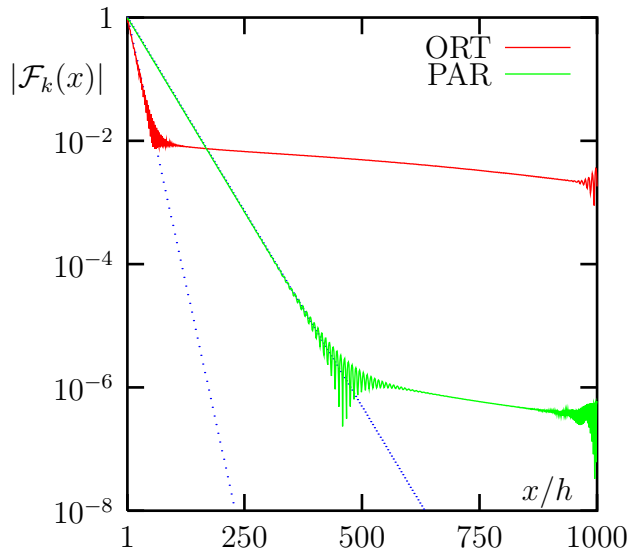,bbllx=0bp,bblly=500bp,bburx=596bp,bbury=725bp,clip=t}}
\caption{Propagation of a SP in an ordered chain of $N=1000$ nanospheres for
  orthogonal (ORT) and parallel (PAR) polarization of oscillations
  with respect to the chain. Parameters: $\omega=\omega_{\mathrm{F}}$,
  $\gamma/\omega_{\mathrm{F}}=0.002$, $\lambda=10h$, $h=4a$.}
\label{fig:ssp_ord}
\end{figure}

The absolute value of the normalized SP Green's function

\begin{equation}  
\label{F_def}
{\mathcal F}_k(x_n) = \frac{{\mathcal D}_k(x_n,x_1)}{{\mathcal D}_k(x_1,x_1)}
\end{equation}

\noindent 
in a chain of $N=1000$ nanospheres is shown in Fig.~\ref{fig:ssp_ord}
as a function of $x$ (sampled at $x=x_{n}$) for two orthogonal
polarizations. Here the frequency of SP was taken to be exactly equal
to the Frohlich frequency $\omega_{\mathrm{F}}$ and the Drude
relaxation constant was $\gamma =0.002\omega_{\mathrm{F}}$. It can be
seen from the figure that two different SP are excited in the system.
The first is the ordinary (quasistatic) SP that decays exponentially
as $\exp (-x/\ell )$, where $\ell $ is defined by (\ref{ell_def}).
Corresponding asymptotes are shown by dotted lines. Note that the
quasistatic SP in a finite chain (with the point of excitation
coinciding with one of the chain ends) is very well described by the
exponential decay formula, even though the latter was obtained for
infinite chains. When the amplitude of the ordinary SP becomes
sufficiently small, there is a crossover to the extraordinary SP. The
decay rate of the extraordinary SP is much slower. We have also
confirmed by inspecting the real and imaginary parts of ${\mathcal
  D}_{k}(x_{n},x_{1})$ (data not shown) that it oscillates at the
spatial frequency corresponding to the ordinary SP in the
fast-decaying segments of the curves shown in Fig.~\ref{fig:ssp_ord}
and with the spatial frequency that corresponds to the extraordinary
SP in the slow-decaying segments.

Mathematically, the relatively slow decay of the extraordinary SP can
be understood as follows. First, note the the exponential decay of the
ordinary SP is, in fact, the result of superposition of an infinite
number of plane waves whose wave numbers are in the interval $\Delta
q\propto \delta$. The corresponding wave packet decays spatially on
scales $\ell \propto 1/\Delta q$. However, in the case of
extraordinary SP, $\Delta q$ can not be defined since the
corresponding resonance is non-Lorentizian. It can be, however, stated
that the extraordinary SP is a superposition of plane waves whose wave
numbers are very close to $k$. Since, however, the wavenumbers can
still slightly deviate from $k$, some spatial decay at large distances
can still occur.

The conclusion we can make so far is that the ordinary SP experiences
exponential decay along the chain due to Ohmic losses in metal. This
decay is very accurately described by the quasi-particle pole
approximation. The extraordinary SP has, initially, much smaller
amplitude than the ordinary one. This is because the peaks in
Fig.~\ref{fig:disp} are very narrow. The quasi-particle pole
approximation is invalid for the extraordinary SP and its decay is
much less affected by Ohmic losses. As a result, the extraordinary SP
decays at a much slower rate and, at sufficiently large propagation
distance, begins to dominate. We also note that the extraordinary SP
can be excited even for longitudinally-polarized SP, although its
amplitude is smaller by some four orders of magnitude than for the
case of transverse oscillations.

\begin{figure}[tbp]
\centerline{\psfig{file=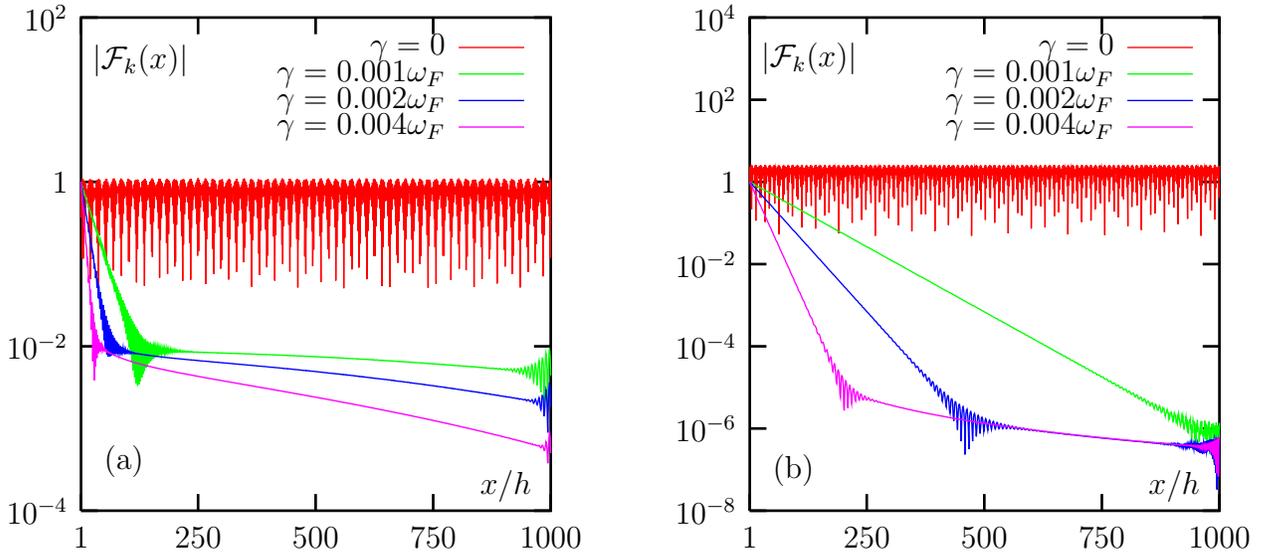,bbllx=0bp,bblly=500bp,bburx=596bp,bbury=725bp,clip=t}}
\caption{Same as in Fig.~\ref{fig:ssp_ord_gamma} for different
  ratios $\gamma/\omega_{\mathrm{F}}$ and for SP
  polarized orthogonally (a) and parallel (b) to the chain.}
\label{fig:ssp_ord_gamma}
\end{figure}

In Fig.~\ref{fig:ssp_ord_gamma}, we illustrate the influence of Ohmic
losses on SP propagation. Here we plot $\vert {\mathcal F}_k(x) \vert$
as a function of $x$ (sampled at $x=x_n$) for
$\omega=\omega_{\mathrm{F}}$ and different values of the ratio
$\gamma/\omega_{\mathrm{F}}$. First, in the absence of absorption
($\gamma=0$), the ordinary SP propagates along the chain without
decay. Once we introduce absorption, the ordinary SP decays
exponentially with the characteristic length scale $\ell$ given by
(\ref{ell_def}). Note that, for the specific metal permeability model
(\ref{Drude}), $\ell\propto \omega_{\mathrm{F}}/\gamma$. Some
dependence of the rate of decay of the extraordinary SP on the ratio
$\gamma/\omega_{\mathrm{F}}$ is visible in the case of orthogonal
polarization [Fig.~\ref{fig:ssp_ord_gamma}(a)]. However, when the
polarization is longitudinal [Fig.~\ref{fig:ssp_ord_gamma}(b)], decay
of the extraordinary SP is dominated by radiative losses. In
particular, the slow-decaying segments of the curves for
$\gamma/\omega_{\mathrm{F}}=0.002$ and
$\gamma/\omega_{\mathrm{F}}=0.004$ in Fig.~\ref{fig:ssp_ord_gamma}(b)
coincide with high precision.

\begin{figure}[tbp]
\centerline{\psfig{file=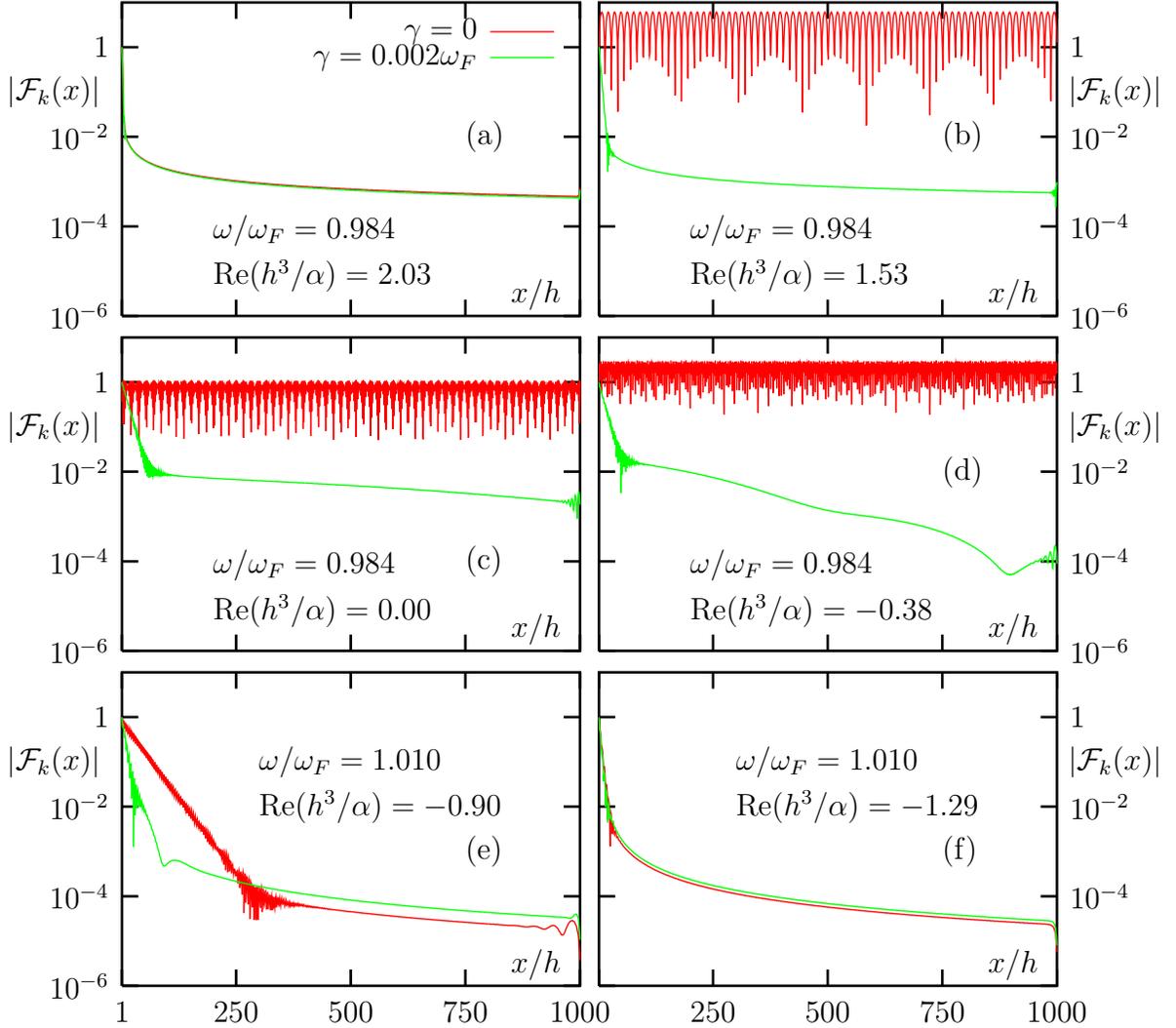,bbllx=0bp,bblly=300bp,bburx=596bp,bbury=725bp,clip=t}}
\caption{Same as in Fig.~\ref{fig:ssp_ord_gamma} for different
  ratios $\omega/\omega_{\mathrm{F}}$ and $\gamma/
  \omega_{\mathrm{F}}$, as indicated, and for SP polarized
  orthogonally to the chain.}
\label{fig:omega}
\end{figure}

Next, we study SP propagation for different values of the ratio $
\omega/\omega_{\mathrm{F}}$. As noted above, we assume the parameters
$ kh=\omega h/c=0.2\pi$ and $h/a=4$ to be fixed. Thus,
$\omega/\omega_{\mathrm{F}}$ can vary either due to a change in
$\omega_{\mathrm{F}}$ or due to a simultaneous change in $\omega$, $h$
and $a$ such that $\omega h = \mathrm{const}$ and
$h/a=\mathrm{const}$. It follows from Fig.~\ref{fig:disp}(a) that, for
the selected set of parameters, the ordinary plasmon can be excited
for $-0.89 < h^3\mathrm{Re}(1/\alpha) <1.56$. This corresponds to
$\omega/\omega_{\mathrm{F}}$ lying in the interval
$0.988<\omega/\omega_{\mathrm{F}} < 1.007$. In Fig.~\ref{fig:omega},
we illustrate propagation of SP excitations for some values of
$\omega/\omega_{\mathrm{F}}$ inside this interval, exactly at the
lower and upper bounds of this interval, and slightly outside of the
interval. Results are shown for two values of absorption strength:
$\gamma/\omega_{\mathrm{F}}=0$ and $\gamma/\omega_{
  \mathrm{F}}=0.002$.

First, consider the two cases when $\omega/\omega_{\mathrm{F}}$ is
outside of the interval where the ordinary SP can be excited:
$\omega/\omega_{ \mathrm{F}} = 0.984$ and $\omega/\omega_{\mathrm{F}}
= 1.010$ (correspondingly, $h^3\mathrm{Re}(1/\alpha) = 2.03$ and
$h^3\mathrm{Re}(1/\alpha) = -1.29$), shown in
Fig.~\ref{fig:omega}(a,f). In the case $\omega/\omega_{\mathrm{F}} =
0.984$, the ordinary SP exhibits very fast spatial decay, which is
characteristic for the noninteracting limit when $G_k(n,n_0) \propto
\delta_{n,n_0}$. After the initial decay of the ordinary SP, the
extraordinary SP becomes dominating. The extraordinary SP decays
slowly by means of radiative losses. It is interesting to note that
decay of the extraordinary SP is almost unaffected by Ohmic losses in
metal, although a noticeable dependence on $\gamma$ appears if we
further increase the parameter $\gamma/\omega_{\mathrm{F}}$ by the
factor of $10$ (data not shown). A qualitatively similar behavior is
obtained at $\omega/\omega_{\mathrm{F}}=1.010$. However, the decay of
extraordinary SP in this case is a little faster. Paradoxically, the
curve corresponding to $\gamma/\omega_{\mathrm{F}}=0.002$ is slightly
higher than the curve corresponding to $\gamma=0$ in
Fig.~\ref{fig:omega}(f) (similar peculiarity is seen in
Fig.~\ref{fig:omega}(e)).

When the ratio $\omega/\omega_{\mathrm{F}}$ is inside the interval
where the ordinary SP can be excited [Fig.~\ref{fig:omega}(c,d)], SP
propagation is strongly influenced by absorptive losses. In the
absence of such losses, the ordinary SP propagates along the chain
indefinitely and dominates the extraordinary SP. However, in the
presence of even small absorption, the ordinary SP decays
exponentially so that, at sufficiently large propagation distances,
the extraordinary SP starts to dominate. A qualitatively similar
picture is also obtained for the borderline case
$\omega/\omega_{\mathrm{F}}=0.984$ [Fig.~\ref{fig:omega}(b)]. In the
second borderline case [Fig.~\ref{fig:omega}(e)], SP propagation is
more complicated. The wave numbers of both ordinary and extraordinary
SPs in this case are close to $k$, so that both can experience
radiative decay, as is evident in the case of zero absorption.

To conclude this section, we note that propagation of SP excitations
in long periodic chains can be characterized by exponential decay.
This decay is caused either by absorptive or by radiative losses. At
small propagation distances, energy is transported by the ordinary SP
excitation, if the ordinary SP can be excited (e.g., if
$\mathrm{Re}(1/\alpha)$ is inside the appropriate interval). However,
at sufficiently large propagation distances, there is a cross over to
transport by means of the extraordinary SP. In this case, propagation
is mediated by far-zone interaction and is characterized by slow,
radiative decay which is affected by absorptive losses only weakly.

We note that exponential decay in ordered chains, if exists, is not
caused by Anderson localization, since we have not, so far, introduced
disorder into the system. For example, as was discussed in
section~\ref{sec:disp}, a linear superposition of delocalized plane
wave modes of the form (\ref{d_n_Fur}) can exhibit exponential decay
with the characteristic length (\ref{ell_def}). The irreversible
exponential decay is, in fact, obtained because the delocalized modes
form a truly continuous spectrum (are indexed by a continuous variable
$q$). Of course, any superposition of \emph{discrete} delocalized
modes would result in Poincare recurrences.

\section{Propagation in Disordered Chains}
\label{sec:prop_dord}

The ordinary (quasistatic) SP propagates in ordered chains without
radiative losses due to the perfect periodicity of the lattice.
However, once this periodicity is broken, the quasistatic SP can
experience radiative losses and spatial decay even in the absence of
absorption. Dependence of the radiative quality of SP modes in finite
one-dimensional chains on disorder strength was studied in
Ref.~\onlinecite{burin_04_1}. It was shown that position disorder
tends to decrease the radiative quality factor of initially
non-radiating (``bound'') modes. In this section we study how the
disorder influences propagation of the SP along the chain and take a
separate look at the ordinary and extraordinary SPs. We also consider
two types of disorder: off-diagonal and diagonal. Off-diagonal
disorder is disorder in particle positions while all particles are
identical.  Diagonal disorder arises due to differences in particle
properties, even if the particle positions are perfectly ordered.

The exponential decay due to Ohmic losses in the material can mask the
effects of disorder. Therefore, we assume in this section that the
nanospheres are non-absorbing, i.e., set $\gamma_n=0$. Physically,
absence of absorption can be realized by embedding the chain of
nanoparticles in a transparent dielectric host medium with positive
gain~\cite{sudarkin_89_1,bergman_03_1,avrutsky_04_1,lawandy_04_1}.
Such medium has dielectric permeability $\epsilon_{\mathrm{h}}$ with
positive real and negative imaginary parts. The gain can be tuned so
that the effective permeability of inclusions
$\epsilon/\epsilon_{\mathrm{h}}$ is purely real and negative. We also
work in the regime $kh=0.2\pi $, $h=4a$.

\subsection{Off-Diagonal Disorder}
\label{subsec:offdiag}

Off-diagonal disorder is disorder in particle position. It is called
``off-diagonal'' because it affects only off-diagonal elements of the
interaction matrix $G_k(x_n,x_{n^{\prime}})$. In the simulations shown
below, coordinates of particles in a disordered chain were taken to be
$ x_n=h(n+\xi_n)$ where $\xi_n$ is a random variable evenly
distributed in the interval $[-A,A]$. The random numbers $\xi_n$ are
taken to be mathematically independent. Therefore, the disorder is
uncorrelated.

\begin{figure}[tbp]
\centerline{\psfig{file=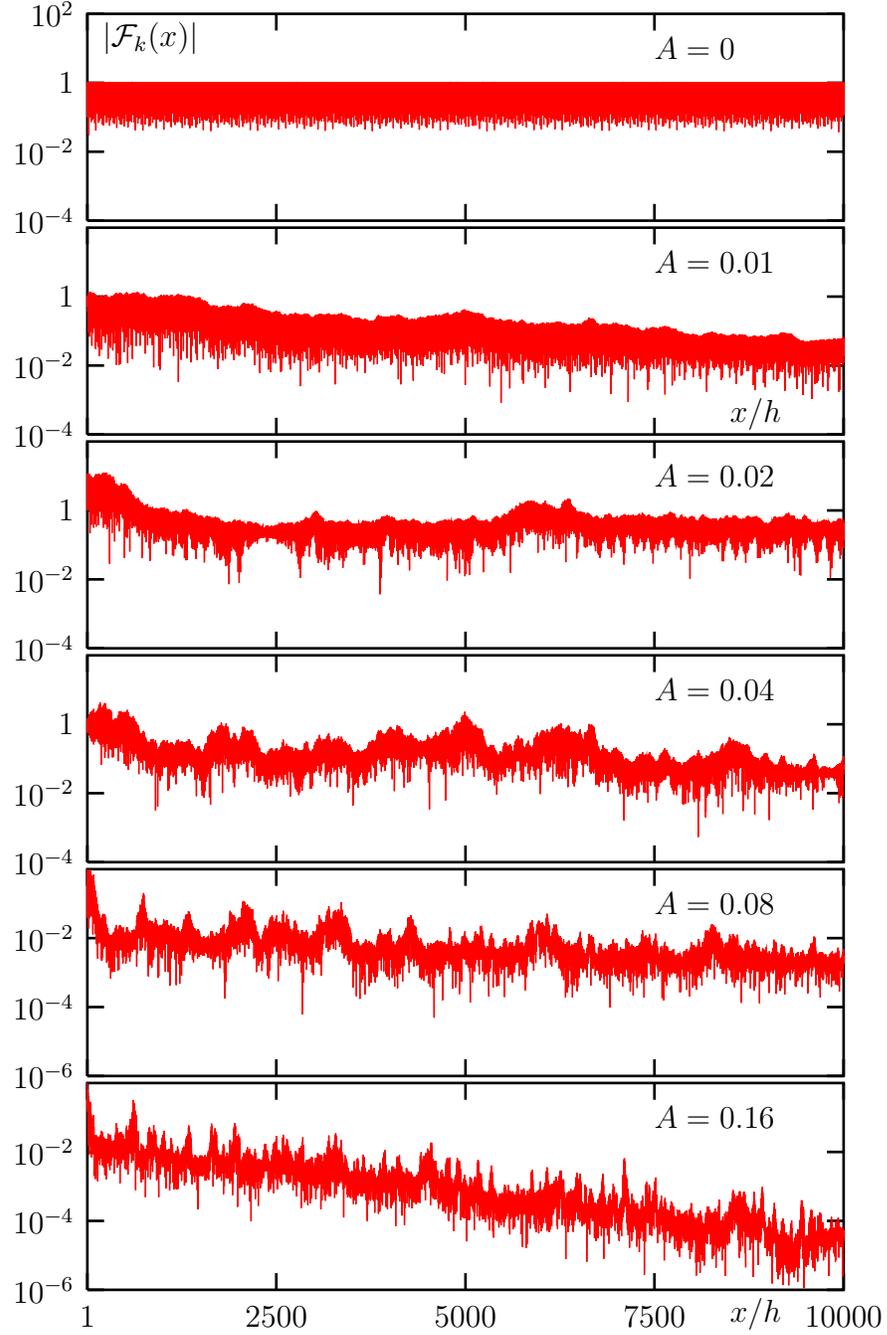,bbllx=100bp,bblly=200bp,bburx=496.bp,bbury=750bp,clip=t}}
\caption{Propagation of SP in a chain of $N=10,000$ non-absorbing ($ 
  \gamma=0$) nanospheres for different levels of \emph{off-diagonal}
  disorder $ A$, as indicated. SP polarization is orthogonal to the
  chain. Other parameters: $kh=0.2\pi$, $\omega/\omega_{\mathrm{F}
  }=1$, $h/a=4$.}
\label{fig:ssp_dord_ort_a1}
\end{figure}

The first example concerns propagation in an non-absorbing chain of
$N=10,000$ nanospheres excited exactly at the Frohlich frequency
$\omega = \omega_{\mathrm{F}}$. Numerical results for different levels
of disorder are illustrated in Fig.~\ref{fig:ssp_dord_ort_a1} where we
plot $\vert {\mathcal F}_k(x) \vert$ as a function of $x$ (sampled at
$x=x_n$).

\begin{figure}[tbp]
\centerline{\psfig{file=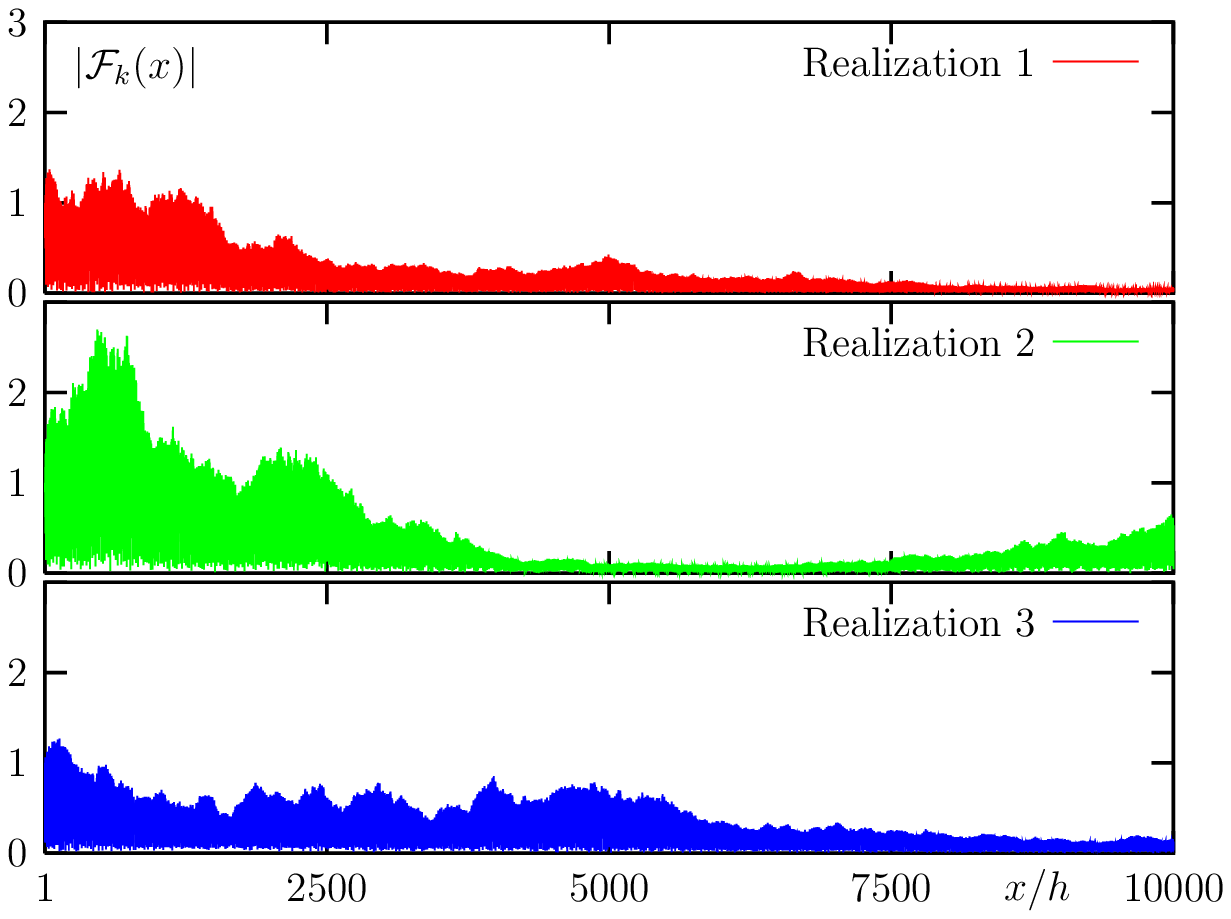,bbllx=0bp,bblly=450bp,bburx=596bp,bbury=710bp,clip=t}}
\caption{Propagation of SPs in a chain with off-diagonal disorder of
  amplitude $A=0.01$ for different random realizations of disorder.
  Other parameters same as in Fig.~\ref{fig:ssp_dord_ort_a1}.}
\label{fig:ssp_dord_ort_a1=0.01_R}
\end{figure}

\begin{figure}[tbp]
\centerline{\psfig{file=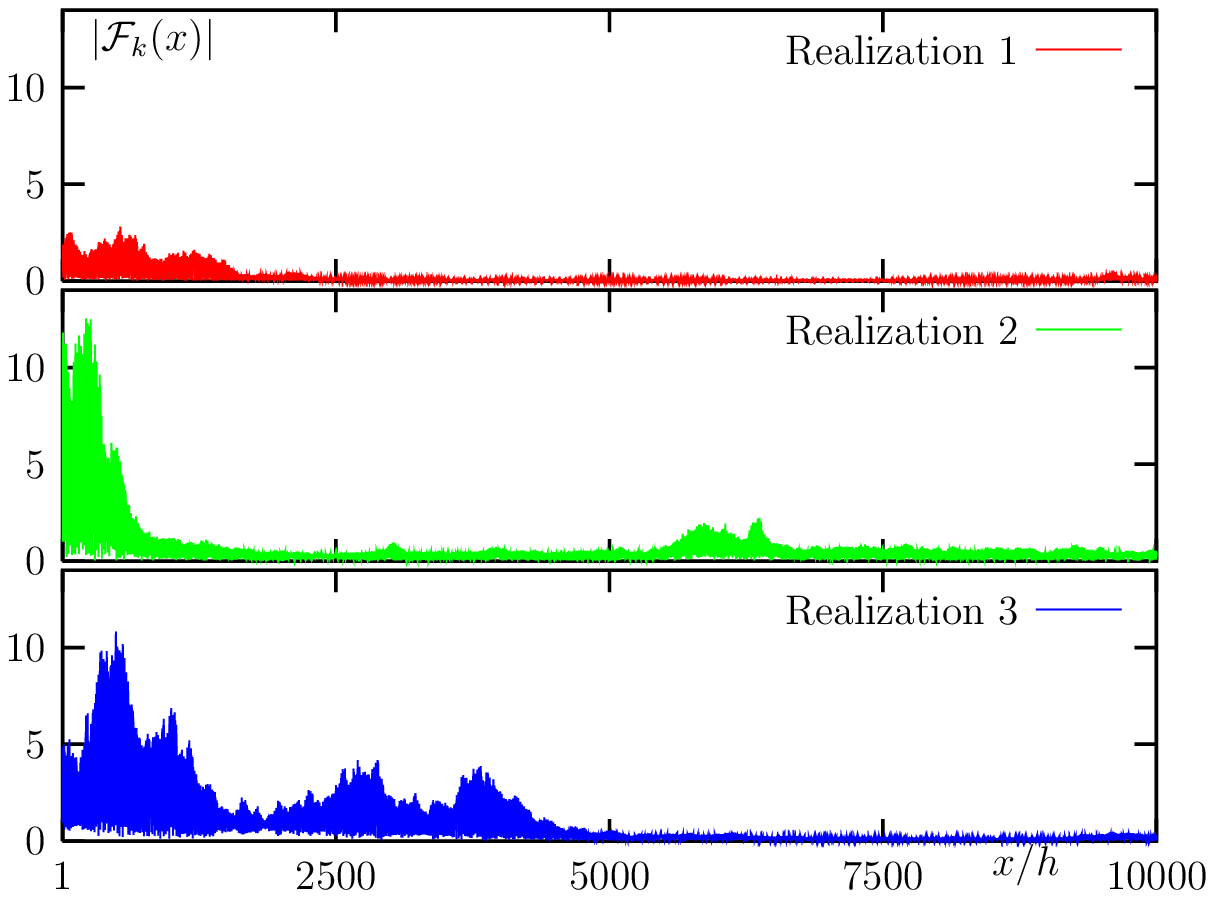,bbllx=0bp,bblly=450bp,bburx=596bp,bbury=710bp,clip=t}}
\caption{Same as in Fig.~\ref{fig:ssp_dord_ort_a1=0.01_R}, but for $ 
  A=0.02$.}
\label{fig:ssp_dord_ort_a1=0.02_R}
\end{figure}

One obvious conclusion that can be made from inspection of
Fig.~\ref{fig:ssp_dord_ort_a1} is that disorder causes spatial decay
of SP. Since the system has no absorption, energy is lost to
radiation. However, the exact law of decay strongly depends on
particular realization of disorder. In
Figs.~\ref{fig:ssp_dord_ort_a1=0.01_R}
and~\ref{fig:ssp_dord_ort_a1=0.02_R}, we plot the function
$\vert{\mathcal F}_k(x) \vert$ for $A=0.01$ and $A=0.02$,
respectively, and for three different realizations of disorder
(without the use of logarithmic scale). Giant fluctuations in the
amplitude of transmitted SP are quite apparent. In all cases, after
some initial growth, the amplitude decays, although some random
recurrences (due to re-excitation) can take place. However, the
amplitude $\vert{\mathcal F}_k(x) \vert$ at a given site $x=x_n$
strongly depends on realization of disorder and can be very far from
its ensemble average, even at very large propagation distances.
Therefore, the function ${\mathcal F}_k(x)$ appears to be \emph{not
  self-averaging}. In particular, any particular realization of the
intensity ${\mathcal I}_k(x)=\vert {\mathcal F}_k(x) \vert^2$ does not
satisfy either the radiative transport equation or the diffusion
equation. The absence of transport and self-averaging is
characteristic for Anderson localization. We note that the
localization discussed here is an interference (radiative) effect
which is different from localization of quasistatic polarization modes
studied in Refs.~\onlinecite{stockman_01_1,genov_05_1}. Localization
properties of the eigenmodes in disordered chains is a subject of
separate investigation and will be reported elsewhere.

\begin{figure}[tbp]
\centerline{\input{ssp_dord_ort_a1_o=0.984.tex}}
\caption{Same as in Fig.~\ref{fig:ssp_dord_ort_a1} but for
  $\omega/\omega_{\mathrm{F}} = 0.984$ and for different levels $A$ of
  off-diagonal disorder, as indicated. Note that the curves for $A=0$
  and $ A=0.01$ and $A=0.04$ are indistinguishable.}
\label{fig:ssp_dord_ort_a1_o=0.984}
\end{figure}
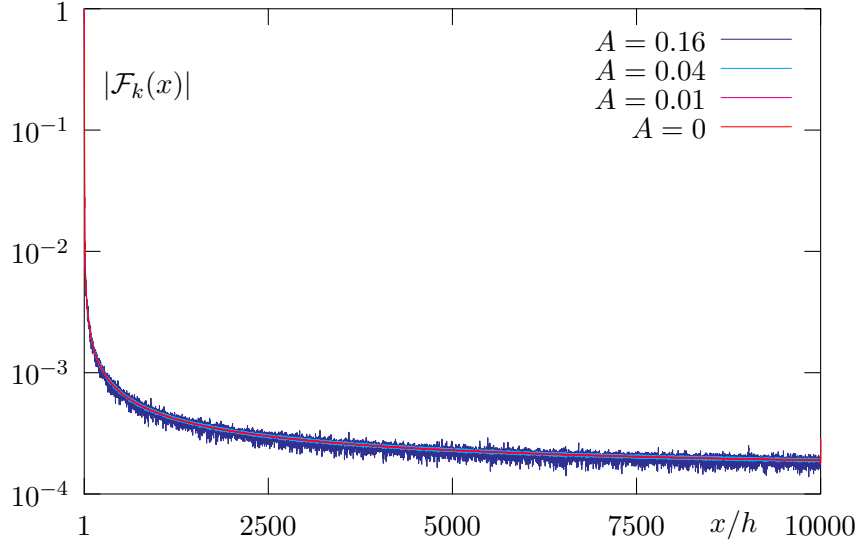

The situation is complicated by the presence of two types of SP
excitations.  One can argue that localization properties of these two
types of SPs might be different. To investigate this possibility, we
perform two types of numerical experiments. First, we repeat
simulations illustrated in Fig.~\ref{fig:ssp_dord_ort_a1} but for
$\omega/\omega_{\mathrm{F}}=0.948$. In this case, the ordinary SP is
not excited in ordered chains (see Fig.~\ref{fig:omega}(a)). Results
are shown in Fig.~\ref{fig:ssp_dord_ort_a1_o=0.984} . It can be
concluded from the figure that off-diagonal disorder does not result
in additional decay of the extraordinary SP. In fact, the curves with
$A=0$, $A=0.1$ and $A=0.04$ are indistinguishable. This should be
contrasted with the case $\omega/\omega_{\mathrm{F}}=1$, when, at the
level of disorder $A=0.02$, spatial decay is already well manifested.
The conclusion one can make is that the extraordinary SP excitations
do not experience localization due to uncorrelated off-diagonal
disorder in the chain. Therefore, spatial decay seen in
Figs.~\ref{fig:ssp_dord_ort_a1} and~\ref{fig:ssp_dord_ort_a1=0.02_R}
is decay of the ordinary SP. At very large propagation distances,
${\mathcal F}_k(x)$ continues to oscillate around the baseline of the
extraordinary SP.

The second numerical experiment that can elucidate the influence of
disorder on ordinary and extraordinary SPs is simulation of an
extinction measurement when the chain is excited, instead of a
nera-field tip, by a plane wave $E_0 \exp (iqx)$. Namely, we will look
at the dependence of the specific (per one nanosphere) extinction
cross section $\sigma_e$ on $q$. There are two different
experimental setups that can be used to measure $\sigma_e(q)$. When
$q$ is in the interval $0\leq q<k$, this experiment can be carried out
simply by varying the angle between the incident beam and the chain.
However, values $q>k$ are not accessible in this experiment. In this
case, the chain can be placed on a dielectric substrate and excited by
evanescent wave originating due to the total internal reflection of
the incident beam. The maximum longitudinal wave number of the
evanescent wave is $nk$, where $n>1$ is the refractive index of the
substrate. We note that it is not realistically possible to access all
wave numbers up to $q=\pi /h$ in this way because, in the particular
case $kh=0.2\pi $, this would require the refractive index $n=5$.
Refractive indices of such magnitude are not achievable in the optical
range. However, in a numerical simulation, we can assume that a
hypothetical transparent substrate with $n=5$ exists. Besides, all
wave numbers $q$ in the first Brillouin zone of the lattice can be
accessible for a different choice of parameters (particularly, for
larger values of $kh$).

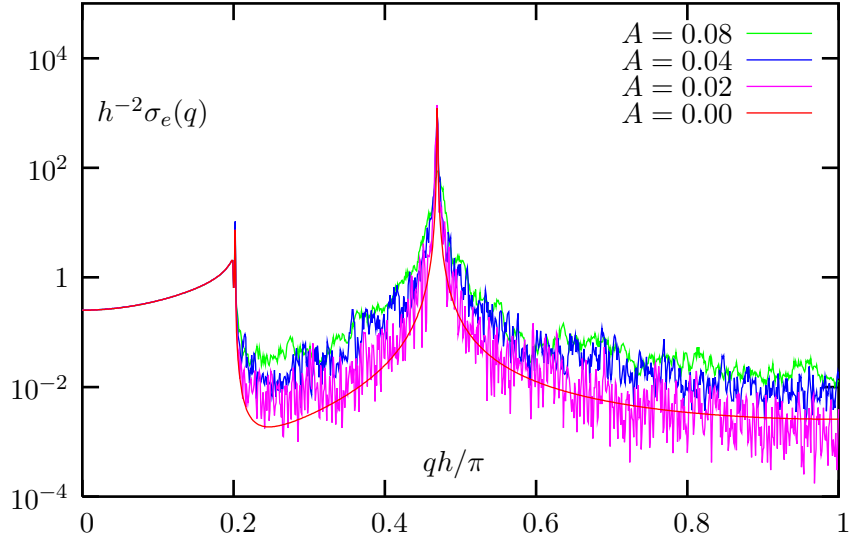
\begin{figure}[tbp]
\centerline{\input{qspect_a1.tex}}
\caption{Specific extinction $\sigma_e$ as a function of lateral
  wave number of incident wave, $q$, for different levels of
  off-diagonal disorder, $A$.}
\label{fig:qspect_a1}
\end{figure}

The specific extinction $\sigma_e(q)$ is given by the following formula:

\begin{equation}  
\label{sigma_e_def}
\sigma_e(q) = \frac{4\pi k}{N\vert E_0 \vert^2} \sum_{n=1}^N E_0^*\exp(-i q
x_n) d_n \ ,
\end{equation}

\noindent 
where $d_n$ is the solution to Eq.~(\ref{CDE}) with the right-hand
side $E_n=E_0 \exp(i q x_n)$. In Fig.~\ref{fig:qspect_a1}, we plot the
dimensionless quantity $h^{-2}\sigma_e(q)$ as a function of $q$ for
transverse oscillations in partially disordered non-absorbing chains.
Two peaks are clearly visible in the spectrum. The first peak at
$q=q_1\approx 0.47\pi/h$ corresponds to excitation of the ordinary SP.
The second peak at $ q=q_2\approx k$ corresponds to excitation of the
extraordinary SP. In infinite, periodic and non-absorbing chains, the
spectrum has a simple pole at $q=q_1$.~\cite{fn4} In the finite chain
with $N=10,000$, the singularity is replaced by a very sharp maximum.
Introduction of even slight disorder tends to further broaden and
randomize this peak. Obviously, this broadening, as well as the
randomization of the spectrum in the vicinity of $ q=q_1$, result in
spatial decay of the ordinary SP excitated by a near-field tip.
However, the peak corresponding to the extraordinary SP is almost
unaffected by disorder. Consequently, the extraordinary SP does not
experience localization and related spatial decay when off-diagonal
disorder is introduced into the system.

The physical reason why the extraordinary SP is not affected by
disorder is quite straightforward. This SP is mediated by
electromagnetic waves in the far (radiation) zone that arrive at a
given nanosphere from all other nanospheres. The synchronism condition
(that all these secondary waves arrive in phase) is not affected by
disorder as long as the displacement amplitudes $\xi_n$ are small
compared to the wavelength. In the numerical examples of this section,
wavelength is ten times larger than the inter-particle spacing, which
is, in turn, much larger than the displacement amplitudes.
Additionally, effects of disorder are expected to be averaged if the
amplitudes $\xi_n$ are mathematically independent because the
scattered field at a given nanosphere is a sum of very large number of
secondary waves. In contrast, the ordinary SP is mediated by
near-field interactions which are very sensitive to even slight
displacements of nanospheres.  Besides, the scattered field at a given
nanosphere (when the ordinary SP propagates in the chain) is a rapidly
converging sum of secondary fields, so that only a few terms in this
sum are important and no effective averaging takes place.

\subsection{Diagonal Disorder}
\label{subsec:diag}

Diagonal disorder is disorder in the properties of nanospheres. There
are several possibilities for introducing such disorder.

First, the spheres can be polydisperse, i.e., have different radiuses $a_n$.
This is, however, not a truly diagonal disorder. Indeed, polarizability of $ 
n $-th nanosphere can be written in this case as $f_n \langle \alpha \rangle$
where $f_n = (a_n /\langle a \rangle)^3$ and $\langle a \rangle$ is the
average radius (note that the radiative correction to $1/\alpha_n$ can be
included into the dipole sum $S(k,q)$, so that this analysis remains valid
even when this correction is important). The factors $f_n$ are all
positive-definite which allows one to introduce a simple transformation of
Eq.~(\ref{CDE}) which removes the diagonal disorder~\cite{perminov_04_1}.
Then the disorder becomes effectively off-diagonal. More importantly, one
can retain a well-defined spectral parameter of the theory, $1/\langle
\alpha \rangle$. We, therefore, do not consider polydispersity in this
section.

The second possibility is variation of the absorptive parameter $\gamma$.
This effect that can be practically important. Yet, we are interested in
propagation in the absence of Ohmic losses and, therefore, set $\gamma_n=0$.

The third, and the most fundamental, reason for diagonal disorder is
variation of the Frohlich frequency of nanospheres. Namely, we take
the Frohlich frequency of the $n$-th particle to be
$\omega_{\mathrm{F} n}=\langle \omega_{\mathrm{F}} \rangle (1 +
\xi_n)$, where $\xi_n$ are statistically independent random variables
evenly distributed in the interval $[-A,A]$ and $\omega/ \langle
\omega_{\mathrm{F}} \rangle = 1$. The factors $f_n$ in this case are
no longer positive-definite and the transformation of
Ref.~\onlinecite{perminov_04_1} can not be applied. The most profound
consequence of introducing the diagonal disorder is that the spectral
parameter such as $1/\alpha$ is no longer well defined. As long as the
disorder amplitude is relatively small, one can view $1/\langle \alpha
\rangle$ as an approximate spectral parameter. However, as the
amplitude of disorder increases, this approach becomes invalid. We
have seen in section~\ref{sec:prop_ord} that variation of the ratio
$\omega/\omega_{\mathrm{F}}$ in the interval $0.988 <
\omega/\omega_{\mathrm{F}} < 1.007$ can result in dramatic changes in
the way an SP excitation propagates along the chain. For
$\omega/\omega_{\mathrm{F}}$ outside of this interval, ordinary SP
could not be effectively excited. However, in
section~\ref{sec:prop_ord}, variation of $\omega/\omega_{\mathrm{F}}$
applied to all nanospheres simultaneously. We now introduce random
uncorrelated variation of this ratio for each individual nanosphere.
The effects of such disorder are difficult to predict theoretically.
We can, however, expect that these effects become dramatic for
$A\gtrsim 0.01$ since, in this case, the ratio
$\omega/\omega_{\mathrm{F} n}$ can be outside of the interval $[0.988,
1.007]$.

\begin{figure}[tbp]
\centerline{\psfig{file=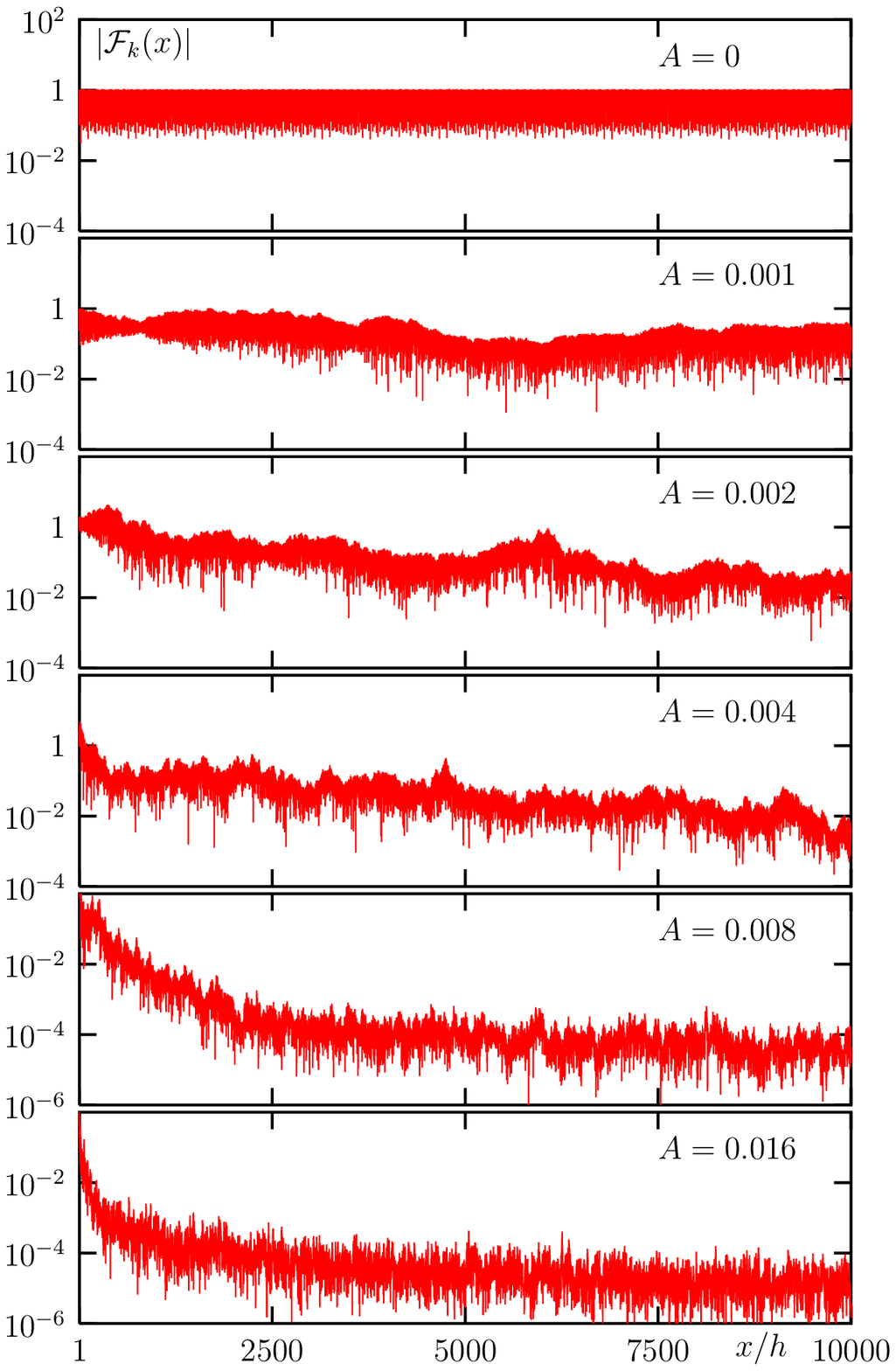,bbllx=100bp,bblly=200bp,bburx=496.bp,bbury=750bp,clip=t}}
\caption{Same as in Fig.~\ref{fig:ssp_dord_ort_a1}, but for
  different levels $A$ of \emph{diagonal} disorder.}
\label{fig:ssp_dord_ort_a3}
\end{figure}

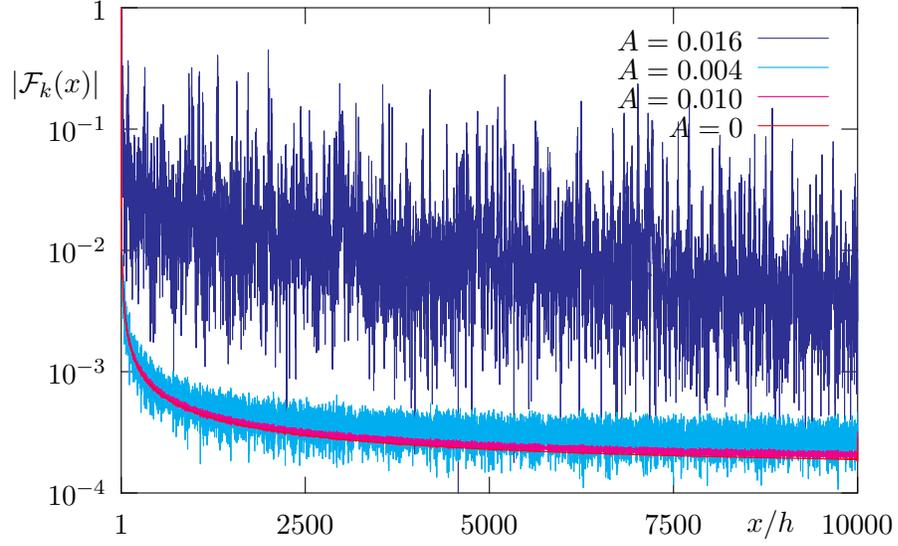
\begin{figure}[tbp]
\centerline{\input{ssp_dord_ort_a3_o=0.984.tex}}
\caption{Same as in Fig.~\ref{fig:ssp_dord_ort_a3} but for $ 
  \omega/\omega_{\mathrm{F}}=0.984$ and for different levels $A$ of
  off-diagonal disorder, as indicated. Note that the curves for $A=0$
  and $ A=0.01$ are indistinguishable.}
\label{fig:ssp_dord_ort_a3_o=0.984}
\end{figure}

We first adduce in Fig.~\ref{fig:ssp_dord_ort_a3} the same
dependencies as shown in Fig.~\ref{fig:ssp_dord_ort_a1} but for
different levels of \emph{diagonal} disorder. The results appear to
be, qualitatively, quite similar, although in the case $A=0.016$,
there is no visible trend for $x/h>3000$. We then investigate whether
the extraordinary SP remains insensitive to diagonal disorder. Data
analogous to those shown in Fig.~\ref{fig:ssp_dord_ort_a1_o=0.984},
but for diagonal disorder, are presented in
Fig.~\ref{fig:ssp_dord_ort_a3_o=0.984}. It can be seen that the
influence of diagonal disorder on the extraordinary SP is stronger
than that of the off-diagonal disorder. When the disorder amplitude
$A$ exceeds $0.01$, the influence becomes quite dramatic. But
paradoxically, at $A=0.016$, the average decay rate is much slower
than for $A=0.008$, although the amplitude of fluctuations is much
larger.

\begin{figure}[tbp]
\centerline{\input{qspect_a3.tex}}
\caption{Specific extinction $\sigma_e$ as a function of lateral
  wave number of incident wave, $q$, for different levels of
  off-diagonal disorder, $A$.}
\label{fig:qspect_a3}
\end{figure}
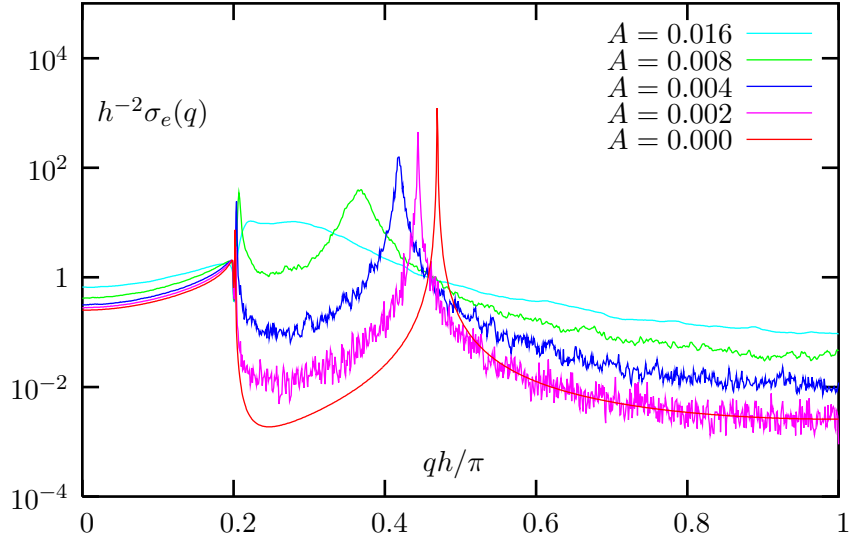

To see why this happens, we look at the extinction spectra
$\sigma_e(q)$.  These are plotted in Fig.~\ref{fig:qspect_a3}. The
fundamental difference between the off-diagonal and diagonal disorder
is clearly revealed by comparing this figure to
Fig.~\ref{fig:qspect_a1}. Namely, the effect of diagonal disorder is
not only to broaden and randomize the peak at $q=q_1$, but also to
shift it towards the peak corresponding to the extraordinary SP.  At
sufficiently large levels of disorder ($A \geq 0.016$), the separate
peaks disappear and a broad structure emerges. At this point, ordinary
and extraordinary SP can no longer be distinguished. Correspondingly,
the ordinary and extraordinary SP are effectively mixed in the
$A=0.016$ curve in Fig.~\ref{fig:ssp_dord_ort_a3_o=0.984}, while fo
smaller amplitudes of the diagonal disorder, the extraordinary SP is
excited predominatlly.

\section{Summary}
\label{sec:summary}

We have considered surface plasmon (SP) propagation in a linear chain
of metal nanoparticles. Computer simulations reveal the existence of
two types of plasmons: ordinary (quasistatic) and extraordinary
(non-quasistatic) SPs.

The ordinary SP is characterized by short-range interaction of
nanospheres in a chain. The retardation effects are inessential for
its existence and properties. The ordinary SP behaves as a quasistatic
excitation. The ordinary SP can not radiate into the far zone in
perfectly periodic chains because its wave number is larger than the
wavenumber $k=\omega/c$ of free electromagnetic waves. However, it can
experience decay due to absorptive dissipation in the material.

The second, extraordinary, SP propagates due to long-range (radiation
zone) interaction in a chain. Its excitation is possible due to the
existence of the non-Lorentzian optical resonance in the chain
introduced in Ref.~\onlinecite{markel_05_2}. 
The extraordinary SP may experience some radiative loss but is much
less afected by absorptive dissipation and disorder. As a result, it
can propagate to much larger distances along the chain. The
extraordinary SP can be used to guide energy or information in
all-optical integrated photonic systems.

We have also considered the effects of disorder and localization of
the ordinary and extraordinary SPs. Results of numerical simulations
suggest that even small disorder in the position or properties of
nanoparticles results in localization of the ordinary SP. However, the
extraordinary SP appears to remain delocalized for all types and
levels of disorder considered in the paper.

The authors can be reached at vmarkel@mail.med.upenn.edu (VAM) and
asarychev@ethertronics.com (AKS).

\bibliographystyle{prsty} 
\bibliography{abbrev,master,local}

\end{document}

%% file: disp_re_ort.tex
\begingroup%
  \makeatletter%
  \newcommand{\GNUPLOTspecial}{%
    \@sanitize\catcode`\%=14\relax\special}%
  \setlength{\unitlength}{0.1bp}%
{\GNUPLOTspecial{!
/gnudict 256 dict def
gnudict begin
/Color true def
/Solid true def
/gnulinewidth 5.000 def
/userlinewidth gnulinewidth def
/vshift -33 def
/dl {10 mul} def
/hpt_ 31.5 def
/vpt_ 31.5 def
/hpt hpt_ def
/vpt vpt_ def
/M {moveto} bind def
/L {lineto} bind def
/R {rmoveto} bind def
/V {rlineto} bind def
/vpt2 vpt 2 mul def
/hpt2 hpt 2 mul def
/Lshow { currentpoint stroke M
  0 vshift R show } def
/Rshow { currentpoint stroke M
  dup stringwidth pop neg vshift R show } def
/Cshow { currentpoint stroke M
  dup stringwidth pop -2 div vshift R show } def
/UP { dup vpt_ mul /vpt exch def hpt_ mul /hpt exch def
  /hpt2 hpt 2 mul def /vpt2 vpt 2 mul def } def
/DL { Color {setrgbcolor Solid {pop []} if 0 setdash }
 {pop pop pop Solid {pop []} if 0 setdash} ifelse } def
/BL { stroke userlinewidth 2 mul setlinewidth } def
/AL { stroke userlinewidth 2 div setlinewidth } def
/UL { dup gnulinewidth mul /userlinewidth exch def
      dup 1 lt {pop 1} if 10 mul /udl exch def } def
/PL { stroke userlinewidth setlinewidth } def
/LTb { BL [] 0 0 0 DL } def
/LTa { AL [1 udl mul 2 udl mul] 0 setdash 0 0 0 setrgbcolor } def
/LT0 { PL [] 1 0 0 DL } def
/LT1 { PL [4 dl 2 dl] 0 1 0 DL } def
/LT2 { PL [2 dl 3 dl] 0 0 1 DL } def
/LT3 { PL [1 dl 1.5 dl] 1 0 1 DL } def
/LT4 { PL [5 dl 2 dl 1 dl 2 dl] 0 1 1 DL } def
/LT5 { PL [4 dl 3 dl 1 dl 3 dl] 1 1 0 DL } def
/LT6 { PL [2 dl 2 dl 2 dl 4 dl] 0 0 0 DL } def
/LT7 { PL [2 dl 2 dl 2 dl 2 dl 2 dl 4 dl] 1 0.3 0 DL } def
/LT8 { PL [2 dl 2 dl 2 dl 2 dl 2 dl 2 dl 2 dl 4 dl] 0.5 0.5 0.5 DL } def
/Pnt { stroke [] 0 setdash
   gsave 1 setlinecap M 0 0 V stroke grestore } def
/Dia { stroke [] 0 setdash 2 copy vpt add M
  hpt neg vpt neg V hpt vpt neg V
  hpt vpt V hpt neg vpt V closepath stroke
  Pnt } def
/Pls { stroke [] 0 setdash vpt sub M 0 vpt2 V
  currentpoint stroke M
  hpt neg vpt neg R hpt2 0 V stroke
  } def
/Box { stroke [] 0 setdash 2 copy exch hpt sub exch vpt add M
  0 vpt2 neg V hpt2 0 V 0 vpt2 V
  hpt2 neg 0 V closepath stroke
  Pnt } def
/Crs { stroke [] 0 setdash exch hpt sub exch vpt add M
  hpt2 vpt2 neg V currentpoint stroke M
  hpt2 neg 0 R hpt2 vpt2 V stroke } def
/TriU { stroke [] 0 setdash 2 copy vpt 1.12 mul add M
  hpt neg vpt -1.62 mul V
  hpt 2 mul 0 V
  hpt neg vpt 1.62 mul V closepath stroke
  Pnt  } def
/Star { 2 copy Pls Crs } def
/BoxF { stroke [] 0 setdash exch hpt sub exch vpt add M
  0 vpt2 neg V  hpt2 0 V  0 vpt2 V
  hpt2 neg 0 V  closepath fill } def
/TriUF { stroke [] 0 setdash vpt 1.12 mul add M
  hpt neg vpt -1.62 mul V
  hpt 2 mul 0 V
  hpt neg vpt 1.62 mul V closepath fill } def
/TriD { stroke [] 0 setdash 2 copy vpt 1.12 mul sub M
  hpt neg vpt 1.62 mul V
  hpt 2 mul 0 V
  hpt neg vpt -1.62 mul V closepath stroke
  Pnt  } def
/TriDF { stroke [] 0 setdash vpt 1.12 mul sub M
  hpt neg vpt 1.62 mul V
  hpt 2 mul 0 V
  hpt neg vpt -1.62 mul V closepath fill} def
/DiaF { stroke [] 0 setdash vpt add M
  hpt neg vpt neg V hpt vpt neg V
  hpt vpt V hpt neg vpt V closepath fill } def
/Pent { stroke [] 0 setdash 2 copy gsave
  translate 0 hpt M 4 {72 rotate 0 hpt L} repeat
  closepath stroke grestore Pnt } def
/PentF { stroke [] 0 setdash gsave
  translate 0 hpt M 4 {72 rotate 0 hpt L} repeat
  closepath fill grestore } def
/Circle { stroke [] 0 setdash 2 copy
  hpt 0 360 arc stroke Pnt } def
/CircleF { stroke [] 0 setdash hpt 0 360 arc fill } def
/C0 { BL [] 0 setdash 2 copy moveto vpt 90 450  arc } bind def
/C1 { BL [] 0 setdash 2 copy        moveto
       2 copy  vpt 0 90 arc closepath fill
               vpt 0 360 arc closepath } bind def
/C2 { BL [] 0 setdash 2 copy moveto
       2 copy  vpt 90 180 arc closepath fill
               vpt 0 360 arc closepath } bind def
/C3 { BL [] 0 setdash 2 copy moveto
       2 copy  vpt 0 180 arc closepath fill
               vpt 0 360 arc closepath } bind def
/C4 { BL [] 0 setdash 2 copy moveto
       2 copy  vpt 180 270 arc closepath fill
               vpt 0 360 arc closepath } bind def
/C5 { BL [] 0 setdash 2 copy moveto
       2 copy  vpt 0 90 arc
       2 copy moveto
       2 copy  vpt 180 270 arc closepath fill
               vpt 0 360 arc } bind def
/C6 { BL [] 0 setdash 2 copy moveto
      2 copy  vpt 90 270 arc closepath fill
              vpt 0 360 arc closepath } bind def
/C7 { BL [] 0 setdash 2 copy moveto
      2 copy  vpt 0 270 arc closepath fill
              vpt 0 360 arc closepath } bind def
/C8 { BL [] 0 setdash 2 copy moveto
      2 copy vpt 270 360 arc closepath fill
              vpt 0 360 arc closepath } bind def
/C9 { BL [] 0 setdash 2 copy moveto
      2 copy  vpt 270 450 arc closepath fill
              vpt 0 360 arc closepath } bind def
/C10 { BL [] 0 setdash 2 copy 2 copy moveto vpt 270 360 arc closepath fill
       2 copy moveto
       2 copy vpt 90 180 arc closepath fill
               vpt 0 360 arc closepath } bind def
/C11 { BL [] 0 setdash 2 copy moveto
       2 copy  vpt 0 180 arc closepath fill
       2 copy moveto
       2 copy  vpt 270 360 arc closepath fill
               vpt 0 360 arc closepath } bind def
/C12 { BL [] 0 setdash 2 copy moveto
       2 copy  vpt 180 360 arc closepath fill
               vpt 0 360 arc closepath } bind def
/C13 { BL [] 0 setdash  2 copy moveto
       2 copy  vpt 0 90 arc closepath fill
       2 copy moveto
       2 copy  vpt 180 360 arc closepath fill
               vpt 0 360 arc closepath } bind def
/C14 { BL [] 0 setdash 2 copy moveto
       2 copy  vpt 90 360 arc closepath fill
               vpt 0 360 arc } bind def
/C15 { BL [] 0 setdash 2 copy vpt 0 360 arc closepath fill
               vpt 0 360 arc closepath } bind def
/Rec   { newpath 4 2 roll moveto 1 index 0 rlineto 0 exch rlineto
       neg 0 rlineto closepath } bind def
/Square { dup Rec } bind def
/Bsquare { vpt sub exch vpt sub exch vpt2 Square } bind def
/S0 { BL [] 0 setdash 2 copy moveto 0 vpt rlineto BL Bsquare } bind def
/S1 { BL [] 0 setdash 2 copy vpt Square fill Bsquare } bind def
/S2 { BL [] 0 setdash 2 copy exch vpt sub exch vpt Square fill Bsquare } bind def
/S3 { BL [] 0 setdash 2 copy exch vpt sub exch vpt2 vpt Rec fill Bsquare } bind def
/S4 { BL [] 0 setdash 2 copy exch vpt sub exch vpt sub vpt Square fill Bsquare } bind def
/S5 { BL [] 0 setdash 2 copy 2 copy vpt Square fill
       exch vpt sub exch vpt sub vpt Square fill Bsquare } bind def
/S6 { BL [] 0 setdash 2 copy exch vpt sub exch vpt sub vpt vpt2 Rec fill Bsquare } bind def
/S7 { BL [] 0 setdash 2 copy exch vpt sub exch vpt sub vpt vpt2 Rec fill
       2 copy vpt Square fill
       Bsquare } bind def
/S8 { BL [] 0 setdash 2 copy vpt sub vpt Square fill Bsquare } bind def
/S9 { BL [] 0 setdash 2 copy vpt sub vpt vpt2 Rec fill Bsquare } bind def
/S10 { BL [] 0 setdash 2 copy vpt sub vpt Square fill 2 copy exch vpt sub exch vpt Square fill
       Bsquare } bind def
/S11 { BL [] 0 setdash 2 copy vpt sub vpt Square fill 2 copy exch vpt sub exch vpt2 vpt Rec fill
       Bsquare } bind def
/S12 { BL [] 0 setdash 2 copy exch vpt sub exch vpt sub vpt2 vpt Rec fill Bsquare } bind def
/S13 { BL [] 0 setdash 2 copy exch vpt sub exch vpt sub vpt2 vpt Rec fill
       2 copy vpt Square fill Bsquare } bind def
/S14 { BL [] 0 setdash 2 copy exch vpt sub exch vpt sub vpt2 vpt Rec fill
       2 copy exch vpt sub exch vpt Square fill Bsquare } bind def
/S15 { BL [] 0 setdash 2 copy Bsquare fill Bsquare } bind def
/D0 { gsave translate 45 rotate 0 0 S0 stroke grestore } bind def
/D1 { gsave translate 45 rotate 0 0 S1 stroke grestore } bind def
/D2 { gsave translate 45 rotate 0 0 S2 stroke grestore } bind def
/D3 { gsave translate 45 rotate 0 0 S3 stroke grestore } bind def
/D4 { gsave translate 45 rotate 0 0 S4 stroke grestore } bind def
/D5 { gsave translate 45 rotate 0 0 S5 stroke grestore } bind def
/D6 { gsave translate 45 rotate 0 0 S6 stroke grestore } bind def
/D7 { gsave translate 45 rotate 0 0 S7 stroke grestore } bind def
/D8 { gsave translate 45 rotate 0 0 S8 stroke grestore } bind def
/D9 { gsave translate 45 rotate 0 0 S9 stroke grestore } bind def
/D10 { gsave translate 45 rotate 0 0 S10 stroke grestore } bind def
/D11 { gsave translate 45 rotate 0 0 S11 stroke grestore } bind def
/D12 { gsave translate 45 rotate 0 0 S12 stroke grestore } bind def
/D13 { gsave translate 45 rotate 0 0 S13 stroke grestore } bind def
/D14 { gsave translate 45 rotate 0 0 S14 stroke grestore } bind def
/D15 { gsave translate 45 rotate 0 0 S15 stroke grestore } bind def
/DiaE { stroke [] 0 setdash vpt add M
  hpt neg vpt neg V hpt vpt neg V
  hpt vpt V hpt neg vpt V closepath stroke } def
/BoxE { stroke [] 0 setdash exch hpt sub exch vpt add M
  0 vpt2 neg V hpt2 0 V 0 vpt2 V
  hpt2 neg 0 V closepath stroke } def
/TriUE { stroke [] 0 setdash vpt 1.12 mul add M
  hpt neg vpt -1.62 mul V
  hpt 2 mul 0 V
  hpt neg vpt 1.62 mul V closepath stroke } def
/TriDE { stroke [] 0 setdash vpt 1.12 mul sub M
  hpt neg vpt 1.62 mul V
  hpt 2 mul 0 V
  hpt neg vpt -1.62 mul V closepath stroke } def
/PentE { stroke [] 0 setdash gsave
  translate 0 hpt M 4 {72 rotate 0 hpt L} repeat
  closepath stroke grestore } def
/CircE { stroke [] 0 setdash 
  hpt 0 360 arc stroke } def
/Opaque { gsave closepath 1 setgray fill grestore 0 setgray closepath } def
/DiaW { stroke [] 0 setdash vpt add M
  hpt neg vpt neg V hpt vpt neg V
  hpt vpt V hpt neg vpt V Opaque stroke } def
/BoxW { stroke [] 0 setdash exch hpt sub exch vpt add M
  0 vpt2 neg V hpt2 0 V 0 vpt2 V
  hpt2 neg 0 V Opaque stroke } def
/TriUW { stroke [] 0 setdash vpt 1.12 mul add M
  hpt neg vpt -1.62 mul V
  hpt 2 mul 0 V
  hpt neg vpt 1.62 mul V Opaque stroke } def
/TriDW { stroke [] 0 setdash vpt 1.12 mul sub M
  hpt neg vpt 1.62 mul V
  hpt 2 mul 0 V
  hpt neg vpt -1.62 mul V Opaque stroke } def
/PentW { stroke [] 0 setdash gsave
  translate 0 hpt M 4 {72 rotate 0 hpt L} repeat
  Opaque stroke grestore } def
/CircW { stroke [] 0 setdash 
  hpt 0 360 arc Opaque stroke } def
/BoxFill { gsave Rec 1 setgray fill grestore } def
/Symbol-Oblique /Symbol findfont [1 0 .167 1 0 0] makefont
dup length dict begin {1 index /FID eq {pop pop} {def} ifelse} forall
currentdict end definefont pop
end
}}%
\begin{picture}(2520,2160)(0,0)%
{\GNUPLOTspecial{"
gnudict begin
gsave
0 0 translate
0.100 0.100 scale
0 setgray
newpath
1.000 UL
LTb
250 200 M
63 0 V
2057 0 R
-63 0 V
250 665 M
63 0 V
2057 0 R
-63 0 V
250 1130 M
63 0 V
2057 0 R
-63 0 V
250 1595 M
63 0 V
2057 0 R
-63 0 V
250 2060 M
63 0 V
2057 0 R
-63 0 V
250 200 M
0 63 V
0 1797 R
0 -63 V
674 200 M
0 63 V
0 1797 R
0 -63 V
1098 200 M
0 63 V
0 1797 R
0 -63 V
1522 200 M
0 63 V
0 1797 R
0 -63 V
1946 200 M
0 63 V
0 1797 R
0 -63 V
2370 200 M
0 63 V
0 1797 R
0 -63 V
1.000 UL
LTb
250 200 M
2120 0 V
0 1860 V
-2120 0 V
250 200 L
1.000 UL
LT4
250 1130 M
2120 0 V
1.000 UL
LT3
2023 1947 M
247 0 V
250 340 M
2 0 V
2 0 V
2 0 V
2 0 V
3 0 V
2 1 V
2 0 V
2 0 V
2 0 V
2 0 V
2 0 V
2 0 V
3 0 V
2 0 V
2 0 V
2 0 V
2 1 V
2 0 V
2 0 V
2 0 V
3 0 V
2 0 V
2 1 V
2 0 V
2 0 V
2 0 V
2 0 V
2 1 V
2 0 V
3 0 V
2 0 V
2 1 V
2 0 V
2 0 V
2 1 V
2 0 V
2 0 V
3 1 V
2 0 V
2 0 V
2 1 V
2 0 V
2 0 V
2 1 V
2 0 V
3 0 V
2 1 V
2 0 V
2 1 V
2 0 V
2 0 V
2 1 V
2 0 V
2 1 V
3 0 V
2 1 V
2 0 V
2 1 V
2 0 V
2 1 V
2 0 V
2 1 V
3 0 V
2 1 V
2 0 V
2 1 V
2 1 V
2 0 V
2 1 V
2 0 V
3 1 V
2 1 V
2 0 V
2 1 V
2 1 V
2 0 V
2 1 V
2 1 V
2 0 V
3 1 V
2 1 V
2 0 V
2 1 V
2 1 V
2 1 V
2 0 V
2 1 V
3 1 V
2 1 V
2 0 V
2 1 V
2 1 V
2 1 V
2 1 V
2 1 V
3 0 V
2 1 V
2 1 V
2 1 V
2 1 V
2 1 V
2 1 V
2 1 V
2 1 V
3 0 V
2 1 V
2 1 V
2 1 V
2 1 V
2 1 V
2 1 V
2 1 V
3 1 V
2 1 V
2 1 V
2 1 V
2 2 V
2 1 V
2 1 V
2 1 V
3 1 V
2 1 V
2 1 V
2 1 V
2 1 V
2 2 V
2 1 V
2 1 V
2 1 V
3 1 V
2 1 V
2 2 V
2 1 V
2 1 V
2 1 V
2 2 V
2 1 V
3 1 V
2 2 V
2 1 V
2 1 V
2 2 V
2 1 V
2 1 V
2 2 V
3 1 V
2 1 V
2 2 V
2 1 V
2 2 V
2 1 V
2 2 V
2 1 V
2 2 V
3 1 V
2 2 V
2 1 V
2 2 V
2 1 V
2 2 V
2 1 V
2 2 V
3 1 V
2 2 V
2 2 V
2 1 V
2 2 V
2 2 V
2 1 V
2 2 V
3 2 V
2 1 V
2 2 V
2 2 V
2 2 V
2 1 V
2 2 V
2 2 V
2 2 V
3 2 V
2 1 V
2 2 V
2 2 V
2 2 V
2 2 V
2 2 V
2 2 V
3 2 V
2 2 V
2 2 V
2 2 V
2 2 V
2 2 V
2 2 V
2 2 V
3 2 V
2 2 V
2 2 V
2 2 V
2 2 V
2 2 V
2 3 V
2 2 V
2 2 V
3 2 V
2 2 V
2 3 V
2 2 V
2 2 V
2 2 V
2 3 V
2 2 V
3 2 V
2 3 V
2 2 V
2 3 V
2 2 V
2 2 V
2 3 V
2 2 V
3 3 V
2 2 V
2 3 V
2 3 V
2 2 V
2 3 V
2 2 V
2 3 V
2 3 V
3 2 V
2 3 V
2 3 V
2 2 V
2 3 V
2 3 V
2 3 V
2 3 V
3 2 V
2 3 V
2 3 V
2 3 V
2 3 V
2 3 V
2 3 V
2 3 V
3 3 V
2 3 V
2 3 V
2 3 V
2 4 V
2 3 V
2 3 V
2 3 V
2 3 V
3 4 V
2 3 V
2 3 V
2 4 V
2 3 V
2 3 V
2 4 V
2 3 V
3 4 V
2 3 V
2 4 V
2 3 V
2 4 V
2 4 V
2 3 V
2 4 V
3 4 V
2 3 V
2 4 V
2 4 V
2 4 V
2 4 V
2 4 V
2 4 V
2 4 V
3 4 V
2 4 V
2 4 V
2 4 V
2 4 V
2 5 V
2 4 V
2 4 V
3 4 V
2 5 V
2 4 V
2 5 V
2 4 V
2 5 V
2 4 V
2 5 V
3 5 V
2 4 V
2 5 V
2 5 V
2 5 V
2 5 V
2 5 V
2 5 V
2 5 V
3 5 V
2 5 V
2 5 V
2 6 V
2 5 V
2 5 V
2 6 V
2 5 V
3 6 V
2 6 V
2 5 V
2 6 V
2 6 V
2 6 V
2 6 V
2 6 V
3 6 V
2 6 V
2 7 V
2 6 V
2 6 V
2 7 V
2 7 V
2 6 V
2 7 V
3 7 V
2 7 V
2 7 V
2 7 V
2 7 V
2 8 V
2 7 V
2 8 V
3 7 V
2 8 V
2 8 V
2 8 V
2 8 V
2 9 V
2 8 V
2 9 V
3 8 V
2 9 V
2 9 V
2 9 V
2 10 V
2 9 V
2 10 V
2 10 V
2 10 V
3 10 V
2 10 V
2 11 V
2 11 V
2 11 V
2 11 V
2 12 V
2 12 V
3 12 V
2 13 V
2 12 V
2 14 V
2 13 V
2 14 V
2 14 V
2 15 V
3 15 V
2 15 V
2 17 V
2 16 V
2 17 V
2 18 V
2 19 V
2 19 V
2 20 V
3 21 V
2 22 V
2 23 V
2 24 V
2 26 V
2 27 V
2 29 V
2 30 V
3 33 V
2 36 V
2 38 V
2 43 V
2 47 V
2 53 V
1 26 V
31 0 R
2 -38 V
2 -34 V
currentpoint stroke M
2 -31 V
2 -29 V
3 -25 V
2 -24 V
2 -21 V
2 -20 V
2 -19 V
2 -17 V
2 -17 V
2 -15 V
3 -15 V
2 -13 V
2 -13 V
2 -13 V
2 -11 V
2 -11 V
2 -11 V
2 -10 V
2 -10 V
3 -9 V
2 -9 V
2 -9 V
2 -8 V
2 -8 V
2 -8 V
2 -7 V
2 -7 V
3 -7 V
2 -6 V
2 -7 V
2 -6 V
2 -6 V
2 -6 V
2 -5 V
2 -6 V
3 -5 V
2 -5 V
2 -5 V
2 -5 V
2 -5 V
2 -4 V
2 -4 V
2 -5 V
2 -4 V
3 -4 V
2 -4 V
2 -4 V
2 -3 V
2 -4 V
2 -4 V
2 -3 V
2 -3 V
3 -4 V
2 -3 V
2 -3 V
2 -3 V
2 -3 V
2 -3 V
2 -3 V
2 -3 V
3 -2 V
2 -3 V
2 -2 V
2 -3 V
2 -2 V
2 -3 V
2 -2 V
2 -2 V
2 -3 V
3 -2 V
2 -2 V
2 -2 V
2 -2 V
2 -2 V
2 -2 V
2 -2 V
2 -2 V
3 -2 V
2 -1 V
2 -2 V
2 -2 V
2 -2 V
2 -1 V
2 -2 V
2 -1 V
3 -2 V
2 -1 V
2 -2 V
2 -1 V
2 -2 V
2 -1 V
2 -1 V
2 -2 V
2 -1 V
3 -1 V
2 -1 V
2 -2 V
2 -1 V
2 -1 V
2 -1 V
2 -1 V
2 -1 V
3 -1 V
2 -1 V
2 -1 V
2 -1 V
2 -1 V
2 -1 V
2 -1 V
2 -1 V
3 -1 V
2 -1 V
2 -1 V
2 -1 V
2 0 V
2 -1 V
2 -1 V
2 -1 V
2 0 V
3 -1 V
2 -1 V
2 -1 V
2 0 V
2 -1 V
2 -1 V
2 0 V
2 -1 V
3 0 V
2 -1 V
2 -1 V
2 0 V
2 -1 V
2 0 V
2 -1 V
2 0 V
3 -1 V
2 0 V
2 -1 V
2 0 V
2 0 V
2 -1 V
2 0 V
2 -1 V
2 0 V
3 0 V
2 -1 V
2 0 V
2 -1 V
2 0 V
2 0 V
2 -1 V
2 0 V
3 0 V
2 0 V
2 -1 V
2 0 V
2 0 V
2 0 V
2 -1 V
2 0 V
3 0 V
2 0 V
2 -1 V
2 0 V
2 0 V
2 0 V
2 0 V
2 -1 V
2 0 V
3 0 V
2 0 V
2 0 V
2 0 V
2 0 V
2 0 V
2 -1 V
2 0 V
3 0 V
2 0 V
2 0 V
2 0 V
2 0 V
2 0 V
2 0 V
2 0 V
3 0 V
2 0 V
2 0 V
2 0 V
2 0 V
2 -1 V
2 0 V
2 0 V
2 0 V
3 0 V
2 0 V
2 0 V
2 1 V
2 0 V
2 0 V
2 0 V
2 0 V
3 0 V
2 0 V
2 0 V
2 0 V
2 0 V
2 0 V
2 0 V
2 0 V
3 0 V
2 0 V
2 0 V
2 0 V
2 0 V
2 1 V
2 0 V
2 0 V
2 0 V
3 0 V
2 0 V
2 0 V
2 0 V
2 0 V
2 1 V
2 0 V
2 0 V
3 0 V
2 0 V
2 0 V
2 0 V
2 1 V
2 0 V
2 0 V
2 0 V
3 0 V
2 0 V
2 0 V
2 1 V
2 0 V
2 0 V
2 0 V
2 0 V
2 1 V
3 0 V
2 0 V
2 0 V
2 0 V
2 0 V
2 1 V
2 0 V
2 0 V
3 0 V
2 0 V
2 1 V
2 0 V
2 0 V
2 0 V
2 0 V
2 1 V
3 0 V
2 0 V
2 0 V
2 1 V
2 0 V
2 0 V
2 0 V
2 0 V
2 1 V
3 0 V
2 0 V
2 0 V
2 1 V
2 0 V
2 0 V
2 0 V
2 0 V
3 1 V
2 0 V
2 0 V
2 0 V
2 1 V
2 0 V
2 0 V
2 0 V
3 1 V
2 0 V
2 0 V
2 0 V
2 1 V
2 0 V
2 0 V
2 0 V
2 1 V
3 0 V
2 0 V
2 0 V
2 1 V
2 0 V
2 0 V
2 1 V
2 0 V
3 0 V
2 0 V
2 1 V
2 0 V
2 0 V
2 0 V
2 1 V
2 0 V
3 0 V
2 0 V
2 1 V
2 0 V
2 0 V
2 1 V
2 0 V
2 0 V
2 0 V
3 1 V
2 0 V
2 0 V
2 0 V
2 1 V
2 0 V
2 0 V
2 0 V
3 1 V
2 0 V
2 0 V
2 1 V
2 0 V
2 0 V
2 0 V
2 1 V
3 0 V
2 0 V
2 0 V
2 1 V
2 0 V
2 0 V
2 1 V
2 0 V
2 0 V
3 0 V
2 1 V
2 0 V
2 0 V
2 0 V
2 1 V
2 0 V
2 0 V
3 1 V
2 0 V
2 0 V
2 0 V
2 1 V
2 0 V
2 0 V
2 0 V
3 1 V
2 0 V
2 0 V
2 1 V
2 0 V
2 0 V
2 0 V
2 1 V
2 0 V
3 0 V
2 0 V
2 1 V
2 0 V
2 0 V
2 0 V
2 1 V
2 0 V
3 0 V
2 0 V
2 1 V
2 0 V
2 0 V
2 1 V
2 0 V
2 0 V
3 0 V
2 1 V
2 0 V
2 0 V
2 0 V
2 1 V
2 0 V
2 0 V
2 0 V
3 1 V
2 0 V
2 0 V
2 0 V
2 1 V
2 0 V
currentpoint stroke M
2 0 V
2 0 V
3 1 V
2 0 V
2 0 V
2 0 V
2 0 V
2 1 V
2 0 V
2 0 V
3 0 V
2 1 V
2 0 V
2 0 V
2 0 V
2 1 V
2 0 V
2 0 V
2 0 V
3 1 V
2 0 V
2 0 V
2 0 V
2 0 V
2 1 V
2 0 V
2 0 V
3 0 V
2 1 V
2 0 V
2 0 V
2 0 V
2 0 V
2 1 V
2 0 V
3 0 V
2 0 V
2 1 V
2 0 V
2 0 V
2 0 V
2 0 V
2 1 V
2 0 V
3 0 V
2 0 V
2 0 V
2 1 V
2 0 V
2 0 V
2 0 V
2 0 V
3 1 V
2 0 V
2 0 V
2 0 V
2 0 V
2 0 V
2 1 V
2 0 V
3 0 V
2 0 V
2 0 V
2 1 V
2 0 V
2 0 V
2 0 V
2 0 V
2 0 V
3 1 V
2 0 V
2 0 V
2 0 V
2 0 V
2 1 V
2 0 V
2 0 V
3 0 V
2 0 V
2 0 V
2 0 V
2 1 V
2 0 V
2 0 V
2 0 V
3 0 V
2 0 V
2 1 V
2 0 V
2 0 V
2 0 V
2 0 V
2 0 V
2 0 V
3 1 V
2 0 V
2 0 V
2 0 V
2 0 V
2 0 V
2 0 V
2 1 V
3 0 V
2 0 V
2 0 V
2 0 V
2 0 V
2 0 V
2 0 V
2 1 V
3 0 V
2 0 V
2 0 V
2 0 V
2 0 V
2 0 V
2 0 V
2 0 V
2 1 V
3 0 V
2 0 V
2 0 V
2 0 V
2 0 V
2 0 V
2 0 V
2 0 V
3 0 V
2 1 V
2 0 V
2 0 V
2 0 V
2 0 V
2 0 V
2 0 V
3 0 V
2 0 V
2 0 V
2 0 V
2 0 V
2 1 V
2 0 V
2 0 V
2 0 V
3 0 V
2 0 V
2 0 V
2 0 V
2 0 V
2 0 V
2 0 V
2 0 V
3 0 V
2 0 V
2 0 V
2 1 V
2 0 V
2 0 V
2 0 V
2 0 V
3 0 V
2 0 V
2 0 V
2 0 V
2 0 V
2 0 V
2 0 V
2 0 V
2 0 V
3 0 V
2 0 V
2 0 V
2 0 V
2 0 V
2 0 V
2 0 V
2 0 V
3 0 V
2 0 V
2 0 V
2 0 V
2 0 V
2 0 V
2 0 V
2 0 V
3 0 V
2 0 V
2 0 V
2 0 V
2 0 V
1.000 UL
LT2
2023 1847 M
247 0 V
250 570 M
2 0 V
2 0 V
2 0 V
2 0 V
3 0 V
2 0 V
2 0 V
2 0 V
2 1 V
2 0 V
2 0 V
2 0 V
3 0 V
2 0 V
2 0 V
2 1 V
2 0 V
2 0 V
2 0 V
2 0 V
3 1 V
2 0 V
2 0 V
2 0 V
2 1 V
2 0 V
2 0 V
2 1 V
2 0 V
3 0 V
2 1 V
2 0 V
2 0 V
2 1 V
2 0 V
2 1 V
2 0 V
3 0 V
2 1 V
2 0 V
2 1 V
2 0 V
2 1 V
2 0 V
2 1 V
3 0 V
2 1 V
2 0 V
2 1 V
2 1 V
2 0 V
2 1 V
2 0 V
2 1 V
3 1 V
2 0 V
2 1 V
2 1 V
2 0 V
2 1 V
2 1 V
2 1 V
3 0 V
2 1 V
2 1 V
2 1 V
2 1 V
2 0 V
2 1 V
2 1 V
3 1 V
2 1 V
2 1 V
2 1 V
2 1 V
2 1 V
2 1 V
2 0 V
2 1 V
3 1 V
2 1 V
2 2 V
2 1 V
2 1 V
2 1 V
2 1 V
2 1 V
3 1 V
2 1 V
2 1 V
2 2 V
2 1 V
2 1 V
2 1 V
2 2 V
3 1 V
2 1 V
2 1 V
2 2 V
2 1 V
2 1 V
2 2 V
2 1 V
2 2 V
3 1 V
2 2 V
2 1 V
2 2 V
2 1 V
2 2 V
2 1 V
2 2 V
3 1 V
2 2 V
2 2 V
2 1 V
2 2 V
2 2 V
2 2 V
2 1 V
3 2 V
2 2 V
2 2 V
2 2 V
2 2 V
2 2 V
2 2 V
2 2 V
2 2 V
3 2 V
2 2 V
2 2 V
2 2 V
2 2 V
2 3 V
2 2 V
2 2 V
3 3 V
2 2 V
2 2 V
2 3 V
2 2 V
2 3 V
2 2 V
2 3 V
3 3 V
2 2 V
2 3 V
2 3 V
2 3 V
2 3 V
2 3 V
2 3 V
2 3 V
3 3 V
2 3 V
2 3 V
2 4 V
2 3 V
2 3 V
2 4 V
2 4 V
3 3 V
2 4 V
2 4 V
2 4 V
2 4 V
2 4 V
2 5 V
2 4 V
3 5 V
2 4 V
2 5 V
2 5 V
2 5 V
2 5 V
2 6 V
2 6 V
2 5 V
3 7 V
2 6 V
2 7 V
2 7 V
2 7 V
2 8 V
2 8 V
2 9 V
3 9 V
2 10 V
2 11 V
2 11 V
2 13 V
2 15 V
2 16 V
2 20 V
3 23 V
2 29 V
2 40 V
2 66 V
2 501 V
2 -493 V
2 -61 V
2 -34 V
2 -24 V
3 -18 V
2 -14 V
2 -12 V
2 -9 V
2 -9 V
2 -7 V
2 -7 V
2 -5 V
3 -6 V
2 -4 V
2 -4 V
2 -4 V
2 -4 V
2 -3 V
2 -3 V
2 -2 V
3 -3 V
2 -2 V
2 -2 V
2 -2 V
2 -1 V
2 -2 V
2 -2 V
2 -1 V
2 -1 V
3 -1 V
2 -1 V
2 -1 V
2 -1 V
2 -1 V
2 -1 V
2 0 V
2 -1 V
3 -1 V
2 0 V
2 -1 V
2 0 V
2 0 V
2 -1 V
2 0 V
2 0 V
3 0 V
2 0 V
2 0 V
2 0 V
2 0 V
2 0 V
2 0 V
2 0 V
2 0 V
3 0 V
2 0 V
2 0 V
2 1 V
2 0 V
2 0 V
2 0 V
2 1 V
3 0 V
2 0 V
2 1 V
2 0 V
2 0 V
2 1 V
2 0 V
2 1 V
3 0 V
2 1 V
2 0 V
2 1 V
2 0 V
2 1 V
2 0 V
2 1 V
2 1 V
3 0 V
2 1 V
2 0 V
2 1 V
2 1 V
2 0 V
2 1 V
2 1 V
3 0 V
2 1 V
2 1 V
2 0 V
2 1 V
2 1 V
2 1 V
2 0 V
3 1 V
2 1 V
2 1 V
2 1 V
2 0 V
2 1 V
2 1 V
2 1 V
2 1 V
3 0 V
2 1 V
2 1 V
2 1 V
2 1 V
2 1 V
2 1 V
2 0 V
3 1 V
2 1 V
2 1 V
2 1 V
2 1 V
2 1 V
2 1 V
2 1 V
3 1 V
2 1 V
2 1 V
2 1 V
2 0 V
2 1 V
2 1 V
2 1 V
2 1 V
3 1 V
2 1 V
2 1 V
2 1 V
2 1 V
2 1 V
2 1 V
2 1 V
3 1 V
2 1 V
2 1 V
2 1 V
2 1 V
2 1 V
2 1 V
2 1 V
3 1 V
2 1 V
2 1 V
2 2 V
2 1 V
2 1 V
2 1 V
2 1 V
2 1 V
3 1 V
2 1 V
2 1 V
2 1 V
2 1 V
2 1 V
2 1 V
2 1 V
3 1 V
2 2 V
2 1 V
2 1 V
2 1 V
2 1 V
2 1 V
2 1 V
3 1 V
2 1 V
2 1 V
2 1 V
2 2 V
2 1 V
2 1 V
2 1 V
2 1 V
3 1 V
2 1 V
2 1 V
2 1 V
2 2 V
2 1 V
2 1 V
2 1 V
3 1 V
2 1 V
2 1 V
2 1 V
2 2 V
2 1 V
2 1 V
2 1 V
3 1 V
2 1 V
currentpoint stroke M
2 1 V
2 2 V
2 1 V
2 1 V
2 1 V
2 1 V
2 1 V
3 1 V
2 2 V
2 1 V
2 1 V
2 1 V
2 1 V
2 1 V
2 1 V
3 2 V
2 1 V
2 1 V
2 1 V
2 1 V
2 1 V
2 2 V
2 1 V
3 1 V
2 1 V
2 1 V
2 1 V
2 1 V
2 2 V
2 1 V
2 1 V
2 1 V
3 1 V
2 1 V
2 2 V
2 1 V
2 1 V
2 1 V
2 1 V
2 1 V
3 2 V
2 1 V
2 1 V
2 1 V
2 1 V
2 1 V
2 2 V
2 1 V
3 1 V
2 1 V
2 1 V
2 1 V
2 2 V
2 1 V
2 1 V
2 1 V
2 1 V
3 1 V
2 2 V
2 1 V
2 1 V
2 1 V
2 1 V
2 1 V
2 1 V
3 2 V
2 1 V
2 1 V
2 1 V
2 1 V
2 1 V
2 2 V
2 1 V
3 1 V
2 1 V
2 1 V
2 1 V
2 2 V
2 1 V
2 1 V
2 1 V
2 1 V
3 1 V
2 1 V
2 2 V
2 1 V
2 1 V
2 1 V
2 1 V
2 1 V
3 2 V
2 1 V
2 1 V
2 1 V
2 1 V
2 1 V
2 1 V
2 2 V
3 1 V
2 1 V
2 1 V
2 1 V
2 1 V
2 1 V
2 2 V
2 1 V
2 1 V
3 1 V
2 1 V
2 1 V
2 1 V
2 2 V
2 1 V
2 1 V
2 1 V
3 1 V
2 1 V
2 1 V
2 1 V
2 2 V
2 1 V
2 1 V
2 1 V
3 1 V
2 1 V
2 1 V
2 1 V
2 2 V
2 1 V
2 1 V
2 1 V
2 1 V
3 1 V
2 1 V
2 1 V
2 1 V
2 2 V
2 1 V
2 1 V
2 1 V
3 1 V
2 1 V
2 1 V
2 1 V
2 1 V
2 2 V
2 1 V
2 1 V
3 1 V
2 1 V
2 1 V
2 1 V
2 1 V
2 1 V
2 1 V
2 1 V
2 2 V
3 1 V
2 1 V
2 1 V
2 1 V
2 1 V
2 1 V
2 1 V
2 1 V
3 1 V
2 1 V
2 1 V
2 1 V
2 2 V
2 1 V
2 1 V
2 1 V
3 1 V
2 1 V
2 1 V
2 1 V
2 1 V
2 1 V
2 1 V
2 1 V
2 1 V
3 1 V
2 1 V
2 1 V
2 1 V
2 1 V
2 1 V
2 2 V
2 1 V
3 1 V
2 1 V
2 1 V
2 1 V
2 1 V
2 1 V
2 1 V
2 1 V
3 1 V
2 1 V
2 1 V
2 1 V
2 1 V
2 1 V
2 1 V
2 1 V
2 1 V
3 1 V
2 1 V
2 1 V
2 1 V
2 1 V
2 1 V
2 1 V
2 1 V
3 1 V
2 1 V
2 1 V
2 1 V
2 1 V
2 1 V
2 1 V
2 1 V
3 1 V
2 1 V
2 1 V
2 1 V
2 1 V
2 1 V
2 1 V
2 1 V
2 1 V
3 1 V
2 0 V
2 1 V
2 1 V
2 1 V
2 1 V
2 1 V
2 1 V
3 1 V
2 1 V
2 1 V
2 1 V
2 1 V
2 1 V
2 1 V
2 1 V
3 1 V
2 1 V
2 1 V
2 0 V
2 1 V
2 1 V
2 1 V
2 1 V
2 1 V
3 1 V
2 1 V
2 1 V
2 1 V
2 1 V
2 0 V
2 1 V
2 1 V
3 1 V
2 1 V
2 1 V
2 1 V
2 1 V
2 1 V
2 0 V
2 1 V
3 1 V
2 1 V
2 1 V
2 1 V
2 1 V
2 1 V
2 0 V
2 1 V
2 1 V
3 1 V
2 1 V
2 1 V
2 1 V
2 0 V
2 1 V
2 1 V
2 1 V
3 1 V
2 1 V
2 1 V
2 0 V
2 1 V
2 1 V
2 1 V
2 1 V
3 1 V
2 0 V
2 1 V
2 1 V
2 1 V
2 1 V
2 0 V
2 1 V
2 1 V
3 1 V
2 1 V
2 0 V
2 1 V
2 1 V
2 1 V
2 1 V
2 0 V
3 1 V
2 1 V
2 1 V
2 0 V
2 1 V
2 1 V
2 1 V
2 1 V
3 0 V
2 1 V
2 1 V
2 1 V
2 0 V
2 1 V
2 1 V
2 1 V
2 0 V
3 1 V
2 1 V
2 1 V
2 0 V
2 1 V
2 1 V
2 1 V
2 0 V
3 1 V
2 1 V
2 0 V
2 1 V
2 1 V
2 1 V
2 0 V
2 1 V
3 1 V
2 0 V
2 1 V
2 1 V
2 1 V
2 0 V
2 1 V
2 1 V
2 0 V
3 1 V
2 1 V
2 0 V
2 1 V
2 1 V
2 0 V
2 1 V
2 1 V
3 0 V
2 1 V
2 1 V
2 0 V
2 1 V
2 1 V
2 0 V
2 1 V
3 1 V
2 0 V
2 1 V
2 0 V
2 1 V
2 1 V
2 0 V
2 1 V
2 1 V
3 0 V
2 1 V
2 0 V
2 1 V
2 1 V
2 0 V
2 1 V
2 0 V
3 1 V
2 1 V
2 0 V
2 1 V
2 0 V
2 1 V
2 1 V
2 0 V
3 1 V
2 0 V
currentpoint stroke M
2 1 V
2 0 V
2 1 V
2 1 V
2 0 V
2 1 V
2 0 V
3 1 V
2 0 V
2 1 V
2 0 V
2 1 V
2 1 V
2 0 V
2 1 V
3 0 V
2 1 V
2 0 V
2 1 V
2 0 V
2 1 V
2 0 V
2 1 V
3 0 V
2 1 V
2 0 V
2 1 V
2 0 V
2 1 V
2 0 V
2 1 V
2 0 V
3 1 V
2 0 V
2 1 V
2 0 V
2 1 V
2 0 V
2 1 V
2 0 V
3 1 V
2 0 V
2 1 V
2 0 V
2 0 V
2 1 V
2 0 V
2 1 V
3 0 V
2 1 V
2 0 V
2 1 V
2 0 V
2 0 V
2 1 V
2 0 V
2 1 V
3 0 V
2 0 V
2 1 V
2 0 V
2 1 V
2 0 V
2 1 V
2 0 V
3 0 V
2 1 V
2 0 V
2 0 V
2 1 V
2 0 V
2 1 V
2 0 V
3 0 V
2 1 V
2 0 V
2 0 V
2 1 V
2 0 V
2 1 V
2 0 V
2 0 V
3 1 V
2 0 V
2 0 V
2 1 V
2 0 V
2 0 V
2 1 V
2 0 V
3 0 V
2 1 V
2 0 V
2 0 V
2 1 V
2 0 V
2 0 V
2 0 V
3 1 V
2 0 V
2 0 V
2 1 V
2 0 V
2 0 V
2 1 V
2 0 V
2 0 V
3 0 V
2 1 V
2 0 V
2 0 V
2 0 V
2 1 V
2 0 V
2 0 V
3 1 V
2 0 V
2 0 V
2 0 V
2 0 V
2 1 V
2 0 V
2 0 V
3 0 V
2 1 V
2 0 V
2 0 V
2 0 V
2 1 V
2 0 V
2 0 V
2 0 V
3 0 V
2 1 V
2 0 V
2 0 V
2 0 V
2 0 V
2 1 V
2 0 V
3 0 V
2 0 V
2 0 V
2 0 V
2 1 V
2 0 V
2 0 V
2 0 V
3 0 V
2 0 V
2 1 V
2 0 V
2 0 V
2 0 V
2 0 V
2 0 V
2 0 V
3 1 V
2 0 V
2 0 V
2 0 V
2 0 V
2 0 V
2 0 V
2 0 V
3 1 V
2 0 V
2 0 V
2 0 V
2 0 V
2 0 V
2 0 V
2 0 V
3 0 V
2 0 V
2 0 V
2 1 V
2 0 V
2 0 V
2 0 V
2 0 V
2 0 V
3 0 V
2 0 V
2 0 V
2 0 V
2 0 V
2 0 V
2 0 V
2 0 V
3 0 V
2 0 V
2 0 V
2 0 V
2 0 V
2 0 V
2 0 V
2 0 V
3 0 V
2 1 V
2 0 V
2 0 V
2 0 V
1.000 UL
LT1
2023 1747 M
247 0 V
250 586 M
2 0 V
2 0 V
2 0 V
2 0 V
3 0 V
2 1 V
2 0 V
2 0 V
2 0 V
2 0 V
2 0 V
2 0 V
3 1 V
2 0 V
2 0 V
2 0 V
2 0 V
2 1 V
2 0 V
2 0 V
3 1 V
2 0 V
2 0 V
2 1 V
2 0 V
2 0 V
2 1 V
2 0 V
2 1 V
3 0 V
2 0 V
2 1 V
2 0 V
2 1 V
2 0 V
2 1 V
2 1 V
3 0 V
2 1 V
2 0 V
2 1 V
2 1 V
2 0 V
2 1 V
2 1 V
3 1 V
2 0 V
2 1 V
2 1 V
2 1 V
2 1 V
2 0 V
2 1 V
2 1 V
3 1 V
2 1 V
2 1 V
2 1 V
2 1 V
2 1 V
2 1 V
2 1 V
3 2 V
2 1 V
2 1 V
2 1 V
2 2 V
2 1 V
2 1 V
2 2 V
3 1 V
2 2 V
2 1 V
2 2 V
2 1 V
2 2 V
2 2 V
2 2 V
2 2 V
3 2 V
2 2 V
2 2 V
2 2 V
2 2 V
2 3 V
2 2 V
2 3 V
3 3 V
2 3 V
2 3 V
2 4 V
2 4 V
2 4 V
2 5 V
2 5 V
3 7 V
2 8 V
2 11 V
2 17 V
2 126 V
2 -122 V
2 -14 V
2 -8 V
2 -5 V
3 -3 V
2 -3 V
2 -2 V
2 -2 V
2 -1 V
2 -1 V
2 -1 V
2 -1 V
3 0 V
2 -1 V
2 0 V
2 0 V
2 0 V
2 0 V
2 0 V
2 0 V
3 0 V
2 1 V
2 0 V
2 0 V
2 1 V
2 0 V
2 1 V
2 0 V
2 0 V
3 1 V
2 1 V
2 0 V
2 1 V
2 0 V
2 1 V
2 1 V
2 0 V
3 1 V
2 1 V
2 0 V
2 1 V
2 1 V
2 1 V
2 1 V
2 0 V
3 1 V
2 1 V
2 1 V
2 1 V
2 0 V
2 1 V
2 1 V
2 1 V
2 1 V
3 1 V
2 1 V
2 1 V
2 1 V
2 1 V
2 1 V
2 1 V
2 1 V
3 1 V
2 0 V
2 1 V
2 1 V
2 1 V
2 1 V
2 2 V
2 1 V
3 1 V
2 1 V
2 1 V
2 1 V
2 1 V
2 1 V
2 1 V
2 1 V
2 1 V
3 1 V
2 1 V
2 1 V
2 1 V
2 2 V
2 1 V
2 1 V
2 1 V
3 1 V
2 1 V
2 1 V
2 1 V
2 2 V
2 1 V
2 1 V
2 1 V
3 1 V
2 1 V
2 1 V
2 2 V
2 1 V
2 1 V
2 1 V
2 1 V
2 2 V
3 1 V
2 1 V
2 1 V
2 1 V
2 2 V
2 1 V
2 1 V
2 1 V
3 1 V
2 2 V
2 1 V
2 1 V
2 1 V
2 2 V
2 1 V
2 1 V
3 1 V
2 2 V
2 1 V
2 1 V
2 1 V
2 2 V
2 1 V
2 1 V
2 1 V
3 2 V
2 1 V
2 1 V
2 2 V
2 1 V
2 1 V
2 1 V
2 2 V
3 1 V
2 1 V
2 2 V
2 1 V
2 1 V
2 2 V
2 1 V
2 1 V
3 1 V
2 2 V
2 1 V
2 1 V
2 2 V
2 1 V
2 1 V
2 2 V
2 1 V
3 1 V
2 2 V
2 1 V
2 1 V
2 2 V
2 1 V
2 1 V
2 2 V
3 1 V
2 1 V
2 2 V
2 1 V
2 1 V
2 2 V
2 1 V
2 2 V
3 1 V
2 1 V
2 2 V
2 1 V
2 1 V
2 2 V
2 1 V
2 1 V
2 2 V
3 1 V
2 2 V
2 1 V
2 1 V
2 2 V
2 1 V
2 1 V
2 2 V
3 1 V
2 2 V
2 1 V
2 1 V
2 2 V
2 1 V
2 1 V
2 2 V
3 1 V
2 2 V
2 1 V
2 1 V
2 2 V
2 1 V
2 2 V
2 1 V
2 1 V
3 2 V
2 1 V
2 1 V
2 2 V
2 1 V
2 2 V
2 1 V
2 1 V
3 2 V
2 1 V
2 2 V
2 1 V
2 1 V
2 2 V
2 1 V
2 2 V
3 1 V
2 2 V
2 1 V
2 1 V
2 2 V
2 1 V
2 2 V
2 1 V
2 1 V
3 2 V
2 1 V
2 2 V
2 1 V
2 1 V
2 2 V
2 1 V
2 2 V
3 1 V
2 1 V
2 2 V
2 1 V
2 2 V
2 1 V
2 2 V
2 1 V
3 1 V
2 2 V
2 1 V
2 2 V
2 1 V
2 1 V
2 2 V
2 1 V
2 2 V
3 1 V
2 1 V
2 2 V
2 1 V
2 2 V
2 1 V
2 2 V
2 1 V
3 1 V
2 2 V
2 1 V
2 2 V
2 1 V
2 1 V
2 2 V
2 1 V
3 2 V
2 1 V
2 1 V
2 2 V
2 1 V
2 2 V
2 1 V
2 2 V
2 1 V
3 1 V
2 2 V
2 1 V
2 2 V
2 1 V
2 1 V
2 2 V
2 1 V
3 2 V
2 1 V
2 1 V
2 2 V
2 1 V
2 2 V
2 1 V
2 1 V
3 2 V
2 1 V
currentpoint stroke M
2 2 V
2 1 V
2 2 V
2 1 V
2 1 V
2 2 V
2 1 V
3 2 V
2 1 V
2 1 V
2 2 V
2 1 V
2 1 V
2 2 V
2 1 V
3 2 V
2 1 V
2 1 V
2 2 V
2 1 V
2 2 V
2 1 V
2 1 V
3 2 V
2 1 V
2 2 V
2 1 V
2 1 V
2 2 V
2 1 V
2 1 V
2 2 V
3 1 V
2 2 V
2 1 V
2 1 V
2 2 V
2 1 V
2 2 V
2 1 V
3 1 V
2 2 V
2 1 V
2 1 V
2 2 V
2 1 V
2 2 V
2 1 V
3 1 V
2 2 V
2 1 V
2 1 V
2 2 V
2 1 V
2 1 V
2 2 V
2 1 V
3 1 V
2 2 V
2 1 V
2 2 V
2 1 V
2 1 V
2 2 V
2 1 V
3 1 V
2 2 V
2 1 V
2 1 V
2 2 V
2 1 V
2 1 V
2 2 V
3 1 V
2 1 V
2 2 V
2 1 V
2 1 V
2 2 V
2 1 V
2 1 V
2 2 V
3 1 V
2 1 V
2 2 V
2 1 V
2 1 V
2 2 V
2 1 V
2 1 V
3 2 V
2 1 V
2 1 V
2 2 V
2 1 V
2 1 V
2 1 V
2 2 V
3 1 V
2 1 V
2 2 V
2 1 V
2 1 V
2 2 V
2 1 V
2 1 V
2 2 V
3 1 V
2 1 V
2 1 V
2 2 V
2 1 V
2 1 V
2 2 V
2 1 V
3 1 V
2 1 V
2 2 V
2 1 V
2 1 V
2 1 V
2 2 V
2 1 V
3 1 V
2 2 V
2 1 V
2 1 V
2 1 V
2 2 V
2 1 V
2 1 V
2 1 V
3 2 V
2 1 V
2 1 V
2 1 V
2 2 V
2 1 V
2 1 V
2 1 V
3 2 V
2 1 V
2 1 V
2 1 V
2 2 V
2 1 V
2 1 V
2 1 V
3 1 V
2 2 V
2 1 V
2 1 V
2 1 V
2 2 V
2 1 V
2 1 V
2 1 V
3 1 V
2 2 V
2 1 V
2 1 V
2 1 V
2 1 V
2 2 V
2 1 V
3 1 V
2 1 V
2 1 V
2 2 V
2 1 V
2 1 V
2 1 V
2 1 V
3 1 V
2 2 V
2 1 V
2 1 V
2 1 V
2 1 V
2 1 V
2 2 V
2 1 V
3 1 V
2 1 V
2 1 V
2 1 V
2 2 V
2 1 V
2 1 V
2 1 V
3 1 V
2 1 V
2 1 V
2 2 V
2 1 V
2 1 V
2 1 V
2 1 V
3 1 V
2 1 V
2 2 V
2 1 V
2 1 V
2 1 V
2 1 V
2 1 V
2 1 V
3 1 V
2 1 V
2 2 V
2 1 V
2 1 V
2 1 V
2 1 V
2 1 V
3 1 V
2 1 V
2 1 V
2 1 V
2 1 V
2 2 V
2 1 V
2 1 V
3 1 V
2 1 V
2 1 V
2 1 V
2 1 V
2 1 V
2 1 V
2 1 V
2 1 V
3 1 V
2 1 V
2 1 V
2 1 V
2 2 V
2 1 V
2 1 V
2 1 V
3 1 V
2 1 V
2 1 V
2 1 V
2 1 V
2 1 V
2 1 V
2 1 V
3 1 V
2 1 V
2 1 V
2 1 V
2 1 V
2 1 V
2 1 V
2 1 V
2 1 V
3 1 V
2 1 V
2 1 V
2 1 V
2 1 V
2 1 V
2 1 V
2 1 V
3 1 V
2 1 V
2 1 V
2 1 V
2 1 V
2 1 V
2 1 V
2 1 V
3 1 V
2 1 V
2 0 V
2 1 V
2 1 V
2 1 V
2 1 V
2 1 V
2 1 V
3 1 V
2 1 V
2 1 V
2 1 V
2 1 V
2 1 V
2 1 V
2 1 V
3 0 V
2 1 V
2 1 V
2 1 V
2 1 V
2 1 V
2 1 V
2 1 V
3 1 V
2 1 V
2 0 V
2 1 V
2 1 V
2 1 V
2 1 V
2 1 V
2 1 V
3 1 V
2 0 V
2 1 V
2 1 V
2 1 V
2 1 V
2 1 V
2 1 V
3 0 V
2 1 V
2 1 V
2 1 V
2 1 V
2 1 V
2 1 V
2 0 V
3 1 V
2 1 V
2 1 V
2 1 V
2 0 V
2 1 V
2 1 V
2 1 V
2 1 V
3 1 V
2 0 V
2 1 V
2 1 V
2 1 V
2 1 V
2 0 V
2 1 V
3 1 V
2 1 V
2 0 V
2 1 V
2 1 V
2 1 V
2 1 V
2 0 V
3 1 V
2 1 V
2 1 V
2 0 V
2 1 V
2 1 V
2 1 V
2 0 V
2 1 V
3 1 V
2 1 V
2 0 V
2 1 V
2 1 V
2 0 V
2 1 V
2 1 V
3 1 V
2 0 V
2 1 V
2 1 V
2 0 V
2 1 V
2 1 V
2 1 V
3 0 V
2 1 V
2 1 V
2 0 V
2 1 V
2 1 V
2 0 V
2 1 V
2 1 V
3 0 V
2 1 V
2 1 V
2 0 V
2 1 V
2 1 V
2 0 V
2 1 V
3 1 V
2 0 V
2 1 V
2 0 V
2 1 V
2 1 V
2 0 V
2 1 V
3 1 V
2 0 V
currentpoint stroke M
2 1 V
2 0 V
2 1 V
2 1 V
2 0 V
2 1 V
2 0 V
3 1 V
2 1 V
2 0 V
2 1 V
2 0 V
2 1 V
2 1 V
2 0 V
3 1 V
2 0 V
2 1 V
2 0 V
2 1 V
2 1 V
2 0 V
2 1 V
3 0 V
2 1 V
2 0 V
2 1 V
2 0 V
2 1 V
2 0 V
2 1 V
2 1 V
3 0 V
2 1 V
2 0 V
2 1 V
2 0 V
2 1 V
2 0 V
2 1 V
3 0 V
2 1 V
2 0 V
2 1 V
2 0 V
2 1 V
2 0 V
2 1 V
3 0 V
2 0 V
2 1 V
2 0 V
2 1 V
2 0 V
2 1 V
2 0 V
2 1 V
3 0 V
2 1 V
2 0 V
2 0 V
2 1 V
2 0 V
2 1 V
2 0 V
3 1 V
2 0 V
2 0 V
2 1 V
2 0 V
2 1 V
2 0 V
2 0 V
3 1 V
2 0 V
2 1 V
2 0 V
2 0 V
2 1 V
2 0 V
2 1 V
2 0 V
3 0 V
2 1 V
2 0 V
2 0 V
2 1 V
2 0 V
2 1 V
2 0 V
3 0 V
2 1 V
2 0 V
2 0 V
2 1 V
2 0 V
2 0 V
2 1 V
3 0 V
2 0 V
2 1 V
2 0 V
2 0 V
2 0 V
2 1 V
2 0 V
2 0 V
3 1 V
2 0 V
2 0 V
2 1 V
2 0 V
2 0 V
2 0 V
2 1 V
3 0 V
2 0 V
2 0 V
2 1 V
2 0 V
2 0 V
2 1 V
2 0 V
3 0 V
2 0 V
2 0 V
2 1 V
2 0 V
2 0 V
2 0 V
2 1 V
2 0 V
3 0 V
2 0 V
2 1 V
2 0 V
2 0 V
2 0 V
2 0 V
2 1 V
3 0 V
2 0 V
2 0 V
2 0 V
2 0 V
2 1 V
2 0 V
2 0 V
3 0 V
2 0 V
2 1 V
2 0 V
2 0 V
2 0 V
2 0 V
2 0 V
2 0 V
3 1 V
2 0 V
2 0 V
2 0 V
2 0 V
2 0 V
2 0 V
2 0 V
3 1 V
2 0 V
2 0 V
2 0 V
2 0 V
2 0 V
2 0 V
2 0 V
3 0 V
2 1 V
2 0 V
2 0 V
2 0 V
2 0 V
2 0 V
2 0 V
2 0 V
3 0 V
2 0 V
2 0 V
2 0 V
2 0 V
2 0 V
2 0 V
2 0 V
3 1 V
2 0 V
2 0 V
2 0 V
2 0 V
2 0 V
2 0 V
2 0 V
3 0 V
2 0 V
2 0 V
2 0 V
2 0 V
1.000 UL
LT0
2023 1647 M
247 0 V
250 571 M
2 0 V
2 0 V
2 0 V
2 0 V
3 0 V
2 0 V
2 1 V
2 0 V
2 0 V
2 0 V
2 0 V
2 1 V
3 0 V
2 0 V
2 0 V
2 1 V
2 0 V
2 0 V
2 1 V
2 0 V
3 0 V
2 1 V
2 0 V
2 0 V
2 1 V
2 0 V
2 1 V
2 0 V
2 1 V
3 0 V
2 0 V
2 1 V
2 0 V
2 1 V
2 0 V
2 1 V
2 1 V
3 0 V
2 1 V
2 0 V
2 1 V
2 0 V
2 1 V
2 0 V
2 1 V
3 1 V
2 0 V
2 1 V
2 1 V
2 0 V
2 1 V
2 1 V
2 0 V
2 1 V
3 1 V
2 0 V
2 1 V
2 1 V
2 0 V
2 1 V
2 1 V
2 1 V
3 0 V
2 1 V
2 1 V
2 1 V
2 1 V
2 0 V
2 1 V
2 1 V
3 1 V
2 1 V
2 0 V
2 1 V
2 1 V
2 1 V
2 1 V
2 1 V
2 0 V
3 1 V
2 1 V
2 1 V
2 1 V
2 1 V
2 1 V
2 1 V
2 1 V
3 1 V
2 0 V
2 1 V
2 1 V
2 1 V
2 1 V
2 1 V
2 1 V
3 1 V
2 1 V
2 1 V
2 1 V
2 1 V
2 1 V
2 1 V
2 1 V
2 1 V
3 1 V
2 1 V
2 1 V
2 1 V
2 1 V
2 1 V
2 1 V
2 1 V
3 1 V
2 1 V
2 1 V
2 2 V
2 1 V
2 1 V
2 1 V
2 1 V
3 1 V
2 1 V
2 1 V
2 1 V
2 1 V
2 1 V
2 2 V
2 1 V
2 1 V
3 1 V
2 1 V
2 1 V
2 1 V
2 2 V
2 1 V
2 1 V
2 1 V
3 1 V
2 1 V
2 2 V
2 1 V
2 1 V
2 1 V
2 1 V
2 1 V
3 2 V
2 1 V
2 1 V
2 1 V
2 1 V
2 2 V
2 1 V
2 1 V
2 1 V
3 2 V
2 1 V
2 1 V
2 1 V
2 1 V
2 2 V
2 1 V
2 1 V
3 1 V
2 2 V
2 1 V
2 1 V
2 1 V
2 2 V
2 1 V
2 1 V
3 2 V
2 1 V
2 1 V
2 1 V
2 2 V
2 1 V
2 1 V
2 2 V
2 1 V
3 1 V
2 1 V
2 2 V
2 1 V
2 1 V
2 2 V
2 1 V
2 1 V
3 2 V
2 1 V
2 1 V
2 2 V
2 1 V
2 1 V
2 2 V
2 1 V
3 1 V
2 2 V
2 1 V
2 1 V
2 2 V
2 1 V
2 1 V
2 2 V
2 1 V
3 1 V
2 2 V
2 1 V
2 1 V
2 2 V
2 1 V
2 2 V
2 1 V
3 1 V
2 2 V
2 1 V
2 1 V
2 2 V
2 1 V
2 2 V
2 1 V
3 1 V
2 2 V
2 1 V
2 2 V
2 1 V
2 1 V
2 2 V
2 1 V
2 2 V
3 1 V
2 1 V
2 2 V
2 1 V
2 2 V
2 1 V
2 1 V
2 2 V
3 1 V
2 2 V
2 1 V
2 2 V
2 1 V
2 1 V
2 2 V
2 1 V
3 2 V
2 1 V
2 2 V
2 1 V
2 1 V
2 2 V
2 1 V
2 2 V
2 1 V
3 2 V
2 1 V
2 1 V
2 2 V
2 1 V
2 2 V
2 1 V
2 2 V
3 1 V
2 2 V
2 1 V
2 2 V
2 1 V
2 1 V
2 2 V
2 1 V
3 2 V
2 1 V
2 2 V
2 1 V
2 2 V
2 1 V
2 2 V
2 1 V
2 1 V
3 2 V
2 1 V
2 2 V
2 1 V
2 2 V
2 1 V
2 2 V
2 1 V
3 2 V
2 1 V
2 2 V
2 1 V
2 2 V
2 1 V
2 1 V
2 2 V
3 1 V
2 2 V
2 1 V
2 2 V
2 1 V
2 2 V
2 1 V
2 2 V
2 1 V
3 2 V
2 1 V
2 2 V
2 1 V
2 2 V
2 1 V
2 2 V
2 1 V
3 2 V
2 1 V
2 2 V
2 1 V
2 1 V
2 2 V
2 1 V
2 2 V
3 1 V
2 2 V
2 1 V
2 2 V
2 1 V
2 2 V
2 1 V
2 2 V
2 1 V
3 2 V
2 1 V
2 2 V
2 1 V
2 2 V
2 1 V
2 2 V
2 1 V
3 2 V
2 1 V
2 2 V
2 1 V
2 2 V
2 1 V
2 1 V
2 2 V
3 1 V
2 2 V
2 1 V
2 2 V
2 1 V
2 2 V
2 1 V
2 2 V
2 1 V
3 2 V
2 1 V
2 2 V
2 1 V
2 2 V
2 1 V
2 2 V
2 1 V
3 2 V
2 1 V
2 2 V
2 1 V
2 2 V
2 1 V
2 1 V
2 2 V
3 1 V
2 2 V
2 1 V
2 2 V
2 1 V
2 2 V
2 1 V
2 2 V
2 1 V
3 2 V
2 1 V
2 2 V
2 1 V
2 2 V
2 1 V
2 1 V
2 2 V
3 1 V
2 2 V
2 1 V
2 2 V
2 1 V
2 2 V
2 1 V
2 2 V
3 1 V
2 2 V
currentpoint stroke M
2 1 V
2 1 V
2 2 V
2 1 V
2 2 V
2 1 V
2 2 V
3 1 V
2 2 V
2 1 V
2 2 V
2 1 V
2 1 V
2 2 V
2 1 V
3 2 V
2 1 V
2 2 V
2 1 V
2 2 V
2 1 V
2 1 V
2 2 V
3 1 V
2 2 V
2 1 V
2 2 V
2 1 V
2 2 V
2 1 V
2 1 V
2 2 V
3 1 V
2 2 V
2 1 V
2 2 V
2 1 V
2 1 V
2 2 V
2 1 V
3 2 V
2 1 V
2 2 V
2 1 V
2 1 V
2 2 V
2 1 V
2 2 V
3 1 V
2 1 V
2 2 V
2 1 V
2 2 V
2 1 V
2 1 V
2 2 V
2 1 V
3 2 V
2 1 V
2 1 V
2 2 V
2 1 V
2 2 V
2 1 V
2 1 V
3 2 V
2 1 V
2 2 V
2 1 V
2 1 V
2 2 V
2 1 V
2 2 V
3 1 V
2 1 V
2 2 V
2 1 V
2 1 V
2 2 V
2 1 V
2 2 V
2 1 V
3 1 V
2 2 V
2 1 V
2 1 V
2 2 V
2 1 V
2 2 V
2 1 V
3 1 V
2 2 V
2 1 V
2 1 V
2 2 V
2 1 V
2 1 V
2 2 V
3 1 V
2 1 V
2 2 V
2 1 V
2 1 V
2 2 V
2 1 V
2 1 V
2 2 V
3 1 V
2 1 V
2 2 V
2 1 V
2 1 V
2 2 V
2 1 V
2 1 V
3 2 V
2 1 V
2 1 V
2 2 V
2 1 V
2 1 V
2 2 V
2 1 V
3 1 V
2 1 V
2 2 V
2 1 V
2 1 V
2 2 V
2 1 V
2 1 V
2 2 V
3 1 V
2 1 V
2 1 V
2 2 V
2 1 V
2 1 V
2 1 V
2 2 V
3 1 V
2 1 V
2 2 V
2 1 V
2 1 V
2 1 V
2 2 V
2 1 V
3 1 V
2 1 V
2 2 V
2 1 V
2 1 V
2 1 V
2 2 V
2 1 V
2 1 V
3 1 V
2 2 V
2 1 V
2 1 V
2 1 V
2 2 V
2 1 V
2 1 V
3 1 V
2 1 V
2 2 V
2 1 V
2 1 V
2 1 V
2 2 V
2 1 V
3 1 V
2 1 V
2 1 V
2 2 V
2 1 V
2 1 V
2 1 V
2 1 V
2 2 V
3 1 V
2 1 V
2 1 V
2 1 V
2 2 V
2 1 V
2 1 V
2 1 V
3 1 V
2 1 V
2 2 V
2 1 V
2 1 V
2 1 V
2 1 V
2 1 V
3 2 V
2 1 V
2 1 V
2 1 V
2 1 V
2 1 V
2 1 V
2 2 V
2 1 V
3 1 V
2 1 V
2 1 V
2 1 V
2 1 V
2 1 V
2 2 V
2 1 V
3 1 V
2 1 V
2 1 V
2 1 V
2 1 V
2 1 V
2 2 V
2 1 V
3 1 V
2 1 V
2 1 V
2 1 V
2 1 V
2 1 V
2 1 V
2 1 V
2 1 V
3 2 V
2 1 V
2 1 V
2 1 V
2 1 V
2 1 V
2 1 V
2 1 V
3 1 V
2 1 V
2 1 V
2 1 V
2 1 V
2 1 V
2 1 V
2 1 V
3 1 V
2 2 V
2 1 V
2 1 V
2 1 V
2 1 V
2 1 V
2 1 V
2 1 V
3 1 V
2 1 V
2 1 V
2 1 V
2 1 V
2 1 V
2 1 V
2 1 V
3 1 V
2 1 V
2 1 V
2 1 V
2 1 V
2 1 V
2 1 V
2 1 V
3 1 V
2 1 V
2 1 V
2 1 V
2 1 V
2 1 V
2 1 V
2 0 V
2 1 V
3 1 V
2 1 V
2 1 V
2 1 V
2 1 V
2 1 V
2 1 V
2 1 V
3 1 V
2 1 V
2 1 V
2 1 V
2 1 V
2 1 V
2 0 V
2 1 V
3 1 V
2 1 V
2 1 V
2 1 V
2 1 V
2 1 V
2 1 V
2 1 V
2 1 V
3 0 V
2 1 V
2 1 V
2 1 V
2 1 V
2 1 V
2 1 V
2 1 V
3 0 V
2 1 V
2 1 V
2 1 V
2 1 V
2 1 V
2 1 V
2 0 V
3 1 V
2 1 V
2 1 V
2 1 V
2 1 V
2 0 V
2 1 V
2 1 V
2 1 V
3 1 V
2 1 V
2 0 V
2 1 V
2 1 V
2 1 V
2 1 V
2 0 V
3 1 V
2 1 V
2 1 V
2 1 V
2 0 V
2 1 V
2 1 V
2 1 V
3 1 V
2 0 V
2 1 V
2 1 V
2 1 V
2 0 V
2 1 V
2 1 V
2 1 V
3 0 V
2 1 V
2 1 V
2 1 V
2 0 V
2 1 V
2 1 V
2 1 V
3 0 V
2 1 V
2 1 V
2 1 V
2 0 V
2 1 V
2 1 V
2 0 V
3 1 V
2 1 V
2 1 V
2 0 V
2 1 V
2 1 V
2 0 V
2 1 V
2 1 V
3 0 V
2 1 V
2 1 V
2 0 V
2 1 V
2 1 V
2 0 V
2 1 V
3 1 V
2 0 V
2 1 V
2 1 V
2 0 V
2 1 V
2 1 V
2 0 V
3 1 V
2 0 V
currentpoint stroke M
2 1 V
2 1 V
2 0 V
2 1 V
2 1 V
2 0 V
2 1 V
3 0 V
2 1 V
2 1 V
2 0 V
2 1 V
2 0 V
2 1 V
2 1 V
3 0 V
2 1 V
2 0 V
2 1 V
2 0 V
2 1 V
2 1 V
2 0 V
3 1 V
2 0 V
2 1 V
2 0 V
2 1 V
2 0 V
2 1 V
2 1 V
2 0 V
3 1 V
2 0 V
2 1 V
2 0 V
2 1 V
2 0 V
2 1 V
2 0 V
3 1 V
2 0 V
2 1 V
2 0 V
2 1 V
2 0 V
2 1 V
2 0 V
3 1 V
2 0 V
2 1 V
2 0 V
2 1 V
2 0 V
2 1 V
2 0 V
2 1 V
3 0 V
2 0 V
2 1 V
2 0 V
2 1 V
2 0 V
2 1 V
2 0 V
3 1 V
2 0 V
2 0 V
2 1 V
2 0 V
2 1 V
2 0 V
2 0 V
3 1 V
2 0 V
2 1 V
2 0 V
2 0 V
2 1 V
2 0 V
2 1 V
2 0 V
3 0 V
2 1 V
2 0 V
2 1 V
2 0 V
2 0 V
2 1 V
2 0 V
3 0 V
2 1 V
2 0 V
2 0 V
2 1 V
2 0 V
2 0 V
2 1 V
3 0 V
2 0 V
2 1 V
2 0 V
2 0 V
2 1 V
2 0 V
2 0 V
2 1 V
3 0 V
2 0 V
2 0 V
2 1 V
2 0 V
2 0 V
2 1 V
2 0 V
3 0 V
2 0 V
2 1 V
2 0 V
2 0 V
2 1 V
2 0 V
2 0 V
3 0 V
2 1 V
2 0 V
2 0 V
2 0 V
2 1 V
2 0 V
2 0 V
2 0 V
3 0 V
2 1 V
2 0 V
2 0 V
2 0 V
2 1 V
2 0 V
2 0 V
3 0 V
2 0 V
2 0 V
2 1 V
2 0 V
2 0 V
2 0 V
2 0 V
3 1 V
2 0 V
2 0 V
2 0 V
2 0 V
2 0 V
2 1 V
2 0 V
2 0 V
3 0 V
2 0 V
2 0 V
2 0 V
2 1 V
2 0 V
2 0 V
2 0 V
3 0 V
2 0 V
2 0 V
2 0 V
2 0 V
2 1 V
2 0 V
2 0 V
3 0 V
2 0 V
2 0 V
2 0 V
2 0 V
2 0 V
2 0 V
2 0 V
2 1 V
3 0 V
2 0 V
2 0 V
2 0 V
2 0 V
2 0 V
2 0 V
2 0 V
3 0 V
2 0 V
2 0 V
2 0 V
2 0 V
2 0 V
2 0 V
2 0 V
3 0 V
2 0 V
2 0 V
2 0 V
2 0 V
stroke
grestore
end
showpage
}}%
\put(1973,1647){\makebox(0,0)[r]{QS}}%
\put(1973,1747){\makebox(0,0)[r]{$kh/\pi=0.1$}}%
\put(1973,1847){\makebox(0,0)[r]{$kh/\pi=0.2$}}%
\put(1973,1947){\makebox(0,0)[r]{$kh/\pi=0.4$}}%
\put(1946,665){\makebox(0,0)[l]{(a)}}%
\put(282,1874){\makebox(0,0)[l]{$h^3{\rm Re}S^{\perp}(k,q)$}}%
\put(1999,270){\makebox(0,0)[l]{$qh/\pi$}}%
\put(2370,100){\makebox(0,0){ 1}}%
\put(1946,100){\makebox(0,0){ 0.8}}%
\put(1522,100){\makebox(0,0){ 0.6}}%
\put(1098,100){\makebox(0,0){ 0.4}}%
\put(674,100){\makebox(0,0){ 0.2}}%
\put(250,100){\makebox(0,0){ 0}}%
\put(200,2060){\makebox(0,0)[r]{ 4}}%
\put(200,1595){\makebox(0,0)[r]{ 2}}%
\put(200,1130){\makebox(0,0)[r]{ 0}}%
\put(200,665){\makebox(0,0)[r]{-2}}%
\put(200,200){\makebox(0,0)[r]{-4}}%
\end{picture}%
\endgroup
 

%% file: disp_re_par.tex
\begingroup%
  \makeatletter%
  \newcommand{\GNUPLOTspecial}{%
    \@sanitize\catcode`\%=14\relax\special}%
  \setlength{\unitlength}{0.1bp}%
{\GNUPLOTspecial{!
/gnudict 256 dict def
gnudict begin
/Color true def
/Solid true def
/gnulinewidth 5.000 def
/userlinewidth gnulinewidth def
/vshift -33 def
/dl {10 mul} def
/hpt_ 31.5 def
/vpt_ 31.5 def
/hpt hpt_ def
/vpt vpt_ def
/M {moveto} bind def
/L {lineto} bind def
/R {rmoveto} bind def
/V {rlineto} bind def
/vpt2 vpt 2 mul def
/hpt2 hpt 2 mul def
/Lshow { currentpoint stroke M
  0 vshift R show } def
/Rshow { currentpoint stroke M
  dup stringwidth pop neg vshift R show } def
/Cshow { currentpoint stroke M
  dup stringwidth pop -2 div vshift R show } def
/UP { dup vpt_ mul /vpt exch def hpt_ mul /hpt exch def
  /hpt2 hpt 2 mul def /vpt2 vpt 2 mul def } def
/DL { Color {setrgbcolor Solid {pop []} if 0 setdash }
 {pop pop pop Solid {pop []} if 0 setdash} ifelse } def
/BL { stroke userlinewidth 2 mul setlinewidth } def
/AL { stroke userlinewidth 2 div setlinewidth } def
/UL { dup gnulinewidth mul /userlinewidth exch def
      dup 1 lt {pop 1} if 10 mul /udl exch def } def
/PL { stroke userlinewidth setlinewidth } def
/LTb { BL [] 0 0 0 DL } def
/LTa { AL [1 udl mul 2 udl mul] 0 setdash 0 0 0 setrgbcolor } def
/LT0 { PL [] 1 0 0 DL } def
/LT1 { PL [4 dl 2 dl] 0 1 0 DL } def
/LT2 { PL [2 dl 3 dl] 0 0 1 DL } def
/LT3 { PL [1 dl 1.5 dl] 1 0 1 DL } def
/LT4 { PL [5 dl 2 dl 1 dl 2 dl] 0 1 1 DL } def
/LT5 { PL [4 dl 3 dl 1 dl 3 dl] 1 1 0 DL } def
/LT6 { PL [2 dl 2 dl 2 dl 4 dl] 0 0 0 DL } def
/LT7 { PL [2 dl 2 dl 2 dl 2 dl 2 dl 4 dl] 1 0.3 0 DL } def
/LT8 { PL [2 dl 2 dl 2 dl 2 dl 2 dl 2 dl 2 dl 4 dl] 0.5 0.5 0.5 DL } def
/Pnt { stroke [] 0 setdash
   gsave 1 setlinecap M 0 0 V stroke grestore } def
/Dia { stroke [] 0 setdash 2 copy vpt add M
  hpt neg vpt neg V hpt vpt neg V
  hpt vpt V hpt neg vpt V closepath stroke
  Pnt } def
/Pls { stroke [] 0 setdash vpt sub M 0 vpt2 V
  currentpoint stroke M
  hpt neg vpt neg R hpt2 0 V stroke
  } def
/Box { stroke [] 0 setdash 2 copy exch hpt sub exch vpt add M
  0 vpt2 neg V hpt2 0 V 0 vpt2 V
  hpt2 neg 0 V closepath stroke
  Pnt } def
/Crs { stroke [] 0 setdash exch hpt sub exch vpt add M
  hpt2 vpt2 neg V currentpoint stroke M
  hpt2 neg 0 R hpt2 vpt2 V stroke } def
/TriU { stroke [] 0 setdash 2 copy vpt 1.12 mul add M
  hpt neg vpt -1.62 mul V
  hpt 2 mul 0 V
  hpt neg vpt 1.62 mul V closepath stroke
  Pnt  } def
/Star { 2 copy Pls Crs } def
/BoxF { stroke [] 0 setdash exch hpt sub exch vpt add M
  0 vpt2 neg V  hpt2 0 V  0 vpt2 V
  hpt2 neg 0 V  closepath fill } def
/TriUF { stroke [] 0 setdash vpt 1.12 mul add M
  hpt neg vpt -1.62 mul V
  hpt 2 mul 0 V
  hpt neg vpt 1.62 mul V closepath fill } def
/TriD { stroke [] 0 setdash 2 copy vpt 1.12 mul sub M
  hpt neg vpt 1.62 mul V
  hpt 2 mul 0 V
  hpt neg vpt -1.62 mul V closepath stroke
  Pnt  } def
/TriDF { stroke [] 0 setdash vpt 1.12 mul sub M
  hpt neg vpt 1.62 mul V
  hpt 2 mul 0 V
  hpt neg vpt -1.62 mul V closepath fill} def
/DiaF { stroke [] 0 setdash vpt add M
  hpt neg vpt neg V hpt vpt neg V
  hpt vpt V hpt neg vpt V closepath fill } def
/Pent { stroke [] 0 setdash 2 copy gsave
  translate 0 hpt M 4 {72 rotate 0 hpt L} repeat
  closepath stroke grestore Pnt } def
/PentF { stroke [] 0 setdash gsave
  translate 0 hpt M 4 {72 rotate 0 hpt L} repeat
  closepath fill grestore } def
/Circle { stroke [] 0 setdash 2 copy
  hpt 0 360 arc stroke Pnt } def
/CircleF { stroke [] 0 setdash hpt 0 360 arc fill } def
/C0 { BL [] 0 setdash 2 copy moveto vpt 90 450  arc } bind def
/C1 { BL [] 0 setdash 2 copy        moveto
       2 copy  vpt 0 90 arc closepath fill
               vpt 0 360 arc closepath } bind def
/C2 { BL [] 0 setdash 2 copy moveto
       2 copy  vpt 90 180 arc closepath fill
               vpt 0 360 arc closepath } bind def
/C3 { BL [] 0 setdash 2 copy moveto
       2 copy  vpt 0 180 arc closepath fill
               vpt 0 360 arc closepath } bind def
/C4 { BL [] 0 setdash 2 copy moveto
       2 copy  vpt 180 270 arc closepath fill
               vpt 0 360 arc closepath } bind def
/C5 { BL [] 0 setdash 2 copy moveto
       2 copy  vpt 0 90 arc
       2 copy moveto
       2 copy  vpt 180 270 arc closepath fill
               vpt 0 360 arc } bind def
/C6 { BL [] 0 setdash 2 copy moveto
      2 copy  vpt 90 270 arc closepath fill
              vpt 0 360 arc closepath } bind def
/C7 { BL [] 0 setdash 2 copy moveto
      2 copy  vpt 0 270 arc closepath fill
              vpt 0 360 arc closepath } bind def
/C8 { BL [] 0 setdash 2 copy moveto
      2 copy vpt 270 360 arc closepath fill
              vpt 0 360 arc closepath } bind def
/C9 { BL [] 0 setdash 2 copy moveto
      2 copy  vpt 270 450 arc closepath fill
              vpt 0 360 arc closepath } bind def
/C10 { BL [] 0 setdash 2 copy 2 copy moveto vpt 270 360 arc closepath fill
       2 copy moveto
       2 copy vpt 90 180 arc closepath fill
               vpt 0 360 arc closepath } bind def
/C11 { BL [] 0 setdash 2 copy moveto
       2 copy  vpt 0 180 arc closepath fill
       2 copy moveto
       2 copy  vpt 270 360 arc closepath fill
               vpt 0 360 arc closepath } bind def
/C12 { BL [] 0 setdash 2 copy moveto
       2 copy  vpt 180 360 arc closepath fill
               vpt 0 360 arc closepath } bind def
/C13 { BL [] 0 setdash  2 copy moveto
       2 copy  vpt 0 90 arc closepath fill
       2 copy moveto
       2 copy  vpt 180 360 arc closepath fill
               vpt 0 360 arc closepath } bind def
/C14 { BL [] 0 setdash 2 copy moveto
       2 copy  vpt 90 360 arc closepath fill
               vpt 0 360 arc } bind def
/C15 { BL [] 0 setdash 2 copy vpt 0 360 arc closepath fill
               vpt 0 360 arc closepath } bind def
/Rec   { newpath 4 2 roll moveto 1 index 0 rlineto 0 exch rlineto
       neg 0 rlineto closepath } bind def
/Square { dup Rec } bind def
/Bsquare { vpt sub exch vpt sub exch vpt2 Square } bind def
/S0 { BL [] 0 setdash 2 copy moveto 0 vpt rlineto BL Bsquare } bind def
/S1 { BL [] 0 setdash 2 copy vpt Square fill Bsquare } bind def
/S2 { BL [] 0 setdash 2 copy exch vpt sub exch vpt Square fill Bsquare } bind def
/S3 { BL [] 0 setdash 2 copy exch vpt sub exch vpt2 vpt Rec fill Bsquare } bind def
/S4 { BL [] 0 setdash 2 copy exch vpt sub exch vpt sub vpt Square fill Bsquare } bind def
/S5 { BL [] 0 setdash 2 copy 2 copy vpt Square fill
       exch vpt sub exch vpt sub vpt Square fill Bsquare } bind def
/S6 { BL [] 0 setdash 2 copy exch vpt sub exch vpt sub vpt vpt2 Rec fill Bsquare } bind def
/S7 { BL [] 0 setdash 2 copy exch vpt sub exch vpt sub vpt vpt2 Rec fill
       2 copy vpt Square fill
       Bsquare } bind def
/S8 { BL [] 0 setdash 2 copy vpt sub vpt Square fill Bsquare } bind def
/S9 { BL [] 0 setdash 2 copy vpt sub vpt vpt2 Rec fill Bsquare } bind def
/S10 { BL [] 0 setdash 2 copy vpt sub vpt Square fill 2 copy exch vpt sub exch vpt Square fill
       Bsquare } bind def
/S11 { BL [] 0 setdash 2 copy vpt sub vpt Square fill 2 copy exch vpt sub exch vpt2 vpt Rec fill
       Bsquare } bind def
/S12 { BL [] 0 setdash 2 copy exch vpt sub exch vpt sub vpt2 vpt Rec fill Bsquare } bind def
/S13 { BL [] 0 setdash 2 copy exch vpt sub exch vpt sub vpt2 vpt Rec fill
       2 copy vpt Square fill Bsquare } bind def
/S14 { BL [] 0 setdash 2 copy exch vpt sub exch vpt sub vpt2 vpt Rec fill
       2 copy exch vpt sub exch vpt Square fill Bsquare } bind def
/S15 { BL [] 0 setdash 2 copy Bsquare fill Bsquare } bind def
/D0 { gsave translate 45 rotate 0 0 S0 stroke grestore } bind def
/D1 { gsave translate 45 rotate 0 0 S1 stroke grestore } bind def
/D2 { gsave translate 45 rotate 0 0 S2 stroke grestore } bind def
/D3 { gsave translate 45 rotate 0 0 S3 stroke grestore } bind def
/D4 { gsave translate 45 rotate 0 0 S4 stroke grestore } bind def
/D5 { gsave translate 45 rotate 0 0 S5 stroke grestore } bind def
/D6 { gsave translate 45 rotate 0 0 S6 stroke grestore } bind def
/D7 { gsave translate 45 rotate 0 0 S7 stroke grestore } bind def
/D8 { gsave translate 45 rotate 0 0 S8 stroke grestore } bind def
/D9 { gsave translate 45 rotate 0 0 S9 stroke grestore } bind def
/D10 { gsave translate 45 rotate 0 0 S10 stroke grestore } bind def
/D11 { gsave translate 45 rotate 0 0 S11 stroke grestore } bind def
/D12 { gsave translate 45 rotate 0 0 S12 stroke grestore } bind def
/D13 { gsave translate 45 rotate 0 0 S13 stroke grestore } bind def
/D14 { gsave translate 45 rotate 0 0 S14 stroke grestore } bind def
/D15 { gsave translate 45 rotate 0 0 S15 stroke grestore } bind def
/DiaE { stroke [] 0 setdash vpt add M
  hpt neg vpt neg V hpt vpt neg V
  hpt vpt V hpt neg vpt V closepath stroke } def
/BoxE { stroke [] 0 setdash exch hpt sub exch vpt add M
  0 vpt2 neg V hpt2 0 V 0 vpt2 V
  hpt2 neg 0 V closepath stroke } def
/TriUE { stroke [] 0 setdash vpt 1.12 mul add M
  hpt neg vpt -1.62 mul V
  hpt 2 mul 0 V
  hpt neg vpt 1.62 mul V closepath stroke } def
/TriDE { stroke [] 0 setdash vpt 1.12 mul sub M
  hpt neg vpt 1.62 mul V
  hpt 2 mul 0 V
  hpt neg vpt -1.62 mul V closepath stroke } def
/PentE { stroke [] 0 setdash gsave
  translate 0 hpt M 4 {72 rotate 0 hpt L} repeat
  closepath stroke grestore } def
/CircE { stroke [] 0 setdash 
  hpt 0 360 arc stroke } def
/Opaque { gsave closepath 1 setgray fill grestore 0 setgray closepath } def
/DiaW { stroke [] 0 setdash vpt add M
  hpt neg vpt neg V hpt vpt neg V
  hpt vpt V hpt neg vpt V Opaque stroke } def
/BoxW { stroke [] 0 setdash exch hpt sub exch vpt add M
  0 vpt2 neg V hpt2 0 V 0 vpt2 V
  hpt2 neg 0 V Opaque stroke } def
/TriUW { stroke [] 0 setdash vpt 1.12 mul add M
  hpt neg vpt -1.62 mul V
  hpt 2 mul 0 V
  hpt neg vpt 1.62 mul V Opaque stroke } def
/TriDW { stroke [] 0 setdash vpt 1.12 mul sub M
  hpt neg vpt 1.62 mul V
  hpt 2 mul 0 V
  hpt neg vpt -1.62 mul V Opaque stroke } def
/PentW { stroke [] 0 setdash gsave
  translate 0 hpt M 4 {72 rotate 0 hpt L} repeat
  Opaque stroke grestore } def
/CircW { stroke [] 0 setdash 
  hpt 0 360 arc Opaque stroke } def
/BoxFill { gsave Rec 1 setgray fill grestore } def
/Symbol-Oblique /Symbol findfont [1 0 .167 1 0 0] makefont
dup length dict begin {1 index /FID eq {pop pop} {def} ifelse} forall
currentdict end definefont pop
end
}}%
\begin{picture}(2520,2160)(0,0)%
{\GNUPLOTspecial{"
gnudict begin
gsave
0 0 translate
0.100 0.100 scale
0 setgray
newpath
1.000 UL
LTb
250 466 M
63 0 V
2057 0 R
-63 0 V
250 731 M
63 0 V
2057 0 R
-63 0 V
250 997 M
63 0 V
2057 0 R
-63 0 V
250 1263 M
63 0 V
2057 0 R
-63 0 V
250 1529 M
63 0 V
2057 0 R
-63 0 V
250 1794 M
63 0 V
2057 0 R
-63 0 V
250 2060 M
63 0 V
2057 0 R
-63 0 V
250 200 M
0 63 V
0 1797 R
0 -63 V
674 200 M
0 63 V
0 1797 R
0 -63 V
1098 200 M
0 63 V
0 1797 R
0 -63 V
1522 200 M
0 63 V
0 1797 R
0 -63 V
1946 200 M
0 63 V
0 1797 R
0 -63 V
2370 200 M
0 63 V
0 1797 R
0 -63 V
1.000 UL
LTb
250 200 M
2120 0 V
0 1860 V
-2120 0 V
250 200 L
1.000 UL
LT4
250 997 M
2120 0 V
1.000 UL
LT3
2023 1947 M
247 0 V
250 1764 M
2 0 V
2 0 V
2 0 V
2 0 V
3 0 V
2 0 V
2 0 V
2 0 V
2 0 V
2 0 V
2 0 V
2 0 V
3 0 V
2 -1 V
2 0 V
2 0 V
2 0 V
2 0 V
2 0 V
2 0 V
3 0 V
2 0 V
2 0 V
2 0 V
2 0 V
2 0 V
2 -1 V
2 0 V
2 0 V
3 0 V
2 0 V
2 0 V
2 0 V
2 0 V
2 0 V
2 -1 V
2 0 V
3 0 V
2 0 V
2 0 V
2 0 V
2 0 V
2 -1 V
2 0 V
2 0 V
3 0 V
2 0 V
2 0 V
2 -1 V
2 0 V
2 0 V
2 0 V
2 0 V
2 -1 V
3 0 V
2 0 V
2 0 V
2 0 V
2 -1 V
2 0 V
2 0 V
2 0 V
3 -1 V
2 0 V
2 0 V
2 0 V
2 -1 V
2 0 V
2 0 V
2 0 V
3 -1 V
2 0 V
2 0 V
2 0 V
2 -1 V
2 0 V
2 0 V
2 -1 V
2 0 V
3 0 V
2 0 V
2 -1 V
2 0 V
2 0 V
2 -1 V
2 0 V
2 0 V
3 -1 V
2 0 V
2 0 V
2 -1 V
2 0 V
2 0 V
2 -1 V
2 0 V
3 -1 V
2 0 V
2 0 V
2 -1 V
2 0 V
2 0 V
2 -1 V
2 0 V
2 -1 V
3 0 V
2 0 V
2 -1 V
2 0 V
2 -1 V
2 0 V
2 -1 V
2 0 V
3 0 V
2 -1 V
2 0 V
2 -1 V
2 0 V
2 -1 V
2 0 V
2 -1 V
3 0 V
2 -1 V
2 0 V
2 -1 V
2 0 V
2 -1 V
2 0 V
2 -1 V
2 0 V
3 -1 V
2 0 V
2 -1 V
2 0 V
2 -1 V
2 0 V
2 -1 V
2 0 V
3 -1 V
2 0 V
2 -1 V
2 -1 V
2 0 V
2 -1 V
2 0 V
2 -1 V
3 -1 V
2 0 V
2 -1 V
2 0 V
2 -1 V
2 -1 V
2 0 V
2 -1 V
2 0 V
3 -1 V
2 -1 V
2 0 V
2 -1 V
2 -1 V
2 0 V
2 -1 V
2 -1 V
3 0 V
2 -1 V
2 -1 V
2 0 V
2 -1 V
2 -1 V
2 -1 V
2 0 V
3 -1 V
2 -1 V
2 0 V
2 -1 V
2 -1 V
2 -1 V
2 0 V
2 -1 V
2 -1 V
3 -1 V
2 0 V
2 -1 V
2 -1 V
2 -1 V
2 -1 V
2 0 V
2 -1 V
3 -1 V
2 -1 V
2 -1 V
2 -1 V
2 0 V
2 -1 V
2 -1 V
2 -1 V
3 -1 V
2 -1 V
2 -1 V
2 0 V
2 -1 V
2 -1 V
2 -1 V
2 -1 V
2 -1 V
3 -1 V
2 -1 V
2 -1 V
2 -1 V
2 -1 V
2 -1 V
2 -1 V
2 -1 V
3 -1 V
2 -1 V
2 0 V
2 -1 V
2 -1 V
2 -2 V
2 -1 V
2 -1 V
3 -1 V
2 -1 V
2 -1 V
2 -1 V
2 -1 V
2 -1 V
2 -1 V
2 -1 V
2 -1 V
3 -1 V
2 -1 V
2 -1 V
2 -1 V
2 -2 V
2 -1 V
2 -1 V
2 -1 V
3 -1 V
2 -1 V
2 -2 V
2 -1 V
2 -1 V
2 -1 V
2 -1 V
2 -1 V
3 -2 V
2 -1 V
2 -1 V
2 -1 V
2 -2 V
2 -1 V
2 -1 V
2 -1 V
2 -2 V
3 -1 V
2 -1 V
2 -2 V
2 -1 V
2 -1 V
2 -2 V
2 -1 V
2 -1 V
3 -2 V
2 -1 V
2 -2 V
2 -1 V
2 -1 V
2 -2 V
2 -1 V
2 -2 V
3 -1 V
2 -1 V
2 -2 V
2 -1 V
2 -2 V
2 -1 V
2 -2 V
2 -1 V
2 -2 V
3 -1 V
2 -2 V
2 -2 V
2 -1 V
2 -2 V
2 -1 V
2 -2 V
2 -2 V
3 -1 V
2 -2 V
2 -1 V
2 -2 V
2 -2 V
2 -1 V
2 -2 V
2 -2 V
3 -2 V
2 -1 V
2 -2 V
2 -2 V
2 -2 V
2 -1 V
2 -2 V
2 -2 V
2 -2 V
3 -2 V
2 -1 V
2 -2 V
2 -2 V
2 -2 V
2 -2 V
2 -2 V
2 -2 V
3 -2 V
2 -2 V
2 -2 V
2 -2 V
2 -2 V
2 -2 V
2 -2 V
2 -2 V
3 -2 V
2 -2 V
2 -2 V
2 -2 V
2 -2 V
2 -2 V
2 -3 V
2 -2 V
2 -2 V
3 -2 V
2 -2 V
2 -3 V
2 -2 V
2 -2 V
2 -3 V
2 -2 V
2 -2 V
3 -3 V
2 -2 V
2 -2 V
2 -3 V
2 -2 V
2 -3 V
2 -2 V
2 -3 V
3 -2 V
2 -3 V
2 -3 V
2 -2 V
2 -3 V
2 -2 V
2 -3 V
2 -3 V
2 -3 V
3 -2 V
2 -3 V
2 -3 V
2 -3 V
2 -3 V
2 -3 V
2 -3 V
2 -3 V
3 -3 V
2 -3 V
2 -3 V
2 -3 V
2 -3 V
2 -3 V
2 -3 V
2 -4 V
3 -3 V
2 -3 V
2 -4 V
2 -3 V
2 -3 V
2 -4 V
2 -4 V
2 -3 V
2 -4 V
3 -4 V
2 -3 V
2 -4 V
2 -4 V
2 -4 V
2 -4 V
2 -4 V
2 -4 V
3 -5 V
2 -4 V
2 -5 V
2 -4 V
2 -5 V
2 -5 V
2 -5 V
2 -5 V
3 -6 V
2 -5 V
currentpoint stroke M
2 -6 V
2 -7 V
2 -8 V
2 -8 V
2 -7 V
2 -6 V
2 -6 V
3 -5 V
2 -6 V
2 -5 V
2 -5 V
2 -5 V
2 -5 V
2 -5 V
2 -4 V
3 -5 V
2 -4 V
2 -5 V
2 -4 V
2 -4 V
2 -4 V
2 -5 V
2 -4 V
3 -4 V
2 -4 V
2 -4 V
2 -4 V
2 -4 V
2 -4 V
2 -4 V
2 -3 V
2 -4 V
3 -4 V
2 -4 V
2 -3 V
2 -4 V
2 -4 V
2 -3 V
2 -4 V
2 -3 V
3 -4 V
2 -3 V
2 -4 V
2 -3 V
2 -4 V
2 -3 V
2 -4 V
2 -3 V
3 -3 V
2 -4 V
2 -3 V
2 -3 V
2 -4 V
2 -3 V
2 -3 V
2 -3 V
2 -4 V
3 -3 V
2 -3 V
2 -3 V
2 -3 V
2 -4 V
2 -3 V
2 -3 V
2 -3 V
3 -3 V
2 -3 V
2 -3 V
2 -3 V
2 -3 V
2 -3 V
2 -3 V
2 -3 V
3 -3 V
2 -3 V
2 -3 V
2 -3 V
2 -3 V
2 -3 V
2 -3 V
2 -3 V
2 -3 V
3 -3 V
2 -2 V
2 -3 V
2 -3 V
2 -3 V
2 -3 V
2 -3 V
2 -2 V
3 -3 V
2 -3 V
2 -3 V
2 -3 V
2 -2 V
2 -3 V
2 -3 V
2 -3 V
3 -2 V
2 -3 V
2 -3 V
2 -2 V
2 -3 V
2 -3 V
2 -2 V
2 -3 V
2 -3 V
3 -2 V
2 -3 V
2 -3 V
2 -2 V
2 -3 V
2 -2 V
2 -3 V
2 -3 V
3 -2 V
2 -3 V
2 -2 V
2 -3 V
2 -2 V
2 -3 V
2 -2 V
2 -3 V
3 -2 V
2 -3 V
2 -2 V
2 -3 V
2 -2 V
2 -3 V
2 -2 V
2 -3 V
2 -2 V
3 -3 V
2 -2 V
2 -3 V
2 -2 V
2 -2 V
2 -3 V
2 -2 V
2 -3 V
3 -2 V
2 -2 V
2 -3 V
2 -2 V
2 -3 V
2 -2 V
2 -2 V
2 -3 V
3 -2 V
2 -2 V
2 -3 V
2 -2 V
2 -2 V
2 -2 V
2 -3 V
2 -2 V
2 -2 V
3 -3 V
2 -2 V
2 -2 V
2 -2 V
2 -3 V
2 -2 V
2 -2 V
2 -2 V
3 -3 V
2 -2 V
2 -2 V
2 -2 V
2 -2 V
2 -3 V
2 -2 V
2 -2 V
3 -2 V
2 -2 V
2 -2 V
2 -3 V
2 -2 V
2 -2 V
2 -2 V
2 -2 V
2 -2 V
3 -2 V
2 -3 V
2 -2 V
2 -2 V
2 -2 V
2 -2 V
2 -2 V
2 -2 V
3 -2 V
2 -2 V
2 -2 V
2 -2 V
2 -2 V
2 -2 V
2 -3 V
2 -2 V
3 -2 V
2 -2 V
2 -2 V
2 -2 V
2 -2 V
2 -2 V
2 -2 V
2 -2 V
2 -2 V
3 -2 V
2 -2 V
2 -2 V
2 -1 V
2 -2 V
2 -2 V
2 -2 V
2 -2 V
3 -2 V
2 -2 V
2 -2 V
2 -2 V
2 -2 V
2 -2 V
2 -2 V
2 -2 V
3 -1 V
2 -2 V
2 -2 V
2 -2 V
2 -2 V
2 -2 V
2 -2 V
2 -2 V
2 -1 V
3 -2 V
2 -2 V
2 -2 V
2 -2 V
2 -2 V
2 -1 V
2 -2 V
2 -2 V
3 -2 V
2 -2 V
2 -1 V
2 -2 V
2 -2 V
2 -2 V
2 -1 V
2 -2 V
3 -2 V
2 -2 V
2 -1 V
2 -2 V
2 -2 V
2 -2 V
2 -1 V
2 -2 V
2 -2 V
3 -2 V
2 -1 V
2 -2 V
2 -2 V
2 -1 V
2 -2 V
2 -2 V
2 -1 V
3 -2 V
2 -2 V
2 -1 V
2 -2 V
2 -2 V
2 -1 V
2 -2 V
2 -2 V
3 -1 V
2 -2 V
2 -1 V
2 -2 V
2 -2 V
2 -1 V
2 -2 V
2 -1 V
2 -2 V
3 -2 V
2 -1 V
2 -2 V
2 -1 V
2 -2 V
2 -1 V
2 -2 V
2 -2 V
3 -1 V
2 -2 V
2 -1 V
2 -2 V
2 -1 V
2 -2 V
2 -1 V
2 -2 V
3 -1 V
2 -2 V
2 -1 V
2 -2 V
2 -1 V
2 -2 V
2 -1 V
2 -2 V
2 -1 V
3 -2 V
2 -1 V
2 -1 V
2 -2 V
2 -1 V
2 -2 V
2 -1 V
2 -2 V
3 -1 V
2 -1 V
2 -2 V
2 -1 V
2 -2 V
2 -1 V
2 -1 V
2 -2 V
3 -1 V
2 -2 V
2 -1 V
2 -1 V
2 -2 V
2 -1 V
2 -1 V
2 -2 V
2 -1 V
3 -1 V
2 -2 V
2 -1 V
2 -1 V
2 -2 V
2 -1 V
2 -1 V
2 -1 V
3 -2 V
2 -1 V
2 -1 V
2 -2 V
2 -1 V
2 -1 V
2 -1 V
2 -2 V
3 -1 V
2 -1 V
2 -1 V
2 -2 V
2 -1 V
2 -1 V
2 -1 V
2 -2 V
2 -1 V
3 -1 V
2 -1 V
2 -1 V
2 -2 V
2 -1 V
2 -1 V
2 -1 V
2 -1 V
3 -1 V
2 -2 V
2 -1 V
2 -1 V
2 -1 V
2 -1 V
2 -1 V
2 -1 V
3 -2 V
2 -1 V
2 -1 V
2 -1 V
2 -1 V
2 -1 V
2 -1 V
2 -1 V
2 -1 V
3 -1 V
2 -2 V
2 -1 V
2 -1 V
2 -1 V
2 -1 V
2 -1 V
2 -1 V
3 -1 V
2 -1 V
2 -1 V
2 -1 V
2 -1 V
2 -1 V
2 -1 V
2 -1 V
3 -1 V
2 -1 V
currentpoint stroke M
2 -1 V
2 -1 V
2 -1 V
2 -1 V
2 -1 V
2 -1 V
2 -1 V
3 -1 V
2 -1 V
2 -1 V
2 -1 V
2 -1 V
2 -1 V
2 -1 V
2 -1 V
3 0 V
2 -1 V
2 -1 V
2 -1 V
2 -1 V
2 -1 V
2 -1 V
2 -1 V
3 -1 V
2 -1 V
2 0 V
2 -1 V
2 -1 V
2 -1 V
2 -1 V
2 -1 V
2 -1 V
3 0 V
2 -1 V
2 -1 V
2 -1 V
2 -1 V
2 -1 V
2 0 V
2 -1 V
3 -1 V
2 -1 V
2 -1 V
2 0 V
2 -1 V
2 -1 V
2 -1 V
2 0 V
3 -1 V
2 -1 V
2 -1 V
2 0 V
2 -1 V
2 -1 V
2 -1 V
2 0 V
2 -1 V
3 -1 V
2 0 V
2 -1 V
2 -1 V
2 -1 V
2 0 V
2 -1 V
2 -1 V
3 0 V
2 -1 V
2 -1 V
2 0 V
2 -1 V
2 -1 V
2 0 V
2 -1 V
3 -1 V
2 0 V
2 -1 V
2 0 V
2 -1 V
2 -1 V
2 0 V
2 -1 V
2 0 V
3 -1 V
2 -1 V
2 0 V
2 -1 V
2 0 V
2 -1 V
2 0 V
2 -1 V
3 -1 V
2 0 V
2 -1 V
2 0 V
2 -1 V
2 0 V
2 -1 V
2 0 V
3 -1 V
2 0 V
2 -1 V
2 0 V
2 -1 V
2 0 V
2 -1 V
2 0 V
2 -1 V
3 0 V
2 -1 V
2 0 V
2 -1 V
2 0 V
2 -1 V
2 0 V
2 0 V
3 -1 V
2 0 V
2 -1 V
2 0 V
2 0 V
2 -1 V
2 0 V
2 -1 V
3 0 V
2 0 V
2 -1 V
2 0 V
2 -1 V
2 0 V
2 0 V
2 -1 V
2 0 V
3 0 V
2 -1 V
2 0 V
2 0 V
2 -1 V
2 0 V
2 0 V
2 -1 V
3 0 V
2 0 V
2 -1 V
2 0 V
2 0 V
2 0 V
2 -1 V
2 0 V
3 0 V
2 -1 V
2 0 V
2 0 V
2 0 V
2 -1 V
2 0 V
2 0 V
2 0 V
3 0 V
2 -1 V
2 0 V
2 0 V
2 0 V
2 0 V
2 -1 V
2 0 V
3 0 V
2 0 V
2 0 V
2 -1 V
2 0 V
2 0 V
2 0 V
2 0 V
3 0 V
2 0 V
2 -1 V
2 0 V
2 0 V
2 0 V
2 0 V
2 0 V
2 0 V
3 0 V
2 -1 V
2 0 V
2 0 V
2 0 V
2 0 V
2 0 V
2 0 V
3 0 V
2 0 V
2 0 V
2 0 V
2 0 V
2 0 V
2 0 V
2 0 V
3 0 V
2 0 V
2 0 V
2 0 V
2 0 V
1.000 UL
LT2
2023 1847 M
247 0 V
250 1738 M
2 0 V
2 0 V
2 0 V
2 0 V
3 0 V
2 0 V
2 0 V
2 0 V
2 0 V
2 0 V
2 0 V
2 -1 V
3 0 V
2 0 V
2 0 V
2 0 V
2 0 V
2 0 V
2 0 V
2 -1 V
3 0 V
2 0 V
2 0 V
2 0 V
2 0 V
2 -1 V
2 0 V
2 0 V
2 0 V
3 0 V
2 -1 V
2 0 V
2 0 V
2 0 V
2 -1 V
2 0 V
2 0 V
3 -1 V
2 0 V
2 0 V
2 0 V
2 -1 V
2 0 V
2 0 V
2 -1 V
3 0 V
2 0 V
2 -1 V
2 0 V
2 -1 V
2 0 V
2 0 V
2 -1 V
2 0 V
3 -1 V
2 0 V
2 -1 V
2 0 V
2 0 V
2 -1 V
2 0 V
2 -1 V
3 0 V
2 -1 V
2 0 V
2 -1 V
2 0 V
2 -1 V
2 0 V
2 -1 V
3 -1 V
2 0 V
2 -1 V
2 0 V
2 -1 V
2 -1 V
2 0 V
2 -1 V
2 0 V
3 -1 V
2 -1 V
2 0 V
2 -1 V
2 -1 V
2 0 V
2 -1 V
2 -1 V
3 0 V
2 -1 V
2 -1 V
2 -1 V
2 0 V
2 -1 V
2 -1 V
2 -1 V
3 -1 V
2 0 V
2 -1 V
2 -1 V
2 -1 V
2 -1 V
2 0 V
2 -1 V
2 -1 V
3 -1 V
2 -1 V
2 -1 V
2 -1 V
2 -1 V
2 -1 V
2 -1 V
2 -1 V
3 -1 V
2 -1 V
2 0 V
2 -1 V
2 -2 V
2 -1 V
2 -1 V
2 -1 V
3 -1 V
2 -1 V
2 -1 V
2 -1 V
2 -1 V
2 -1 V
2 -1 V
2 -1 V
2 -2 V
3 -1 V
2 -1 V
2 -1 V
2 -1 V
2 -2 V
2 -1 V
2 -1 V
2 -1 V
3 -2 V
2 -1 V
2 -1 V
2 -1 V
2 -2 V
2 -1 V
2 -2 V
2 -1 V
3 -1 V
2 -2 V
2 -1 V
2 -2 V
2 -1 V
2 -2 V
2 -1 V
2 -2 V
2 -1 V
3 -2 V
2 -1 V
2 -2 V
2 -1 V
2 -2 V
2 -2 V
2 -1 V
2 -2 V
3 -2 V
2 -1 V
2 -2 V
2 -2 V
2 -2 V
2 -1 V
2 -2 V
2 -2 V
3 -2 V
2 -2 V
2 -2 V
2 -2 V
2 -2 V
2 -2 V
2 -2 V
2 -2 V
2 -2 V
3 -2 V
2 -2 V
2 -3 V
2 -2 V
2 -2 V
2 -2 V
2 -3 V
2 -2 V
3 -3 V
2 -2 V
2 -3 V
2 -2 V
2 -3 V
2 -3 V
2 -3 V
2 -3 V
3 -3 V
2 -3 V
2 -3 V
2 -4 V
2 -5 V
2 -4 V
2 -4 V
2 -3 V
2 -4 V
3 -3 V
2 -3 V
2 -3 V
2 -3 V
2 -3 V
2 -3 V
2 -3 V
2 -3 V
3 -2 V
2 -3 V
2 -3 V
2 -2 V
2 -3 V
2 -3 V
2 -2 V
2 -3 V
3 -2 V
2 -3 V
2 -3 V
2 -2 V
2 -3 V
2 -2 V
2 -3 V
2 -2 V
2 -2 V
3 -3 V
2 -2 V
2 -3 V
2 -2 V
2 -3 V
2 -2 V
2 -2 V
2 -3 V
3 -2 V
2 -2 V
2 -3 V
2 -2 V
2 -2 V
2 -3 V
2 -2 V
2 -2 V
3 -3 V
2 -2 V
2 -2 V
2 -3 V
2 -2 V
2 -2 V
2 -3 V
2 -2 V
2 -2 V
3 -2 V
2 -3 V
2 -2 V
2 -2 V
2 -2 V
2 -3 V
2 -2 V
2 -2 V
3 -2 V
2 -3 V
2 -2 V
2 -2 V
2 -2 V
2 -2 V
2 -3 V
2 -2 V
3 -2 V
2 -2 V
2 -2 V
2 -3 V
2 -2 V
2 -2 V
2 -2 V
2 -2 V
2 -2 V
3 -3 V
2 -2 V
2 -2 V
2 -2 V
2 -2 V
2 -2 V
2 -3 V
2 -2 V
3 -2 V
2 -2 V
2 -2 V
2 -2 V
2 -3 V
2 -2 V
2 -2 V
2 -2 V
3 -2 V
2 -2 V
2 -2 V
2 -2 V
2 -3 V
2 -2 V
2 -2 V
2 -2 V
2 -2 V
3 -2 V
2 -2 V
2 -2 V
2 -2 V
2 -3 V
2 -2 V
2 -2 V
2 -2 V
3 -2 V
2 -2 V
2 -2 V
2 -2 V
2 -2 V
2 -2 V
2 -3 V
2 -2 V
3 -2 V
2 -2 V
2 -2 V
2 -2 V
2 -2 V
2 -2 V
2 -2 V
2 -2 V
2 -2 V
3 -2 V
2 -2 V
2 -2 V
2 -2 V
2 -3 V
2 -2 V
2 -2 V
2 -2 V
3 -2 V
2 -2 V
2 -2 V
2 -2 V
2 -2 V
2 -2 V
2 -2 V
2 -2 V
3 -2 V
2 -2 V
2 -2 V
2 -2 V
2 -2 V
2 -2 V
2 -2 V
2 -2 V
2 -2 V
3 -2 V
2 -2 V
2 -2 V
2 -2 V
2 -2 V
2 -2 V
2 -2 V
2 -2 V
3 -2 V
2 -2 V
2 -2 V
2 -2 V
2 -2 V
2 -2 V
2 -2 V
2 -2 V
3 -2 V
2 -2 V
2 -2 V
2 -2 V
2 -2 V
2 -2 V
2 -2 V
2 -2 V
2 -2 V
3 -2 V
2 -2 V
2 -2 V
2 -2 V
2 -2 V
2 -2 V
2 -2 V
2 -2 V
3 -2 V
2 -2 V
2 -2 V
2 -2 V
2 -2 V
2 -2 V
2 -1 V
2 -2 V
3 -2 V
2 -2 V
currentpoint stroke M
2 -2 V
2 -2 V
2 -2 V
2 -2 V
2 -2 V
2 -2 V
2 -2 V
3 -2 V
2 -2 V
2 -2 V
2 -1 V
2 -2 V
2 -2 V
2 -2 V
2 -2 V
3 -2 V
2 -2 V
2 -2 V
2 -2 V
2 -2 V
2 -2 V
2 -1 V
2 -2 V
3 -2 V
2 -2 V
2 -2 V
2 -2 V
2 -2 V
2 -2 V
2 -1 V
2 -2 V
2 -2 V
3 -2 V
2 -2 V
2 -2 V
2 -2 V
2 -2 V
2 -1 V
2 -2 V
2 -2 V
3 -2 V
2 -2 V
2 -2 V
2 -2 V
2 -1 V
2 -2 V
2 -2 V
2 -2 V
3 -2 V
2 -2 V
2 -1 V
2 -2 V
2 -2 V
2 -2 V
2 -2 V
2 -2 V
2 -1 V
3 -2 V
2 -2 V
2 -2 V
2 -2 V
2 -1 V
2 -2 V
2 -2 V
2 -2 V
3 -2 V
2 -1 V
2 -2 V
2 -2 V
2 -2 V
2 -2 V
2 -1 V
2 -2 V
3 -2 V
2 -2 V
2 -2 V
2 -1 V
2 -2 V
2 -2 V
2 -2 V
2 -1 V
2 -2 V
3 -2 V
2 -2 V
2 -1 V
2 -2 V
2 -2 V
2 -2 V
2 -1 V
2 -2 V
3 -2 V
2 -2 V
2 -1 V
2 -2 V
2 -2 V
2 -2 V
2 -1 V
2 -2 V
3 -2 V
2 -2 V
2 -1 V
2 -2 V
2 -2 V
2 -1 V
2 -2 V
2 -2 V
2 -2 V
3 -1 V
2 -2 V
2 -2 V
2 -1 V
2 -2 V
2 -2 V
2 -1 V
2 -2 V
3 -2 V
2 -1 V
2 -2 V
2 -2 V
2 -1 V
2 -2 V
2 -2 V
2 -1 V
3 -2 V
2 -2 V
2 -1 V
2 -2 V
2 -2 V
2 -1 V
2 -2 V
2 -2 V
2 -1 V
3 -2 V
2 -2 V
2 -1 V
2 -2 V
2 -1 V
2 -2 V
2 -2 V
2 -1 V
3 -2 V
2 -2 V
2 -1 V
2 -2 V
2 -1 V
2 -2 V
2 -2 V
2 -1 V
3 -2 V
2 -1 V
2 -2 V
2 -2 V
2 -1 V
2 -2 V
2 -1 V
2 -2 V
2 -1 V
3 -2 V
2 -2 V
2 -1 V
2 -2 V
2 -1 V
2 -2 V
2 -1 V
2 -2 V
3 -1 V
2 -2 V
2 -2 V
2 -1 V
2 -2 V
2 -1 V
2 -2 V
2 -1 V
3 -2 V
2 -1 V
2 -2 V
2 -1 V
2 -2 V
2 -1 V
2 -2 V
2 -1 V
2 -2 V
3 -1 V
2 -2 V
2 -1 V
2 -2 V
2 -1 V
2 -2 V
2 -1 V
2 -2 V
3 -1 V
2 -2 V
2 -1 V
2 -2 V
2 -1 V
2 -2 V
2 -1 V
2 -1 V
3 -2 V
2 -1 V
2 -2 V
2 -1 V
2 -2 V
2 -1 V
2 -2 V
2 -1 V
2 -1 V
3 -2 V
2 -1 V
2 -2 V
2 -1 V
2 -2 V
2 -1 V
2 -1 V
2 -2 V
3 -1 V
2 -2 V
2 -1 V
2 -1 V
2 -2 V
2 -1 V
2 -2 V
2 -1 V
3 -1 V
2 -2 V
2 -1 V
2 -1 V
2 -2 V
2 -1 V
2 -1 V
2 -2 V
2 -1 V
3 -2 V
2 -1 V
2 -1 V
2 -2 V
2 -1 V
2 -1 V
2 -2 V
2 -1 V
3 -1 V
2 -2 V
2 -1 V
2 -1 V
2 -2 V
2 -1 V
2 -1 V
2 -1 V
3 -2 V
2 -1 V
2 -1 V
2 -2 V
2 -1 V
2 -1 V
2 -1 V
2 -2 V
2 -1 V
3 -1 V
2 -2 V
2 -1 V
2 -1 V
2 -1 V
2 -2 V
2 -1 V
2 -1 V
3 -1 V
2 -2 V
2 -1 V
2 -1 V
2 -1 V
2 -2 V
2 -1 V
2 -1 V
3 -1 V
2 -1 V
2 -2 V
2 -1 V
2 -1 V
2 -1 V
2 -1 V
2 -2 V
2 -1 V
3 -1 V
2 -1 V
2 -1 V
2 -2 V
2 -1 V
2 -1 V
2 -1 V
2 -1 V
3 -1 V
2 -2 V
2 -1 V
2 -1 V
2 -1 V
2 -1 V
2 -1 V
2 -1 V
3 -2 V
2 -1 V
2 -1 V
2 -1 V
2 -1 V
2 -1 V
2 -1 V
2 -1 V
2 -2 V
3 -1 V
2 -1 V
2 -1 V
2 -1 V
2 -1 V
2 -1 V
2 -1 V
2 -1 V
3 -1 V
2 -1 V
2 -2 V
2 -1 V
2 -1 V
2 -1 V
2 -1 V
2 -1 V
3 -1 V
2 -1 V
2 -1 V
2 -1 V
2 -1 V
2 -1 V
2 -1 V
2 -1 V
2 -1 V
3 -1 V
2 -1 V
2 -1 V
2 -1 V
2 -1 V
2 -1 V
2 -1 V
2 -1 V
3 -1 V
2 -1 V
2 -1 V
2 -1 V
2 -1 V
2 -1 V
2 -1 V
2 -1 V
3 -1 V
2 -1 V
2 -1 V
2 -1 V
2 -1 V
2 -1 V
2 -1 V
2 -1 V
2 -1 V
3 0 V
2 -1 V
2 -1 V
2 -1 V
2 -1 V
2 -1 V
2 -1 V
2 -1 V
3 -1 V
2 -1 V
2 -1 V
2 0 V
2 -1 V
2 -1 V
2 -1 V
2 -1 V
3 -1 V
2 -1 V
2 -1 V
2 -1 V
2 0 V
2 -1 V
2 -1 V
2 -1 V
2 -1 V
3 -1 V
2 0 V
2 -1 V
2 -1 V
2 -1 V
2 -1 V
2 -1 V
2 0 V
3 -1 V
2 -1 V
2 -1 V
2 -1 V
2 0 V
2 -1 V
2 -1 V
2 -1 V
3 -1 V
2 0 V
currentpoint stroke M
2 -1 V
2 -1 V
2 -1 V
2 0 V
2 -1 V
2 -1 V
2 -1 V
3 0 V
2 -1 V
2 -1 V
2 -1 V
2 0 V
2 -1 V
2 -1 V
2 -1 V
3 0 V
2 -1 V
2 -1 V
2 0 V
2 -1 V
2 -1 V
2 -1 V
2 0 V
3 -1 V
2 -1 V
2 0 V
2 -1 V
2 -1 V
2 0 V
2 -1 V
2 -1 V
2 0 V
3 -1 V
2 -1 V
2 0 V
2 -1 V
2 -1 V
2 0 V
2 -1 V
2 0 V
3 -1 V
2 -1 V
2 0 V
2 -1 V
2 -1 V
2 0 V
2 -1 V
2 0 V
3 -1 V
2 -1 V
2 0 V
2 -1 V
2 0 V
2 -1 V
2 0 V
2 -1 V
2 -1 V
3 0 V
2 -1 V
2 0 V
2 -1 V
2 0 V
2 -1 V
2 0 V
2 -1 V
3 -1 V
2 0 V
2 -1 V
2 0 V
2 -1 V
2 0 V
2 -1 V
2 0 V
3 -1 V
2 0 V
2 -1 V
2 0 V
2 -1 V
2 0 V
2 -1 V
2 0 V
2 -1 V
3 0 V
2 0 V
2 -1 V
2 0 V
2 -1 V
2 0 V
2 -1 V
2 0 V
3 -1 V
2 0 V
2 0 V
2 -1 V
2 0 V
2 -1 V
2 0 V
2 -1 V
3 0 V
2 0 V
2 -1 V
2 0 V
2 -1 V
2 0 V
2 0 V
2 -1 V
2 0 V
3 0 V
2 -1 V
2 0 V
2 -1 V
2 0 V
2 0 V
2 -1 V
2 0 V
3 0 V
2 -1 V
2 0 V
2 0 V
2 -1 V
2 0 V
2 0 V
2 -1 V
3 0 V
2 0 V
2 0 V
2 -1 V
2 0 V
2 0 V
2 -1 V
2 0 V
2 0 V
3 0 V
2 -1 V
2 0 V
2 0 V
2 -1 V
2 0 V
2 0 V
2 0 V
3 -1 V
2 0 V
2 0 V
2 0 V
2 0 V
2 -1 V
2 0 V
2 0 V
3 0 V
2 -1 V
2 0 V
2 0 V
2 0 V
2 0 V
2 -1 V
2 0 V
2 0 V
3 0 V
2 0 V
2 0 V
2 -1 V
2 0 V
2 0 V
2 0 V
2 0 V
3 0 V
2 -1 V
2 0 V
2 0 V
2 0 V
2 0 V
2 0 V
2 0 V
3 0 V
2 -1 V
2 0 V
2 0 V
2 0 V
2 0 V
2 0 V
2 0 V
2 0 V
3 0 V
2 0 V
2 0 V
2 0 V
2 -1 V
2 0 V
2 0 V
2 0 V
3 0 V
2 0 V
2 0 V
2 0 V
2 0 V
2 0 V
2 0 V
2 0 V
3 0 V
2 0 V
2 0 V
2 0 V
2 0 V
1.000 UL
LT1
2023 1747 M
247 0 V
250 1679 M
2 0 V
2 0 V
2 0 V
2 0 V
3 0 V
2 0 V
2 0 V
2 0 V
2 0 V
2 0 V
2 0 V
2 0 V
3 0 V
2 -1 V
2 0 V
2 0 V
2 0 V
2 0 V
2 -1 V
2 0 V
3 0 V
2 0 V
2 -1 V
2 0 V
2 0 V
2 0 V
2 -1 V
2 0 V
2 0 V
3 -1 V
2 0 V
2 0 V
2 -1 V
2 0 V
2 -1 V
2 0 V
2 0 V
3 -1 V
2 0 V
2 -1 V
2 0 V
2 -1 V
2 0 V
2 -1 V
2 0 V
3 -1 V
2 0 V
2 -1 V
2 0 V
2 -1 V
2 -1 V
2 0 V
2 -1 V
2 -1 V
3 0 V
2 -1 V
2 -1 V
2 0 V
2 -1 V
2 -1 V
2 -1 V
2 0 V
3 -1 V
2 -1 V
2 -1 V
2 -1 V
2 0 V
2 -1 V
2 -1 V
2 -1 V
3 -1 V
2 -1 V
2 -1 V
2 -1 V
2 -1 V
2 -1 V
2 -1 V
2 -1 V
2 -1 V
3 -1 V
2 -1 V
2 -2 V
2 -1 V
2 -1 V
2 -1 V
2 -1 V
2 -2 V
3 -1 V
2 -1 V
2 -2 V
2 -1 V
2 -2 V
2 -1 V
2 -2 V
2 -1 V
3 -2 V
2 -2 V
2 -2 V
2 -2 V
2 -2 V
2 -3 V
2 -2 V
2 -2 V
2 -1 V
3 -2 V
2 -2 V
2 -2 V
2 -2 V
2 -1 V
2 -2 V
2 -2 V
2 -1 V
3 -2 V
2 -2 V
2 -1 V
2 -2 V
2 -2 V
2 -1 V
2 -2 V
2 -2 V
3 -1 V
2 -2 V
2 -1 V
2 -2 V
2 -2 V
2 -1 V
2 -2 V
2 -1 V
2 -2 V
3 -2 V
2 -1 V
2 -2 V
2 -1 V
2 -2 V
2 -2 V
2 -1 V
2 -2 V
3 -1 V
2 -2 V
2 -2 V
2 -1 V
2 -2 V
2 -1 V
2 -2 V
2 -2 V
3 -1 V
2 -2 V
2 -1 V
2 -2 V
2 -2 V
2 -1 V
2 -2 V
2 -1 V
2 -2 V
3 -2 V
2 -1 V
2 -2 V
2 -1 V
2 -2 V
2 -2 V
2 -1 V
2 -2 V
3 -2 V
2 -1 V
2 -2 V
2 -1 V
2 -2 V
2 -2 V
2 -1 V
2 -2 V
3 -2 V
2 -1 V
2 -2 V
2 -2 V
2 -1 V
2 -2 V
2 -2 V
2 -1 V
2 -2 V
3 -2 V
2 -1 V
2 -2 V
2 -1 V
2 -2 V
2 -2 V
2 -1 V
2 -2 V
3 -2 V
2 -1 V
2 -2 V
2 -2 V
2 -2 V
2 -1 V
2 -2 V
2 -2 V
3 -1 V
2 -2 V
2 -2 V
2 -1 V
2 -2 V
2 -2 V
2 -1 V
2 -2 V
2 -2 V
3 -1 V
2 -2 V
2 -2 V
2 -2 V
2 -1 V
2 -2 V
2 -2 V
2 -1 V
3 -2 V
2 -2 V
2 -1 V
2 -2 V
2 -2 V
2 -2 V
2 -1 V
2 -2 V
3 -2 V
2 -1 V
2 -2 V
2 -2 V
2 -2 V
2 -1 V
2 -2 V
2 -2 V
2 -2 V
3 -1 V
2 -2 V
2 -2 V
2 -1 V
2 -2 V
2 -2 V
2 -2 V
2 -1 V
3 -2 V
2 -2 V
2 -2 V
2 -1 V
2 -2 V
2 -2 V
2 -2 V
2 -1 V
3 -2 V
2 -2 V
2 -2 V
2 -1 V
2 -2 V
2 -2 V
2 -2 V
2 -1 V
2 -2 V
3 -2 V
2 -2 V
2 -1 V
2 -2 V
2 -2 V
2 -2 V
2 -1 V
2 -2 V
3 -2 V
2 -2 V
2 -1 V
2 -2 V
2 -2 V
2 -2 V
2 -1 V
2 -2 V
3 -2 V
2 -2 V
2 -1 V
2 -2 V
2 -2 V
2 -2 V
2 -1 V
2 -2 V
2 -2 V
3 -2 V
2 -2 V
2 -1 V
2 -2 V
2 -2 V
2 -2 V
2 -1 V
2 -2 V
3 -2 V
2 -2 V
2 -1 V
2 -2 V
2 -2 V
2 -2 V
2 -2 V
2 -1 V
3 -2 V
2 -2 V
2 -2 V
2 -1 V
2 -2 V
2 -2 V
2 -2 V
2 -2 V
2 -1 V
3 -2 V
2 -2 V
2 -2 V
2 -1 V
2 -2 V
2 -2 V
2 -2 V
2 -1 V
3 -2 V
2 -2 V
2 -2 V
2 -2 V
2 -1 V
2 -2 V
2 -2 V
2 -2 V
3 -1 V
2 -2 V
2 -2 V
2 -2 V
2 -2 V
2 -1 V
2 -2 V
2 -2 V
2 -2 V
3 -1 V
2 -2 V
2 -2 V
2 -2 V
2 -1 V
2 -2 V
2 -2 V
2 -2 V
3 -2 V
2 -1 V
2 -2 V
2 -2 V
2 -2 V
2 -1 V
2 -2 V
2 -2 V
3 -2 V
2 -1 V
2 -2 V
2 -2 V
2 -2 V
2 -1 V
2 -2 V
2 -2 V
2 -2 V
3 -2 V
2 -1 V
2 -2 V
2 -2 V
2 -2 V
2 -1 V
2 -2 V
2 -2 V
3 -2 V
2 -1 V
2 -2 V
2 -2 V
2 -2 V
2 -1 V
2 -2 V
2 -2 V
3 -2 V
2 -1 V
2 -2 V
2 -2 V
2 -2 V
2 -1 V
2 -2 V
2 -2 V
2 -2 V
3 -1 V
2 -2 V
2 -2 V
2 -2 V
2 -1 V
2 -2 V
2 -2 V
2 -1 V
3 -2 V
2 -2 V
2 -2 V
2 -1 V
2 -2 V
2 -2 V
2 -2 V
2 -1 V
3 -2 V
2 -2 V
currentpoint stroke M
2 -2 V
2 -1 V
2 -2 V
2 -2 V
2 -1 V
2 -2 V
2 -2 V
3 -2 V
2 -1 V
2 -2 V
2 -2 V
2 -1 V
2 -2 V
2 -2 V
2 -2 V
3 -1 V
2 -2 V
2 -2 V
2 -1 V
2 -2 V
2 -2 V
2 -2 V
2 -1 V
3 -2 V
2 -2 V
2 -1 V
2 -2 V
2 -2 V
2 -1 V
2 -2 V
2 -2 V
2 -1 V
3 -2 V
2 -2 V
2 -2 V
2 -1 V
2 -2 V
2 -2 V
2 -1 V
2 -2 V
3 -2 V
2 -1 V
2 -2 V
2 -2 V
2 -1 V
2 -2 V
2 -2 V
2 -1 V
3 -2 V
2 -2 V
2 -1 V
2 -2 V
2 -2 V
2 -1 V
2 -2 V
2 -2 V
2 -1 V
3 -2 V
2 -2 V
2 -1 V
2 -2 V
2 -2 V
2 -1 V
2 -2 V
2 -1 V
3 -2 V
2 -2 V
2 -1 V
2 -2 V
2 -2 V
2 -1 V
2 -2 V
2 -2 V
3 -1 V
2 -2 V
2 -1 V
2 -2 V
2 -2 V
2 -1 V
2 -2 V
2 -1 V
2 -2 V
3 -2 V
2 -1 V
2 -2 V
2 -2 V
2 -1 V
2 -2 V
2 -1 V
2 -2 V
3 -2 V
2 -1 V
2 -2 V
2 -1 V
2 -2 V
2 -1 V
2 -2 V
2 -2 V
3 -1 V
2 -2 V
2 -1 V
2 -2 V
2 -2 V
2 -1 V
2 -2 V
2 -1 V
2 -2 V
3 -1 V
2 -2 V
2 -1 V
2 -2 V
2 -2 V
2 -1 V
2 -2 V
2 -1 V
3 -2 V
2 -1 V
2 -2 V
2 -1 V
2 -2 V
2 -1 V
2 -2 V
2 -2 V
3 -1 V
2 -2 V
2 -1 V
2 -2 V
2 -1 V
2 -2 V
2 -1 V
2 -2 V
2 -1 V
3 -2 V
2 -1 V
2 -2 V
2 -1 V
2 -2 V
2 -1 V
2 -2 V
2 -1 V
3 -2 V
2 -1 V
2 -2 V
2 -1 V
2 -2 V
2 -1 V
2 -2 V
2 -1 V
3 -2 V
2 -1 V
2 -2 V
2 -1 V
2 -1 V
2 -2 V
2 -1 V
2 -2 V
2 -1 V
3 -2 V
2 -1 V
2 -2 V
2 -1 V
2 -2 V
2 -1 V
2 -1 V
2 -2 V
3 -1 V
2 -2 V
2 -1 V
2 -2 V
2 -1 V
2 -1 V
2 -2 V
2 -1 V
3 -2 V
2 -1 V
2 -2 V
2 -1 V
2 -1 V
2 -2 V
2 -1 V
2 -2 V
2 -1 V
3 -1 V
2 -2 V
2 -1 V
2 -2 V
2 -1 V
2 -1 V
2 -2 V
2 -1 V
3 -1 V
2 -2 V
2 -1 V
2 -2 V
2 -1 V
2 -1 V
2 -2 V
2 -1 V
3 -1 V
2 -2 V
2 -1 V
2 -1 V
2 -2 V
2 -1 V
2 -1 V
2 -2 V
2 -1 V
3 -1 V
2 -2 V
2 -1 V
2 -1 V
2 -2 V
2 -1 V
2 -1 V
2 -2 V
3 -1 V
2 -1 V
2 -1 V
2 -2 V
2 -1 V
2 -1 V
2 -2 V
2 -1 V
3 -1 V
2 -2 V
2 -1 V
2 -1 V
2 -1 V
2 -2 V
2 -1 V
2 -1 V
2 -1 V
3 -2 V
2 -1 V
2 -1 V
2 -1 V
2 -2 V
2 -1 V
2 -1 V
2 -1 V
3 -2 V
2 -1 V
2 -1 V
2 -1 V
2 -2 V
2 -1 V
2 -1 V
2 -1 V
3 -1 V
2 -2 V
2 -1 V
2 -1 V
2 -1 V
2 -2 V
2 -1 V
2 -1 V
2 -1 V
3 -1 V
2 -1 V
2 -2 V
2 -1 V
2 -1 V
2 -1 V
2 -1 V
2 -2 V
3 -1 V
2 -1 V
2 -1 V
2 -1 V
2 -1 V
2 -1 V
2 -2 V
2 -1 V
3 -1 V
2 -1 V
2 -1 V
2 -1 V
2 -1 V
2 -2 V
2 -1 V
2 -1 V
2 -1 V
3 -1 V
2 -1 V
2 -1 V
2 -1 V
2 -2 V
2 -1 V
2 -1 V
2 -1 V
3 -1 V
2 -1 V
2 -1 V
2 -1 V
2 -1 V
2 -1 V
2 -1 V
2 -1 V
3 -2 V
2 -1 V
2 -1 V
2 -1 V
2 -1 V
2 -1 V
2 -1 V
2 -1 V
2 -1 V
3 -1 V
2 -1 V
2 -1 V
2 -1 V
2 -1 V
2 -1 V
2 -1 V
2 -1 V
3 -1 V
2 -1 V
2 -1 V
2 -1 V
2 -1 V
2 -1 V
2 -1 V
2 -1 V
3 -1 V
2 -1 V
2 -1 V
2 -1 V
2 -1 V
2 -1 V
2 -1 V
2 -1 V
2 -1 V
3 -1 V
2 -1 V
2 -1 V
2 -1 V
2 -1 V
2 -1 V
2 -1 V
2 -1 V
3 0 V
2 -1 V
2 -1 V
2 -1 V
2 -1 V
2 -1 V
2 -1 V
2 -1 V
3 -1 V
2 -1 V
2 -1 V
2 0 V
2 -1 V
2 -1 V
2 -1 V
2 -1 V
2 -1 V
3 -1 V
2 -1 V
2 -1 V
2 0 V
2 -1 V
2 -1 V
2 -1 V
2 -1 V
3 -1 V
2 -1 V
2 0 V
2 -1 V
2 -1 V
2 -1 V
2 -1 V
2 -1 V
3 0 V
2 -1 V
2 -1 V
2 -1 V
2 -1 V
2 0 V
2 -1 V
2 -1 V
2 -1 V
3 -1 V
2 0 V
2 -1 V
2 -1 V
2 -1 V
2 -1 V
2 0 V
2 -1 V
3 -1 V
2 -1 V
2 0 V
2 -1 V
2 -1 V
2 -1 V
2 0 V
2 -1 V
3 -1 V
2 -1 V
currentpoint stroke M
2 0 V
2 -1 V
2 -1 V
2 -1 V
2 0 V
2 -1 V
2 -1 V
3 0 V
2 -1 V
2 -1 V
2 0 V
2 -1 V
2 -1 V
2 -1 V
2 0 V
3 -1 V
2 -1 V
2 0 V
2 -1 V
2 -1 V
2 0 V
2 -1 V
2 -1 V
3 0 V
2 -1 V
2 -1 V
2 0 V
2 -1 V
2 0 V
2 -1 V
2 -1 V
2 0 V
3 -1 V
2 -1 V
2 0 V
2 -1 V
2 0 V
2 -1 V
2 -1 V
2 0 V
3 -1 V
2 0 V
2 -1 V
2 -1 V
2 0 V
2 -1 V
2 0 V
2 -1 V
3 0 V
2 -1 V
2 -1 V
2 0 V
2 -1 V
2 0 V
2 -1 V
2 0 V
2 -1 V
3 0 V
2 -1 V
2 0 V
2 -1 V
2 0 V
2 -1 V
2 0 V
2 -1 V
3 0 V
2 -1 V
2 0 V
2 -1 V
2 0 V
2 -1 V
2 0 V
2 -1 V
3 0 V
2 -1 V
2 0 V
2 -1 V
2 0 V
2 -1 V
2 0 V
2 -1 V
2 0 V
3 0 V
2 -1 V
2 0 V
2 -1 V
2 0 V
2 -1 V
2 0 V
2 0 V
3 -1 V
2 0 V
2 -1 V
2 0 V
2 0 V
2 -1 V
2 0 V
2 -1 V
3 0 V
2 0 V
2 -1 V
2 0 V
2 -1 V
2 0 V
2 0 V
2 -1 V
2 0 V
3 0 V
2 -1 V
2 0 V
2 0 V
2 -1 V
2 0 V
2 0 V
2 -1 V
3 0 V
2 0 V
2 -1 V
2 0 V
2 0 V
2 -1 V
2 0 V
2 0 V
3 0 V
2 -1 V
2 0 V
2 0 V
2 -1 V
2 0 V
2 0 V
2 0 V
2 -1 V
3 0 V
2 0 V
2 0 V
2 -1 V
2 0 V
2 0 V
2 0 V
2 -1 V
3 0 V
2 0 V
2 0 V
2 -1 V
2 0 V
2 0 V
2 0 V
2 0 V
3 -1 V
2 0 V
2 0 V
2 0 V
2 0 V
2 -1 V
2 0 V
2 0 V
2 0 V
3 0 V
2 0 V
2 -1 V
2 0 V
2 0 V
2 0 V
2 0 V
2 0 V
3 0 V
2 -1 V
2 0 V
2 0 V
2 0 V
2 0 V
2 0 V
2 0 V
3 0 V
2 -1 V
2 0 V
2 0 V
2 0 V
2 0 V
2 0 V
2 0 V
2 0 V
3 0 V
2 0 V
2 0 V
2 0 V
2 -1 V
2 0 V
2 0 V
2 0 V
3 0 V
2 0 V
2 0 V
2 0 V
2 0 V
2 0 V
2 0 V
2 0 V
3 0 V
2 0 V
2 0 V
2 0 V
2 0 V
1.000 UL
LT0
2023 1647 M
247 0 V
250 1636 M
2 0 V
2 0 V
2 0 V
2 0 V
3 0 V
2 -1 V
2 0 V
2 0 V
2 0 V
2 0 V
2 -1 V
2 0 V
3 0 V
2 0 V
2 -1 V
2 0 V
2 0 V
2 -1 V
2 0 V
2 -1 V
3 0 V
2 0 V
2 -1 V
2 0 V
2 -1 V
2 0 V
2 -1 V
2 0 V
2 -1 V
3 0 V
2 -1 V
2 0 V
2 -1 V
2 0 V
2 -1 V
2 -1 V
2 0 V
3 -1 V
2 0 V
2 -1 V
2 -1 V
2 0 V
2 -1 V
2 -1 V
2 0 V
3 -1 V
2 -1 V
2 -1 V
2 0 V
2 -1 V
2 -1 V
2 -1 V
2 0 V
2 -1 V
3 -1 V
2 -1 V
2 0 V
2 -1 V
2 -1 V
2 -1 V
2 -1 V
2 -1 V
3 -1 V
2 0 V
2 -1 V
2 -1 V
2 -1 V
2 -1 V
2 -1 V
2 -1 V
3 -1 V
2 -1 V
2 -1 V
2 -1 V
2 -1 V
2 0 V
2 -1 V
2 -1 V
2 -1 V
3 -1 V
2 -1 V
2 -1 V
2 -1 V
2 -1 V
2 -1 V
2 -2 V
2 -1 V
3 -1 V
2 -1 V
2 -1 V
2 -1 V
2 -1 V
2 -1 V
2 -1 V
2 -1 V
3 -1 V
2 -1 V
2 -2 V
2 -1 V
2 -1 V
2 -1 V
2 -1 V
2 -1 V
2 -1 V
3 -2 V
2 -1 V
2 -1 V
2 -1 V
2 -1 V
2 -1 V
2 -2 V
2 -1 V
3 -1 V
2 -1 V
2 -1 V
2 -2 V
2 -1 V
2 -1 V
2 -1 V
2 -2 V
3 -1 V
2 -1 V
2 -1 V
2 -2 V
2 -1 V
2 -1 V
2 -1 V
2 -2 V
2 -1 V
3 -1 V
2 -2 V
2 -1 V
2 -1 V
2 -1 V
2 -2 V
2 -1 V
2 -1 V
3 -2 V
2 -1 V
2 -1 V
2 -2 V
2 -1 V
2 -1 V
2 -2 V
2 -1 V
3 -2 V
2 -1 V
2 -1 V
2 -2 V
2 -1 V
2 -1 V
2 -2 V
2 -1 V
2 -2 V
3 -1 V
2 -1 V
2 -2 V
2 -1 V
2 -2 V
2 -1 V
2 -1 V
2 -2 V
3 -1 V
2 -2 V
2 -1 V
2 -2 V
2 -1 V
2 -2 V
2 -1 V
2 -1 V
3 -2 V
2 -1 V
2 -2 V
2 -1 V
2 -2 V
2 -1 V
2 -2 V
2 -1 V
2 -2 V
3 -1 V
2 -2 V
2 -1 V
2 -2 V
2 -1 V
2 -2 V
2 -1 V
2 -2 V
3 -1 V
2 -2 V
2 -1 V
2 -2 V
2 -1 V
2 -2 V
2 -1 V
2 -2 V
3 -1 V
2 -2 V
2 -1 V
2 -2 V
2 -1 V
2 -2 V
2 -2 V
2 -1 V
2 -2 V
3 -1 V
2 -2 V
2 -1 V
2 -2 V
2 -1 V
2 -2 V
2 -2 V
2 -1 V
3 -2 V
2 -1 V
2 -2 V
2 -1 V
2 -2 V
2 -2 V
2 -1 V
2 -2 V
3 -1 V
2 -2 V
2 -2 V
2 -1 V
2 -2 V
2 -1 V
2 -2 V
2 -2 V
2 -1 V
3 -2 V
2 -1 V
2 -2 V
2 -2 V
2 -1 V
2 -2 V
2 -1 V
2 -2 V
3 -2 V
2 -1 V
2 -2 V
2 -2 V
2 -1 V
2 -2 V
2 -1 V
2 -2 V
3 -2 V
2 -1 V
2 -2 V
2 -2 V
2 -1 V
2 -2 V
2 -1 V
2 -2 V
2 -2 V
3 -1 V
2 -2 V
2 -2 V
2 -1 V
2 -2 V
2 -2 V
2 -1 V
2 -2 V
3 -2 V
2 -1 V
2 -2 V
2 -2 V
2 -1 V
2 -2 V
2 -2 V
2 -1 V
3 -2 V
2 -2 V
2 -1 V
2 -2 V
2 -2 V
2 -1 V
2 -2 V
2 -2 V
2 -1 V
3 -2 V
2 -2 V
2 -1 V
2 -2 V
2 -2 V
2 -1 V
2 -2 V
2 -2 V
3 -1 V
2 -2 V
2 -2 V
2 -1 V
2 -2 V
2 -2 V
2 -1 V
2 -2 V
3 -2 V
2 -1 V
2 -2 V
2 -2 V
2 -2 V
2 -1 V
2 -2 V
2 -2 V
2 -1 V
3 -2 V
2 -2 V
2 -1 V
2 -2 V
2 -2 V
2 -1 V
2 -2 V
2 -2 V
3 -1 V
2 -2 V
2 -2 V
2 -2 V
2 -1 V
2 -2 V
2 -2 V
2 -1 V
3 -2 V
2 -2 V
2 -1 V
2 -2 V
2 -2 V
2 -1 V
2 -2 V
2 -2 V
2 -2 V
3 -1 V
2 -2 V
2 -2 V
2 -1 V
2 -2 V
2 -2 V
2 -1 V
2 -2 V
3 -2 V
2 -2 V
2 -1 V
2 -2 V
2 -2 V
2 -1 V
2 -2 V
2 -2 V
3 -1 V
2 -2 V
2 -2 V
2 -1 V
2 -2 V
2 -2 V
2 -2 V
2 -1 V
2 -2 V
3 -2 V
2 -1 V
2 -2 V
2 -2 V
2 -1 V
2 -2 V
2 -2 V
2 -1 V
3 -2 V
2 -2 V
2 -2 V
2 -1 V
2 -2 V
2 -2 V
2 -1 V
2 -2 V
3 -2 V
2 -1 V
2 -2 V
2 -2 V
2 -1 V
2 -2 V
2 -2 V
2 -1 V
2 -2 V
3 -2 V
2 -1 V
2 -2 V
2 -2 V
2 -1 V
2 -2 V
2 -2 V
2 -2 V
3 -1 V
2 -2 V
2 -2 V
2 -1 V
2 -2 V
2 -2 V
2 -1 V
2 -2 V
3 -2 V
2 -1 V
currentpoint stroke M
2 -2 V
2 -2 V
2 -1 V
2 -2 V
2 -2 V
2 -1 V
2 -2 V
3 -2 V
2 -1 V
2 -2 V
2 -2 V
2 -1 V
2 -2 V
2 -1 V
2 -2 V
3 -2 V
2 -1 V
2 -2 V
2 -2 V
2 -1 V
2 -2 V
2 -2 V
2 -1 V
3 -2 V
2 -2 V
2 -1 V
2 -2 V
2 -2 V
2 -1 V
2 -2 V
2 -1 V
2 -2 V
3 -2 V
2 -1 V
2 -2 V
2 -2 V
2 -1 V
2 -2 V
2 -2 V
2 -1 V
3 -2 V
2 -1 V
2 -2 V
2 -2 V
2 -1 V
2 -2 V
2 -2 V
2 -1 V
3 -2 V
2 -1 V
2 -2 V
2 -2 V
2 -1 V
2 -2 V
2 -1 V
2 -2 V
2 -2 V
3 -1 V
2 -2 V
2 -1 V
2 -2 V
2 -2 V
2 -1 V
2 -2 V
2 -1 V
3 -2 V
2 -2 V
2 -1 V
2 -2 V
2 -1 V
2 -2 V
2 -2 V
2 -1 V
3 -2 V
2 -1 V
2 -2 V
2 -1 V
2 -2 V
2 -2 V
2 -1 V
2 -2 V
2 -1 V
3 -2 V
2 -1 V
2 -2 V
2 -1 V
2 -2 V
2 -2 V
2 -1 V
2 -2 V
3 -1 V
2 -2 V
2 -1 V
2 -2 V
2 -1 V
2 -2 V
2 -2 V
2 -1 V
3 -2 V
2 -1 V
2 -2 V
2 -1 V
2 -2 V
2 -1 V
2 -2 V
2 -1 V
2 -2 V
3 -1 V
2 -2 V
2 -1 V
2 -2 V
2 -1 V
2 -2 V
2 -1 V
2 -2 V
3 -1 V
2 -2 V
2 -1 V
2 -2 V
2 -1 V
2 -2 V
2 -1 V
2 -2 V
3 -1 V
2 -2 V
2 -1 V
2 -2 V
2 -1 V
2 -2 V
2 -1 V
2 -2 V
2 -1 V
3 -2 V
2 -1 V
2 -2 V
2 -1 V
2 -2 V
2 -1 V
2 -2 V
2 -1 V
3 -1 V
2 -2 V
2 -1 V
2 -2 V
2 -1 V
2 -2 V
2 -1 V
2 -2 V
3 -1 V
2 -1 V
2 -2 V
2 -1 V
2 -2 V
2 -1 V
2 -2 V
2 -1 V
2 -1 V
3 -2 V
2 -1 V
2 -2 V
2 -1 V
2 -2 V
2 -1 V
2 -1 V
2 -2 V
3 -1 V
2 -2 V
2 -1 V
2 -1 V
2 -2 V
2 -1 V
2 -2 V
2 -1 V
3 -1 V
2 -2 V
2 -1 V
2 -1 V
2 -2 V
2 -1 V
2 -2 V
2 -1 V
2 -1 V
3 -2 V
2 -1 V
2 -1 V
2 -2 V
2 -1 V
2 -1 V
2 -2 V
2 -1 V
3 -1 V
2 -2 V
2 -1 V
2 -1 V
2 -2 V
2 -1 V
2 -1 V
2 -2 V
3 -1 V
2 -1 V
2 -2 V
2 -1 V
2 -1 V
2 -2 V
2 -1 V
2 -1 V
2 -2 V
3 -1 V
2 -1 V
2 -1 V
2 -2 V
2 -1 V
2 -1 V
2 -2 V
2 -1 V
3 -1 V
2 -1 V
2 -2 V
2 -1 V
2 -1 V
2 -1 V
2 -2 V
2 -1 V
3 -1 V
2 -2 V
2 -1 V
2 -1 V
2 -1 V
2 -2 V
2 -1 V
2 -1 V
2 -1 V
3 -1 V
2 -2 V
2 -1 V
2 -1 V
2 -1 V
2 -2 V
2 -1 V
2 -1 V
3 -1 V
2 -1 V
2 -2 V
2 -1 V
2 -1 V
2 -1 V
2 -1 V
2 -2 V
3 -1 V
2 -1 V
2 -1 V
2 -1 V
2 -2 V
2 -1 V
2 -1 V
2 -1 V
2 -1 V
3 -1 V
2 -2 V
2 -1 V
2 -1 V
2 -1 V
2 -1 V
2 -1 V
2 -1 V
3 -2 V
2 -1 V
2 -1 V
2 -1 V
2 -1 V
2 -1 V
2 -1 V
2 -2 V
3 -1 V
2 -1 V
2 -1 V
2 -1 V
2 -1 V
2 -1 V
2 -1 V
2 -1 V
2 -1 V
3 -2 V
2 -1 V
2 -1 V
2 -1 V
2 -1 V
2 -1 V
2 -1 V
2 -1 V
3 -1 V
2 -1 V
2 -1 V
2 -1 V
2 -1 V
2 -2 V
2 -1 V
2 -1 V
3 -1 V
2 -1 V
2 -1 V
2 -1 V
2 -1 V
2 -1 V
2 -1 V
2 -1 V
2 -1 V
3 -1 V
2 -1 V
2 -1 V
2 -1 V
2 -1 V
2 -1 V
2 -1 V
2 -1 V
3 -1 V
2 -1 V
2 -1 V
2 -1 V
2 -1 V
2 -1 V
2 -1 V
2 -1 V
3 -1 V
2 -1 V
2 -1 V
2 -1 V
2 -1 V
2 -1 V
2 -1 V
2 0 V
2 -1 V
3 -1 V
2 -1 V
2 -1 V
2 -1 V
2 -1 V
2 -1 V
2 -1 V
2 -1 V
3 -1 V
2 -1 V
2 -1 V
2 0 V
2 -1 V
2 -1 V
2 -1 V
2 -1 V
3 -1 V
2 -1 V
2 -1 V
2 -1 V
2 0 V
2 -1 V
2 -1 V
2 -1 V
2 -1 V
3 -1 V
2 -1 V
2 -1 V
2 0 V
2 -1 V
2 -1 V
2 -1 V
2 -1 V
3 -1 V
2 0 V
2 -1 V
2 -1 V
2 -1 V
2 -1 V
2 -1 V
2 0 V
3 -1 V
2 -1 V
2 -1 V
2 -1 V
2 0 V
2 -1 V
2 -1 V
2 -1 V
2 0 V
3 -1 V
2 -1 V
2 -1 V
2 -1 V
2 0 V
2 -1 V
2 -1 V
2 -1 V
3 0 V
2 -1 V
2 -1 V
2 -1 V
2 0 V
2 -1 V
2 -1 V
2 -1 V
3 0 V
2 -1 V
currentpoint stroke M
2 -1 V
2 0 V
2 -1 V
2 -1 V
2 -1 V
2 0 V
2 -1 V
3 -1 V
2 0 V
2 -1 V
2 -1 V
2 0 V
2 -1 V
2 -1 V
2 0 V
3 -1 V
2 -1 V
2 0 V
2 -1 V
2 -1 V
2 0 V
2 -1 V
2 -1 V
3 0 V
2 -1 V
2 -1 V
2 0 V
2 -1 V
2 0 V
2 -1 V
2 -1 V
2 0 V
3 -1 V
2 0 V
2 -1 V
2 -1 V
2 0 V
2 -1 V
2 0 V
2 -1 V
3 -1 V
2 0 V
2 -1 V
2 0 V
2 -1 V
2 0 V
2 -1 V
2 -1 V
3 0 V
2 -1 V
2 0 V
2 -1 V
2 0 V
2 -1 V
2 0 V
2 -1 V
2 0 V
3 -1 V
2 -1 V
2 0 V
2 -1 V
2 0 V
2 -1 V
2 0 V
2 -1 V
3 0 V
2 -1 V
2 0 V
2 0 V
2 -1 V
2 0 V
2 -1 V
2 0 V
3 -1 V
2 0 V
2 -1 V
2 0 V
2 -1 V
2 0 V
2 -1 V
2 0 V
2 0 V
3 -1 V
2 0 V
2 -1 V
2 0 V
2 -1 V
2 0 V
2 0 V
2 -1 V
3 0 V
2 -1 V
2 0 V
2 0 V
2 -1 V
2 0 V
2 -1 V
2 0 V
3 0 V
2 -1 V
2 0 V
2 0 V
2 -1 V
2 0 V
2 -1 V
2 0 V
2 0 V
3 -1 V
2 0 V
2 0 V
2 -1 V
2 0 V
2 0 V
2 -1 V
2 0 V
3 0 V
2 -1 V
2 0 V
2 0 V
2 0 V
2 -1 V
2 0 V
2 0 V
3 -1 V
2 0 V
2 0 V
2 0 V
2 -1 V
2 0 V
2 0 V
2 -1 V
2 0 V
3 0 V
2 0 V
2 -1 V
2 0 V
2 0 V
2 0 V
2 -1 V
2 0 V
3 0 V
2 0 V
2 0 V
2 -1 V
2 0 V
2 0 V
2 0 V
2 0 V
3 -1 V
2 0 V
2 0 V
2 0 V
2 0 V
2 -1 V
2 0 V
2 0 V
2 0 V
3 0 V
2 0 V
2 -1 V
2 0 V
2 0 V
2 0 V
2 0 V
2 0 V
3 0 V
2 -1 V
2 0 V
2 0 V
2 0 V
2 0 V
2 0 V
2 0 V
3 0 V
2 -1 V
2 0 V
2 0 V
2 0 V
2 0 V
2 0 V
2 0 V
2 0 V
3 0 V
2 0 V
2 0 V
2 0 V
2 0 V
2 -1 V
2 0 V
2 0 V
3 0 V
2 0 V
2 0 V
2 0 V
2 0 V
2 0 V
2 0 V
2 0 V
3 0 V
2 0 V
2 0 V
2 0 V
2 0 V
stroke
grestore
end
showpage
}}%
\put(1973,1647){\makebox(0,0)[r]{QS}}%
\put(1973,1747){\makebox(0,0)[r]{$kh/\pi=0.1$}}%
\put(1973,1847){\makebox(0,0)[r]{$kh/\pi=0.2$}}%
\put(1973,1947){\makebox(0,0)[r]{$kh/\pi=0.4$}}%
\put(1946,731){\makebox(0,0)[l]{(b)}}%
\put(282,1927){\makebox(0,0)[l]{$h^3{\rm Re}S^{\parallel}(k,q)$}}%
\put(1575,270){\makebox(0,0)[l]{$qh/\pi$}}%
\put(2370,100){\makebox(0,0){ 1}}%
\put(1946,100){\makebox(0,0){ 0.8}}%
\put(1522,100){\makebox(0,0){ 0.6}}%
\put(1098,100){\makebox(0,0){ 0.4}}%
\put(674,100){\makebox(0,0){ 0.2}}%
\put(250,100){\makebox(0,0){ 0}}%
\put(200,2060){\makebox(0,0)[r]{ 8}}%
\put(200,1794){\makebox(0,0)[r]{ 6}}%
\put(200,1529){\makebox(0,0)[r]{ 4}}%
\put(200,1263){\makebox(0,0)[r]{ 2}}%
\put(200,997){\makebox(0,0)[r]{ 0}}%
\put(200,731){\makebox(0,0)[r]{-2}}%
\put(200,466){\makebox(0,0)[r]{-4}}%
\end{picture}%
\endgroup
 

%% file: disp_im_ort.tex
\begingroup%
  \makeatletter%
  \newcommand{\GNUPLOTspecial}{%
    \@sanitize\catcode`\%=14\relax\special}%
  \setlength{\unitlength}{0.1bp}%
{\GNUPLOTspecial{!
/gnudict 256 dict def
gnudict begin
/Color true def
/Solid true def
/gnulinewidth 5.000 def
/userlinewidth gnulinewidth def
/vshift -33 def
/dl {10 mul} def
/hpt_ 31.5 def
/vpt_ 31.5 def
/hpt hpt_ def
/vpt vpt_ def
/M {moveto} bind def
/L {lineto} bind def
/R {rmoveto} bind def
/V {rlineto} bind def
/vpt2 vpt 2 mul def
/hpt2 hpt 2 mul def
/Lshow { currentpoint stroke M
  0 vshift R show } def
/Rshow { currentpoint stroke M
  dup stringwidth pop neg vshift R show } def
/Cshow { currentpoint stroke M
  dup stringwidth pop -2 div vshift R show } def
/UP { dup vpt_ mul /vpt exch def hpt_ mul /hpt exch def
  /hpt2 hpt 2 mul def /vpt2 vpt 2 mul def } def
/DL { Color {setrgbcolor Solid {pop []} if 0 setdash }
 {pop pop pop Solid {pop []} if 0 setdash} ifelse } def
/BL { stroke userlinewidth 2 mul setlinewidth } def
/AL { stroke userlinewidth 2 div setlinewidth } def
/UL { dup gnulinewidth mul /userlinewidth exch def
      dup 1 lt {pop 1} if 10 mul /udl exch def } def
/PL { stroke userlinewidth setlinewidth } def
/LTb { BL [] 0 0 0 DL } def
/LTa { AL [1 udl mul 2 udl mul] 0 setdash 0 0 0 setrgbcolor } def
/LT0 { PL [] 1 0 0 DL } def
/LT1 { PL [4 dl 2 dl] 0 1 0 DL } def
/LT2 { PL [2 dl 3 dl] 0 0 1 DL } def
/LT3 { PL [1 dl 1.5 dl] 1 0 1 DL } def
/LT4 { PL [5 dl 2 dl 1 dl 2 dl] 0 1 1 DL } def
/LT5 { PL [4 dl 3 dl 1 dl 3 dl] 1 1 0 DL } def
/LT6 { PL [2 dl 2 dl 2 dl 4 dl] 0 0 0 DL } def
/LT7 { PL [2 dl 2 dl 2 dl 2 dl 2 dl 4 dl] 1 0.3 0 DL } def
/LT8 { PL [2 dl 2 dl 2 dl 2 dl 2 dl 2 dl 2 dl 4 dl] 0.5 0.5 0.5 DL } def
/Pnt { stroke [] 0 setdash
   gsave 1 setlinecap M 0 0 V stroke grestore } def
/Dia { stroke [] 0 setdash 2 copy vpt add M
  hpt neg vpt neg V hpt vpt neg V
  hpt vpt V hpt neg vpt V closepath stroke
  Pnt } def
/Pls { stroke [] 0 setdash vpt sub M 0 vpt2 V
  currentpoint stroke M
  hpt neg vpt neg R hpt2 0 V stroke
  } def
/Box { stroke [] 0 setdash 2 copy exch hpt sub exch vpt add M
  0 vpt2 neg V hpt2 0 V 0 vpt2 V
  hpt2 neg 0 V closepath stroke
  Pnt } def
/Crs { stroke [] 0 setdash exch hpt sub exch vpt add M
  hpt2 vpt2 neg V currentpoint stroke M
  hpt2 neg 0 R hpt2 vpt2 V stroke } def
/TriU { stroke [] 0 setdash 2 copy vpt 1.12 mul add M
  hpt neg vpt -1.62 mul V
  hpt 2 mul 0 V
  hpt neg vpt 1.62 mul V closepath stroke
  Pnt  } def
/Star { 2 copy Pls Crs } def
/BoxF { stroke [] 0 setdash exch hpt sub exch vpt add M
  0 vpt2 neg V  hpt2 0 V  0 vpt2 V
  hpt2 neg 0 V  closepath fill } def
/TriUF { stroke [] 0 setdash vpt 1.12 mul add M
  hpt neg vpt -1.62 mul V
  hpt 2 mul 0 V
  hpt neg vpt 1.62 mul V closepath fill } def
/TriD { stroke [] 0 setdash 2 copy vpt 1.12 mul sub M
  hpt neg vpt 1.62 mul V
  hpt 2 mul 0 V
  hpt neg vpt -1.62 mul V closepath stroke
  Pnt  } def
/TriDF { stroke [] 0 setdash vpt 1.12 mul sub M
  hpt neg vpt 1.62 mul V
  hpt 2 mul 0 V
  hpt neg vpt -1.62 mul V closepath fill} def
/DiaF { stroke [] 0 setdash vpt add M
  hpt neg vpt neg V hpt vpt neg V
  hpt vpt V hpt neg vpt V closepath fill } def
/Pent { stroke [] 0 setdash 2 copy gsave
  translate 0 hpt M 4 {72 rotate 0 hpt L} repeat
  closepath stroke grestore Pnt } def
/PentF { stroke [] 0 setdash gsave
  translate 0 hpt M 4 {72 rotate 0 hpt L} repeat
  closepath fill grestore } def
/Circle { stroke [] 0 setdash 2 copy
  hpt 0 360 arc stroke Pnt } def
/CircleF { stroke [] 0 setdash hpt 0 360 arc fill } def
/C0 { BL [] 0 setdash 2 copy moveto vpt 90 450  arc } bind def
/C1 { BL [] 0 setdash 2 copy        moveto
       2 copy  vpt 0 90 arc closepath fill
               vpt 0 360 arc closepath } bind def
/C2 { BL [] 0 setdash 2 copy moveto
       2 copy  vpt 90 180 arc closepath fill
               vpt 0 360 arc closepath } bind def
/C3 { BL [] 0 setdash 2 copy moveto
       2 copy  vpt 0 180 arc closepath fill
               vpt 0 360 arc closepath } bind def
/C4 { BL [] 0 setdash 2 copy moveto
       2 copy  vpt 180 270 arc closepath fill
               vpt 0 360 arc closepath } bind def
/C5 { BL [] 0 setdash 2 copy moveto
       2 copy  vpt 0 90 arc
       2 copy moveto
       2 copy  vpt 180 270 arc closepath fill
               vpt 0 360 arc } bind def
/C6 { BL [] 0 setdash 2 copy moveto
      2 copy  vpt 90 270 arc closepath fill
              vpt 0 360 arc closepath } bind def
/C7 { BL [] 0 setdash 2 copy moveto
      2 copy  vpt 0 270 arc closepath fill
              vpt 0 360 arc closepath } bind def
/C8 { BL [] 0 setdash 2 copy moveto
      2 copy vpt 270 360 arc closepath fill
              vpt 0 360 arc closepath } bind def
/C9 { BL [] 0 setdash 2 copy moveto
      2 copy  vpt 270 450 arc closepath fill
              vpt 0 360 arc closepath } bind def
/C10 { BL [] 0 setdash 2 copy 2 copy moveto vpt 270 360 arc closepath fill
       2 copy moveto
       2 copy vpt 90 180 arc closepath fill
               vpt 0 360 arc closepath } bind def
/C11 { BL [] 0 setdash 2 copy moveto
       2 copy  vpt 0 180 arc closepath fill
       2 copy moveto
       2 copy  vpt 270 360 arc closepath fill
               vpt 0 360 arc closepath } bind def
/C12 { BL [] 0 setdash 2 copy moveto
       2 copy  vpt 180 360 arc closepath fill
               vpt 0 360 arc closepath } bind def
/C13 { BL [] 0 setdash  2 copy moveto
       2 copy  vpt 0 90 arc closepath fill
       2 copy moveto
       2 copy  vpt 180 360 arc closepath fill
               vpt 0 360 arc closepath } bind def
/C14 { BL [] 0 setdash 2 copy moveto
       2 copy  vpt 90 360 arc closepath fill
               vpt 0 360 arc } bind def
/C15 { BL [] 0 setdash 2 copy vpt 0 360 arc closepath fill
               vpt 0 360 arc closepath } bind def
/Rec   { newpath 4 2 roll moveto 1 index 0 rlineto 0 exch rlineto
       neg 0 rlineto closepath } bind def
/Square { dup Rec } bind def
/Bsquare { vpt sub exch vpt sub exch vpt2 Square } bind def
/S0 { BL [] 0 setdash 2 copy moveto 0 vpt rlineto BL Bsquare } bind def
/S1 { BL [] 0 setdash 2 copy vpt Square fill Bsquare } bind def
/S2 { BL [] 0 setdash 2 copy exch vpt sub exch vpt Square fill Bsquare } bind def
/S3 { BL [] 0 setdash 2 copy exch vpt sub exch vpt2 vpt Rec fill Bsquare } bind def
/S4 { BL [] 0 setdash 2 copy exch vpt sub exch vpt sub vpt Square fill Bsquare } bind def
/S5 { BL [] 0 setdash 2 copy 2 copy vpt Square fill
       exch vpt sub exch vpt sub vpt Square fill Bsquare } bind def
/S6 { BL [] 0 setdash 2 copy exch vpt sub exch vpt sub vpt vpt2 Rec fill Bsquare } bind def
/S7 { BL [] 0 setdash 2 copy exch vpt sub exch vpt sub vpt vpt2 Rec fill
       2 copy vpt Square fill
       Bsquare } bind def
/S8 { BL [] 0 setdash 2 copy vpt sub vpt Square fill Bsquare } bind def
/S9 { BL [] 0 setdash 2 copy vpt sub vpt vpt2 Rec fill Bsquare } bind def
/S10 { BL [] 0 setdash 2 copy vpt sub vpt Square fill 2 copy exch vpt sub exch vpt Square fill
       Bsquare } bind def
/S11 { BL [] 0 setdash 2 copy vpt sub vpt Square fill 2 copy exch vpt sub exch vpt2 vpt Rec fill
       Bsquare } bind def
/S12 { BL [] 0 setdash 2 copy exch vpt sub exch vpt sub vpt2 vpt Rec fill Bsquare } bind def
/S13 { BL [] 0 setdash 2 copy exch vpt sub exch vpt sub vpt2 vpt Rec fill
       2 copy vpt Square fill Bsquare } bind def
/S14 { BL [] 0 setdash 2 copy exch vpt sub exch vpt sub vpt2 vpt Rec fill
       2 copy exch vpt sub exch vpt Square fill Bsquare } bind def
/S15 { BL [] 0 setdash 2 copy Bsquare fill Bsquare } bind def
/D0 { gsave translate 45 rotate 0 0 S0 stroke grestore } bind def
/D1 { gsave translate 45 rotate 0 0 S1 stroke grestore } bind def
/D2 { gsave translate 45 rotate 0 0 S2 stroke grestore } bind def
/D3 { gsave translate 45 rotate 0 0 S3 stroke grestore } bind def
/D4 { gsave translate 45 rotate 0 0 S4 stroke grestore } bind def
/D5 { gsave translate 45 rotate 0 0 S5 stroke grestore } bind def
/D6 { gsave translate 45 rotate 0 0 S6 stroke grestore } bind def
/D7 { gsave translate 45 rotate 0 0 S7 stroke grestore } bind def
/D8 { gsave translate 45 rotate 0 0 S8 stroke grestore } bind def
/D9 { gsave translate 45 rotate 0 0 S9 stroke grestore } bind def
/D10 { gsave translate 45 rotate 0 0 S10 stroke grestore } bind def
/D11 { gsave translate 45 rotate 0 0 S11 stroke grestore } bind def
/D12 { gsave translate 45 rotate 0 0 S12 stroke grestore } bind def
/D13 { gsave translate 45 rotate 0 0 S13 stroke grestore } bind def
/D14 { gsave translate 45 rotate 0 0 S14 stroke grestore } bind def
/D15 { gsave translate 45 rotate 0 0 S15 stroke grestore } bind def
/DiaE { stroke [] 0 setdash vpt add M
  hpt neg vpt neg V hpt vpt neg V
  hpt vpt V hpt neg vpt V closepath stroke } def
/BoxE { stroke [] 0 setdash exch hpt sub exch vpt add M
  0 vpt2 neg V hpt2 0 V 0 vpt2 V
  hpt2 neg 0 V closepath stroke } def
/TriUE { stroke [] 0 setdash vpt 1.12 mul add M
  hpt neg vpt -1.62 mul V
  hpt 2 mul 0 V
  hpt neg vpt 1.62 mul V closepath stroke } def
/TriDE { stroke [] 0 setdash vpt 1.12 mul sub M
  hpt neg vpt 1.62 mul V
  hpt 2 mul 0 V
  hpt neg vpt -1.62 mul V closepath stroke } def
/PentE { stroke [] 0 setdash gsave
  translate 0 hpt M 4 {72 rotate 0 hpt L} repeat
  closepath stroke grestore } def
/CircE { stroke [] 0 setdash 
  hpt 0 360 arc stroke } def
/Opaque { gsave closepath 1 setgray fill grestore 0 setgray closepath } def
/DiaW { stroke [] 0 setdash vpt add M
  hpt neg vpt neg V hpt vpt neg V
  hpt vpt V hpt neg vpt V Opaque stroke } def
/BoxW { stroke [] 0 setdash exch hpt sub exch vpt add M
  0 vpt2 neg V hpt2 0 V 0 vpt2 V
  hpt2 neg 0 V Opaque stroke } def
/TriUW { stroke [] 0 setdash vpt 1.12 mul add M
  hpt neg vpt -1.62 mul V
  hpt 2 mul 0 V
  hpt neg vpt 1.62 mul V Opaque stroke } def
/TriDW { stroke [] 0 setdash vpt 1.12 mul sub M
  hpt neg vpt 1.62 mul V
  hpt 2 mul 0 V
  hpt neg vpt -1.62 mul V Opaque stroke } def
/PentW { stroke [] 0 setdash gsave
  translate 0 hpt M 4 {72 rotate 0 hpt L} repeat
  Opaque stroke grestore } def
/CircW { stroke [] 0 setdash 
  hpt 0 360 arc Opaque stroke } def
/BoxFill { gsave Rec 1 setgray fill grestore } def
/Symbol-Oblique /Symbol findfont [1 0 .167 1 0 0] makefont
dup length dict begin {1 index /FID eq {pop pop} {def} ifelse} forall
currentdict end definefont pop
end
}}%
\begin{picture}(2520,2160)(0,0)%
{\GNUPLOTspecial{"
gnudict begin
gsave
0 0 translate
0.100 0.100 scale
0 setgray
newpath
1.000 UL
LTb
300 443 M
63 0 V
2007 0 R
-63 0 V
300 847 M
63 0 V
2007 0 R
-63 0 V
300 1251 M
63 0 V
2007 0 R
-63 0 V
300 1656 M
63 0 V
2007 0 R
-63 0 V
300 2060 M
63 0 V
2007 0 R
-63 0 V
300 200 M
0 63 V
0 1797 R
0 -63 V
714 200 M
0 63 V
0 1797 R
0 -63 V
1128 200 M
0 63 V
0 1797 R
0 -63 V
1542 200 M
0 63 V
0 1797 R
0 -63 V
1956 200 M
0 63 V
0 1797 R
0 -63 V
2370 200 M
0 63 V
0 1797 R
0 -63 V
1.000 UL
LTb
300 200 M
2070 0 V
0 1860 V
-2070 0 V
300 200 L
1.000 UL
LT4
300 443 M
2070 0 V
1.000 UL
LT1
2023 1947 M
247 0 V
300 1049 M
2 0 V
2 0 V
2 1 V
2 0 V
2 1 V
2 0 V
2 1 V
3 1 V
2 1 V
2 1 V
2 1 V
2 2 V
2 1 V
2 2 V
2 2 V
2 2 V
2 2 V
2 2 V
2 2 V
2 2 V
2 3 V
3 2 V
2 3 V
2 3 V
2 3 V
2 3 V
2 3 V
2 4 V
2 3 V
2 4 V
2 3 V
2 4 V
2 4 V
2 4 V
2 4 V
3 5 V
2 4 V
2 5 V
2 4 V
2 5 V
2 5 V
2 5 V
2 5 V
2 5 V
2 6 V
2 5 V
2 6 V
2 6 V
2 6 V
3 6 V
2 6 V
2 6 V
2 6 V
2 7 V
2 7 V
2 6 V
2 7 V
2 7 V
2 7 V
2 7 V
2 8 V
2 7 V
2 8 V
2 7 V
3 8 V
2 8 V
2 8 V
2 8 V
2 9 V
2 8 V
2 9 V
2 8 V
2 9 V
2 9 V
2 9 V
2 9 V
2 10 V
2 9 V
3 9 V
2 10 V
2 10 V
2 10 V
2 10 V
2 10 V
2 10 V
2 10 V
2 11 V
2 11 V
2 10 V
2 11 V
2 11 V
2 11 V
3 11 V
2 12 V
2 11 V
2 11 V
2 12 V
2 11 V
2 10 V
2 -591 V
2 -603 V
2 -2 V
2 0 V
2 -1 V
2 0 V
2 0 V
2 0 V
3 0 V
2 0 V
2 0 V
2 0 V
2 0 V
2 0 V
2 0 V
2 0 V
2 0 V
2 0 V
2 0 V
2 0 V
2 0 V
2 0 V
3 0 V
2 0 V
2 0 V
2 0 V
2 0 V
2 0 V
2 0 V
2 0 V
2 0 V
2 0 V
2 0 V
2 0 V
2 0 V
2 0 V
3 0 V
2 0 V
2 0 V
2 0 V
2 0 V
2 0 V
2 0 V
2 0 V
2 0 V
2 0 V
2 0 V
2 0 V
2 0 V
2 0 V
3 0 V
2 0 V
2 0 V
2 0 V
2 0 V
2 0 V
2 0 V
2 0 V
2 0 V
2 0 V
2 0 V
2 0 V
2 0 V
2 0 V
2 0 V
3 0 V
2 0 V
2 0 V
2 0 V
2 0 V
2 0 V
2 0 V
2 0 V
2 0 V
2 0 V
2 0 V
2 0 V
2 0 V
2 0 V
3 0 V
2 0 V
2 0 V
2 0 V
2 0 V
2 0 V
2 0 V
2 0 V
2 0 V
2 0 V
2 0 V
2 0 V
2 0 V
2 0 V
3 0 V
2 0 V
2 0 V
2 0 V
2 0 V
2 0 V
2 0 V
2 0 V
2 0 V
2 0 V
2 0 V
2 0 V
2 0 V
2 0 V
2 0 V
3 0 V
2 0 V
2 0 V
2 0 V
2 0 V
2 0 V
2 0 V
2 0 V
2 0 V
2 0 V
2 0 V
2 0 V
2 0 V
2 0 V
3 0 V
2 0 V
2 0 V
2 0 V
2 0 V
2 0 V
2 0 V
2 0 V
2 0 V
2 0 V
2 0 V
2 0 V
2 0 V
2 0 V
3 0 V
2 0 V
2 0 V
2 0 V
2 0 V
2 0 V
2 0 V
2 0 V
2 0 V
2 0 V
2 0 V
2 0 V
2 0 V
2 0 V
3 0 V
2 0 V
2 0 V
2 0 V
2 0 V
2 0 V
2 0 V
2 0 V
2 0 V
2 0 V
2 0 V
2 0 V
2 0 V
2 0 V
2 0 V
3 0 V
2 0 V
2 0 V
2 0 V
2 0 V
2 0 V
2 0 V
2 0 V
2 0 V
2 0 V
2 0 V
2 0 V
2 0 V
2 0 V
3 0 V
2 0 V
2 0 V
2 0 V
2 0 V
2 0 V
2 0 V
2 0 V
2 0 V
2 0 V
2 0 V
2 0 V
2 0 V
2 0 V
3 0 V
2 0 V
2 0 V
2 0 V
2 0 V
2 0 V
2 0 V
2 0 V
2 0 V
2 0 V
2 0 V
2 0 V
2 0 V
2 0 V
2 0 V
3 0 V
2 0 V
2 0 V
2 0 V
2 0 V
2 0 V
2 0 V
2 0 V
2 0 V
2 0 V
2 0 V
2 0 V
2 0 V
2 0 V
3 0 V
2 0 V
2 0 V
2 0 V
2 0 V
2 0 V
2 0 V
2 0 V
2 0 V
2 0 V
2 0 V
2 0 V
2 0 V
2 0 V
3 0 V
2 0 V
2 0 V
2 0 V
2 0 V
2 0 V
2 0 V
2 0 V
2 0 V
2 0 V
2 0 V
2 0 V
2 0 V
2 0 V
3 0 V
2 0 V
2 0 V
2 0 V
2 0 V
2 0 V
2 0 V
2 0 V
2 0 V
2 0 V
2 0 V
2 0 V
2 0 V
2 0 V
2 0 V
3 0 V
2 0 V
2 0 V
2 0 V
2 0 V
2 0 V
2 0 V
2 0 V
2 0 V
2 0 V
2 0 V
2 0 V
2 0 V
2 0 V
3 0 V
2 0 V
2 0 V
2 0 V
2 0 V
2 0 V
2 0 V
2 0 V
2 0 V
2 0 V
2 0 V
2 0 V
2 0 V
2 0 V
3 0 V
2 0 V
2 0 V
2 0 V
2 0 V
currentpoint stroke M
2 0 V
2 0 V
2 0 V
2 0 V
2 0 V
2 0 V
2 0 V
2 0 V
2 0 V
2 0 V
3 0 V
2 0 V
2 0 V
2 0 V
2 0 V
2 0 V
2 0 V
2 0 V
2 0 V
2 0 V
2 0 V
2 0 V
2 0 V
2 0 V
3 0 V
2 0 V
2 0 V
2 0 V
2 0 V
2 0 V
2 0 V
2 0 V
2 0 V
2 0 V
2 0 V
2 0 V
2 0 V
2 0 V
3 0 V
2 0 V
2 0 V
2 0 V
2 0 V
2 0 V
2 0 V
2 0 V
2 0 V
2 0 V
2 0 V
2 0 V
2 0 V
2 0 V
3 0 V
2 0 V
2 0 V
2 0 V
2 0 V
2 0 V
2 0 V
2 0 V
2 0 V
2 0 V
2 0 V
2 0 V
2 0 V
2 0 V
2 0 V
3 0 V
2 0 V
2 0 V
2 0 V
2 0 V
2 0 V
2 0 V
2 0 V
2 0 V
2 0 V
2 0 V
2 0 V
2 0 V
2 0 V
3 0 V
2 0 V
2 0 V
2 0 V
2 0 V
2 0 V
2 0 V
2 0 V
2 0 V
2 0 V
2 0 V
2 0 V
2 0 V
2 0 V
3 0 V
2 0 V
2 0 V
2 0 V
2 0 V
2 0 V
2 0 V
2 0 V
2 0 V
2 0 V
2 0 V
2 0 V
2 0 V
2 0 V
2 0 V
3 0 V
2 0 V
2 0 V
2 0 V
2 0 V
2 0 V
2 0 V
2 0 V
2 0 V
2 0 V
2 0 V
2 0 V
2 0 V
2 0 V
3 0 V
2 0 V
2 0 V
2 0 V
2 0 V
2 0 V
2 0 V
2 0 V
2 0 V
2 0 V
2 0 V
2 0 V
2 0 V
2 0 V
3 0 V
2 0 V
2 0 V
2 0 V
2 0 V
2 0 V
2 0 V
2 0 V
2 0 V
2 0 V
2 0 V
2 0 V
2 0 V
2 0 V
3 0 V
2 0 V
2 0 V
2 0 V
2 0 V
2 0 V
2 0 V
2 0 V
2 0 V
2 0 V
2 0 V
2 0 V
2 0 V
2 0 V
2 0 V
3 0 V
2 0 V
2 0 V
2 0 V
2 0 V
2 0 V
2 0 V
2 0 V
2 0 V
2 0 V
2 0 V
2 0 V
2 0 V
2 0 V
3 0 V
2 0 V
2 0 V
2 0 V
2 0 V
2 0 V
2 0 V
2 0 V
2 0 V
2 0 V
2 0 V
2 0 V
2 0 V
2 0 V
3 0 V
2 0 V
2 0 V
2 0 V
2 0 V
2 0 V
2 0 V
2 0 V
2 0 V
2 0 V
2 0 V
2 0 V
2 0 V
2 0 V
2 0 V
3 0 V
2 0 V
2 0 V
2 0 V
2 0 V
2 0 V
2 0 V
2 0 V
2 0 V
2 0 V
2 0 V
2 0 V
2 0 V
2 0 V
3 0 V
2 0 V
2 0 V
2 0 V
2 0 V
2 0 V
2 0 V
2 0 V
2 0 V
2 0 V
2 0 V
2 0 V
2 0 V
2 0 V
3 0 V
2 0 V
2 0 V
2 0 V
2 0 V
2 0 V
2 0 V
2 0 V
2 0 V
2 0 V
2 0 V
2 0 V
2 0 V
2 0 V
3 0 V
2 0 V
2 0 V
2 0 V
2 0 V
2 0 V
2 0 V
2 0 V
2 0 V
2 0 V
2 0 V
2 0 V
2 0 V
2 0 V
2 0 V
3 0 V
2 0 V
2 0 V
2 0 V
2 0 V
2 0 V
2 0 V
2 0 V
2 0 V
2 0 V
2 0 V
2 0 V
2 0 V
2 0 V
3 0 V
2 0 V
2 0 V
2 0 V
2 0 V
2 0 V
2 0 V
2 0 V
2 0 V
2 0 V
2 0 V
2 0 V
2 0 V
2 0 V
3 0 V
2 0 V
2 0 V
2 0 V
2 0 V
2 0 V
2 0 V
2 0 V
2 0 V
2 0 V
2 0 V
2 0 V
2 0 V
2 0 V
2 0 V
3 0 V
2 0 V
2 0 V
2 0 V
2 0 V
2 0 V
2 0 V
2 0 V
2 0 V
2 0 V
2 0 V
2 0 V
2 0 V
2 0 V
3 0 V
2 0 V
2 0 V
2 0 V
2 0 V
2 0 V
2 0 V
2 0 V
2 0 V
2 0 V
2 0 V
2 0 V
2 0 V
2 0 V
3 0 V
2 0 V
2 0 V
2 0 V
2 0 V
2 0 V
2 0 V
2 0 V
2 0 V
2 0 V
2 0 V
2 0 V
2 0 V
2 0 V
3 0 V
2 0 V
2 0 V
2 0 V
2 0 V
2 0 V
2 0 V
2 0 V
2 0 V
2 0 V
2 0 V
2 0 V
2 0 V
2 0 V
2 0 V
3 0 V
2 0 V
2 0 V
2 0 V
2 0 V
2 0 V
2 0 V
2 0 V
2 0 V
2 0 V
2 0 V
2 0 V
2 0 V
2 0 V
3 0 V
2 0 V
2 0 V
2 0 V
2 0 V
2 0 V
2 0 V
2 0 V
2 0 V
2 0 V
2 0 V
2 0 V
2 0 V
2 0 V
3 0 V
2 0 V
2 0 V
2 0 V
2 0 V
currentpoint stroke M
2 0 V
2 0 V
2 0 V
2 0 V
2 0 V
2 0 V
2 0 V
2 0 V
2 0 V
2 0 V
3 0 V
2 0 V
2 0 V
2 0 V
2 0 V
2 0 V
2 0 V
2 0 V
2 0 V
2 0 V
2 0 V
2 0 V
2 0 V
2 0 V
3 0 V
2 0 V
2 0 V
2 0 V
2 0 V
2 0 V
2 0 V
2 0 V
2 0 V
2 0 V
2 0 V
2 0 V
2 0 V
2 0 V
3 0 V
2 0 V
2 0 V
2 0 V
2 0 V
2 0 V
2 0 V
2 0 V
2 0 V
2 0 V
2 0 V
2 0 V
2 0 V
2 0 V
3 0 V
2 0 V
2 0 V
2 0 V
2 0 V
2 0 V
2 0 V
2 0 V
2 0 V
2 0 V
2 0 V
2 0 V
2 0 V
2 0 V
2 0 V
3 0 V
2 0 V
2 0 V
2 0 V
2 0 V
2 0 V
2 0 V
2 0 V
2 0 V
2 0 V
2 0 V
2 0 V
2 0 V
2 0 V
3 0 V
2 0 V
2 0 V
2 0 V
2 0 V
2 0 V
2 0 V
2 0 V
2 0 V
2 0 V
2 0 V
2 0 V
2 0 V
2 0 V
3 0 V
2 0 V
2 0 V
2 0 V
2 0 V
2 0 V
2 0 V
2 0 V
2 0 V
2 0 V
2 0 V
2 0 V
2 0 V
2 0 V
2 0 V
3 0 V
2 0 V
2 0 V
2 0 V
2 0 V
2 0 V
2 0 V
2 0 V
2 0 V
2 0 V
2 0 V
2 0 V
2 0 V
2 0 V
3 0 V
2 0 V
2 0 V
2 0 V
2 0 V
2 0 V
2 0 V
2 0 V
2 0 V
2 0 V
2 0 V
2 0 V
2 0 V
2 0 V
3 0 V
2 0 V
2 0 V
2 0 V
2 0 V
2 0 V
2 0 V
2 0 V
2 0 V
2 0 V
2 0 V
2 0 V
2 0 V
2 0 V
3 0 V
2 0 V
2 0 V
2 0 V
2 0 V
2 0 V
2 0 V
2 0 V
2 0 V
2 0 V
2 0 V
2 0 V
2 0 V
2 0 V
2 0 V
3 0 V
2 0 V
2 0 V
2 0 V
2 0 V
2 0 V
2 0 V
2 0 V
2 0 V
2 0 V
2 0 V
2 0 V
2 0 V
2 0 V
3 0 V
2 0 V
2 0 V
2 0 V
2 0 V
2 0 V
2 0 V
2 0 V
2 0 V
2 0 V
2 0 V
2 0 V
2 0 V
2 0 V
3 0 V
2 0 V
2 0 V
2 0 V
2 0 V
2 0 V
2 0 V
2 0 V
1.000 UL
LT2
2023 1847 M
247 0 V
300 746 M
2 0 V
2 0 V
2 0 V
2 0 V
2 0 V
2 0 V
2 0 V
3 0 V
2 0 V
2 1 V
2 0 V
2 0 V
2 0 V
2 0 V
2 1 V
2 0 V
2 0 V
2 0 V
2 1 V
2 0 V
2 0 V
3 1 V
2 0 V
2 0 V
2 1 V
2 0 V
2 0 V
2 1 V
2 0 V
2 1 V
2 0 V
2 1 V
2 0 V
2 1 V
2 0 V
3 1 V
2 0 V
2 1 V
2 0 V
2 1 V
2 1 V
2 0 V
2 1 V
2 1 V
2 0 V
2 1 V
2 1 V
2 0 V
2 1 V
3 1 V
2 1 V
2 0 V
2 1 V
2 1 V
2 1 V
2 1 V
2 0 V
2 1 V
2 1 V
2 1 V
2 1 V
2 1 V
2 1 V
2 1 V
3 1 V
2 1 V
2 1 V
2 1 V
2 1 V
2 1 V
2 1 V
2 1 V
2 1 V
2 1 V
2 1 V
2 2 V
2 1 V
2 1 V
3 1 V
2 1 V
2 2 V
2 1 V
2 1 V
2 1 V
2 2 V
2 1 V
2 1 V
2 2 V
2 1 V
2 1 V
2 2 V
2 1 V
3 1 V
2 2 V
2 1 V
2 2 V
2 1 V
2 2 V
2 1 V
2 2 V
2 1 V
2 2 V
2 1 V
2 2 V
2 1 V
2 2 V
2 2 V
3 1 V
2 2 V
2 2 V
2 1 V
2 2 V
2 2 V
2 1 V
2 2 V
2 2 V
2 2 V
2 1 V
2 2 V
2 2 V
2 2 V
3 2 V
2 2 V
2 1 V
2 2 V
2 2 V
2 2 V
2 2 V
2 2 V
2 2 V
2 2 V
2 2 V
2 2 V
2 2 V
2 2 V
3 2 V
2 2 V
2 2 V
2 2 V
2 2 V
2 3 V
2 2 V
2 2 V
2 2 V
2 2 V
2 2 V
2 3 V
2 2 V
2 2 V
3 2 V
2 3 V
2 2 V
2 2 V
2 3 V
2 2 V
2 2 V
2 3 V
2 2 V
2 2 V
2 3 V
2 2 V
2 3 V
2 2 V
2 3 V
3 2 V
2 3 V
2 2 V
2 3 V
2 2 V
2 3 V
2 3 V
2 2 V
2 3 V
2 2 V
2 3 V
2 3 V
2 2 V
2 3 V
3 3 V
2 2 V
2 3 V
2 3 V
2 3 V
2 2 V
2 3 V
2 3 V
2 3 V
2 3 V
2 3 V
2 2 V
2 3 V
2 3 V
3 3 V
2 3 V
2 3 V
2 3 V
2 3 V
2 2 V
2 3 V
2 -299 V
2 -302 V
2 -1 V
2 0 V
2 0 V
2 0 V
2 0 V
2 0 V
3 0 V
2 0 V
2 0 V
2 0 V
2 0 V
2 0 V
2 0 V
2 0 V
2 0 V
2 0 V
2 0 V
2 0 V
2 0 V
2 0 V
3 0 V
2 0 V
2 0 V
2 0 V
2 0 V
2 0 V
2 0 V
2 0 V
2 0 V
2 0 V
2 0 V
2 0 V
2 0 V
2 0 V
3 0 V
2 0 V
2 0 V
2 0 V
2 0 V
2 0 V
2 0 V
2 0 V
2 0 V
2 0 V
2 0 V
2 0 V
2 0 V
2 0 V
3 0 V
2 0 V
2 0 V
2 0 V
2 0 V
2 0 V
2 0 V
2 0 V
2 0 V
2 0 V
2 0 V
2 0 V
2 0 V
2 0 V
2 0 V
3 0 V
2 0 V
2 0 V
2 0 V
2 0 V
2 0 V
2 0 V
2 0 V
2 0 V
2 0 V
2 0 V
2 0 V
2 0 V
2 0 V
3 0 V
2 0 V
2 0 V
2 0 V
2 0 V
2 0 V
2 0 V
2 0 V
2 0 V
2 0 V
2 0 V
2 0 V
2 0 V
2 0 V
3 0 V
2 0 V
2 0 V
2 0 V
2 0 V
2 0 V
2 0 V
2 0 V
2 0 V
2 0 V
2 0 V
2 0 V
2 0 V
2 0 V
2 0 V
3 0 V
2 0 V
2 0 V
2 0 V
2 0 V
2 0 V
2 0 V
2 0 V
2 0 V
2 0 V
2 0 V
2 0 V
2 0 V
2 0 V
3 0 V
2 0 V
2 0 V
2 0 V
2 0 V
2 0 V
2 0 V
2 0 V
2 0 V
2 0 V
2 0 V
2 0 V
2 0 V
2 0 V
3 0 V
2 0 V
2 0 V
2 0 V
2 0 V
2 0 V
2 0 V
2 0 V
2 0 V
2 0 V
2 0 V
2 0 V
2 0 V
2 0 V
3 0 V
2 0 V
2 0 V
2 0 V
2 0 V
2 0 V
2 0 V
2 0 V
2 0 V
2 0 V
2 0 V
2 0 V
2 0 V
2 0 V
2 0 V
3 0 V
2 0 V
2 0 V
2 0 V
2 0 V
2 0 V
2 0 V
2 0 V
2 0 V
2 0 V
2 0 V
2 0 V
2 0 V
2 0 V
3 0 V
2 0 V
2 0 V
2 0 V
2 0 V
2 0 V
2 0 V
2 0 V
2 0 V
2 0 V
2 0 V
2 0 V
2 0 V
2 0 V
3 0 V
2 0 V
2 0 V
2 0 V
2 0 V
currentpoint stroke M
2 0 V
2 0 V
2 0 V
2 0 V
2 0 V
2 0 V
2 0 V
2 0 V
2 0 V
2 0 V
3 0 V
2 0 V
2 0 V
2 0 V
2 0 V
2 0 V
2 0 V
2 0 V
2 0 V
2 0 V
2 0 V
2 0 V
2 0 V
2 0 V
3 0 V
2 0 V
2 0 V
2 0 V
2 0 V
2 0 V
2 0 V
2 0 V
2 0 V
2 0 V
2 0 V
2 0 V
2 0 V
2 0 V
3 0 V
2 0 V
2 0 V
2 0 V
2 0 V
2 0 V
2 0 V
2 0 V
2 0 V
2 0 V
2 0 V
2 0 V
2 0 V
2 0 V
3 0 V
2 0 V
2 0 V
2 0 V
2 0 V
2 0 V
2 0 V
2 0 V
2 0 V
2 0 V
2 0 V
2 0 V
2 0 V
2 0 V
2 0 V
3 0 V
2 0 V
2 0 V
2 0 V
2 0 V
2 0 V
2 0 V
2 0 V
2 0 V
2 0 V
2 0 V
2 0 V
2 0 V
2 0 V
3 0 V
2 0 V
2 0 V
2 0 V
2 0 V
2 0 V
2 0 V
2 0 V
2 0 V
2 0 V
2 0 V
2 0 V
2 0 V
2 0 V
3 0 V
2 0 V
2 0 V
2 0 V
2 0 V
2 0 V
2 0 V
2 0 V
2 0 V
2 0 V
2 0 V
2 0 V
2 0 V
2 0 V
2 0 V
3 0 V
2 0 V
2 0 V
2 0 V
2 0 V
2 0 V
2 0 V
2 0 V
2 0 V
2 0 V
2 0 V
2 0 V
2 0 V
2 0 V
3 0 V
2 0 V
2 0 V
2 0 V
2 0 V
2 0 V
2 0 V
2 0 V
2 0 V
2 0 V
2 0 V
2 0 V
2 0 V
2 0 V
3 0 V
2 0 V
2 0 V
2 0 V
2 0 V
2 0 V
2 0 V
2 0 V
2 0 V
2 0 V
2 0 V
2 0 V
2 0 V
2 0 V
3 0 V
2 0 V
2 0 V
2 0 V
2 0 V
2 0 V
2 0 V
2 0 V
2 0 V
2 0 V
2 0 V
2 0 V
2 0 V
2 0 V
2 0 V
3 0 V
2 0 V
2 0 V
2 0 V
2 0 V
2 0 V
2 0 V
2 0 V
2 0 V
2 0 V
2 0 V
2 0 V
2 0 V
2 0 V
3 0 V
2 0 V
2 0 V
2 0 V
2 0 V
2 0 V
2 0 V
2 0 V
2 0 V
2 0 V
2 0 V
2 0 V
2 0 V
2 0 V
3 0 V
2 0 V
2 0 V
2 0 V
2 0 V
2 0 V
2 0 V
2 0 V
2 0 V
2 0 V
2 0 V
2 0 V
2 0 V
2 0 V
2 0 V
3 0 V
2 0 V
2 0 V
2 0 V
2 0 V
2 0 V
2 0 V
2 0 V
2 0 V
2 0 V
2 0 V
2 0 V
2 0 V
2 0 V
3 0 V
2 0 V
2 0 V
2 0 V
2 0 V
2 0 V
2 0 V
2 0 V
2 0 V
2 0 V
2 0 V
2 0 V
2 0 V
2 0 V
3 0 V
2 0 V
2 0 V
2 0 V
2 0 V
2 0 V
2 0 V
2 0 V
2 0 V
2 0 V
2 0 V
2 0 V
2 0 V
2 0 V
3 0 V
2 0 V
2 0 V
2 0 V
2 0 V
2 0 V
2 0 V
2 0 V
2 0 V
2 0 V
2 0 V
2 0 V
2 0 V
2 0 V
2 0 V
3 0 V
2 0 V
2 0 V
2 0 V
2 0 V
2 0 V
2 0 V
2 0 V
2 0 V
2 0 V
2 0 V
2 0 V
2 0 V
2 0 V
3 0 V
2 0 V
2 0 V
2 0 V
2 0 V
2 0 V
2 0 V
2 0 V
2 0 V
2 0 V
2 0 V
2 0 V
2 0 V
2 0 V
3 0 V
2 0 V
2 0 V
2 0 V
2 0 V
2 0 V
2 0 V
2 0 V
2 0 V
2 0 V
2 0 V
2 0 V
2 0 V
2 0 V
2 0 V
3 0 V
2 0 V
2 0 V
2 0 V
2 0 V
2 0 V
2 0 V
2 0 V
2 0 V
2 0 V
2 0 V
2 0 V
2 0 V
2 0 V
3 0 V
2 0 V
2 0 V
2 0 V
2 0 V
2 0 V
2 0 V
2 0 V
2 0 V
2 0 V
2 0 V
2 0 V
2 0 V
2 0 V
3 0 V
2 0 V
2 0 V
2 0 V
2 0 V
2 0 V
2 0 V
2 0 V
2 0 V
2 0 V
2 0 V
2 0 V
2 0 V
2 0 V
3 0 V
2 0 V
2 0 V
2 0 V
2 0 V
2 0 V
2 0 V
2 0 V
2 0 V
2 0 V
2 0 V
2 0 V
2 0 V
2 0 V
2 0 V
3 0 V
2 0 V
2 0 V
2 0 V
2 0 V
2 0 V
2 0 V
2 0 V
2 0 V
2 0 V
2 0 V
2 0 V
2 0 V
2 0 V
3 0 V
2 0 V
2 0 V
2 0 V
2 0 V
2 0 V
2 0 V
2 0 V
2 0 V
2 0 V
2 0 V
2 0 V
2 0 V
2 0 V
3 0 V
2 0 V
2 0 V
2 0 V
2 0 V
currentpoint stroke M
2 0 V
2 0 V
2 0 V
2 0 V
2 0 V
2 0 V
2 0 V
2 0 V
2 0 V
2 0 V
3 0 V
2 0 V
2 0 V
2 0 V
2 0 V
2 0 V
2 0 V
2 0 V
2 0 V
2 0 V
2 0 V
2 0 V
2 0 V
2 0 V
3 0 V
2 0 V
2 0 V
2 0 V
2 0 V
2 0 V
2 0 V
2 0 V
2 0 V
2 0 V
2 0 V
2 0 V
2 0 V
2 0 V
3 0 V
2 0 V
2 0 V
2 0 V
2 0 V
2 0 V
2 0 V
2 0 V
2 0 V
2 0 V
2 0 V
2 0 V
2 0 V
2 0 V
3 0 V
2 0 V
2 0 V
2 0 V
2 0 V
2 0 V
2 0 V
2 0 V
2 0 V
2 0 V
2 0 V
2 0 V
2 0 V
2 0 V
2 0 V
3 0 V
2 0 V
2 0 V
2 0 V
2 0 V
2 0 V
2 0 V
2 0 V
2 0 V
2 0 V
2 0 V
2 0 V
2 0 V
2 0 V
3 0 V
2 0 V
2 0 V
2 0 V
2 0 V
2 0 V
2 0 V
2 0 V
2 0 V
2 0 V
2 0 V
2 0 V
2 0 V
2 0 V
3 0 V
2 0 V
2 0 V
2 0 V
2 0 V
2 0 V
2 0 V
2 0 V
2 0 V
2 0 V
2 0 V
2 0 V
2 0 V
2 0 V
2 0 V
3 0 V
2 0 V
2 0 V
2 0 V
2 0 V
2 0 V
2 0 V
2 0 V
2 0 V
2 0 V
2 0 V
2 0 V
2 0 V
2 0 V
3 0 V
2 0 V
2 0 V
2 0 V
2 0 V
2 0 V
2 0 V
2 0 V
2 0 V
2 0 V
2 0 V
2 0 V
2 0 V
2 0 V
3 0 V
2 0 V
2 0 V
2 0 V
2 0 V
2 0 V
2 0 V
2 0 V
2 0 V
2 0 V
2 0 V
2 0 V
2 0 V
2 0 V
3 0 V
2 0 V
2 0 V
2 0 V
2 0 V
2 0 V
2 0 V
2 0 V
2 0 V
2 0 V
2 0 V
2 0 V
2 0 V
2 0 V
2 0 V
3 0 V
2 0 V
2 0 V
2 0 V
2 0 V
2 0 V
2 0 V
2 0 V
2 0 V
2 0 V
2 0 V
2 0 V
2 0 V
2 0 V
3 0 V
2 0 V
2 0 V
2 0 V
2 0 V
2 0 V
2 0 V
2 0 V
2 0 V
2 0 V
2 0 V
2 0 V
2 0 V
2 0 V
3 0 V
2 0 V
2 0 V
2 0 V
2 0 V
2 0 V
2 0 V
2 0 V
1.000 UL
LT3
2023 1747 M
247 0 V
300 594 M
2 0 V
2 0 V
2 0 V
2 0 V
2 0 V
2 0 V
2 0 V
3 0 V
2 0 V
2 0 V
2 0 V
2 0 V
2 0 V
2 0 V
2 0 V
2 0 V
2 1 V
2 0 V
2 0 V
2 0 V
2 0 V
3 0 V
2 0 V
2 0 V
2 0 V
2 0 V
2 0 V
2 0 V
2 0 V
2 0 V
2 0 V
2 0 V
2 0 V
2 0 V
2 0 V
3 0 V
2 1 V
2 0 V
2 0 V
2 0 V
2 0 V
2 0 V
2 0 V
2 0 V
2 0 V
2 0 V
2 0 V
2 0 V
2 1 V
3 0 V
2 0 V
2 0 V
2 0 V
2 0 V
2 0 V
2 0 V
2 0 V
2 0 V
2 1 V
2 0 V
2 0 V
2 0 V
2 0 V
2 0 V
3 0 V
2 0 V
2 0 V
2 1 V
2 0 V
2 0 V
2 0 V
2 0 V
2 0 V
2 0 V
2 1 V
2 0 V
2 0 V
2 0 V
3 0 V
2 0 V
2 0 V
2 1 V
2 0 V
2 0 V
2 0 V
2 0 V
2 0 V
2 1 V
2 0 V
2 0 V
2 0 V
2 0 V
3 0 V
2 1 V
2 0 V
2 0 V
2 0 V
2 0 V
2 1 V
2 0 V
2 0 V
2 0 V
2 0 V
2 0 V
2 1 V
2 0 V
2 0 V
3 0 V
2 0 V
2 1 V
2 0 V
2 0 V
2 0 V
2 1 V
2 0 V
2 0 V
2 0 V
2 0 V
2 1 V
2 0 V
2 0 V
3 0 V
2 1 V
2 0 V
2 0 V
2 0 V
2 1 V
2 0 V
2 0 V
2 0 V
2 0 V
2 1 V
2 0 V
2 0 V
2 1 V
3 0 V
2 0 V
2 0 V
2 1 V
2 0 V
2 0 V
2 0 V
2 1 V
2 0 V
2 0 V
2 0 V
2 1 V
2 0 V
2 0 V
3 1 V
2 0 V
2 0 V
2 0 V
2 1 V
2 0 V
2 0 V
2 1 V
2 0 V
2 0 V
2 0 V
2 1 V
2 0 V
2 0 V
2 1 V
3 0 V
2 0 V
2 1 V
2 0 V
2 0 V
2 1 V
2 0 V
2 0 V
2 1 V
2 0 V
2 0 V
2 1 V
2 0 V
2 0 V
3 1 V
2 0 V
2 0 V
2 1 V
2 0 V
2 0 V
2 1 V
2 0 V
2 0 V
2 1 V
2 0 V
2 0 V
2 1 V
2 0 V
3 1 V
2 0 V
2 0 V
2 1 V
2 0 V
2 0 V
2 1 V
2 0 V
2 1 V
2 0 V
2 0 V
2 1 V
2 0 V
2 0 V
2 1 V
3 0 V
2 1 V
2 0 V
2 0 V
2 1 V
2 0 V
2 1 V
2 0 V
2 0 V
2 1 V
2 0 V
2 1 V
2 0 V
2 1 V
3 0 V
2 0 V
2 1 V
2 0 V
2 1 V
2 0 V
2 0 V
2 1 V
2 0 V
2 1 V
2 0 V
2 1 V
2 0 V
2 1 V
3 0 V
2 0 V
2 1 V
2 0 V
2 1 V
2 0 V
2 1 V
2 0 V
2 1 V
2 0 V
2 1 V
2 0 V
2 1 V
2 0 V
3 0 V
2 1 V
2 0 V
2 1 V
2 0 V
2 1 V
2 0 V
2 1 V
2 0 V
2 1 V
2 0 V
2 1 V
2 0 V
2 1 V
2 0 V
3 1 V
2 0 V
2 1 V
2 0 V
2 1 V
2 0 V
2 1 V
2 0 V
2 1 V
2 0 V
2 1 V
2 0 V
2 1 V
2 0 V
3 1 V
2 1 V
2 0 V
2 1 V
2 0 V
2 1 V
2 0 V
2 1 V
2 0 V
2 1 V
2 0 V
2 1 V
2 0 V
2 1 V
3 1 V
2 0 V
2 1 V
2 0 V
2 1 V
2 0 V
2 1 V
2 1 V
2 0 V
2 1 V
2 0 V
2 1 V
2 0 V
2 1 V
2 1 V
3 0 V
2 1 V
2 0 V
2 1 V
2 0 V
2 1 V
2 1 V
2 0 V
2 1 V
2 0 V
2 1 V
2 1 V
2 0 V
2 1 V
3 0 V
2 1 V
2 1 V
2 0 V
2 1 V
2 1 V
2 0 V
2 1 V
2 0 V
2 1 V
2 1 V
2 0 V
2 1 V
2 1 V
3 0 V
2 1 V
2 0 V
2 1 V
2 1 V
2 0 V
2 1 V
2 1 V
2 0 V
2 1 V
2 1 V
2 0 V
2 1 V
2 1 V
3 0 V
2 1 V
2 1 V
2 0 V
2 1 V
2 1 V
2 0 V
2 1 V
2 1 V
2 0 V
2 1 V
2 1 V
2 0 V
2 1 V
2 1 V
3 0 V
2 1 V
2 1 V
2 1 V
2 0 V
2 1 V
2 1 V
2 0 V
2 1 V
2 1 V
2 0 V
2 1 V
2 1 V
2 1 V
3 0 V
2 1 V
2 1 V
2 0 V
2 1 V
2 1 V
2 1 V
2 0 V
2 1 V
2 1 V
2 1 V
2 0 V
2 1 V
2 1 V
3 0 V
2 1 V
2 1 V
2 1 V
2 0 V
currentpoint stroke M
2 1 V
2 0 V
2 -150 V
2 -151 V
2 0 V
2 0 V
2 0 V
2 0 V
2 0 V
2 0 V
3 0 V
2 0 V
2 0 V
2 0 V
2 0 V
2 0 V
2 0 V
2 0 V
2 0 V
2 0 V
2 0 V
2 0 V
2 0 V
2 0 V
3 0 V
2 0 V
2 0 V
2 0 V
2 0 V
2 0 V
2 0 V
2 0 V
2 0 V
2 0 V
2 0 V
2 0 V
2 0 V
2 0 V
3 0 V
2 0 V
2 0 V
2 0 V
2 0 V
2 0 V
2 0 V
2 0 V
2 0 V
2 0 V
2 0 V
2 0 V
2 0 V
2 0 V
3 0 V
2 0 V
2 0 V
2 0 V
2 0 V
2 0 V
2 0 V
2 0 V
2 0 V
2 0 V
2 0 V
2 0 V
2 0 V
2 0 V
2 0 V
3 0 V
2 0 V
2 0 V
2 0 V
2 0 V
2 0 V
2 0 V
2 0 V
2 0 V
2 0 V
2 0 V
2 0 V
2 0 V
2 0 V
3 0 V
2 0 V
2 0 V
2 0 V
2 0 V
2 0 V
2 0 V
2 0 V
2 0 V
2 0 V
2 0 V
2 0 V
2 0 V
2 0 V
3 0 V
2 0 V
2 0 V
2 0 V
2 0 V
2 0 V
2 0 V
2 0 V
2 0 V
2 0 V
2 0 V
2 0 V
2 0 V
2 0 V
2 0 V
3 0 V
2 0 V
2 0 V
2 0 V
2 0 V
2 0 V
2 0 V
2 0 V
2 0 V
2 0 V
2 0 V
2 0 V
2 0 V
2 0 V
3 0 V
2 0 V
2 0 V
2 0 V
2 0 V
2 0 V
2 0 V
2 0 V
2 0 V
2 0 V
2 0 V
2 0 V
2 0 V
2 0 V
3 0 V
2 0 V
2 0 V
2 0 V
2 0 V
2 0 V
2 0 V
2 0 V
2 0 V
2 0 V
2 0 V
2 0 V
2 0 V
2 0 V
3 0 V
2 0 V
2 0 V
2 0 V
2 0 V
2 0 V
2 0 V
2 0 V
2 0 V
2 0 V
2 0 V
2 0 V
2 0 V
2 0 V
2 0 V
3 0 V
2 0 V
2 0 V
2 0 V
2 0 V
2 0 V
2 0 V
2 0 V
2 0 V
2 0 V
2 0 V
2 0 V
2 0 V
2 0 V
3 0 V
2 0 V
2 0 V
2 0 V
2 0 V
2 0 V
2 0 V
2 0 V
2 0 V
2 0 V
2 0 V
2 0 V
2 0 V
2 0 V
3 0 V
2 0 V
2 0 V
2 0 V
2 0 V
2 0 V
2 0 V
2 0 V
2 0 V
2 0 V
2 0 V
2 0 V
2 0 V
2 0 V
2 0 V
3 0 V
2 0 V
2 0 V
2 0 V
2 0 V
2 0 V
2 0 V
2 0 V
2 0 V
2 0 V
2 0 V
2 0 V
2 0 V
2 0 V
3 0 V
2 0 V
2 0 V
2 0 V
2 0 V
2 0 V
2 0 V
2 0 V
2 0 V
2 0 V
2 0 V
2 0 V
2 0 V
2 0 V
3 0 V
2 0 V
2 0 V
2 0 V
2 0 V
2 0 V
2 0 V
2 0 V
2 0 V
2 0 V
2 0 V
2 0 V
2 0 V
2 0 V
3 0 V
2 0 V
2 0 V
2 0 V
2 0 V
2 0 V
2 0 V
2 0 V
2 0 V
2 0 V
2 0 V
2 0 V
2 0 V
2 0 V
2 0 V
3 0 V
2 0 V
2 0 V
2 0 V
2 0 V
2 0 V
2 0 V
2 0 V
2 0 V
2 0 V
2 0 V
2 0 V
2 0 V
2 0 V
3 0 V
2 0 V
2 0 V
2 0 V
2 0 V
2 0 V
2 0 V
2 0 V
2 0 V
2 0 V
2 0 V
2 0 V
2 0 V
2 0 V
3 0 V
2 0 V
2 0 V
2 0 V
2 0 V
2 0 V
2 0 V
2 0 V
2 0 V
2 0 V
2 0 V
2 0 V
2 0 V
2 0 V
2 0 V
3 0 V
2 0 V
2 0 V
2 0 V
2 0 V
2 0 V
2 0 V
2 0 V
2 0 V
2 0 V
2 0 V
2 0 V
2 0 V
2 0 V
3 0 V
2 0 V
2 0 V
2 0 V
2 0 V
2 0 V
2 0 V
2 0 V
2 0 V
2 0 V
2 0 V
2 0 V
2 0 V
2 0 V
3 0 V
2 0 V
2 0 V
2 0 V
2 0 V
2 0 V
2 0 V
2 0 V
2 0 V
2 0 V
2 0 V
2 0 V
2 0 V
2 0 V
3 0 V
2 0 V
2 0 V
2 0 V
2 0 V
2 0 V
2 0 V
2 0 V
2 0 V
2 0 V
2 0 V
2 0 V
2 0 V
2 0 V
2 0 V
3 0 V
2 0 V
2 0 V
2 0 V
2 0 V
2 0 V
2 0 V
2 0 V
2 0 V
2 0 V
2 0 V
2 0 V
2 0 V
2 0 V
3 0 V
2 0 V
2 0 V
2 0 V
2 0 V
2 0 V
2 0 V
2 0 V
2 0 V
2 0 V
2 0 V
2 0 V
2 0 V
2 0 V
3 0 V
2 0 V
2 0 V
2 0 V
2 0 V
currentpoint stroke M
2 0 V
2 0 V
2 0 V
2 0 V
2 0 V
2 0 V
2 0 V
2 0 V
2 0 V
2 0 V
3 0 V
2 0 V
2 0 V
2 0 V
2 0 V
2 0 V
2 0 V
2 0 V
2 0 V
2 0 V
2 0 V
2 0 V
2 0 V
2 0 V
3 0 V
2 0 V
2 0 V
2 0 V
2 0 V
2 0 V
2 0 V
2 0 V
2 0 V
2 0 V
2 0 V
2 0 V
2 0 V
2 0 V
3 0 V
2 0 V
2 0 V
2 0 V
2 0 V
2 0 V
2 0 V
2 0 V
2 0 V
2 0 V
2 0 V
2 0 V
2 0 V
2 0 V
3 0 V
2 0 V
2 0 V
2 0 V
2 0 V
2 0 V
2 0 V
2 0 V
2 0 V
2 0 V
2 0 V
2 0 V
2 0 V
2 0 V
2 0 V
3 0 V
2 0 V
2 0 V
2 0 V
2 0 V
2 0 V
2 0 V
2 0 V
2 0 V
2 0 V
2 0 V
2 0 V
2 0 V
2 0 V
3 0 V
2 0 V
2 0 V
2 0 V
2 0 V
2 0 V
2 0 V
2 0 V
2 0 V
2 0 V
2 0 V
2 0 V
2 0 V
2 0 V
3 0 V
2 0 V
2 0 V
2 0 V
2 0 V
2 0 V
2 0 V
2 0 V
2 0 V
2 0 V
2 0 V
2 0 V
2 0 V
2 0 V
2 0 V
3 0 V
2 0 V
2 0 V
2 0 V
2 0 V
2 0 V
2 0 V
2 0 V
2 0 V
2 0 V
2 0 V
2 0 V
2 0 V
2 0 V
3 0 V
2 0 V
2 0 V
2 0 V
2 0 V
2 0 V
2 0 V
2 0 V
2 0 V
2 0 V
2 0 V
2 0 V
2 0 V
2 0 V
3 0 V
2 0 V
2 0 V
2 0 V
2 0 V
2 0 V
2 0 V
2 0 V
2 0 V
2 0 V
2 0 V
2 0 V
2 0 V
2 0 V
3 0 V
2 0 V
2 0 V
2 0 V
2 0 V
2 0 V
2 0 V
2 0 V
2 0 V
2 0 V
2 0 V
2 0 V
2 0 V
2 0 V
2 0 V
3 0 V
2 0 V
2 0 V
2 0 V
2 0 V
2 0 V
2 0 V
2 0 V
2 0 V
2 0 V
2 0 V
2 0 V
2 0 V
2 0 V
3 0 V
2 0 V
2 0 V
2 0 V
2 0 V
2 0 V
2 0 V
2 0 V
2 0 V
2 0 V
2 0 V
2 0 V
2 0 V
2 0 V
3 0 V
2 0 V
2 0 V
2 0 V
2 0 V
2 0 V
2 0 V
2 0 V
stroke
grestore
end
showpage
}}%
\put(1973,1747){\makebox(0,0)[r]{$kh/\pi=0.4$}}%
\put(1973,1847){\makebox(0,0)[r]{$kh/\pi=0.2$}}%
\put(1973,1947){\makebox(0,0)[r]{$kh/\pi=0.1$}}%
\put(1956,847){\makebox(0,0)[l]{(c)}}%
\put(331,1858){\makebox(0,0)[l]{$Q^{\perp}(k,q)$}}%
\put(2008,281){\makebox(0,0)[l]{$qh/\pi$}}%
\put(2370,100){\makebox(0,0){ 1}}%
\put(1956,100){\makebox(0,0){ 0.8}}%
\put(1542,100){\makebox(0,0){ 0.6}}%
\put(1128,100){\makebox(0,0){ 0.4}}%
\put(714,100){\makebox(0,0){ 0.2}}%
\put(300,100){\makebox(0,0){ 0}}%
\put(250,2060){\makebox(0,0)[r]{ 20}}%
\put(250,1656){\makebox(0,0)[r]{ 15}}%
\put(250,1251){\makebox(0,0)[r]{ 10}}%
\put(250,847){\makebox(0,0)[r]{ 5}}%
\put(250,443){\makebox(0,0)[r]{ 0}}%
\end{picture}%
\endgroup
 

%% file: disp_im_par.tex
\begingroup%
  \makeatletter%
  \newcommand{\GNUPLOTspecial}{%
    \@sanitize\catcode`\%=14\relax\special}%
  \setlength{\unitlength}{0.1bp}%
{\GNUPLOTspecial{!
/gnudict 256 dict def
gnudict begin
/Color true def
/Solid true def
/gnulinewidth 5.000 def
/userlinewidth gnulinewidth def
/vshift -33 def
/dl {10 mul} def
/hpt_ 31.5 def
/vpt_ 31.5 def
/hpt hpt_ def
/vpt vpt_ def
/M {moveto} bind def
/L {lineto} bind def
/R {rmoveto} bind def
/V {rlineto} bind def
/vpt2 vpt 2 mul def
/hpt2 hpt 2 mul def
/Lshow { currentpoint stroke M
  0 vshift R show } def
/Rshow { currentpoint stroke M
  dup stringwidth pop neg vshift R show } def
/Cshow { currentpoint stroke M
  dup stringwidth pop -2 div vshift R show } def
/UP { dup vpt_ mul /vpt exch def hpt_ mul /hpt exch def
  /hpt2 hpt 2 mul def /vpt2 vpt 2 mul def } def
/DL { Color {setrgbcolor Solid {pop []} if 0 setdash }
 {pop pop pop Solid {pop []} if 0 setdash} ifelse } def
/BL { stroke userlinewidth 2 mul setlinewidth } def
/AL { stroke userlinewidth 2 div setlinewidth } def
/UL { dup gnulinewidth mul /userlinewidth exch def
      dup 1 lt {pop 1} if 10 mul /udl exch def } def
/PL { stroke userlinewidth setlinewidth } def
/LTb { BL [] 0 0 0 DL } def
/LTa { AL [1 udl mul 2 udl mul] 0 setdash 0 0 0 setrgbcolor } def
/LT0 { PL [] 1 0 0 DL } def
/LT1 { PL [4 dl 2 dl] 0 1 0 DL } def
/LT2 { PL [2 dl 3 dl] 0 0 1 DL } def
/LT3 { PL [1 dl 1.5 dl] 1 0 1 DL } def
/LT4 { PL [5 dl 2 dl 1 dl 2 dl] 0 1 1 DL } def
/LT5 { PL [4 dl 3 dl 1 dl 3 dl] 1 1 0 DL } def
/LT6 { PL [2 dl 2 dl 2 dl 4 dl] 0 0 0 DL } def
/LT7 { PL [2 dl 2 dl 2 dl 2 dl 2 dl 4 dl] 1 0.3 0 DL } def
/LT8 { PL [2 dl 2 dl 2 dl 2 dl 2 dl 2 dl 2 dl 4 dl] 0.5 0.5 0.5 DL } def
/Pnt { stroke [] 0 setdash
   gsave 1 setlinecap M 0 0 V stroke grestore } def
/Dia { stroke [] 0 setdash 2 copy vpt add M
  hpt neg vpt neg V hpt vpt neg V
  hpt vpt V hpt neg vpt V closepath stroke
  Pnt } def
/Pls { stroke [] 0 setdash vpt sub M 0 vpt2 V
  currentpoint stroke M
  hpt neg vpt neg R hpt2 0 V stroke
  } def
/Box { stroke [] 0 setdash 2 copy exch hpt sub exch vpt add M
  0 vpt2 neg V hpt2 0 V 0 vpt2 V
  hpt2 neg 0 V closepath stroke
  Pnt } def
/Crs { stroke [] 0 setdash exch hpt sub exch vpt add M
  hpt2 vpt2 neg V currentpoint stroke M
  hpt2 neg 0 R hpt2 vpt2 V stroke } def
/TriU { stroke [] 0 setdash 2 copy vpt 1.12 mul add M
  hpt neg vpt -1.62 mul V
  hpt 2 mul 0 V
  hpt neg vpt 1.62 mul V closepath stroke
  Pnt  } def
/Star { 2 copy Pls Crs } def
/BoxF { stroke [] 0 setdash exch hpt sub exch vpt add M
  0 vpt2 neg V  hpt2 0 V  0 vpt2 V
  hpt2 neg 0 V  closepath fill } def
/TriUF { stroke [] 0 setdash vpt 1.12 mul add M
  hpt neg vpt -1.62 mul V
  hpt 2 mul 0 V
  hpt neg vpt 1.62 mul V closepath fill } def
/TriD { stroke [] 0 setdash 2 copy vpt 1.12 mul sub M
  hpt neg vpt 1.62 mul V
  hpt 2 mul 0 V
  hpt neg vpt -1.62 mul V closepath stroke
  Pnt  } def
/TriDF { stroke [] 0 setdash vpt 1.12 mul sub M
  hpt neg vpt 1.62 mul V
  hpt 2 mul 0 V
  hpt neg vpt -1.62 mul V closepath fill} def
/DiaF { stroke [] 0 setdash vpt add M
  hpt neg vpt neg V hpt vpt neg V
  hpt vpt V hpt neg vpt V closepath fill } def
/Pent { stroke [] 0 setdash 2 copy gsave
  translate 0 hpt M 4 {72 rotate 0 hpt L} repeat
  closepath stroke grestore Pnt } def
/PentF { stroke [] 0 setdash gsave
  translate 0 hpt M 4 {72 rotate 0 hpt L} repeat
  closepath fill grestore } def
/Circle { stroke [] 0 setdash 2 copy
  hpt 0 360 arc stroke Pnt } def
/CircleF { stroke [] 0 setdash hpt 0 360 arc fill } def
/C0 { BL [] 0 setdash 2 copy moveto vpt 90 450  arc } bind def
/C1 { BL [] 0 setdash 2 copy        moveto
       2 copy  vpt 0 90 arc closepath fill
               vpt 0 360 arc closepath } bind def
/C2 { BL [] 0 setdash 2 copy moveto
       2 copy  vpt 90 180 arc closepath fill
               vpt 0 360 arc closepath } bind def
/C3 { BL [] 0 setdash 2 copy moveto
       2 copy  vpt 0 180 arc closepath fill
               vpt 0 360 arc closepath } bind def
/C4 { BL [] 0 setdash 2 copy moveto
       2 copy  vpt 180 270 arc closepath fill
               vpt 0 360 arc closepath } bind def
/C5 { BL [] 0 setdash 2 copy moveto
       2 copy  vpt 0 90 arc
       2 copy moveto
       2 copy  vpt 180 270 arc closepath fill
               vpt 0 360 arc } bind def
/C6 { BL [] 0 setdash 2 copy moveto
      2 copy  vpt 90 270 arc closepath fill
              vpt 0 360 arc closepath } bind def
/C7 { BL [] 0 setdash 2 copy moveto
      2 copy  vpt 0 270 arc closepath fill
              vpt 0 360 arc closepath } bind def
/C8 { BL [] 0 setdash 2 copy moveto
      2 copy vpt 270 360 arc closepath fill
              vpt 0 360 arc closepath } bind def
/C9 { BL [] 0 setdash 2 copy moveto
      2 copy  vpt 270 450 arc closepath fill
              vpt 0 360 arc closepath } bind def
/C10 { BL [] 0 setdash 2 copy 2 copy moveto vpt 270 360 arc closepath fill
       2 copy moveto
       2 copy vpt 90 180 arc closepath fill
               vpt 0 360 arc closepath } bind def
/C11 { BL [] 0 setdash 2 copy moveto
       2 copy  vpt 0 180 arc closepath fill
       2 copy moveto
       2 copy  vpt 270 360 arc closepath fill
               vpt 0 360 arc closepath } bind def
/C12 { BL [] 0 setdash 2 copy moveto
       2 copy  vpt 180 360 arc closepath fill
               vpt 0 360 arc closepath } bind def
/C13 { BL [] 0 setdash  2 copy moveto
       2 copy  vpt 0 90 arc closepath fill
       2 copy moveto
       2 copy  vpt 180 360 arc closepath fill
               vpt 0 360 arc closepath } bind def
/C14 { BL [] 0 setdash 2 copy moveto
       2 copy  vpt 90 360 arc closepath fill
               vpt 0 360 arc } bind def
/C15 { BL [] 0 setdash 2 copy vpt 0 360 arc closepath fill
               vpt 0 360 arc closepath } bind def
/Rec   { newpath 4 2 roll moveto 1 index 0 rlineto 0 exch rlineto
       neg 0 rlineto closepath } bind def
/Square { dup Rec } bind def
/Bsquare { vpt sub exch vpt sub exch vpt2 Square } bind def
/S0 { BL [] 0 setdash 2 copy moveto 0 vpt rlineto BL Bsquare } bind def
/S1 { BL [] 0 setdash 2 copy vpt Square fill Bsquare } bind def
/S2 { BL [] 0 setdash 2 copy exch vpt sub exch vpt Square fill Bsquare } bind def
/S3 { BL [] 0 setdash 2 copy exch vpt sub exch vpt2 vpt Rec fill Bsquare } bind def
/S4 { BL [] 0 setdash 2 copy exch vpt sub exch vpt sub vpt Square fill Bsquare } bind def
/S5 { BL [] 0 setdash 2 copy 2 copy vpt Square fill
       exch vpt sub exch vpt sub vpt Square fill Bsquare } bind def
/S6 { BL [] 0 setdash 2 copy exch vpt sub exch vpt sub vpt vpt2 Rec fill Bsquare } bind def
/S7 { BL [] 0 setdash 2 copy exch vpt sub exch vpt sub vpt vpt2 Rec fill
       2 copy vpt Square fill
       Bsquare } bind def
/S8 { BL [] 0 setdash 2 copy vpt sub vpt Square fill Bsquare } bind def
/S9 { BL [] 0 setdash 2 copy vpt sub vpt vpt2 Rec fill Bsquare } bind def
/S10 { BL [] 0 setdash 2 copy vpt sub vpt Square fill 2 copy exch vpt sub exch vpt Square fill
       Bsquare } bind def
/S11 { BL [] 0 setdash 2 copy vpt sub vpt Square fill 2 copy exch vpt sub exch vpt2 vpt Rec fill
       Bsquare } bind def
/S12 { BL [] 0 setdash 2 copy exch vpt sub exch vpt sub vpt2 vpt Rec fill Bsquare } bind def
/S13 { BL [] 0 setdash 2 copy exch vpt sub exch vpt sub vpt2 vpt Rec fill
       2 copy vpt Square fill Bsquare } bind def
/S14 { BL [] 0 setdash 2 copy exch vpt sub exch vpt sub vpt2 vpt Rec fill
       2 copy exch vpt sub exch vpt Square fill Bsquare } bind def
/S15 { BL [] 0 setdash 2 copy Bsquare fill Bsquare } bind def
/D0 { gsave translate 45 rotate 0 0 S0 stroke grestore } bind def
/D1 { gsave translate 45 rotate 0 0 S1 stroke grestore } bind def
/D2 { gsave translate 45 rotate 0 0 S2 stroke grestore } bind def
/D3 { gsave translate 45 rotate 0 0 S3 stroke grestore } bind def
/D4 { gsave translate 45 rotate 0 0 S4 stroke grestore } bind def
/D5 { gsave translate 45 rotate 0 0 S5 stroke grestore } bind def
/D6 { gsave translate 45 rotate 0 0 S6 stroke grestore } bind def
/D7 { gsave translate 45 rotate 0 0 S7 stroke grestore } bind def
/D8 { gsave translate 45 rotate 0 0 S8 stroke grestore } bind def
/D9 { gsave translate 45 rotate 0 0 S9 stroke grestore } bind def
/D10 { gsave translate 45 rotate 0 0 S10 stroke grestore } bind def
/D11 { gsave translate 45 rotate 0 0 S11 stroke grestore } bind def
/D12 { gsave translate 45 rotate 0 0 S12 stroke grestore } bind def
/D13 { gsave translate 45 rotate 0 0 S13 stroke grestore } bind def
/D14 { gsave translate 45 rotate 0 0 S14 stroke grestore } bind def
/D15 { gsave translate 45 rotate 0 0 S15 stroke grestore } bind def
/DiaE { stroke [] 0 setdash vpt add M
  hpt neg vpt neg V hpt vpt neg V
  hpt vpt V hpt neg vpt V closepath stroke } def
/BoxE { stroke [] 0 setdash exch hpt sub exch vpt add M
  0 vpt2 neg V hpt2 0 V 0 vpt2 V
  hpt2 neg 0 V closepath stroke } def
/TriUE { stroke [] 0 setdash vpt 1.12 mul add M
  hpt neg vpt -1.62 mul V
  hpt 2 mul 0 V
  hpt neg vpt 1.62 mul V closepath stroke } def
/TriDE { stroke [] 0 setdash vpt 1.12 mul sub M
  hpt neg vpt 1.62 mul V
  hpt 2 mul 0 V
  hpt neg vpt -1.62 mul V closepath stroke } def
/PentE { stroke [] 0 setdash gsave
  translate 0 hpt M 4 {72 rotate 0 hpt L} repeat
  closepath stroke grestore } def
/CircE { stroke [] 0 setdash 
  hpt 0 360 arc stroke } def
/Opaque { gsave closepath 1 setgray fill grestore 0 setgray closepath } def
/DiaW { stroke [] 0 setdash vpt add M
  hpt neg vpt neg V hpt vpt neg V
  hpt vpt V hpt neg vpt V Opaque stroke } def
/BoxW { stroke [] 0 setdash exch hpt sub exch vpt add M
  0 vpt2 neg V hpt2 0 V 0 vpt2 V
  hpt2 neg 0 V Opaque stroke } def
/TriUW { stroke [] 0 setdash vpt 1.12 mul add M
  hpt neg vpt -1.62 mul V
  hpt 2 mul 0 V
  hpt neg vpt 1.62 mul V Opaque stroke } def
/TriDW { stroke [] 0 setdash vpt 1.12 mul sub M
  hpt neg vpt 1.62 mul V
  hpt 2 mul 0 V
  hpt neg vpt -1.62 mul V Opaque stroke } def
/PentW { stroke [] 0 setdash gsave
  translate 0 hpt M 4 {72 rotate 0 hpt L} repeat
  Opaque stroke grestore } def
/CircW { stroke [] 0 setdash 
  hpt 0 360 arc Opaque stroke } def
/BoxFill { gsave Rec 1 setgray fill grestore } def
/Symbol-Oblique /Symbol findfont [1 0 .167 1 0 0] makefont
dup length dict begin {1 index /FID eq {pop pop} {def} ifelse} forall
currentdict end definefont pop
end
}}%
\begin{picture}(2520,2160)(0,0)%
{\GNUPLOTspecial{"
gnudict begin
gsave
0 0 translate
0.100 0.100 scale
0 setgray
newpath
1.000 UL
LTb
300 443 M
63 0 V
2007 0 R
-63 0 V
300 847 M
63 0 V
2007 0 R
-63 0 V
300 1251 M
63 0 V
2007 0 R
-63 0 V
300 1656 M
63 0 V
2007 0 R
-63 0 V
300 2060 M
63 0 V
2007 0 R
-63 0 V
300 200 M
0 63 V
0 1797 R
0 -63 V
714 200 M
0 63 V
0 1797 R
0 -63 V
1128 200 M
0 63 V
0 1797 R
0 -63 V
1542 200 M
0 63 V
0 1797 R
0 -63 V
1956 200 M
0 63 V
0 1797 R
0 -63 V
2370 200 M
0 63 V
0 1797 R
0 -63 V
1.000 UL
LTb
300 200 M
2070 0 V
0 1860 V
-2070 0 V
300 200 L
1.000 UL
LT4
300 443 M
2070 0 V
1.000 UL
LT1
2023 1947 M
247 0 V
300 1656 M
2 0 V
2 -1 V
2 0 V
2 -1 V
2 -1 V
2 -2 V
2 -1 V
3 -2 V
2 -2 V
2 -2 V
2 -3 V
2 -3 V
2 -3 V
2 -3 V
2 -4 V
2 -3 V
2 -4 V
2 -5 V
2 -4 V
2 -5 V
2 -5 V
3 -5 V
2 -6 V
2 -5 V
2 -6 V
2 -6 V
2 -7 V
2 -6 V
2 -7 V
2 -8 V
2 -7 V
2 -8 V
2 -7 V
2 -9 V
2 -8 V
3 -9 V
2 -8 V
2 -10 V
2 -9 V
2 -9 V
2 -10 V
2 -10 V
2 -11 V
2 -10 V
2 -11 V
2 -11 V
2 -11 V
2 -12 V
2 -12 V
3 -12 V
2 -12 V
2 -12 V
2 -13 V
2 -13 V
2 -13 V
2 -14 V
2 -13 V
2 -14 V
2 -15 V
2 -14 V
2 -15 V
2 -15 V
2 -15 V
2 -15 V
3 -16 V
2 -16 V
2 -16 V
2 -16 V
2 -17 V
2 -17 V
2 -17 V
2 -17 V
2 -18 V
2 -18 V
2 -18 V
2 -18 V
2 -19 V
2 -18 V
3 -19 V
2 -20 V
2 -19 V
2 -20 V
2 -20 V
2 -20 V
2 -21 V
2 -21 V
2 -21 V
2 -21 V
2 -21 V
2 -22 V
2 -22 V
2 -22 V
3 -23 V
2 -22 V
2 -23 V
2 -23 V
2 -24 V
2 -23 V
2 -24 V
2 -24 V
2 0 V
2 0 V
2 0 V
2 0 V
2 0 V
2 0 V
2 0 V
3 0 V
2 0 V
2 0 V
2 0 V
2 0 V
2 0 V
2 0 V
2 0 V
2 0 V
2 0 V
2 0 V
2 0 V
2 0 V
2 0 V
3 0 V
2 0 V
2 0 V
2 0 V
2 0 V
2 0 V
2 0 V
2 0 V
2 0 V
2 0 V
2 0 V
2 0 V
2 0 V
2 0 V
3 0 V
2 0 V
2 0 V
2 0 V
2 0 V
2 0 V
2 0 V
2 0 V
2 0 V
2 0 V
2 0 V
2 0 V
2 0 V
2 0 V
3 0 V
2 0 V
2 0 V
2 0 V
2 0 V
2 0 V
2 0 V
2 0 V
2 0 V
2 0 V
2 0 V
2 0 V
2 0 V
2 0 V
2 0 V
3 0 V
2 0 V
2 0 V
2 0 V
2 0 V
2 0 V
2 0 V
2 0 V
2 0 V
2 0 V
2 0 V
2 0 V
2 0 V
2 0 V
3 0 V
2 0 V
2 0 V
2 0 V
2 0 V
2 0 V
2 0 V
2 0 V
2 0 V
2 0 V
2 0 V
2 0 V
2 0 V
2 0 V
3 0 V
2 0 V
2 0 V
2 0 V
2 0 V
2 0 V
2 0 V
2 0 V
2 0 V
2 0 V
2 0 V
2 0 V
2 0 V
2 0 V
2 0 V
3 0 V
2 0 V
2 0 V
2 0 V
2 0 V
2 0 V
2 0 V
2 0 V
2 0 V
2 0 V
2 0 V
2 0 V
2 0 V
2 0 V
3 0 V
2 0 V
2 0 V
2 0 V
2 0 V
2 0 V
2 0 V
2 0 V
2 0 V
2 0 V
2 0 V
2 0 V
2 0 V
2 0 V
3 0 V
2 0 V
2 0 V
2 0 V
2 0 V
2 0 V
2 0 V
2 0 V
2 0 V
2 0 V
2 0 V
2 0 V
2 0 V
2 0 V
3 0 V
2 0 V
2 0 V
2 0 V
2 0 V
2 0 V
2 0 V
2 0 V
2 0 V
2 0 V
2 0 V
2 0 V
2 0 V
2 0 V
2 0 V
3 0 V
2 0 V
2 0 V
2 0 V
2 0 V
2 0 V
2 0 V
2 0 V
2 0 V
2 0 V
2 0 V
2 0 V
2 0 V
2 0 V
3 0 V
2 0 V
2 0 V
2 0 V
2 0 V
2 0 V
2 0 V
2 0 V
2 0 V
2 0 V
2 0 V
2 0 V
2 0 V
2 0 V
3 0 V
2 0 V
2 0 V
2 0 V
2 0 V
2 0 V
2 0 V
2 0 V
2 0 V
2 0 V
2 0 V
2 0 V
2 0 V
2 0 V
2 0 V
3 0 V
2 0 V
2 0 V
2 0 V
2 0 V
2 0 V
2 0 V
2 0 V
2 0 V
2 0 V
2 0 V
2 0 V
2 0 V
2 0 V
3 0 V
2 0 V
2 0 V
2 0 V
2 0 V
2 0 V
2 0 V
2 0 V
2 0 V
2 0 V
2 0 V
2 0 V
2 0 V
2 0 V
3 0 V
2 0 V
2 0 V
2 0 V
2 0 V
2 0 V
2 0 V
2 0 V
2 0 V
2 0 V
2 0 V
2 0 V
2 0 V
2 0 V
3 0 V
2 0 V
2 0 V
2 0 V
2 0 V
2 0 V
2 0 V
2 0 V
2 0 V
2 0 V
2 0 V
2 0 V
2 0 V
2 0 V
2 0 V
3 0 V
2 0 V
2 0 V
2 0 V
2 0 V
2 0 V
2 0 V
2 0 V
2 0 V
2 0 V
2 0 V
2 0 V
2 0 V
2 0 V
3 0 V
2 0 V
2 0 V
2 0 V
2 0 V
2 0 V
2 0 V
2 0 V
2 0 V
2 0 V
2 0 V
2 0 V
2 0 V
2 0 V
3 0 V
2 0 V
2 0 V
2 0 V
2 0 V
currentpoint stroke M
2 0 V
2 0 V
2 0 V
2 0 V
2 0 V
2 0 V
2 0 V
2 0 V
2 0 V
2 0 V
3 0 V
2 0 V
2 0 V
2 0 V
2 0 V
2 0 V
2 0 V
2 0 V
2 0 V
2 0 V
2 0 V
2 0 V
2 0 V
2 0 V
3 0 V
2 0 V
2 0 V
2 0 V
2 0 V
2 0 V
2 0 V
2 0 V
2 0 V
2 0 V
2 0 V
2 0 V
2 0 V
2 0 V
3 0 V
2 0 V
2 0 V
2 0 V
2 0 V
2 0 V
2 0 V
2 0 V
2 0 V
2 0 V
2 0 V
2 0 V
2 0 V
2 0 V
3 0 V
2 0 V
2 0 V
2 0 V
2 0 V
2 0 V
2 0 V
2 0 V
2 0 V
2 0 V
2 0 V
2 0 V
2 0 V
2 0 V
2 0 V
3 0 V
2 0 V
2 0 V
2 0 V
2 0 V
2 0 V
2 0 V
2 0 V
2 0 V
2 0 V
2 0 V
2 0 V
2 0 V
2 0 V
3 0 V
2 0 V
2 0 V
2 0 V
2 0 V
2 0 V
2 0 V
2 0 V
2 0 V
2 0 V
2 0 V
2 0 V
2 0 V
2 0 V
3 0 V
2 0 V
2 0 V
2 0 V
2 0 V
2 0 V
2 0 V
2 0 V
2 0 V
2 0 V
2 0 V
2 0 V
2 0 V
2 0 V
2 0 V
3 0 V
2 0 V
2 0 V
2 0 V
2 0 V
2 0 V
2 0 V
2 0 V
2 0 V
2 0 V
2 0 V
2 0 V
2 0 V
2 0 V
3 0 V
2 0 V
2 0 V
2 0 V
2 0 V
2 0 V
2 0 V
2 0 V
2 0 V
2 0 V
2 0 V
2 0 V
2 0 V
2 0 V
3 0 V
2 0 V
2 0 V
2 0 V
2 0 V
2 0 V
2 0 V
2 0 V
2 0 V
2 0 V
2 0 V
2 0 V
2 0 V
2 0 V
3 0 V
2 0 V
2 0 V
2 0 V
2 0 V
2 0 V
2 0 V
2 0 V
2 0 V
2 0 V
2 0 V
2 0 V
2 0 V
2 0 V
2 0 V
3 0 V
2 0 V
2 0 V
2 0 V
2 0 V
2 0 V
2 0 V
2 0 V
2 0 V
2 0 V
2 0 V
2 0 V
2 0 V
2 0 V
3 0 V
2 0 V
2 0 V
2 0 V
2 0 V
2 0 V
2 0 V
2 0 V
2 0 V
2 0 V
2 0 V
2 0 V
2 0 V
2 0 V
3 0 V
2 0 V
2 0 V
2 0 V
2 0 V
2 0 V
2 0 V
2 0 V
2 0 V
2 0 V
2 0 V
2 0 V
2 0 V
2 0 V
2 0 V
3 0 V
2 0 V
2 0 V
2 0 V
2 0 V
2 0 V
2 0 V
2 0 V
2 0 V
2 0 V
2 0 V
2 0 V
2 0 V
2 0 V
3 0 V
2 0 V
2 0 V
2 0 V
2 0 V
2 0 V
2 0 V
2 0 V
2 0 V
2 0 V
2 0 V
2 0 V
2 0 V
2 0 V
3 0 V
2 0 V
2 0 V
2 0 V
2 0 V
2 0 V
2 0 V
2 0 V
2 0 V
2 0 V
2 0 V
2 0 V
2 0 V
2 0 V
3 0 V
2 0 V
2 0 V
2 0 V
2 0 V
2 0 V
2 0 V
2 0 V
2 0 V
2 0 V
2 0 V
2 0 V
2 0 V
2 0 V
2 0 V
3 0 V
2 0 V
2 0 V
2 0 V
2 0 V
2 0 V
2 0 V
2 0 V
2 0 V
2 0 V
2 0 V
2 0 V
2 0 V
2 0 V
3 0 V
2 0 V
2 0 V
2 0 V
2 0 V
2 0 V
2 0 V
2 0 V
2 0 V
2 0 V
2 0 V
2 0 V
2 0 V
2 0 V
3 0 V
2 0 V
2 0 V
2 0 V
2 0 V
2 0 V
2 0 V
2 0 V
2 0 V
2 0 V
2 0 V
2 0 V
2 0 V
2 0 V
2 0 V
3 0 V
2 0 V
2 0 V
2 0 V
2 0 V
2 0 V
2 0 V
2 0 V
2 0 V
2 0 V
2 0 V
2 0 V
2 0 V
2 0 V
3 0 V
2 0 V
2 0 V
2 0 V
2 0 V
2 0 V
2 0 V
2 0 V
2 0 V
2 0 V
2 0 V
2 0 V
2 0 V
2 0 V
3 0 V
2 0 V
2 0 V
2 0 V
2 0 V
2 0 V
2 0 V
2 0 V
2 0 V
2 0 V
2 0 V
2 0 V
2 0 V
2 0 V
3 0 V
2 0 V
2 0 V
2 0 V
2 0 V
2 0 V
2 0 V
2 0 V
2 0 V
2 0 V
2 0 V
2 0 V
2 0 V
2 0 V
2 0 V
3 0 V
2 0 V
2 0 V
2 0 V
2 0 V
2 0 V
2 0 V
2 0 V
2 0 V
2 0 V
2 0 V
2 0 V
2 0 V
2 0 V
3 0 V
2 0 V
2 0 V
2 0 V
2 0 V
2 0 V
2 0 V
2 0 V
2 0 V
2 0 V
2 0 V
2 0 V
2 0 V
2 0 V
3 0 V
2 0 V
2 0 V
2 0 V
2 0 V
currentpoint stroke M
2 0 V
2 0 V
2 0 V
2 0 V
2 0 V
2 0 V
2 0 V
2 0 V
2 0 V
2 0 V
3 0 V
2 0 V
2 0 V
2 0 V
2 0 V
2 0 V
2 0 V
2 0 V
2 0 V
2 0 V
2 0 V
2 0 V
2 0 V
2 0 V
3 0 V
2 0 V
2 0 V
2 0 V
2 0 V
2 0 V
2 0 V
2 0 V
2 0 V
2 0 V
2 0 V
2 0 V
2 0 V
2 0 V
3 0 V
2 0 V
2 0 V
2 0 V
2 0 V
2 0 V
2 0 V
2 0 V
2 0 V
2 0 V
2 0 V
2 0 V
2 0 V
2 0 V
3 0 V
2 0 V
2 0 V
2 0 V
2 0 V
2 0 V
2 0 V
2 0 V
2 0 V
2 0 V
2 0 V
2 0 V
2 0 V
2 0 V
2 0 V
3 0 V
2 0 V
2 0 V
2 0 V
2 0 V
2 0 V
2 0 V
2 0 V
2 0 V
2 0 V
2 0 V
2 0 V
2 0 V
2 0 V
3 0 V
2 0 V
2 0 V
2 0 V
2 0 V
2 0 V
2 0 V
2 0 V
2 0 V
2 0 V
2 0 V
2 0 V
2 0 V
2 0 V
3 0 V
2 0 V
2 0 V
2 0 V
2 0 V
2 0 V
2 0 V
2 0 V
2 0 V
2 0 V
2 0 V
2 0 V
2 0 V
2 0 V
2 0 V
3 0 V
2 0 V
2 0 V
2 0 V
2 0 V
2 0 V
2 0 V
2 0 V
2 0 V
2 0 V
2 0 V
2 0 V
2 0 V
2 0 V
3 0 V
2 0 V
2 0 V
2 0 V
2 0 V
2 0 V
2 0 V
2 0 V
2 0 V
2 0 V
2 0 V
2 0 V
2 0 V
2 0 V
3 0 V
2 0 V
2 0 V
2 0 V
2 0 V
2 0 V
2 0 V
2 0 V
2 0 V
2 0 V
2 0 V
2 0 V
2 0 V
2 0 V
3 0 V
2 0 V
2 0 V
2 0 V
2 0 V
2 0 V
2 0 V
2 0 V
2 0 V
2 0 V
2 0 V
2 0 V
2 0 V
2 0 V
2 0 V
3 0 V
2 0 V
2 0 V
2 0 V
2 0 V
2 0 V
2 0 V
2 0 V
2 0 V
2 0 V
2 0 V
2 0 V
2 0 V
2 0 V
3 0 V
2 0 V
2 0 V
2 0 V
2 0 V
2 0 V
2 0 V
2 0 V
2 0 V
2 0 V
2 0 V
2 0 V
2 0 V
2 0 V
3 0 V
2 0 V
2 0 V
2 0 V
2 0 V
2 0 V
2 0 V
2 0 V
1.000 UL
LT2
2023 1847 M
247 0 V
300 1049 M
2 0 V
2 0 V
2 0 V
2 0 V
2 0 V
2 0 V
2 -1 V
3 0 V
2 0 V
2 0 V
2 -1 V
2 0 V
2 0 V
2 -1 V
2 0 V
2 -1 V
2 0 V
2 -1 V
2 0 V
2 -1 V
2 -1 V
3 0 V
2 -1 V
2 -1 V
2 0 V
2 -1 V
2 -1 V
2 -1 V
2 -1 V
2 -1 V
2 0 V
2 -1 V
2 -1 V
2 -1 V
2 -1 V
3 -2 V
2 -1 V
2 -1 V
2 -1 V
2 -1 V
2 -1 V
2 -2 V
2 -1 V
2 -1 V
2 -2 V
2 -1 V
2 -1 V
2 -2 V
2 -1 V
3 -2 V
2 -1 V
2 -2 V
2 -1 V
2 -2 V
2 -2 V
2 -1 V
2 -2 V
2 -2 V
2 -2 V
2 -1 V
2 -2 V
2 -2 V
2 -2 V
2 -2 V
3 -2 V
2 -2 V
2 -2 V
2 -2 V
2 -2 V
2 -2 V
2 -2 V
2 -2 V
2 -3 V
2 -2 V
2 -2 V
2 -2 V
2 -3 V
2 -2 V
3 -3 V
2 -2 V
2 -2 V
2 -3 V
2 -2 V
2 -3 V
2 -2 V
2 -3 V
2 -3 V
2 -2 V
2 -3 V
2 -3 V
2 -2 V
2 -3 V
3 -3 V
2 -3 V
2 -3 V
2 -3 V
2 -3 V
2 -2 V
2 -3 V
2 -4 V
2 -3 V
2 -3 V
2 -3 V
2 -3 V
2 -3 V
2 -3 V
2 -3 V
3 -4 V
2 -3 V
2 -3 V
2 -4 V
2 -3 V
2 -3 V
2 -4 V
2 -3 V
2 -4 V
2 -3 V
2 -4 V
2 -4 V
2 -3 V
2 -4 V
3 -4 V
2 -3 V
2 -4 V
2 -4 V
2 -4 V
2 -3 V
2 -4 V
2 -4 V
2 -4 V
2 -4 V
2 -4 V
2 -4 V
2 -4 V
2 -4 V
3 -4 V
2 -4 V
2 -5 V
2 -4 V
2 -4 V
2 -4 V
2 -5 V
2 -4 V
2 -4 V
2 -5 V
2 -4 V
2 -5 V
2 -4 V
2 -5 V
3 -4 V
2 -5 V
2 -4 V
2 -5 V
2 -4 V
2 -5 V
2 -5 V
2 -5 V
2 -4 V
2 -5 V
2 -5 V
2 -5 V
2 -5 V
2 -5 V
2 -5 V
3 -5 V
2 -5 V
2 -5 V
2 -5 V
2 -5 V
2 -5 V
2 -5 V
2 -5 V
2 -6 V
2 -5 V
2 -5 V
2 -6 V
2 -5 V
2 -5 V
3 -6 V
2 -5 V
2 -6 V
2 -5 V
2 -6 V
2 -5 V
2 -6 V
2 -5 V
2 -6 V
2 -6 V
2 -6 V
2 -5 V
2 -6 V
2 -6 V
3 -6 V
2 -6 V
2 -5 V
2 -6 V
2 -6 V
2 -6 V
2 -6 V
2 -6 V
2 0 V
2 0 V
2 0 V
2 0 V
2 0 V
2 0 V
2 0 V
3 0 V
2 0 V
2 0 V
2 0 V
2 0 V
2 0 V
2 0 V
2 0 V
2 0 V
2 0 V
2 0 V
2 0 V
2 0 V
2 0 V
3 0 V
2 0 V
2 0 V
2 0 V
2 0 V
2 0 V
2 0 V
2 0 V
2 0 V
2 0 V
2 0 V
2 0 V
2 0 V
2 0 V
3 0 V
2 0 V
2 0 V
2 0 V
2 0 V
2 0 V
2 0 V
2 0 V
2 0 V
2 0 V
2 0 V
2 0 V
2 0 V
2 0 V
3 0 V
2 0 V
2 0 V
2 0 V
2 0 V
2 0 V
2 0 V
2 0 V
2 0 V
2 0 V
2 0 V
2 0 V
2 0 V
2 0 V
2 0 V
3 0 V
2 0 V
2 0 V
2 0 V
2 0 V
2 0 V
2 0 V
2 0 V
2 0 V
2 0 V
2 0 V
2 0 V
2 0 V
2 0 V
3 0 V
2 0 V
2 0 V
2 0 V
2 0 V
2 0 V
2 0 V
2 0 V
2 0 V
2 0 V
2 0 V
2 0 V
2 0 V
2 0 V
3 0 V
2 0 V
2 0 V
2 0 V
2 0 V
2 0 V
2 0 V
2 0 V
2 0 V
2 0 V
2 0 V
2 0 V
2 0 V
2 0 V
2 0 V
3 0 V
2 0 V
2 0 V
2 0 V
2 0 V
2 0 V
2 0 V
2 0 V
2 0 V
2 0 V
2 0 V
2 0 V
2 0 V
2 0 V
3 0 V
2 0 V
2 0 V
2 0 V
2 0 V
2 0 V
2 0 V
2 0 V
2 0 V
2 0 V
2 0 V
2 0 V
2 0 V
2 0 V
3 0 V
2 0 V
2 0 V
2 0 V
2 0 V
2 0 V
2 0 V
2 0 V
2 0 V
2 0 V
2 0 V
2 0 V
2 0 V
2 0 V
3 0 V
2 0 V
2 0 V
2 0 V
2 0 V
2 0 V
2 0 V
2 0 V
2 0 V
2 0 V
2 0 V
2 0 V
2 0 V
2 0 V
2 0 V
3 0 V
2 0 V
2 0 V
2 0 V
2 0 V
2 0 V
2 0 V
2 0 V
2 0 V
2 0 V
2 0 V
2 0 V
2 0 V
2 0 V
3 0 V
2 0 V
2 0 V
2 0 V
2 0 V
2 0 V
2 0 V
2 0 V
2 0 V
2 0 V
2 0 V
2 0 V
2 0 V
2 0 V
3 0 V
2 0 V
2 0 V
2 0 V
2 0 V
currentpoint stroke M
2 0 V
2 0 V
2 0 V
2 0 V
2 0 V
2 0 V
2 0 V
2 0 V
2 0 V
2 0 V
3 0 V
2 0 V
2 0 V
2 0 V
2 0 V
2 0 V
2 0 V
2 0 V
2 0 V
2 0 V
2 0 V
2 0 V
2 0 V
2 0 V
3 0 V
2 0 V
2 0 V
2 0 V
2 0 V
2 0 V
2 0 V
2 0 V
2 0 V
2 0 V
2 0 V
2 0 V
2 0 V
2 0 V
3 0 V
2 0 V
2 0 V
2 0 V
2 0 V
2 0 V
2 0 V
2 0 V
2 0 V
2 0 V
2 0 V
2 0 V
2 0 V
2 0 V
3 0 V
2 0 V
2 0 V
2 0 V
2 0 V
2 0 V
2 0 V
2 0 V
2 0 V
2 0 V
2 0 V
2 0 V
2 0 V
2 0 V
2 0 V
3 0 V
2 0 V
2 0 V
2 0 V
2 0 V
2 0 V
2 0 V
2 0 V
2 0 V
2 0 V
2 0 V
2 0 V
2 0 V
2 0 V
3 0 V
2 0 V
2 0 V
2 0 V
2 0 V
2 0 V
2 0 V
2 0 V
2 0 V
2 0 V
2 0 V
2 0 V
2 0 V
2 0 V
3 0 V
2 0 V
2 0 V
2 0 V
2 0 V
2 0 V
2 0 V
2 0 V
2 0 V
2 0 V
2 0 V
2 0 V
2 0 V
2 0 V
2 0 V
3 0 V
2 0 V
2 0 V
2 0 V
2 0 V
2 0 V
2 0 V
2 0 V
2 0 V
2 0 V
2 0 V
2 0 V
2 0 V
2 0 V
3 0 V
2 0 V
2 0 V
2 0 V
2 0 V
2 0 V
2 0 V
2 0 V
2 0 V
2 0 V
2 0 V
2 0 V
2 0 V
2 0 V
3 0 V
2 0 V
2 0 V
2 0 V
2 0 V
2 0 V
2 0 V
2 0 V
2 0 V
2 0 V
2 0 V
2 0 V
2 0 V
2 0 V
3 0 V
2 0 V
2 0 V
2 0 V
2 0 V
2 0 V
2 0 V
2 0 V
2 0 V
2 0 V
2 0 V
2 0 V
2 0 V
2 0 V
2 0 V
3 0 V
2 0 V
2 0 V
2 0 V
2 0 V
2 0 V
2 0 V
2 0 V
2 0 V
2 0 V
2 0 V
2 0 V
2 0 V
2 0 V
3 0 V
2 0 V
2 0 V
2 0 V
2 0 V
2 0 V
2 0 V
2 0 V
2 0 V
2 0 V
2 0 V
2 0 V
2 0 V
2 0 V
3 0 V
2 0 V
2 0 V
2 0 V
2 0 V
2 0 V
2 0 V
2 0 V
2 0 V
2 0 V
2 0 V
2 0 V
2 0 V
2 0 V
2 0 V
3 0 V
2 0 V
2 0 V
2 0 V
2 0 V
2 0 V
2 0 V
2 0 V
2 0 V
2 0 V
2 0 V
2 0 V
2 0 V
2 0 V
3 0 V
2 0 V
2 0 V
2 0 V
2 0 V
2 0 V
2 0 V
2 0 V
2 0 V
2 0 V
2 0 V
2 0 V
2 0 V
2 0 V
3 0 V
2 0 V
2 0 V
2 0 V
2 0 V
2 0 V
2 0 V
2 0 V
2 0 V
2 0 V
2 0 V
2 0 V
2 0 V
2 0 V
3 0 V
2 0 V
2 0 V
2 0 V
2 0 V
2 0 V
2 0 V
2 0 V
2 0 V
2 0 V
2 0 V
2 0 V
2 0 V
2 0 V
2 0 V
3 0 V
2 0 V
2 0 V
2 0 V
2 0 V
2 0 V
2 0 V
2 0 V
2 0 V
2 0 V
2 0 V
2 0 V
2 0 V
2 0 V
3 0 V
2 0 V
2 0 V
2 0 V
2 0 V
2 0 V
2 0 V
2 0 V
2 0 V
2 0 V
2 0 V
2 0 V
2 0 V
2 0 V
3 0 V
2 0 V
2 0 V
2 0 V
2 0 V
2 0 V
2 0 V
2 0 V
2 0 V
2 0 V
2 0 V
2 0 V
2 0 V
2 0 V
2 0 V
3 0 V
2 0 V
2 0 V
2 0 V
2 0 V
2 0 V
2 0 V
2 0 V
2 0 V
2 0 V
2 0 V
2 0 V
2 0 V
2 0 V
3 0 V
2 0 V
2 0 V
2 0 V
2 0 V
2 0 V
2 0 V
2 0 V
2 0 V
2 0 V
2 0 V
2 0 V
2 0 V
2 0 V
3 0 V
2 0 V
2 0 V
2 0 V
2 0 V
2 0 V
2 0 V
2 0 V
2 0 V
2 0 V
2 0 V
2 0 V
2 0 V
2 0 V
3 0 V
2 0 V
2 0 V
2 0 V
2 0 V
2 0 V
2 0 V
2 0 V
2 0 V
2 0 V
2 0 V
2 0 V
2 0 V
2 0 V
2 0 V
3 0 V
2 0 V
2 0 V
2 0 V
2 0 V
2 0 V
2 0 V
2 0 V
2 0 V
2 0 V
2 0 V
2 0 V
2 0 V
2 0 V
3 0 V
2 0 V
2 0 V
2 0 V
2 0 V
2 0 V
2 0 V
2 0 V
2 0 V
2 0 V
2 0 V
2 0 V
2 0 V
2 0 V
3 0 V
2 0 V
2 0 V
2 0 V
2 0 V
currentpoint stroke M
2 0 V
2 0 V
2 0 V
2 0 V
2 0 V
2 0 V
2 0 V
2 0 V
2 0 V
2 0 V
3 0 V
2 0 V
2 0 V
2 0 V
2 0 V
2 0 V
2 0 V
2 0 V
2 0 V
2 0 V
2 0 V
2 0 V
2 0 V
2 0 V
3 0 V
2 0 V
2 0 V
2 0 V
2 0 V
2 0 V
2 0 V
2 0 V
2 0 V
2 0 V
2 0 V
2 0 V
2 0 V
2 0 V
3 0 V
2 0 V
2 0 V
2 0 V
2 0 V
2 0 V
2 0 V
2 0 V
2 0 V
2 0 V
2 0 V
2 0 V
2 0 V
2 0 V
3 0 V
2 0 V
2 0 V
2 0 V
2 0 V
2 0 V
2 0 V
2 0 V
2 0 V
2 0 V
2 0 V
2 0 V
2 0 V
2 0 V
2 0 V
3 0 V
2 0 V
2 0 V
2 0 V
2 0 V
2 0 V
2 0 V
2 0 V
2 0 V
2 0 V
2 0 V
2 0 V
2 0 V
2 0 V
3 0 V
2 0 V
2 0 V
2 0 V
2 0 V
2 0 V
2 0 V
2 0 V
2 0 V
2 0 V
2 0 V
2 0 V
2 0 V
2 0 V
3 0 V
2 0 V
2 0 V
2 0 V
2 0 V
2 0 V
2 0 V
2 0 V
2 0 V
2 0 V
2 0 V
2 0 V
2 0 V
2 0 V
2 0 V
3 0 V
2 0 V
2 0 V
2 0 V
2 0 V
2 0 V
2 0 V
2 0 V
2 0 V
2 0 V
2 0 V
2 0 V
2 0 V
2 0 V
3 0 V
2 0 V
2 0 V
2 0 V
2 0 V
2 0 V
2 0 V
2 0 V
2 0 V
2 0 V
2 0 V
2 0 V
2 0 V
2 0 V
3 0 V
2 0 V
2 0 V
2 0 V
2 0 V
2 0 V
2 0 V
2 0 V
2 0 V
2 0 V
2 0 V
2 0 V
2 0 V
2 0 V
3 0 V
2 0 V
2 0 V
2 0 V
2 0 V
2 0 V
2 0 V
2 0 V
2 0 V
2 0 V
2 0 V
2 0 V
2 0 V
2 0 V
2 0 V
3 0 V
2 0 V
2 0 V
2 0 V
2 0 V
2 0 V
2 0 V
2 0 V
2 0 V
2 0 V
2 0 V
2 0 V
2 0 V
2 0 V
3 0 V
2 0 V
2 0 V
2 0 V
2 0 V
2 0 V
2 0 V
2 0 V
2 0 V
2 0 V
2 0 V
2 0 V
2 0 V
2 0 V
3 0 V
2 0 V
2 0 V
2 0 V
2 0 V
2 0 V
2 0 V
2 0 V
1.000 UL
LT3
2023 1747 M
247 0 V
300 746 M
2 0 V
2 0 V
2 0 V
2 0 V
2 0 V
2 0 V
2 0 V
3 0 V
2 0 V
2 0 V
2 0 V
2 0 V
2 0 V
2 -1 V
2 0 V
2 0 V
2 0 V
2 0 V
2 0 V
2 0 V
2 0 V
3 0 V
2 0 V
2 0 V
2 0 V
2 0 V
2 -1 V
2 0 V
2 0 V
2 0 V
2 0 V
2 0 V
2 0 V
2 0 V
2 0 V
3 -1 V
2 0 V
2 0 V
2 0 V
2 0 V
2 0 V
2 0 V
2 -1 V
2 0 V
2 0 V
2 0 V
2 0 V
2 0 V
2 -1 V
3 0 V
2 0 V
2 0 V
2 0 V
2 -1 V
2 0 V
2 0 V
2 0 V
2 -1 V
2 0 V
2 0 V
2 0 V
2 0 V
2 -1 V
2 0 V
3 0 V
2 0 V
2 -1 V
2 0 V
2 0 V
2 0 V
2 -1 V
2 0 V
2 0 V
2 -1 V
2 0 V
2 0 V
2 0 V
2 -1 V
3 0 V
2 0 V
2 -1 V
2 0 V
2 0 V
2 -1 V
2 0 V
2 0 V
2 0 V
2 -1 V
2 0 V
2 0 V
2 -1 V
2 0 V
3 -1 V
2 0 V
2 0 V
2 -1 V
2 0 V
2 0 V
2 -1 V
2 0 V
2 0 V
2 -1 V
2 0 V
2 -1 V
2 0 V
2 0 V
2 -1 V
3 0 V
2 -1 V
2 0 V
2 0 V
2 -1 V
2 0 V
2 -1 V
2 0 V
2 -1 V
2 0 V
2 -1 V
2 0 V
2 0 V
2 -1 V
3 0 V
2 -1 V
2 0 V
2 -1 V
2 0 V
2 -1 V
2 0 V
2 -1 V
2 0 V
2 -1 V
2 0 V
2 -1 V
2 0 V
2 -1 V
3 0 V
2 -1 V
2 0 V
2 -1 V
2 0 V
2 -1 V
2 0 V
2 -1 V
2 0 V
2 -1 V
2 -1 V
2 0 V
2 -1 V
2 0 V
3 -1 V
2 0 V
2 -1 V
2 0 V
2 -1 V
2 -1 V
2 0 V
2 -1 V
2 0 V
2 -1 V
2 -1 V
2 0 V
2 -1 V
2 0 V
2 -1 V
3 -1 V
2 0 V
2 -1 V
2 -1 V
2 0 V
2 -1 V
2 -1 V
2 0 V
2 -1 V
2 -1 V
2 0 V
2 -1 V
2 -1 V
2 0 V
3 -1 V
2 -1 V
2 0 V
2 -1 V
2 -1 V
2 0 V
2 -1 V
2 -1 V
2 0 V
2 -1 V
2 -1 V
2 -1 V
2 0 V
2 -1 V
3 -1 V
2 0 V
2 -1 V
2 -1 V
2 -1 V
2 0 V
2 -1 V
2 -1 V
2 -1 V
2 0 V
2 -1 V
2 -1 V
2 -1 V
2 -1 V
2 0 V
3 -1 V
2 -1 V
2 -1 V
2 -1 V
2 0 V
2 -1 V
2 -1 V
2 -1 V
2 -1 V
2 0 V
2 -1 V
2 -1 V
2 -1 V
2 -1 V
3 -1 V
2 0 V
2 -1 V
2 -1 V
2 -1 V
2 -1 V
2 -1 V
2 -1 V
2 0 V
2 -1 V
2 -1 V
2 -1 V
2 -1 V
2 -1 V
3 -1 V
2 -1 V
2 0 V
2 -1 V
2 -1 V
2 -1 V
2 -1 V
2 -1 V
2 -1 V
2 -1 V
2 -1 V
2 -1 V
2 -1 V
2 -1 V
3 -1 V
2 -1 V
2 0 V
2 -1 V
2 -1 V
2 -1 V
2 -1 V
2 -1 V
2 -1 V
2 -1 V
2 -1 V
2 -1 V
2 -1 V
2 -1 V
2 -1 V
3 -1 V
2 -1 V
2 -1 V
2 -1 V
2 -1 V
2 -1 V
2 -1 V
2 -1 V
2 -1 V
2 -1 V
2 -1 V
2 -2 V
2 -1 V
2 -1 V
3 -1 V
2 -1 V
2 -1 V
2 -1 V
2 -1 V
2 -1 V
2 -1 V
2 -1 V
2 -1 V
2 -1 V
2 -1 V
2 -2 V
2 -1 V
2 -1 V
3 -1 V
2 -1 V
2 -1 V
2 -1 V
2 -1 V
2 -1 V
2 -2 V
2 -1 V
2 -1 V
2 -1 V
2 -1 V
2 -1 V
2 -1 V
2 -2 V
2 -1 V
3 -1 V
2 -1 V
2 -1 V
2 -1 V
2 -2 V
2 -1 V
2 -1 V
2 -1 V
2 -1 V
2 -2 V
2 -1 V
2 -1 V
2 -1 V
2 -1 V
3 -2 V
2 -1 V
2 -1 V
2 -1 V
2 -2 V
2 -1 V
2 -1 V
2 -1 V
2 -2 V
2 -1 V
2 -1 V
2 -1 V
2 -2 V
2 -1 V
3 -1 V
2 -1 V
2 -2 V
2 -1 V
2 -1 V
2 -2 V
2 -1 V
2 -1 V
2 -1 V
2 -2 V
2 -1 V
2 -1 V
2 -2 V
2 -1 V
3 -1 V
2 -2 V
2 -1 V
2 -1 V
2 -2 V
2 -1 V
2 -1 V
2 -2 V
2 -1 V
2 -1 V
2 -2 V
2 -1 V
2 -2 V
2 -1 V
2 -1 V
3 -2 V
2 -1 V
2 -1 V
2 -2 V
2 -1 V
2 -2 V
2 -1 V
2 -1 V
2 -2 V
2 -1 V
2 -2 V
2 -1 V
2 -2 V
2 -1 V
3 -1 V
2 -2 V
2 -1 V
2 -2 V
2 -1 V
2 -2 V
2 -1 V
2 -2 V
2 -1 V
2 -1 V
2 -2 V
2 -1 V
2 -2 V
2 -1 V
3 -2 V
2 -1 V
2 -2 V
2 -1 V
2 -2 V
currentpoint stroke M
2 -1 V
2 -2 V
2 -1 V
2 0 V
2 0 V
2 0 V
2 0 V
2 0 V
2 0 V
2 0 V
3 0 V
2 0 V
2 0 V
2 0 V
2 0 V
2 0 V
2 0 V
2 0 V
2 0 V
2 0 V
2 0 V
2 0 V
2 0 V
2 0 V
3 0 V
2 0 V
2 0 V
2 0 V
2 0 V
2 0 V
2 0 V
2 0 V
2 0 V
2 0 V
2 0 V
2 0 V
2 0 V
2 0 V
3 0 V
2 0 V
2 0 V
2 0 V
2 0 V
2 0 V
2 0 V
2 0 V
2 0 V
2 0 V
2 0 V
2 0 V
2 0 V
2 0 V
3 0 V
2 0 V
2 0 V
2 0 V
2 0 V
2 0 V
2 0 V
2 0 V
2 0 V
2 0 V
2 0 V
2 0 V
2 0 V
2 0 V
2 0 V
3 0 V
2 0 V
2 0 V
2 0 V
2 0 V
2 0 V
2 0 V
2 0 V
2 0 V
2 0 V
2 0 V
2 0 V
2 0 V
2 0 V
3 0 V
2 0 V
2 0 V
2 0 V
2 0 V
2 0 V
2 0 V
2 0 V
2 0 V
2 0 V
2 0 V
2 0 V
2 0 V
2 0 V
3 0 V
2 0 V
2 0 V
2 0 V
2 0 V
2 0 V
2 0 V
2 0 V
2 0 V
2 0 V
2 0 V
2 0 V
2 0 V
2 0 V
2 0 V
3 0 V
2 0 V
2 0 V
2 0 V
2 0 V
2 0 V
2 0 V
2 0 V
2 0 V
2 0 V
2 0 V
2 0 V
2 0 V
2 0 V
3 0 V
2 0 V
2 0 V
2 0 V
2 0 V
2 0 V
2 0 V
2 0 V
2 0 V
2 0 V
2 0 V
2 0 V
2 0 V
2 0 V
3 0 V
2 0 V
2 0 V
2 0 V
2 0 V
2 0 V
2 0 V
2 0 V
2 0 V
2 0 V
2 0 V
2 0 V
2 0 V
2 0 V
3 0 V
2 0 V
2 0 V
2 0 V
2 0 V
2 0 V
2 0 V
2 0 V
2 0 V
2 0 V
2 0 V
2 0 V
2 0 V
2 0 V
2 0 V
3 0 V
2 0 V
2 0 V
2 0 V
2 0 V
2 0 V
2 0 V
2 0 V
2 0 V
2 0 V
2 0 V
2 0 V
2 0 V
2 0 V
3 0 V
2 0 V
2 0 V
2 0 V
2 0 V
2 0 V
2 0 V
2 0 V
2 0 V
2 0 V
2 0 V
2 0 V
2 0 V
2 0 V
3 0 V
2 0 V
2 0 V
2 0 V
2 0 V
2 0 V
2 0 V
2 0 V
2 0 V
2 0 V
2 0 V
2 0 V
2 0 V
2 0 V
2 0 V
3 0 V
2 0 V
2 0 V
2 0 V
2 0 V
2 0 V
2 0 V
2 0 V
2 0 V
2 0 V
2 0 V
2 0 V
2 0 V
2 0 V
3 0 V
2 0 V
2 0 V
2 0 V
2 0 V
2 0 V
2 0 V
2 0 V
2 0 V
2 0 V
2 0 V
2 0 V
2 0 V
2 0 V
3 0 V
2 0 V
2 0 V
2 0 V
2 0 V
2 0 V
2 0 V
2 0 V
2 0 V
2 0 V
2 0 V
2 0 V
2 0 V
2 0 V
3 0 V
2 0 V
2 0 V
2 0 V
2 0 V
2 0 V
2 0 V
2 0 V
2 0 V
2 0 V
2 0 V
2 0 V
2 0 V
2 0 V
2 0 V
3 0 V
2 0 V
2 0 V
2 0 V
2 0 V
2 0 V
2 0 V
2 0 V
2 0 V
2 0 V
2 0 V
2 0 V
2 0 V
2 0 V
3 0 V
2 0 V
2 0 V
2 0 V
2 0 V
2 0 V
2 0 V
2 0 V
2 0 V
2 0 V
2 0 V
2 0 V
2 0 V
2 0 V
3 0 V
2 0 V
2 0 V
2 0 V
2 0 V
2 0 V
2 0 V
2 0 V
2 0 V
2 0 V
2 0 V
2 0 V
2 0 V
2 0 V
2 0 V
3 0 V
2 0 V
2 0 V
2 0 V
2 0 V
2 0 V
2 0 V
2 0 V
2 0 V
2 0 V
2 0 V
2 0 V
2 0 V
2 0 V
3 0 V
2 0 V
2 0 V
2 0 V
2 0 V
2 0 V
2 0 V
2 0 V
2 0 V
2 0 V
2 0 V
2 0 V
2 0 V
2 0 V
3 0 V
2 0 V
2 0 V
2 0 V
2 0 V
2 0 V
2 0 V
2 0 V
2 0 V
2 0 V
2 0 V
2 0 V
2 0 V
2 0 V
3 0 V
2 0 V
2 0 V
2 0 V
2 0 V
2 0 V
2 0 V
2 0 V
2 0 V
2 0 V
2 0 V
2 0 V
2 0 V
2 0 V
2 0 V
3 0 V
2 0 V
2 0 V
2 0 V
2 0 V
2 0 V
2 0 V
2 0 V
2 0 V
2 0 V
2 0 V
2 0 V
2 0 V
2 0 V
3 0 V
2 0 V
2 0 V
2 0 V
2 0 V
2 0 V
2 0 V
2 0 V
2 0 V
2 0 V
2 0 V
2 0 V
2 0 V
2 0 V
3 0 V
2 0 V
2 0 V
2 0 V
2 0 V
currentpoint stroke M
2 0 V
2 0 V
2 0 V
2 0 V
2 0 V
2 0 V
2 0 V
2 0 V
2 0 V
2 0 V
3 0 V
2 0 V
2 0 V
2 0 V
2 0 V
2 0 V
2 0 V
2 0 V
2 0 V
2 0 V
2 0 V
2 0 V
2 0 V
2 0 V
3 0 V
2 0 V
2 0 V
2 0 V
2 0 V
2 0 V
2 0 V
2 0 V
2 0 V
2 0 V
2 0 V
2 0 V
2 0 V
2 0 V
3 0 V
2 0 V
2 0 V
2 0 V
2 0 V
2 0 V
2 0 V
2 0 V
2 0 V
2 0 V
2 0 V
2 0 V
2 0 V
2 0 V
3 0 V
2 0 V
2 0 V
2 0 V
2 0 V
2 0 V
2 0 V
2 0 V
2 0 V
2 0 V
2 0 V
2 0 V
2 0 V
2 0 V
2 0 V
3 0 V
2 0 V
2 0 V
2 0 V
2 0 V
2 0 V
2 0 V
2 0 V
2 0 V
2 0 V
2 0 V
2 0 V
2 0 V
2 0 V
3 0 V
2 0 V
2 0 V
2 0 V
2 0 V
2 0 V
2 0 V
2 0 V
2 0 V
2 0 V
2 0 V
2 0 V
2 0 V
2 0 V
3 0 V
2 0 V
2 0 V
2 0 V
2 0 V
2 0 V
2 0 V
2 0 V
2 0 V
2 0 V
2 0 V
2 0 V
2 0 V
2 0 V
2 0 V
3 0 V
2 0 V
2 0 V
2 0 V
2 0 V
2 0 V
2 0 V
2 0 V
2 0 V
2 0 V
2 0 V
2 0 V
2 0 V
2 0 V
3 0 V
2 0 V
2 0 V
2 0 V
2 0 V
2 0 V
2 0 V
2 0 V
2 0 V
2 0 V
2 0 V
2 0 V
2 0 V
2 0 V
3 0 V
2 0 V
2 0 V
2 0 V
2 0 V
2 0 V
2 0 V
2 0 V
2 0 V
2 0 V
2 0 V
2 0 V
2 0 V
2 0 V
3 0 V
2 0 V
2 0 V
2 0 V
2 0 V
2 0 V
2 0 V
2 0 V
2 0 V
2 0 V
2 0 V
2 0 V
2 0 V
2 0 V
2 0 V
3 0 V
2 0 V
2 0 V
2 0 V
2 0 V
2 0 V
2 0 V
2 0 V
2 0 V
2 0 V
2 0 V
2 0 V
2 0 V
2 0 V
3 0 V
2 0 V
2 0 V
2 0 V
2 0 V
2 0 V
2 0 V
2 0 V
2 0 V
2 0 V
2 0 V
2 0 V
2 0 V
2 0 V
3 0 V
2 0 V
2 0 V
2 0 V
2 0 V
2 0 V
2 0 V
2 0 V
stroke
grestore
end
showpage
}}%
\put(1973,1747){\makebox(0,0)[r]{$kh/\pi=0.4$}}%
\put(1973,1847){\makebox(0,0)[r]{$kh/\pi=0.2$}}%
\put(1973,1947){\makebox(0,0)[r]{$kh/\pi=0.1$}}%
\put(1956,847){\makebox(0,0)[l]{(d)}}%
\put(331,1858){\makebox(0,0)[l]{$Q^{\parallel}(k,q)$}}%
\put(2008,281){\makebox(0,0)[l]{$qh/\pi$}}%
\put(2370,100){\makebox(0,0){ 1}}%
\put(1956,100){\makebox(0,0){ 0.8}}%
\put(1542,100){\makebox(0,0){ 0.6}}%
\put(1128,100){\makebox(0,0){ 0.4}}%
\put(714,100){\makebox(0,0){ 0.2}}%
\put(300,100){\makebox(0,0){ 0}}%
\put(250,2060){\makebox(0,0)[r]{ 20}}%
\put(250,1656){\makebox(0,0)[r]{ 15}}%
\put(250,1251){\makebox(0,0)[r]{ 10}}%
\put(250,847){\makebox(0,0)[r]{ 5}}%
\put(250,443){\makebox(0,0)[r]{ 0}}%
\end{picture}%
\endgroup
 

%% file: ssp_dord_ort_a1_o=0.984.tex
\begin{picture}(0,0)%
\includegraphics{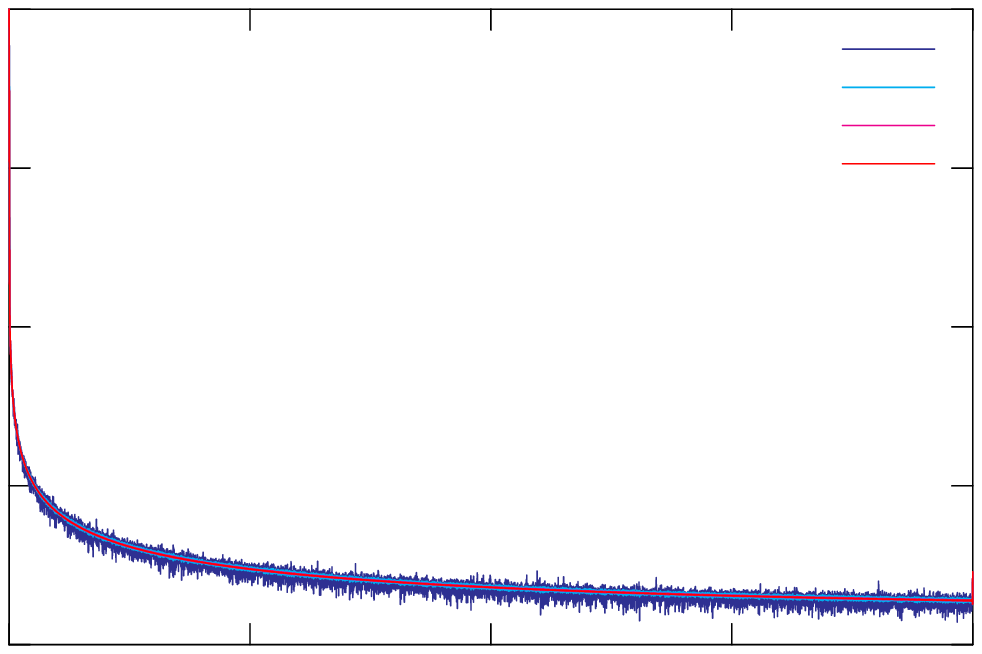}%
\end{picture}%
\setlength{\unitlength}{0.0200bp}%
\begin{picture}(18000,10800)(0,0)%
\put(3025,1100){\makebox(0,0)[r]{\strut{}$10^{-4}$}}%
\put(3025,3387){\makebox(0,0)[r]{\strut{}$10^{-3}$}}%
\put(3025,5675){\makebox(0,0)[r]{\strut{}$10^{-2}$}}%
\put(3025,7962){\makebox(0,0)[r]{\strut{}$10^{-1}$}}%
\put(3025,10250){\makebox(0,0)[r]{\strut{}$1$}}%
\put(3300,550){\makebox(0,0){\strut{}1}}%
\put(6768,550){\makebox(0,0){\strut{}2500}}%
\put(10237,550){\makebox(0,0){\strut{}5000}}%
\put(13706,550){\makebox(0,0){\strut{}7500}}%
\put(17175,550){\makebox(0,0){\strut{}10000}}%
\put(15094,593){\makebox(0,0)[l]{\strut{}$x/h$}}%
\put(3646,8873){\makebox(0,0)[l]{\strut{}$\vert {\mathcal F}_k(x) \vert$}}%
\put(15025,9675){\makebox(0,0)[r]{\strut{}$A=0.16$}}%
\put(15025,9125){\makebox(0,0)[r]{\strut{}$A=0.04$}}%
\put(15025,8575){\makebox(0,0)[r]{\strut{}$A=0.01$}}%
\put(15025,8025){\makebox(0,0)[r]{\strut{}$A=0$}}%
\end{picture}%
 

%% file: qspect_a1.tex
\begingroup%
  \makeatletter%
  \newcommand{\GNUPLOTspecial}{%
    \@sanitize\catcode`\%=14\relax\special}%
  \setlength{\unitlength}{0.1bp}%
{\GNUPLOTspecial{!
/gnudict 256 dict def
gnudict begin
/Color true def
/Solid true def
/gnulinewidth 5.000 def
/userlinewidth gnulinewidth def
/vshift -33 def
/dl {10 mul} def
/hpt_ 31.5 def
/vpt_ 31.5 def
/hpt hpt_ def
/vpt vpt_ def
/M {moveto} bind def
/L {lineto} bind def
/R {rmoveto} bind def
/V {rlineto} bind def
/vpt2 vpt 2 mul def
/hpt2 hpt 2 mul def
/Lshow { currentpoint stroke M
  0 vshift R show } def
/Rshow { currentpoint stroke M
  dup stringwidth pop neg vshift R show } def
/Cshow { currentpoint stroke M
  dup stringwidth pop -2 div vshift R show } def
/UP { dup vpt_ mul /vpt exch def hpt_ mul /hpt exch def
  /hpt2 hpt 2 mul def /vpt2 vpt 2 mul def } def
/DL { Color {setrgbcolor Solid {pop []} if 0 setdash }
 {pop pop pop Solid {pop []} if 0 setdash} ifelse } def
/BL { stroke userlinewidth 2 mul setlinewidth } def
/AL { stroke userlinewidth 2 div setlinewidth } def
/UL { dup gnulinewidth mul /userlinewidth exch def
      dup 1 lt {pop 1} if 10 mul /udl exch def } def
/PL { stroke userlinewidth setlinewidth } def
/LTb { BL [] 0 0 0 DL } def
/LTa { AL [1 udl mul 2 udl mul] 0 setdash 0 0 0 setrgbcolor } def
/LT0 { PL [] 1 0 0 DL } def
/LT1 { PL [4 dl 2 dl] 0 1 0 DL } def
/LT2 { PL [2 dl 3 dl] 0 0 1 DL } def
/LT3 { PL [1 dl 1.5 dl] 1 0 1 DL } def
/LT4 { PL [5 dl 2 dl 1 dl 2 dl] 0 1 1 DL } def
/LT5 { PL [4 dl 3 dl 1 dl 3 dl] 1 1 0 DL } def
/LT6 { PL [2 dl 2 dl 2 dl 4 dl] 0 0 0 DL } def
/LT7 { PL [2 dl 2 dl 2 dl 2 dl 2 dl 4 dl] 1 0.3 0 DL } def
/LT8 { PL [2 dl 2 dl 2 dl 2 dl 2 dl 2 dl 2 dl 4 dl] 0.5 0.5 0.5 DL } def
/Pnt { stroke [] 0 setdash
   gsave 1 setlinecap M 0 0 V stroke grestore } def
/Dia { stroke [] 0 setdash 2 copy vpt add M
  hpt neg vpt neg V hpt vpt neg V
  hpt vpt V hpt neg vpt V closepath stroke
  Pnt } def
/Pls { stroke [] 0 setdash vpt sub M 0 vpt2 V
  currentpoint stroke M
  hpt neg vpt neg R hpt2 0 V stroke
  } def
/Box { stroke [] 0 setdash 2 copy exch hpt sub exch vpt add M
  0 vpt2 neg V hpt2 0 V 0 vpt2 V
  hpt2 neg 0 V closepath stroke
  Pnt } def
/Crs { stroke [] 0 setdash exch hpt sub exch vpt add M
  hpt2 vpt2 neg V currentpoint stroke M
  hpt2 neg 0 R hpt2 vpt2 V stroke } def
/TriU { stroke [] 0 setdash 2 copy vpt 1.12 mul add M
  hpt neg vpt -1.62 mul V
  hpt 2 mul 0 V
  hpt neg vpt 1.62 mul V closepath stroke
  Pnt  } def
/Star { 2 copy Pls Crs } def
/BoxF { stroke [] 0 setdash exch hpt sub exch vpt add M
  0 vpt2 neg V  hpt2 0 V  0 vpt2 V
  hpt2 neg 0 V  closepath fill } def
/TriUF { stroke [] 0 setdash vpt 1.12 mul add M
  hpt neg vpt -1.62 mul V
  hpt 2 mul 0 V
  hpt neg vpt 1.62 mul V closepath fill } def
/TriD { stroke [] 0 setdash 2 copy vpt 1.12 mul sub M
  hpt neg vpt 1.62 mul V
  hpt 2 mul 0 V
  hpt neg vpt -1.62 mul V closepath stroke
  Pnt  } def
/TriDF { stroke [] 0 setdash vpt 1.12 mul sub M
  hpt neg vpt 1.62 mul V
  hpt 2 mul 0 V
  hpt neg vpt -1.62 mul V closepath fill} def
/DiaF { stroke [] 0 setdash vpt add M
  hpt neg vpt neg V hpt vpt neg V
  hpt vpt V hpt neg vpt V closepath fill } def
/Pent { stroke [] 0 setdash 2 copy gsave
  translate 0 hpt M 4 {72 rotate 0 hpt L} repeat
  closepath stroke grestore Pnt } def
/PentF { stroke [] 0 setdash gsave
  translate 0 hpt M 4 {72 rotate 0 hpt L} repeat
  closepath fill grestore } def
/Circle { stroke [] 0 setdash 2 copy
  hpt 0 360 arc stroke Pnt } def
/CircleF { stroke [] 0 setdash hpt 0 360 arc fill } def
/C0 { BL [] 0 setdash 2 copy moveto vpt 90 450  arc } bind def
/C1 { BL [] 0 setdash 2 copy        moveto
       2 copy  vpt 0 90 arc closepath fill
               vpt 0 360 arc closepath } bind def
/C2 { BL [] 0 setdash 2 copy moveto
       2 copy  vpt 90 180 arc closepath fill
               vpt 0 360 arc closepath } bind def
/C3 { BL [] 0 setdash 2 copy moveto
       2 copy  vpt 0 180 arc closepath fill
               vpt 0 360 arc closepath } bind def
/C4 { BL [] 0 setdash 2 copy moveto
       2 copy  vpt 180 270 arc closepath fill
               vpt 0 360 arc closepath } bind def
/C5 { BL [] 0 setdash 2 copy moveto
       2 copy  vpt 0 90 arc
       2 copy moveto
       2 copy  vpt 180 270 arc closepath fill
               vpt 0 360 arc } bind def
/C6 { BL [] 0 setdash 2 copy moveto
      2 copy  vpt 90 270 arc closepath fill
              vpt 0 360 arc closepath } bind def
/C7 { BL [] 0 setdash 2 copy moveto
      2 copy  vpt 0 270 arc closepath fill
              vpt 0 360 arc closepath } bind def
/C8 { BL [] 0 setdash 2 copy moveto
      2 copy vpt 270 360 arc closepath fill
              vpt 0 360 arc closepath } bind def
/C9 { BL [] 0 setdash 2 copy moveto
      2 copy  vpt 270 450 arc closepath fill
              vpt 0 360 arc closepath } bind def
/C10 { BL [] 0 setdash 2 copy 2 copy moveto vpt 270 360 arc closepath fill
       2 copy moveto
       2 copy vpt 90 180 arc closepath fill
               vpt 0 360 arc closepath } bind def
/C11 { BL [] 0 setdash 2 copy moveto
       2 copy  vpt 0 180 arc closepath fill
       2 copy moveto
       2 copy  vpt 270 360 arc closepath fill
               vpt 0 360 arc closepath } bind def
/C12 { BL [] 0 setdash 2 copy moveto
       2 copy  vpt 180 360 arc closepath fill
               vpt 0 360 arc closepath } bind def
/C13 { BL [] 0 setdash  2 copy moveto
       2 copy  vpt 0 90 arc closepath fill
       2 copy moveto
       2 copy  vpt 180 360 arc closepath fill
               vpt 0 360 arc closepath } bind def
/C14 { BL [] 0 setdash 2 copy moveto
       2 copy  vpt 90 360 arc closepath fill
               vpt 0 360 arc } bind def
/C15 { BL [] 0 setdash 2 copy vpt 0 360 arc closepath fill
               vpt 0 360 arc closepath } bind def
/Rec   { newpath 4 2 roll moveto 1 index 0 rlineto 0 exch rlineto
       neg 0 rlineto closepath } bind def
/Square { dup Rec } bind def
/Bsquare { vpt sub exch vpt sub exch vpt2 Square } bind def
/S0 { BL [] 0 setdash 2 copy moveto 0 vpt rlineto BL Bsquare } bind def
/S1 { BL [] 0 setdash 2 copy vpt Square fill Bsquare } bind def
/S2 { BL [] 0 setdash 2 copy exch vpt sub exch vpt Square fill Bsquare } bind def
/S3 { BL [] 0 setdash 2 copy exch vpt sub exch vpt2 vpt Rec fill Bsquare } bind def
/S4 { BL [] 0 setdash 2 copy exch vpt sub exch vpt sub vpt Square fill Bsquare } bind def
/S5 { BL [] 0 setdash 2 copy 2 copy vpt Square fill
       exch vpt sub exch vpt sub vpt Square fill Bsquare } bind def
/S6 { BL [] 0 setdash 2 copy exch vpt sub exch vpt sub vpt vpt2 Rec fill Bsquare } bind def
/S7 { BL [] 0 setdash 2 copy exch vpt sub exch vpt sub vpt vpt2 Rec fill
       2 copy vpt Square fill
       Bsquare } bind def
/S8 { BL [] 0 setdash 2 copy vpt sub vpt Square fill Bsquare } bind def
/S9 { BL [] 0 setdash 2 copy vpt sub vpt vpt2 Rec fill Bsquare } bind def
/S10 { BL [] 0 setdash 2 copy vpt sub vpt Square fill 2 copy exch vpt sub exch vpt Square fill
       Bsquare } bind def
/S11 { BL [] 0 setdash 2 copy vpt sub vpt Square fill 2 copy exch vpt sub exch vpt2 vpt Rec fill
       Bsquare } bind def
/S12 { BL [] 0 setdash 2 copy exch vpt sub exch vpt sub vpt2 vpt Rec fill Bsquare } bind def
/S13 { BL [] 0 setdash 2 copy exch vpt sub exch vpt sub vpt2 vpt Rec fill
       2 copy vpt Square fill Bsquare } bind def
/S14 { BL [] 0 setdash 2 copy exch vpt sub exch vpt sub vpt2 vpt Rec fill
       2 copy exch vpt sub exch vpt Square fill Bsquare } bind def
/S15 { BL [] 0 setdash 2 copy Bsquare fill Bsquare } bind def
/D0 { gsave translate 45 rotate 0 0 S0 stroke grestore } bind def
/D1 { gsave translate 45 rotate 0 0 S1 stroke grestore } bind def
/D2 { gsave translate 45 rotate 0 0 S2 stroke grestore } bind def
/D3 { gsave translate 45 rotate 0 0 S3 stroke grestore } bind def
/D4 { gsave translate 45 rotate 0 0 S4 stroke grestore } bind def
/D5 { gsave translate 45 rotate 0 0 S5 stroke grestore } bind def
/D6 { gsave translate 45 rotate 0 0 S6 stroke grestore } bind def
/D7 { gsave translate 45 rotate 0 0 S7 stroke grestore } bind def
/D8 { gsave translate 45 rotate 0 0 S8 stroke grestore } bind def
/D9 { gsave translate 45 rotate 0 0 S9 stroke grestore } bind def
/D10 { gsave translate 45 rotate 0 0 S10 stroke grestore } bind def
/D11 { gsave translate 45 rotate 0 0 S11 stroke grestore } bind def
/D12 { gsave translate 45 rotate 0 0 S12 stroke grestore } bind def
/D13 { gsave translate 45 rotate 0 0 S13 stroke grestore } bind def
/D14 { gsave translate 45 rotate 0 0 S14 stroke grestore } bind def
/D15 { gsave translate 45 rotate 0 0 S15 stroke grestore } bind def
/DiaE { stroke [] 0 setdash vpt add M
  hpt neg vpt neg V hpt vpt neg V
  hpt vpt V hpt neg vpt V closepath stroke } def
/BoxE { stroke [] 0 setdash exch hpt sub exch vpt add M
  0 vpt2 neg V hpt2 0 V 0 vpt2 V
  hpt2 neg 0 V closepath stroke } def
/TriUE { stroke [] 0 setdash vpt 1.12 mul add M
  hpt neg vpt -1.62 mul V
  hpt 2 mul 0 V
  hpt neg vpt 1.62 mul V closepath stroke } def
/TriDE { stroke [] 0 setdash vpt 1.12 mul sub M
  hpt neg vpt 1.62 mul V
  hpt 2 mul 0 V
  hpt neg vpt -1.62 mul V closepath stroke } def
/PentE { stroke [] 0 setdash gsave
  translate 0 hpt M 4 {72 rotate 0 hpt L} repeat
  closepath stroke grestore } def
/CircE { stroke [] 0 setdash 
  hpt 0 360 arc stroke } def
/Opaque { gsave closepath 1 setgray fill grestore 0 setgray closepath } def
/DiaW { stroke [] 0 setdash vpt add M
  hpt neg vpt neg V hpt vpt neg V
  hpt vpt V hpt neg vpt V Opaque stroke } def
/BoxW { stroke [] 0 setdash exch hpt sub exch vpt add M
  0 vpt2 neg V hpt2 0 V 0 vpt2 V
  hpt2 neg 0 V Opaque stroke } def
/TriUW { stroke [] 0 setdash vpt 1.12 mul add M
  hpt neg vpt -1.62 mul V
  hpt 2 mul 0 V
  hpt neg vpt 1.62 mul V Opaque stroke } def
/TriDW { stroke [] 0 setdash vpt 1.12 mul sub M
  hpt neg vpt 1.62 mul V
  hpt 2 mul 0 V
  hpt neg vpt -1.62 mul V Opaque stroke } def
/PentW { stroke [] 0 setdash gsave
  translate 0 hpt M 4 {72 rotate 0 hpt L} repeat
  Opaque stroke grestore } def
/CircW { stroke [] 0 setdash 
  hpt 0 360 arc Opaque stroke } def
/BoxFill { gsave Rec 1 setgray fill grestore } def
/Symbol-Oblique /Symbol findfont [1 0 .167 1 0 0] makefont
dup length dict begin {1 index /FID eq {pop pop} {def} ifelse} forall
currentdict end definefont pop
end
}}%
\begin{picture}(3600,2160)(0,0)%
{\GNUPLOTspecial{"
gnudict begin
gsave
0 0 translate
0.100 0.100 scale
0 setgray
newpath
1.000 UL
LTb
600 200 M
63 0 V
2787 0 R
-63 0 V
600 613 M
63 0 V
2787 0 R
-63 0 V
600 1027 M
63 0 V
2787 0 R
-63 0 V
600 1440 M
63 0 V
2787 0 R
-63 0 V
600 1853 M
63 0 V
2787 0 R
-63 0 V
600 200 M
0 63 V
0 1797 R
0 -63 V
1170 200 M
0 63 V
0 1797 R
0 -63 V
1740 200 M
0 63 V
0 1797 R
0 -63 V
2310 200 M
0 63 V
0 1797 R
0 -63 V
2880 200 M
0 63 V
0 1797 R
0 -63 V
3450 200 M
0 63 V
0 1797 R
0 -63 V
1.000 UL
LTb
600 200 M
2850 0 V
0 1860 V
-2850 0 V
600 200 L
1.000 UL
LT1
3103 1947 M
247 0 V
600 903 M
3 0 V
3 0 V
3 0 V
2 0 V
3 0 V
3 0 V
3 1 V
3 -1 V
3 1 V
2 -1 V
3 0 V
3 0 V
3 0 V
3 0 V
3 0 V
3 1 V
2 0 V
3 0 V
3 0 V
3 1 V
3 0 V
3 0 V
3 0 V
2 0 V
3 0 V
3 1 V
3 0 V
3 0 V
3 0 V
2 0 V
3 0 V
3 0 V
3 1 V
3 0 V
3 1 V
3 0 V
2 0 V
3 0 V
3 0 V
3 1 V
3 0 V
3 0 V
3 1 V
2 0 V
3 0 V
3 1 V
3 0 V
3 0 V
3 1 V
3 0 V
2 1 V
3 0 V
3 0 V
3 1 V
3 1 V
3 0 V
2 0 V
3 0 V
3 1 V
3 0 V
3 0 V
3 1 V
3 1 V
2 0 V
3 1 V
3 0 V
3 0 V
3 1 V
3 0 V
3 1 V
2 0 V
3 0 V
3 1 V
3 0 V
3 1 V
3 0 V
2 1 V
3 0 V
3 1 V
3 0 V
3 1 V
3 1 V
3 1 V
2 0 V
3 1 V
3 1 V
3 0 V
3 1 V
3 0 V
3 1 V
2 0 V
3 1 V
3 1 V
3 0 V
3 1 V
3 1 V
2 1 V
3 1 V
3 0 V
3 1 V
3 1 V
3 1 V
3 0 V
2 1 V
3 0 V
3 1 V
3 1 V
3 1 V
3 1 V
3 0 V
2 1 V
3 1 V
3 1 V
3 1 V
3 1 V
3 0 V
2 1 V
3 1 V
3 0 V
3 2 V
3 0 V
3 1 V
3 1 V
2 1 V
3 1 V
3 1 V
3 1 V
3 1 V
3 1 V
3 1 V
2 1 V
3 1 V
3 0 V
3 2 V
3 0 V
3 1 V
2 1 V
3 1 V
3 1 V
3 1 V
3 1 V
3 1 V
3 2 V
2 1 V
3 1 V
3 1 V
3 1 V
3 1 V
3 1 V
3 2 V
2 1 V
3 1 V
3 1 V
3 1 V
3 2 V
3 1 V
2 1 V
3 1 V
3 2 V
3 1 V
3 2 V
3 1 V
3 2 V
2 1 V
3 2 V
3 1 V
3 2 V
3 1 V
3 2 V
3 2 V
2 2 V
3 2 V
3 1 V
3 2 V
3 2 V
3 2 V
2 2 V
3 1 V
3 3 V
3 1 V
3 3 V
3 2 V
3 2 V
2 3 V
3 2 V
3 3 V
3 3 V
3 3 V
3 3 V
3 3 V
2 3 V
3 4 V
3 4 V
3 4 V
3 4 V
3 5 V
2 5 V
3 3 V
3 0 V
3 -102 V
3 84 V
3 145 V
3 -245 V
2 -57 V
3 -20 V
3 -26 V
3 -48 V
3 7 V
3 -4 V
3 -23 V
2 -28 V
3 -42 V
3 10 V
3 6 V
3 12 V
3 -17 V
2 -8 V
3 4 V
3 -1 V
3 -31 V
3 -11 V
3 6 V
3 16 V
2 2 V
3 -7 V
3 -6 V
3 9 V
3 16 V
3 11 V
3 15 V
2 10 V
3 -19 V
3 -30 V
3 42 V
3 -23 V
3 -17 V
2 -26 V
3 13 V
3 24 V
3 -26 V
3 19 V
3 -25 V
3 28 V
2 7 V
3 -15 V
3 -18 V
3 -6 V
3 15 V
3 23 V
3 -4 V
2 -3 V
3 -7 V
3 12 V
3 -7 V
3 4 V
3 -9 V
2 1 V
3 -13 V
3 23 V
3 26 V
3 -19 V
3 26 V
3 -20 V
2 4 V
3 5 V
3 -10 V
3 -13 V
3 -10 V
3 8 V
3 35 V
2 13 V
3 8 V
3 6 V
3 6 V
3 8 V
3 -10 V
2 -6 V
3 9 V
3 -27 V
3 -9 V
3 -26 V
3 2 V
3 13 V
2 -8 V
3 17 V
3 -31 V
3 20 V
3 20 V
3 -21 V
3 5 V
2 6 V
3 -9 V
3 -13 V
3 6 V
3 -2 V
3 17 V
2 1 V
3 0 V
3 -13 V
3 1 V
3 23 V
3 -23 V
3 -8 V
2 -13 V
3 16 V
3 -11 V
3 24 V
3 -8 V
3 -3 V
3 -6 V
2 -14 V
3 2 V
3 -21 V
3 8 V
3 -1 V
3 0 V
2 14 V
3 -2 V
3 -12 V
3 -15 V
3 6 V
3 -1 V
3 26 V
2 27 V
3 -14 V
3 -1 V
3 16 V
3 -10 V
3 15 V
3 17 V
2 9 V
3 17 V
3 -27 V
3 12 V
3 11 V
3 10 V
2 -13 V
3 4 V
3 11 V
3 -6 V
3 -14 V
3 3 V
3 4 V
2 17 V
3 -2 V
3 -4 V
3 -20 V
3 2 V
3 -10 V
3 -1 V
2 1 V
3 7 V
3 16 V
3 -7 V
3 3 V
3 14 V
2 39 V
3 -12 V
3 21 V
3 10 V
3 12 V
3 8 V
3 9 V
2 -5 V
3 -2 V
3 -1 V
3 -19 V
3 5 V
3 22 V
3 -28 V
2 13 V
3 -3 V
3 2 V
3 -4 V
3 10 V
3 -9 V
2 -25 V
3 17 V
3 9 V
3 -6 V
3 12 V
3 -8 V
3 22 V
2 15 V
3 -6 V
3 0 V
3 -19 V
3 9 V
3 13 V
3 -8 V
2 21 V
3 -8 V
3 -16 V
3 6 V
3 0 V
3 -16 V
2 7 V
currentpoint stroke M
3 10 V
3 -9 V
3 11 V
3 8 V
3 -1 V
3 6 V
2 -5 V
3 -8 V
3 6 V
3 12 V
3 12 V
3 -7 V
3 15 V
2 -9 V
3 27 V
3 -6 V
3 30 V
3 -11 V
3 2 V
2 18 V
3 -13 V
3 15 V
3 4 V
3 -1 V
3 9 V
3 1 V
2 -12 V
3 1 V
3 22 V
3 -6 V
3 25 V
3 -19 V
3 22 V
2 -13 V
3 8 V
3 -18 V
3 9 V
3 3 V
3 13 V
2 -1 V
3 19 V
3 2 V
3 -13 V
3 -4 V
3 -3 V
3 34 V
2 -1 V
3 37 V
3 6 V
3 15 V
3 18 V
3 -10 V
3 33 V
2 -19 V
3 11 V
3 22 V
3 23 V
3 -1 V
3 23 V
2 9 V
3 18 V
3 -29 V
3 29 V
3 -6 V
3 8 V
3 31 V
2 7 V
3 40 V
3 16 V
3 35 V
3 4 V
3 23 V
3 -1 V
2 3 V
3 -24 V
3 -3 V
3 -29 V
3 -8 V
3 -9 V
2 6 V
3 -43 V
3 -4 V
3 -45 V
3 1 V
3 -21 V
3 -29 V
2 -23 V
3 -20 V
3 -30 V
3 -11 V
3 -35 V
3 -6 V
3 -18 V
2 7 V
3 9 V
3 4 V
3 -15 V
3 9 V
3 14 V
2 -12 V
3 13 V
3 -19 V
3 2 V
3 -8 V
3 9 V
3 -1 V
2 -5 V
3 -7 V
3 -10 V
3 -21 V
3 -19 V
3 16 V
3 -34 V
2 -28 V
3 12 V
3 -28 V
3 2 V
3 -21 V
3 -2 V
2 -4 V
3 27 V
3 -8 V
3 21 V
3 -57 V
3 2 V
3 -8 V
2 5 V
3 19 V
3 28 V
3 -29 V
3 2 V
3 -2 V
3 4 V
2 3 V
3 5 V
3 -18 V
3 5 V
3 -4 V
3 -19 V
2 -21 V
3 -4 V
3 -3 V
3 -16 V
3 16 V
3 5 V
3 5 V
2 6 V
3 25 V
3 -1 V
3 -14 V
3 27 V
3 -6 V
3 3 V
2 -11 V
3 -3 V
3 6 V
3 0 V
3 -33 V
3 -7 V
2 -11 V
3 -35 V
3 -14 V
3 -14 V
3 -5 V
3 -7 V
3 -9 V
2 -5 V
3 4 V
3 -9 V
3 0 V
3 0 V
3 23 V
3 -9 V
2 32 V
3 -7 V
3 -15 V
3 9 V
3 15 V
3 -6 V
2 -6 V
3 -2 V
3 -1 V
3 -12 V
3 8 V
3 -11 V
3 -5 V
2 0 V
3 -19 V
3 -16 V
3 -5 V
3 -21 V
3 -9 V
3 -12 V
2 -10 V
3 2 V
3 -2 V
3 25 V
3 0 V
3 24 V
2 -9 V
3 17 V
3 -17 V
3 22 V
3 -20 V
3 -10 V
3 18 V
2 6 V
3 22 V
3 -13 V
3 -7 V
3 5 V
3 5 V
3 -3 V
2 5 V
3 -8 V
3 1 V
3 7 V
3 15 V
3 -26 V
2 -8 V
3 -1 V
3 -14 V
3 -7 V
3 15 V
3 10 V
3 34 V
2 -22 V
3 26 V
3 -6 V
3 -2 V
3 1 V
3 -8 V
3 12 V
2 32 V
3 -21 V
3 24 V
3 8 V
3 7 V
3 7 V
2 -10 V
3 -13 V
3 -21 V
3 -5 V
3 -12 V
3 -15 V
3 5 V
2 -27 V
3 14 V
3 -5 V
3 35 V
3 -16 V
3 -3 V
3 -2 V
2 7 V
3 -11 V
3 35 V
3 -30 V
3 15 V
3 -17 V
2 2 V
3 -26 V
3 24 V
3 -19 V
3 15 V
3 -1 V
3 19 V
2 -19 V
3 -8 V
3 6 V
3 14 V
3 2 V
3 9 V
3 -7 V
2 4 V
3 -13 V
3 24 V
3 -7 V
3 4 V
3 2 V
2 -3 V
3 16 V
3 -11 V
3 -26 V
3 -7 V
3 26 V
3 -33 V
2 21 V
3 0 V
3 -1 V
3 -2 V
3 3 V
3 16 V
3 -2 V
2 -11 V
3 6 V
3 -19 V
3 0 V
3 15 V
3 -32 V
2 -6 V
3 11 V
3 -29 V
3 13 V
3 17 V
3 0 V
3 -13 V
2 9 V
3 15 V
3 -22 V
3 -12 V
3 3 V
3 14 V
3 -13 V
2 -2 V
3 7 V
3 21 V
3 1 V
3 5 V
3 -8 V
2 -4 V
3 -13 V
3 -3 V
3 -16 V
3 -26 V
3 11 V
3 23 V
2 -22 V
3 -7 V
3 2 V
3 -1 V
3 -5 V
3 1 V
3 -17 V
2 -20 V
3 17 V
3 -37 V
3 31 V
3 -24 V
3 1 V
2 4 V
3 -5 V
3 19 V
3 6 V
3 10 V
3 14 V
3 -30 V
2 16 V
3 -12 V
3 9 V
3 11 V
3 0 V
3 23 V
3 9 V
2 -11 V
3 -2 V
3 14 V
3 -4 V
3 -23 V
3 19 V
2 -30 V
3 11 V
3 1 V
3 10 V
3 0 V
3 -2 V
3 11 V
2 -22 V
3 12 V
3 6 V
3 8 V
3 -2 V
3 7 V
3 -6 V
2 -11 V
3 5 V
3 9 V
3 -18 V
3 15 V
3 -5 V
2 -10 V
3 -5 V
3 -13 V
3 19 V
3 13 V
3 1 V
3 -3 V
2 -7 V
3 23 V
3 -23 V
3 -15 V
3 -9 V
3 -2 V
3 -21 V
2 13 V
3 -21 V
3 22 V
3 -1 V
3 49 V
3 -13 V
2 20 V
currentpoint stroke M
3 -2 V
3 0 V
3 -15 V
3 -20 V
3 -30 V
3 18 V
2 -17 V
3 10 V
3 -2 V
3 16 V
3 16 V
3 22 V
3 1 V
2 2 V
3 -17 V
3 38 V
3 -25 V
3 26 V
3 -40 V
2 16 V
3 0 V
3 -6 V
3 -8 V
3 13 V
3 -35 V
3 7 V
2 -4 V
3 -39 V
3 10 V
3 0 V
3 20 V
3 25 V
3 7 V
2 -29 V
3 19 V
3 -28 V
3 -1 V
3 -14 V
3 -20 V
2 4 V
3 24 V
3 1 V
3 2 V
3 9 V
3 -33 V
3 22 V
2 -18 V
3 -1 V
3 -2 V
3 -16 V
3 -7 V
3 22 V
3 -7 V
2 10 V
3 16 V
3 -18 V
3 11 V
3 -48 V
3 26 V
2 3 V
3 -11 V
3 18 V
3 8 V
3 3 V
3 13 V
3 -1 V
2 2 V
3 18 V
3 -20 V
3 -30 V
3 9 V
3 -5 V
3 -19 V
2 17 V
3 -12 V
3 -3 V
3 36 V
3 12 V
3 -15 V
2 -12 V
3 1 V
3 13 V
3 6 V
3 -7 V
3 -7 V
3 7 V
2 -21 V
3 17 V
3 2 V
3 -17 V
3 6 V
3 5 V
3 -13 V
2 10 V
3 -5 V
3 -10 V
3 -1 V
3 7 V
3 13 V
2 -17 V
3 -7 V
3 -13 V
3 9 V
3 -17 V
3 15 V
3 -1 V
2 -17 V
3 13 V
3 10 V
3 18 V
3 25 V
3 -12 V
3 2 V
2 -7 V
3 8 V
3 22 V
3 -33 V
3 -9 V
3 -25 V
2 18 V
3 -3 V
3 7 V
3 -3 V
3 -12 V
3 29 V
3 -15 V
2 -9 V
3 -19 V
3 28 V
3 5 V
3 -4 V
3 -3 V
3 12 V
2 -15 V
3 3 V
3 10 V
3 -19 V
3 11 V
3 -4 V
2 4 V
3 -11 V
3 33 V
3 5 V
3 28 V
3 -20 V
3 19 V
2 -11 V
3 -6 V
3 -4 V
3 22 V
3 -18 V
3 0 V
3 16 V
2 -8 V
3 -10 V
3 -4 V
3 -10 V
3 -6 V
3 11 V
2 30 V
3 -6 V
3 7 V
3 -9 V
3 9 V
3 -13 V
3 21 V
2 -16 V
3 -17 V
3 -8 V
3 11 V
3 1 V
3 -2 V
3 -3 V
2 9 V
3 -20 V
3 -19 V
3 2 V
3 -6 V
3 -14 V
2 11 V
3 9 V
3 -5 V
3 -41 V
3 25 V
3 3 V
3 4 V
2 10 V
3 -5 V
3 -5 V
3 2 V
3 -4 V
3 6 V
3 0 V
2 -5 V
3 -5 V
3 4 V
3 -9 V
3 -19 V
3 14 V
2 -3 V
3 -6 V
3 17 V
3 -2 V
1.000 UL
LT2
3103 1847 M
247 0 V
600 904 M
3 -1 V
3 0 V
3 0 V
2 1 V
3 0 V
3 -1 V
3 0 V
3 1 V
3 0 V
2 0 V
3 0 V
3 0 V
3 0 V
3 0 V
3 0 V
3 0 V
2 1 V
3 0 V
3 -1 V
3 1 V
3 0 V
3 0 V
3 1 V
2 0 V
3 0 V
3 0 V
3 0 V
3 0 V
3 0 V
2 1 V
3 1 V
3 0 V
3 0 V
3 0 V
3 0 V
3 1 V
2 -1 V
3 0 V
3 1 V
3 1 V
3 -1 V
3 1 V
3 0 V
2 0 V
3 1 V
3 0 V
3 1 V
3 0 V
3 0 V
3 0 V
2 1 V
3 0 V
3 1 V
3 1 V
3 0 V
3 0 V
2 0 V
3 0 V
3 1 V
3 0 V
3 1 V
3 0 V
3 1 V
2 0 V
3 1 V
3 0 V
3 1 V
3 0 V
3 0 V
3 1 V
2 0 V
3 1 V
3 1 V
3 0 V
3 1 V
3 1 V
2 0 V
3 0 V
3 1 V
3 1 V
3 0 V
3 0 V
3 1 V
2 1 V
3 1 V
3 0 V
3 0 V
3 1 V
3 1 V
3 0 V
2 1 V
3 1 V
3 0 V
3 1 V
3 1 V
3 0 V
2 1 V
3 1 V
3 0 V
3 1 V
3 1 V
3 0 V
3 1 V
2 1 V
3 1 V
3 0 V
3 1 V
3 1 V
3 1 V
3 0 V
2 1 V
3 1 V
3 1 V
3 1 V
3 1 V
3 0 V
2 1 V
3 1 V
3 0 V
3 2 V
3 0 V
3 1 V
3 1 V
2 1 V
3 1 V
3 1 V
3 1 V
3 1 V
3 1 V
3 1 V
2 1 V
3 1 V
3 1 V
3 0 V
3 1 V
3 1 V
2 1 V
3 2 V
3 1 V
3 1 V
3 1 V
3 1 V
3 1 V
2 2 V
3 0 V
3 1 V
3 2 V
3 1 V
3 1 V
3 1 V
2 1 V
3 2 V
3 1 V
3 1 V
3 1 V
3 2 V
2 1 V
3 1 V
3 2 V
3 1 V
3 1 V
3 2 V
3 1 V
2 2 V
3 1 V
3 2 V
3 1 V
3 2 V
3 1 V
3 2 V
2 2 V
3 2 V
3 2 V
3 2 V
3 1 V
3 2 V
2 2 V
3 2 V
3 2 V
3 2 V
3 2 V
3 2 V
3 3 V
2 2 V
3 3 V
3 3 V
3 2 V
3 3 V
3 3 V
3 4 V
2 3 V
3 3 V
3 4 V
3 5 V
3 4 V
3 4 V
2 5 V
3 3 V
3 0 V
3 -103 V
3 51 V
3 200 V
3 -278 V
2 -50 V
3 -63 V
3 -41 V
3 -17 V
3 -17 V
3 -42 V
3 -32 V
2 13 V
3 70 V
3 -11 V
3 35 V
3 8 V
3 -13 V
2 -113 V
3 38 V
3 -34 V
3 -18 V
3 -45 V
3 50 V
3 -9 V
2 65 V
3 -69 V
3 6 V
3 69 V
3 -5 V
3 -36 V
3 -129 V
2 60 V
3 41 V
3 -24 V
3 17 V
3 -28 V
3 -4 V
2 10 V
3 17 V
3 5 V
3 -35 V
3 -7 V
3 30 V
3 -3 V
2 -26 V
3 36 V
3 -17 V
3 -17 V
3 3 V
3 2 V
3 -12 V
2 -40 V
3 19 V
3 32 V
3 48 V
3 -72 V
3 -29 V
2 53 V
3 6 V
3 -35 V
3 39 V
3 26 V
3 -38 V
3 -35 V
2 42 V
3 -27 V
3 33 V
3 11 V
3 2 V
3 -15 V
3 -45 V
2 71 V
3 -5 V
3 -29 V
3 13 V
3 31 V
3 5 V
2 -1 V
3 29 V
3 29 V
3 -44 V
3 -87 V
3 117 V
3 37 V
2 -43 V
3 -8 V
3 -9 V
3 19 V
3 -54 V
3 -129 V
3 121 V
2 86 V
3 33 V
3 -15 V
3 9 V
3 9 V
3 -73 V
2 9 V
3 -60 V
3 93 V
3 -77 V
3 -51 V
3 -23 V
3 -13 V
2 66 V
3 14 V
3 41 V
3 50 V
3 -44 V
3 53 V
3 -66 V
2 72 V
3 -31 V
3 -88 V
3 43 V
3 8 V
3 4 V
2 -82 V
3 120 V
3 -35 V
3 -34 V
3 60 V
3 -27 V
3 26 V
2 52 V
3 -18 V
3 -10 V
3 -30 V
3 -70 V
3 65 V
3 29 V
2 -46 V
3 -53 V
3 5 V
3 -4 V
3 -3 V
3 50 V
2 1 V
3 3 V
3 -23 V
3 20 V
3 54 V
3 -10 V
3 -26 V
2 -9 V
3 34 V
3 -17 V
3 -10 V
3 12 V
3 -5 V
3 -14 V
2 9 V
3 42 V
3 56 V
3 -1 V
3 35 V
3 20 V
2 34 V
3 0 V
3 -45 V
3 -2 V
3 -93 V
3 -90 V
3 28 V
2 98 V
3 -27 V
3 -34 V
3 -17 V
3 62 V
3 -1 V
3 46 V
2 63 V
3 -62 V
3 60 V
3 -7 V
3 23 V
3 -6 V
2 2 V
3 -40 V
3 -51 V
3 30 V
3 17 V
3 43 V
3 -47 V
2 27 V
3 -5 V
3 14 V
3 -35 V
3 -27 V
3 30 V
3 -17 V
2 41 V
3 -51 V
3 35 V
3 -52 V
3 16 V
3 6 V
2 -5 V
currentpoint stroke M
3 34 V
3 23 V
3 -62 V
3 -32 V
3 17 V
3 -16 V
2 118 V
3 15 V
3 -10 V
3 41 V
3 -47 V
3 -3 V
3 39 V
2 -23 V
3 64 V
3 -89 V
3 65 V
3 -5 V
3 36 V
2 -21 V
3 -13 V
3 -9 V
3 -116 V
3 53 V
3 21 V
3 0 V
2 -58 V
3 101 V
3 -21 V
3 -37 V
3 36 V
3 -54 V
3 9 V
2 109 V
3 29 V
3 22 V
3 -117 V
3 -11 V
3 31 V
2 65 V
3 -57 V
3 -15 V
3 82 V
3 -60 V
3 36 V
3 36 V
2 0 V
3 12 V
3 -3 V
3 -35 V
3 60 V
3 20 V
3 16 V
2 23 V
3 -9 V
3 2 V
3 -24 V
3 63 V
3 51 V
2 -37 V
3 -19 V
3 76 V
3 14 V
3 -1 V
3 8 V
3 42 V
2 31 V
3 10 V
3 52 V
3 134 V
3 36 V
3 84 V
3 -39 V
2 -144 V
3 -69 V
3 7 V
3 -86 V
3 -58 V
3 -6 V
2 -9 V
3 -58 V
3 27 V
3 -52 V
3 -70 V
3 20 V
3 -23 V
2 114 V
3 -29 V
3 -7 V
3 15 V
3 -125 V
3 -21 V
3 4 V
2 -1 V
3 66 V
3 -83 V
3 -31 V
3 33 V
3 39 V
2 -11 V
3 28 V
3 -43 V
3 45 V
3 -77 V
3 -20 V
3 4 V
2 -42 V
3 48 V
3 -29 V
3 68 V
3 19 V
3 15 V
3 -52 V
2 -132 V
3 82 V
3 -29 V
3 -14 V
3 42 V
3 -39 V
2 -35 V
3 8 V
3 14 V
3 -20 V
3 13 V
3 9 V
3 -57 V
2 84 V
3 2 V
3 -9 V
3 -10 V
3 -28 V
3 36 V
3 -98 V
2 -48 V
3 159 V
3 39 V
3 -22 V
3 -63 V
3 39 V
2 -67 V
3 -9 V
3 46 V
3 -37 V
3 -14 V
3 73 V
3 -76 V
2 -56 V
3 49 V
3 30 V
3 -7 V
3 -23 V
3 -33 V
3 -20 V
2 18 V
3 86 V
3 6 V
3 -1 V
3 -71 V
3 -8 V
2 -53 V
3 28 V
3 62 V
3 -37 V
3 35 V
3 33 V
3 6 V
2 3 V
3 -56 V
3 -109 V
3 -29 V
3 53 V
3 87 V
3 -68 V
2 16 V
3 -40 V
3 -16 V
3 47 V
3 80 V
3 29 V
2 -77 V
3 -89 V
3 0 V
3 -15 V
3 46 V
3 3 V
3 -49 V
2 35 V
3 -8 V
3 -18 V
3 49 V
3 3 V
3 -87 V
3 44 V
2 3 V
3 -112 V
3 62 V
3 35 V
3 14 V
3 37 V
2 -39 V
3 -54 V
3 72 V
3 31 V
3 50 V
3 -37 V
3 -28 V
2 8 V
3 9 V
3 46 V
3 -41 V
3 -42 V
3 12 V
3 -50 V
2 17 V
3 38 V
3 55 V
3 -65 V
3 -16 V
3 -33 V
2 72 V
3 -16 V
3 -19 V
3 20 V
3 0 V
3 -62 V
3 12 V
2 37 V
3 -15 V
3 -24 V
3 -19 V
3 16 V
3 95 V
3 -47 V
2 -6 V
3 25 V
3 -55 V
3 -34 V
3 50 V
3 -7 V
2 -59 V
3 -3 V
3 4 V
3 8 V
3 -14 V
3 70 V
3 18 V
2 35 V
3 33 V
3 -65 V
3 55 V
3 -71 V
3 35 V
3 16 V
2 6 V
3 19 V
3 -65 V
3 47 V
3 -36 V
3 19 V
2 -40 V
3 -41 V
3 3 V
3 46 V
3 -60 V
3 -2 V
3 76 V
2 -37 V
3 -12 V
3 4 V
3 -2 V
3 -6 V
3 18 V
3 -8 V
2 -85 V
3 108 V
3 -46 V
3 38 V
3 52 V
3 -24 V
2 -14 V
3 -7 V
3 28 V
3 -50 V
3 3 V
3 -42 V
3 60 V
2 -56 V
3 66 V
3 63 V
3 28 V
3 -29 V
3 8 V
3 -16 V
2 -60 V
3 55 V
3 -59 V
3 -84 V
3 -6 V
3 102 V
2 -7 V
3 -74 V
3 -25 V
3 23 V
3 67 V
3 38 V
3 34 V
2 5 V
3 -9 V
3 -25 V
3 -59 V
3 45 V
3 -22 V
3 -30 V
2 -58 V
3 -25 V
3 16 V
3 24 V
3 28 V
3 0 V
2 -52 V
3 -14 V
3 66 V
3 27 V
3 -51 V
3 -14 V
3 -4 V
2 -24 V
3 35 V
3 -18 V
3 24 V
3 -44 V
3 -84 V
3 82 V
2 35 V
3 29 V
3 -79 V
3 49 V
3 47 V
3 32 V
2 13 V
3 -7 V
3 -57 V
3 30 V
3 -19 V
3 8 V
3 -28 V
2 -19 V
3 28 V
3 43 V
3 -89 V
3 52 V
3 -25 V
3 4 V
2 -4 V
3 40 V
3 -53 V
3 34 V
3 -38 V
3 -37 V
2 -37 V
3 30 V
3 42 V
3 -74 V
3 70 V
3 25 V
3 -14 V
2 -14 V
3 -10 V
3 69 V
3 31 V
3 28 V
3 37 V
3 -79 V
2 4 V
3 -115 V
3 69 V
3 28 V
3 18 V
3 -13 V
2 -61 V
3 15 V
3 22 V
3 -5 V
3 10 V
3 30 V
3 -7 V
2 -74 V
3 -14 V
3 11 V
3 -18 V
3 7 V
3 62 V
3 -25 V
2 24 V
3 -91 V
3 27 V
3 -3 V
3 -75 V
3 46 V
2 -6 V
currentpoint stroke M
3 59 V
3 -36 V
3 4 V
3 -8 V
3 29 V
3 -5 V
2 45 V
3 -64 V
3 -79 V
3 72 V
3 1 V
3 44 V
3 -68 V
2 31 V
3 -5 V
3 -30 V
3 50 V
3 4 V
3 18 V
2 -22 V
3 55 V
3 -7 V
3 -24 V
3 25 V
3 -95 V
3 -51 V
2 34 V
3 39 V
3 28 V
3 18 V
3 -54 V
3 59 V
3 -34 V
2 -43 V
3 0 V
3 47 V
3 29 V
3 -46 V
3 28 V
2 22 V
3 -33 V
3 -16 V
3 16 V
3 -46 V
3 103 V
3 -4 V
2 -13 V
3 28 V
3 44 V
3 -7 V
3 8 V
3 -45 V
3 -18 V
2 -65 V
3 -11 V
3 23 V
3 17 V
3 2 V
3 39 V
2 -74 V
3 -20 V
3 -18 V
3 2 V
3 -2 V
3 87 V
3 -47 V
2 1 V
3 -31 V
3 81 V
3 -43 V
3 -40 V
3 -46 V
3 8 V
2 38 V
3 40 V
3 5 V
3 17 V
3 16 V
3 -21 V
2 42 V
3 -1 V
3 22 V
3 -57 V
3 59 V
3 -37 V
3 -13 V
2 -59 V
3 33 V
3 26 V
3 -12 V
3 6 V
3 -7 V
3 22 V
2 -42 V
3 -65 V
3 -5 V
3 10 V
3 31 V
3 16 V
2 17 V
3 -42 V
3 -10 V
3 1 V
3 11 V
3 -82 V
3 82 V
2 48 V
3 -28 V
3 -18 V
3 63 V
3 10 V
3 -74 V
3 37 V
2 -28 V
3 3 V
3 47 V
3 3 V
3 -52 V
3 -8 V
2 3 V
3 35 V
3 -45 V
3 96 V
3 -41 V
3 -96 V
3 34 V
2 -17 V
3 1 V
3 10 V
3 64 V
3 -19 V
3 15 V
3 30 V
2 -50 V
3 50 V
3 -56 V
3 -38 V
3 43 V
3 -25 V
2 -3 V
3 79 V
3 -61 V
3 9 V
3 -7 V
3 -4 V
3 -3 V
2 -60 V
3 40 V
3 73 V
3 -79 V
3 110 V
3 -22 V
3 11 V
2 -49 V
3 5 V
3 -10 V
3 -66 V
3 67 V
3 -119 V
2 11 V
3 23 V
3 73 V
3 34 V
3 70 V
3 -50 V
3 -34 V
2 -18 V
3 -82 V
3 152 V
3 -54 V
3 -38 V
3 12 V
3 6 V
2 11 V
3 -36 V
3 7 V
3 5 V
3 -12 V
3 52 V
2 -40 V
3 41 V
3 15 V
3 -23 V
3 10 V
3 39 V
3 -21 V
2 -111 V
3 70 V
3 -7 V
3 40 V
3 1 V
3 -45 V
3 -27 V
2 30 V
3 28 V
3 -85 V
3 28 V
3 -21 V
3 58 V
2 17 V
3 43 V
3 19 V
3 -26 V
1.000 UL
LT3
3103 1747 M
247 0 V
600 903 M
3 0 V
3 0 V
3 0 V
2 0 V
3 0 V
3 0 V
3 0 V
3 0 V
3 1 V
2 0 V
3 0 V
3 0 V
3 0 V
3 0 V
3 0 V
3 0 V
2 0 V
3 0 V
3 1 V
3 0 V
3 0 V
3 0 V
3 0 V
2 0 V
3 0 V
3 1 V
3 0 V
3 0 V
3 0 V
2 0 V
3 1 V
3 0 V
3 0 V
3 0 V
3 1 V
3 0 V
2 0 V
3 0 V
3 1 V
3 0 V
3 1 V
3 0 V
3 0 V
2 0 V
3 1 V
3 0 V
3 0 V
3 1 V
3 0 V
3 0 V
2 0 V
3 1 V
3 0 V
3 1 V
3 0 V
3 0 V
2 1 V
3 0 V
3 1 V
3 0 V
3 1 V
3 0 V
3 1 V
2 0 V
3 0 V
3 1 V
3 0 V
3 1 V
3 0 V
3 1 V
2 0 V
3 1 V
3 0 V
3 1 V
3 1 V
3 0 V
2 0 V
3 1 V
3 1 V
3 0 V
3 1 V
3 0 V
3 1 V
2 0 V
3 1 V
3 1 V
3 0 V
3 1 V
3 1 V
3 0 V
2 1 V
3 0 V
3 1 V
3 1 V
3 0 V
3 1 V
2 1 V
3 1 V
3 0 V
3 1 V
3 1 V
3 0 V
3 1 V
2 1 V
3 0 V
3 1 V
3 1 V
3 1 V
3 0 V
3 1 V
2 1 V
3 1 V
3 1 V
3 0 V
3 1 V
3 1 V
2 1 V
3 1 V
3 1 V
3 1 V
3 0 V
3 1 V
3 1 V
2 1 V
3 1 V
3 1 V
3 1 V
3 1 V
3 1 V
3 1 V
2 1 V
3 1 V
3 1 V
3 1 V
3 1 V
3 1 V
2 1 V
3 1 V
3 1 V
3 1 V
3 1 V
3 1 V
3 1 V
2 1 V
3 1 V
3 1 V
3 2 V
3 1 V
3 1 V
3 1 V
2 1 V
3 2 V
3 1 V
3 1 V
3 1 V
3 2 V
2 1 V
3 1 V
3 2 V
3 1 V
3 1 V
3 2 V
3 1 V
2 2 V
3 1 V
3 2 V
3 1 V
3 2 V
3 2 V
3 1 V
2 2 V
3 2 V
3 1 V
3 2 V
3 2 V
3 2 V
2 2 V
3 2 V
3 2 V
3 2 V
3 2 V
3 3 V
3 2 V
2 2 V
3 3 V
3 3 V
3 2 V
3 3 V
3 3 V
3 4 V
2 3 V
3 4 V
3 4 V
3 4 V
3 4 V
3 4 V
2 5 V
3 4 V
3 -1 V
3 -102 V
3 35 V
3 183 V
3 -275 V
2 -63 V
3 -79 V
3 -59 V
3 16 V
3 -101 V
3 66 V
3 -15 V
2 -27 V
3 -99 V
3 97 V
3 12 V
3 -53 V
3 -88 V
2 88 V
3 -40 V
3 -32 V
3 27 V
3 37 V
3 20 V
3 6 V
2 -45 V
3 -9 V
3 10 V
3 23 V
3 52 V
3 -77 V
3 -59 V
2 -45 V
3 124 V
3 34 V
3 21 V
3 -19 V
3 -144 V
2 -17 V
3 56 V
3 44 V
3 -53 V
3 -83 V
3 101 V
3 39 V
2 -73 V
3 104 V
3 29 V
3 -171 V
3 40 V
3 148 V
3 -10 V
2 -3 V
3 -40 V
3 -70 V
3 -45 V
3 35 V
3 13 V
2 66 V
3 -12 V
3 -1 V
3 17 V
3 -72 V
3 -7 V
3 -46 V
2 106 V
3 -176 V
3 -3 V
3 82 V
3 110 V
3 -33 V
3 34 V
2 -43 V
3 -117 V
3 94 V
3 39 V
3 3 V
3 -104 V
2 137 V
3 14 V
3 44 V
3 -67 V
3 -108 V
3 -2 V
3 119 V
2 -4 V
3 -218 V
3 87 V
3 103 V
3 24 V
3 -102 V
3 -8 V
2 156 V
3 -30 V
3 -41 V
3 -69 V
3 109 V
3 19 V
2 34 V
3 -174 V
3 37 V
3 176 V
3 30 V
3 -66 V
3 -86 V
2 72 V
3 101 V
3 -81 V
3 -33 V
3 -102 V
3 39 V
3 14 V
2 15 V
3 -45 V
3 -85 V
3 -4 V
3 199 V
3 -22 V
2 -116 V
3 -37 V
3 76 V
3 10 V
3 80 V
3 -92 V
3 -108 V
2 98 V
3 -42 V
3 5 V
3 107 V
3 -35 V
3 -58 V
3 -82 V
2 175 V
3 73 V
3 -71 V
3 -59 V
3 -125 V
3 109 V
2 28 V
3 -77 V
3 15 V
3 -25 V
3 205 V
3 -35 V
3 -233 V
2 18 V
3 122 V
3 -1 V
3 -63 V
3 57 V
3 96 V
3 -97 V
2 -7 V
3 61 V
3 24 V
3 22 V
3 -27 V
3 -39 V
2 86 V
3 -81 V
3 52 V
3 -159 V
3 28 V
3 170 V
3 -111 V
2 84 V
3 -68 V
3 -107 V
3 203 V
3 57 V
3 -93 V
3 -102 V
2 90 V
3 145 V
3 -2 V
3 1 V
3 -48 V
3 -68 V
2 -87 V
3 93 V
3 55 V
3 -31 V
3 -89 V
3 -1 V
3 128 V
2 -4 V
3 3 V
3 -58 V
3 -152 V
3 136 V
3 -76 V
3 191 V
2 -57 V
3 63 V
3 -23 V
3 -165 V
3 -33 V
3 238 V
2 -22 V
currentpoint stroke M
3 3 V
3 -91 V
3 -2 V
3 105 V
3 6 V
3 34 V
2 -128 V
3 -57 V
3 36 V
3 88 V
3 -80 V
3 -54 V
3 96 V
2 86 V
3 -119 V
3 30 V
3 -39 V
3 -3 V
3 86 V
2 -111 V
3 24 V
3 41 V
3 38 V
3 68 V
3 20 V
3 -120 V
2 -57 V
3 162 V
3 -32 V
3 -128 V
3 123 V
3 60 V
3 16 V
2 34 V
3 -62 V
3 -76 V
3 117 V
3 107 V
3 -3 V
2 -85 V
3 -70 V
3 -117 V
3 216 V
3 8 V
3 -53 V
3 -7 V
2 48 V
3 60 V
3 68 V
3 -14 V
3 -80 V
3 -241 V
3 261 V
2 131 V
3 -43 V
3 -50 V
3 -240 V
3 34 V
3 204 V
2 51 V
3 -4 V
3 -1 V
3 41 V
3 -18 V
3 44 V
3 73 V
2 29 V
3 20 V
3 151 V
3 93 V
3 16 V
3 123 V
3 -153 V
2 -102 V
3 -151 V
3 41 V
3 -34 V
3 8 V
3 -93 V
2 -42 V
3 24 V
3 62 V
3 -41 V
3 -136 V
3 -144 V
3 31 V
2 173 V
3 22 V
3 -174 V
3 17 V
3 -47 V
3 92 V
3 17 V
2 -5 V
3 -136 V
3 39 V
3 112 V
3 -42 V
3 -76 V
2 -130 V
3 16 V
3 25 V
3 97 V
3 30 V
3 -37 V
3 -143 V
2 10 V
3 27 V
3 84 V
3 -178 V
3 27 V
3 96 V
3 -20 V
2 -98 V
3 38 V
3 66 V
3 -8 V
3 -31 V
3 105 V
2 -67 V
3 -58 V
3 23 V
3 -111 V
3 126 V
3 11 V
3 -104 V
2 125 V
3 -5 V
3 23 V
3 -138 V
3 49 V
3 -46 V
3 20 V
2 44 V
3 0 V
3 -2 V
3 -33 V
3 -211 V
3 182 V
2 88 V
3 46 V
3 -54 V
3 -170 V
3 -71 V
3 140 V
3 68 V
2 -140 V
3 -74 V
3 234 V
3 -114 V
3 142 V
3 -64 V
3 -146 V
2 86 V
3 -9 V
3 30 V
3 14 V
3 -161 V
3 49 V
2 13 V
3 77 V
3 -144 V
3 112 V
3 60 V
3 33 V
3 -15 V
2 -164 V
3 30 V
3 -127 V
3 176 V
3 115 V
3 1 V
3 -87 V
2 -101 V
3 86 V
3 89 V
3 -94 V
3 -16 V
3 42 V
2 -30 V
3 19 V
3 -8 V
3 -1 V
3 -65 V
3 40 V
3 74 V
2 6 V
3 34 V
3 -107 V
3 12 V
3 -19 V
3 6 V
3 -74 V
2 28 V
3 1 V
3 -147 V
3 108 V
3 224 V
3 30 V
2 -62 V
3 -136 V
3 -43 V
3 66 V
3 42 V
3 -28 V
3 -159 V
2 75 V
3 76 V
3 -15 V
3 -60 V
3 33 V
3 -106 V
3 97 V
2 149 V
3 14 V
3 -26 V
3 -133 V
3 -7 V
3 125 V
2 6 V
3 -17 V
3 -41 V
3 -41 V
3 24 V
3 50 V
3 -59 V
2 -128 V
3 38 V
3 148 V
3 82 V
3 -10 V
3 -161 V
3 -76 V
2 50 V
3 49 V
3 33 V
3 12 V
3 -42 V
3 -214 V
2 98 V
3 35 V
3 -32 V
3 85 V
3 43 V
3 -6 V
3 -28 V
2 -7 V
3 -39 V
3 -43 V
3 -45 V
3 46 V
3 -29 V
3 -41 V
2 133 V
3 -28 V
3 -68 V
3 -90 V
3 137 V
3 19 V
2 44 V
3 -141 V
3 100 V
3 37 V
3 82 V
3 -41 V
3 -207 V
2 65 V
3 100 V
3 77 V
3 -121 V
3 -106 V
3 -9 V
3 117 V
2 14 V
3 55 V
3 -202 V
3 116 V
3 -160 V
3 103 V
2 89 V
3 0 V
3 70 V
3 -35 V
3 -163 V
3 16 V
3 137 V
2 -19 V
3 -68 V
3 31 V
3 -31 V
3 90 V
3 20 V
3 105 V
2 -80 V
3 -83 V
3 -6 V
3 19 V
3 22 V
3 -17 V
2 -4 V
3 -226 V
3 174 V
3 62 V
3 -88 V
3 0 V
3 -81 V
2 81 V
3 95 V
3 -13 V
3 -114 V
3 -87 V
3 156 V
3 -3 V
2 -50 V
3 -50 V
3 120 V
3 -116 V
3 -56 V
3 69 V
2 -83 V
3 42 V
3 117 V
3 3 V
3 31 V
3 -74 V
3 61 V
2 -256 V
3 160 V
3 43 V
3 -36 V
3 -10 V
3 -62 V
3 187 V
2 -42 V
3 -95 V
3 -82 V
3 156 V
3 72 V
3 -124 V
2 4 V
3 72 V
3 -194 V
3 65 V
3 92 V
3 51 V
3 21 V
2 -135 V
3 47 V
3 82 V
3 -2 V
3 -59 V
3 22 V
3 -178 V
2 26 V
3 139 V
3 72 V
3 -141 V
3 -58 V
3 136 V
2 133 V
3 14 V
3 -46 V
3 -99 V
3 72 V
3 60 V
3 -47 V
2 -16 V
3 -112 V
3 85 V
3 40 V
3 -202 V
3 20 V
3 111 V
2 59 V
3 -102 V
3 -125 V
3 103 V
3 102 V
3 -84 V
2 -60 V
3 12 V
3 59 V
3 17 V
3 -188 V
3 116 V
3 106 V
2 -36 V
3 87 V
3 -31 V
3 -199 V
3 77 V
3 38 V
3 -103 V
2 5 V
3 141 V
3 -19 V
3 26 V
3 -120 V
3 88 V
2 -60 V
currentpoint stroke M
3 -34 V
3 14 V
3 -95 V
3 63 V
3 167 V
3 63 V
2 -44 V
3 -59 V
3 42 V
3 -50 V
3 -113 V
3 36 V
3 79 V
2 95 V
3 30 V
3 -40 V
3 -45 V
3 -76 V
3 96 V
2 -47 V
3 -68 V
3 59 V
3 -82 V
3 -31 V
3 -6 V
3 26 V
2 57 V
3 7 V
3 -14 V
3 -22 V
3 62 V
3 36 V
3 30 V
2 -96 V
3 -191 V
3 152 V
3 97 V
3 29 V
3 8 V
2 -64 V
3 -15 V
3 48 V
3 11 V
3 -66 V
3 0 V
3 39 V
2 -144 V
3 76 V
3 -105 V
3 64 V
3 44 V
3 54 V
3 -59 V
2 -54 V
3 117 V
3 13 V
3 -201 V
3 32 V
3 54 V
2 -34 V
3 105 V
3 -67 V
3 14 V
3 -63 V
3 -17 V
3 80 V
2 -63 V
3 -39 V
3 211 V
3 -73 V
3 75 V
3 -7 V
3 -128 V
2 -8 V
3 -6 V
3 126 V
3 -59 V
3 41 V
3 -58 V
2 63 V
3 27 V
3 -14 V
3 -163 V
3 10 V
3 81 V
3 -33 V
2 -38 V
3 -74 V
3 105 V
3 100 V
3 -22 V
3 -87 V
3 -89 V
2 90 V
3 144 V
3 2 V
3 -123 V
3 -39 V
3 33 V
2 -70 V
3 72 V
3 -136 V
3 49 V
3 7 V
3 102 V
3 26 V
2 71 V
3 -144 V
3 -110 V
3 99 V
3 39 V
3 -72 V
3 -95 V
2 50 V
3 188 V
3 -113 V
3 -5 V
3 -27 V
3 176 V
2 7 V
3 -35 V
3 -160 V
3 176 V
3 -44 V
3 -12 V
3 -106 V
2 91 V
3 74 V
3 -67 V
3 -111 V
3 -43 V
3 118 V
3 36 V
2 -132 V
3 152 V
3 -64 V
3 -13 V
3 -45 V
3 57 V
2 40 V
3 -30 V
3 77 V
3 -91 V
3 -7 V
3 68 V
3 -56 V
2 -28 V
3 -22 V
3 104 V
3 91 V
3 -39 V
3 -33 V
3 -46 V
2 -165 V
3 79 V
3 -48 V
3 -58 V
3 213 V
3 29 V
2 -106 V
3 48 V
3 -81 V
3 60 V
3 87 V
3 -63 V
3 -106 V
2 135 V
3 61 V
3 15 V
3 11 V
3 -337 V
3 88 V
3 41 V
2 119 V
3 -5 V
3 43 V
3 -66 V
3 41 V
3 -133 V
2 -23 V
3 16 V
3 66 V
3 98 V
3 -91 V
3 -29 V
3 -2 V
2 51 V
3 -70 V
3 -77 V
3 144 V
3 21 V
3 54 V
3 30 V
2 -170 V
3 -42 V
3 61 V
3 -44 V
3 -21 V
3 90 V
2 -70 V
3 -70 V
3 124 V
3 106 V
1.000 UL
LT0
3103 1647 M
247 0 V
600 903 M
3 0 V
3 0 V
3 0 V
2 0 V
3 0 V
3 0 V
3 0 V
3 0 V
3 0 V
2 0 V
3 0 V
3 1 V
3 0 V
3 0 V
3 0 V
3 0 V
2 0 V
3 0 V
3 0 V
3 0 V
3 1 V
3 0 V
3 0 V
2 0 V
3 0 V
3 0 V
3 1 V
3 0 V
3 0 V
2 0 V
3 1 V
3 0 V
3 0 V
3 0 V
3 0 V
3 1 V
2 0 V
3 0 V
3 1 V
3 0 V
3 0 V
3 0 V
3 1 V
2 0 V
3 0 V
3 1 V
3 0 V
3 0 V
3 1 V
3 0 V
2 0 V
3 1 V
3 0 V
3 1 V
3 0 V
3 0 V
2 1 V
3 0 V
3 1 V
3 0 V
3 0 V
3 1 V
3 0 V
2 1 V
3 0 V
3 1 V
3 0 V
3 1 V
3 0 V
3 1 V
2 0 V
3 1 V
3 0 V
3 1 V
3 0 V
3 1 V
2 0 V
3 1 V
3 0 V
3 1 V
3 1 V
3 0 V
3 1 V
2 0 V
3 1 V
3 1 V
3 0 V
3 1 V
3 0 V
3 1 V
2 1 V
3 0 V
3 1 V
3 1 V
3 0 V
3 1 V
2 1 V
3 0 V
3 1 V
3 1 V
3 1 V
3 0 V
3 1 V
2 1 V
3 0 V
3 1 V
3 1 V
3 1 V
3 1 V
3 0 V
2 1 V
3 1 V
3 1 V
3 0 V
3 1 V
3 1 V
2 1 V
3 1 V
3 1 V
3 1 V
3 0 V
3 1 V
3 1 V
2 1 V
3 1 V
3 1 V
3 1 V
3 1 V
3 1 V
3 1 V
2 1 V
3 1 V
3 1 V
3 1 V
3 1 V
3 1 V
2 1 V
3 1 V
3 1 V
3 1 V
3 1 V
3 1 V
3 1 V
2 1 V
3 1 V
3 1 V
3 2 V
3 1 V
3 1 V
3 1 V
2 1 V
3 2 V
3 1 V
3 1 V
3 1 V
3 2 V
2 1 V
3 1 V
3 2 V
3 1 V
3 1 V
3 2 V
3 1 V
2 2 V
3 1 V
3 2 V
3 1 V
3 2 V
3 2 V
3 1 V
2 2 V
3 2 V
3 1 V
3 2 V
3 2 V
3 2 V
2 2 V
3 2 V
3 2 V
3 2 V
3 2 V
3 3 V
3 2 V
2 3 V
3 2 V
3 3 V
3 3 V
3 3 V
3 3 V
3 3 V
2 3 V
3 4 V
3 4 V
3 4 V
3 4 V
3 5 V
2 4 V
3 4 V
3 -1 V
3 -103 V
3 31 V
3 188 V
3 -283 V
2 -102 V
3 -62 V
3 -44 V
3 -35 V
3 -27 V
3 -23 V
3 -19 V
2 -16 V
3 -15 V
3 -12 V
3 -11 V
3 -10 V
3 -9 V
2 -8 V
3 -7 V
3 -6 V
3 -6 V
3 -5 V
3 -5 V
3 -4 V
2 -4 V
3 -3 V
3 -4 V
3 -2 V
3 -3 V
3 -2 V
3 -3 V
2 -1 V
3 -2 V
3 -2 V
3 -1 V
3 -1 V
3 -1 V
2 -1 V
3 -1 V
3 -1 V
3 0 V
3 -1 V
3 0 V
3 -1 V
2 0 V
3 0 V
3 0 V
3 0 V
3 0 V
3 0 V
3 0 V
2 0 V
3 1 V
3 0 V
3 0 V
3 1 V
3 0 V
2 1 V
3 0 V
3 1 V
3 0 V
3 1 V
3 1 V
3 0 V
2 1 V
3 1 V
3 1 V
3 0 V
3 1 V
3 1 V
3 1 V
2 1 V
3 1 V
3 0 V
3 1 V
3 1 V
3 1 V
2 1 V
3 1 V
3 1 V
3 1 V
3 1 V
3 1 V
3 2 V
2 1 V
3 1 V
3 1 V
3 1 V
3 1 V
3 1 V
3 1 V
2 2 V
3 1 V
3 1 V
3 1 V
3 1 V
3 2 V
2 1 V
3 1 V
3 1 V
3 2 V
3 1 V
3 1 V
3 2 V
2 1 V
3 1 V
3 1 V
3 2 V
3 1 V
3 2 V
3 1 V
2 1 V
3 2 V
3 1 V
3 1 V
3 2 V
3 1 V
2 2 V
3 1 V
3 2 V
3 1 V
3 2 V
3 1 V
3 2 V
2 1 V
3 2 V
3 1 V
3 2 V
3 1 V
3 2 V
3 1 V
2 2 V
3 1 V
3 2 V
3 2 V
3 1 V
3 2 V
2 1 V
3 2 V
3 2 V
3 1 V
3 2 V
3 2 V
3 2 V
2 1 V
3 2 V
3 2 V
3 1 V
3 2 V
3 2 V
3 2 V
2 2 V
3 1 V
3 2 V
3 2 V
3 2 V
3 2 V
2 2 V
3 2 V
3 1 V
3 2 V
3 2 V
3 2 V
3 2 V
2 2 V
3 2 V
3 2 V
3 2 V
3 2 V
3 2 V
3 2 V
2 3 V
3 2 V
3 2 V
3 2 V
3 2 V
3 2 V
2 3 V
3 2 V
3 2 V
3 2 V
3 3 V
3 2 V
3 2 V
2 3 V
3 2 V
3 3 V
3 2 V
3 3 V
3 2 V
3 3 V
2 2 V
3 3 V
3 2 V
3 3 V
3 3 V
3 2 V
2 3 V
currentpoint stroke M
3 3 V
3 3 V
3 3 V
3 3 V
3 2 V
3 3 V
2 3 V
3 3 V
3 4 V
3 3 V
3 3 V
3 3 V
3 3 V
2 4 V
3 3 V
3 4 V
3 3 V
3 4 V
3 3 V
2 4 V
3 4 V
3 3 V
3 4 V
3 4 V
3 4 V
3 4 V
2 5 V
3 4 V
3 4 V
3 5 V
3 4 V
3 5 V
3 5 V
2 5 V
3 5 V
3 5 V
3 5 V
3 6 V
3 5 V
2 6 V
3 6 V
3 6 V
3 6 V
3 7 V
3 6 V
3 7 V
2 8 V
3 7 V
3 8 V
3 8 V
3 8 V
3 9 V
3 10 V
2 9 V
3 11 V
3 11 V
3 11 V
3 13 V
3 13 V
2 14 V
3 16 V
3 16 V
3 19 V
3 20 V
3 24 V
3 26 V
2 31 V
3 38 V
3 47 V
3 64 V
3 100 V
3 243 V
3 -112 V
2 -167 V
3 -85 V
3 -57 V
3 -43 V
3 -35 V
3 -29 V
2 -25 V
3 -21 V
3 -20 V
3 -17 V
3 -16 V
3 -15 V
3 -13 V
2 -12 V
3 -12 V
3 -11 V
3 -10 V
3 -10 V
3 -9 V
3 -9 V
2 -8 V
3 -8 V
3 -7 V
3 -7 V
3 -7 V
3 -7 V
2 -6 V
3 -6 V
3 -6 V
3 -6 V
3 -5 V
3 -6 V
3 -5 V
2 -5 V
3 -5 V
3 -4 V
3 -5 V
3 -4 V
3 -5 V
3 -4 V
2 -4 V
3 -4 V
3 -4 V
3 -4 V
3 -3 V
3 -4 V
2 -4 V
3 -3 V
3 -4 V
3 -3 V
3 -3 V
3 -3 V
3 -3 V
2 -3 V
3 -3 V
3 -3 V
3 -3 V
3 -3 V
3 -3 V
3 -3 V
2 -2 V
3 -3 V
3 -3 V
3 -2 V
3 -3 V
3 -2 V
2 -3 V
3 -2 V
3 -2 V
3 -3 V
3 -2 V
3 -2 V
3 -2 V
2 -2 V
3 -3 V
3 -2 V
3 -2 V
3 -2 V
3 -2 V
3 -2 V
2 -2 V
3 -2 V
3 -2 V
3 -2 V
3 -1 V
3 -2 V
2 -2 V
3 -2 V
3 -2 V
3 -1 V
3 -2 V
3 -2 V
3 -1 V
2 -2 V
3 -2 V
3 -1 V
3 -2 V
3 -2 V
3 -1 V
3 -2 V
2 -1 V
3 -2 V
3 -1 V
3 -2 V
3 -1 V
3 -2 V
2 -1 V
3 -1 V
3 -2 V
3 -1 V
3 -2 V
3 -1 V
3 -1 V
2 -2 V
3 -1 V
3 -1 V
3 -1 V
3 -2 V
3 -1 V
3 -1 V
2 -1 V
3 -2 V
3 -1 V
3 -1 V
3 -1 V
3 -1 V
2 -2 V
3 -1 V
3 -1 V
3 -1 V
3 -1 V
3 -1 V
3 -1 V
2 -1 V
3 -2 V
3 -1 V
3 -1 V
3 -1 V
3 -1 V
3 -1 V
2 -1 V
3 -1 V
3 -1 V
3 -1 V
3 -1 V
3 -1 V
2 -1 V
3 -1 V
3 -1 V
3 -1 V
3 -1 V
3 -1 V
3 0 V
2 -1 V
3 -1 V
3 -1 V
3 -1 V
3 -1 V
3 -1 V
3 -1 V
2 -1 V
3 -1 V
3 0 V
3 -1 V
3 -1 V
3 -1 V
2 -1 V
3 -1 V
3 0 V
3 -1 V
3 -1 V
3 -1 V
3 -1 V
2 0 V
3 -1 V
3 -1 V
3 -1 V
3 0 V
3 -1 V
3 -1 V
2 -1 V
3 0 V
3 -1 V
3 -1 V
3 -1 V
3 0 V
2 -1 V
3 -1 V
3 0 V
3 -1 V
3 -1 V
3 -1 V
3 0 V
2 -1 V
3 -1 V
3 0 V
3 -1 V
3 -1 V
3 0 V
3 -1 V
2 0 V
3 -1 V
3 -1 V
3 0 V
3 -1 V
3 -1 V
2 0 V
3 -1 V
3 -1 V
3 0 V
3 -1 V
3 0 V
3 -1 V
2 0 V
3 -1 V
3 -1 V
3 0 V
3 -1 V
3 0 V
3 -1 V
2 -1 V
3 0 V
3 -1 V
3 0 V
3 -1 V
3 0 V
2 -1 V
3 0 V
3 -1 V
3 0 V
3 -1 V
3 0 V
3 -1 V
2 -1 V
3 0 V
3 -1 V
3 0 V
3 -1 V
3 0 V
3 -1 V
2 0 V
3 -1 V
3 0 V
3 -1 V
3 0 V
3 0 V
2 -1 V
3 0 V
3 -1 V
3 0 V
3 -1 V
3 0 V
3 -1 V
2 0 V
3 -1 V
3 0 V
3 -1 V
3 0 V
3 0 V
3 -1 V
2 0 V
3 -1 V
3 0 V
3 -1 V
3 0 V
3 0 V
2 -1 V
3 0 V
3 -1 V
3 0 V
3 -1 V
3 0 V
3 0 V
2 -1 V
3 0 V
3 -1 V
3 0 V
3 0 V
3 -1 V
3 0 V
2 0 V
3 -1 V
3 0 V
3 -1 V
3 0 V
3 0 V
2 -1 V
3 0 V
3 0 V
3 -1 V
3 0 V
3 -1 V
3 0 V
2 0 V
3 -1 V
3 0 V
3 0 V
3 -1 V
3 0 V
3 0 V
2 -1 V
3 0 V
3 0 V
3 -1 V
3 0 V
3 0 V
2 -1 V
3 0 V
3 0 V
3 -1 V
3 0 V
3 0 V
3 -1 V
2 0 V
3 0 V
3 0 V
3 -1 V
3 0 V
3 0 V
3 -1 V
2 0 V
3 0 V
3 -1 V
3 0 V
3 0 V
3 0 V
2 -1 V
currentpoint stroke M
3 0 V
3 0 V
3 -1 V
3 0 V
3 0 V
3 0 V
2 -1 V
3 0 V
3 0 V
3 0 V
3 -1 V
3 0 V
3 0 V
2 0 V
3 -1 V
3 0 V
3 0 V
3 0 V
3 -1 V
2 0 V
3 0 V
3 0 V
3 -1 V
3 0 V
3 0 V
3 0 V
2 -1 V
3 0 V
3 0 V
3 0 V
3 -1 V
3 0 V
3 0 V
2 0 V
3 -1 V
3 0 V
3 0 V
3 0 V
3 0 V
2 -1 V
3 0 V
3 0 V
3 0 V
3 0 V
3 -1 V
3 0 V
2 0 V
3 0 V
3 0 V
3 -1 V
3 0 V
3 0 V
3 0 V
2 0 V
3 -1 V
3 0 V
3 0 V
3 0 V
3 0 V
2 -1 V
3 0 V
3 0 V
3 0 V
3 0 V
3 0 V
3 -1 V
2 0 V
3 0 V
3 0 V
3 0 V
3 -1 V
3 0 V
3 0 V
2 0 V
3 0 V
3 0 V
3 0 V
3 -1 V
3 0 V
2 0 V
3 0 V
3 0 V
3 0 V
3 -1 V
3 0 V
3 0 V
2 0 V
3 0 V
3 0 V
3 0 V
3 -1 V
3 0 V
3 0 V
2 0 V
3 0 V
3 0 V
3 0 V
3 -1 V
3 0 V
2 0 V
3 0 V
3 0 V
3 0 V
3 0 V
3 0 V
3 -1 V
2 0 V
3 0 V
3 0 V
3 0 V
3 0 V
3 0 V
3 0 V
2 0 V
3 -1 V
3 0 V
3 0 V
3 0 V
3 0 V
2 0 V
3 0 V
3 0 V
3 0 V
3 0 V
3 -1 V
3 0 V
2 0 V
3 0 V
3 0 V
3 0 V
3 0 V
3 0 V
3 0 V
2 0 V
3 0 V
3 -1 V
3 0 V
3 0 V
3 0 V
2 0 V
3 0 V
3 0 V
3 0 V
3 0 V
3 0 V
3 0 V
2 0 V
3 0 V
3 0 V
3 -1 V
3 0 V
3 0 V
3 0 V
2 0 V
3 0 V
3 0 V
3 0 V
3 0 V
3 0 V
2 0 V
3 0 V
3 0 V
3 0 V
3 0 V
3 0 V
3 0 V
2 0 V
3 0 V
3 -1 V
3 0 V
3 0 V
3 0 V
3 0 V
2 0 V
3 0 V
3 0 V
3 0 V
3 0 V
3 0 V
2 0 V
3 0 V
3 0 V
3 0 V
3 0 V
3 0 V
3 0 V
2 0 V
3 0 V
3 0 V
3 0 V
3 0 V
3 0 V
3 0 V
2 0 V
3 0 V
3 0 V
3 0 V
3 0 V
3 0 V
2 0 V
3 0 V
3 0 V
3 0 V
stroke
grestore
end
showpage
}}%
\put(3053,1647){\makebox(0,0)[r]{$A=0.00$}}%
\put(3053,1747){\makebox(0,0)[r]{$A=0.02$}}%
\put(3053,1847){\makebox(0,0)[r]{$A=0.04$}}%
\put(3053,1947){\makebox(0,0)[r]{$A=0.08$}}%
\put(657,1647){\makebox(0,0)[l]{$h^{-2}\sigma_e(q)$}}%
\put(1883,324){\makebox(0,0)[l]{$qh/\pi$}}%
\put(3450,100){\makebox(0,0){ 1}}%
\put(2880,100){\makebox(0,0){ 0.8}}%
\put(2310,100){\makebox(0,0){ 0.6}}%
\put(1740,100){\makebox(0,0){ 0.4}}%
\put(1170,100){\makebox(0,0){ 0.2}}%
\put(600,100){\makebox(0,0){ 0}}%
\put(550,1853){\makebox(0,0)[r]{$10^{4}$}}%
\put(550,1440){\makebox(0,0)[r]{$10^{2}$}}%
\put(550,1027){\makebox(0,0)[r]{$1$}}%
\put(550,613){\makebox(0,0)[r]{$10^{-2}$}}%
\put(550,200){\makebox(0,0)[r]{$10^{-4}$}}%
\end{picture}%
\endgroup
 

%% file: ssp_dord_ort_a3_o=0.984.tex
\begin{picture}(0,0)%
\includegraphics{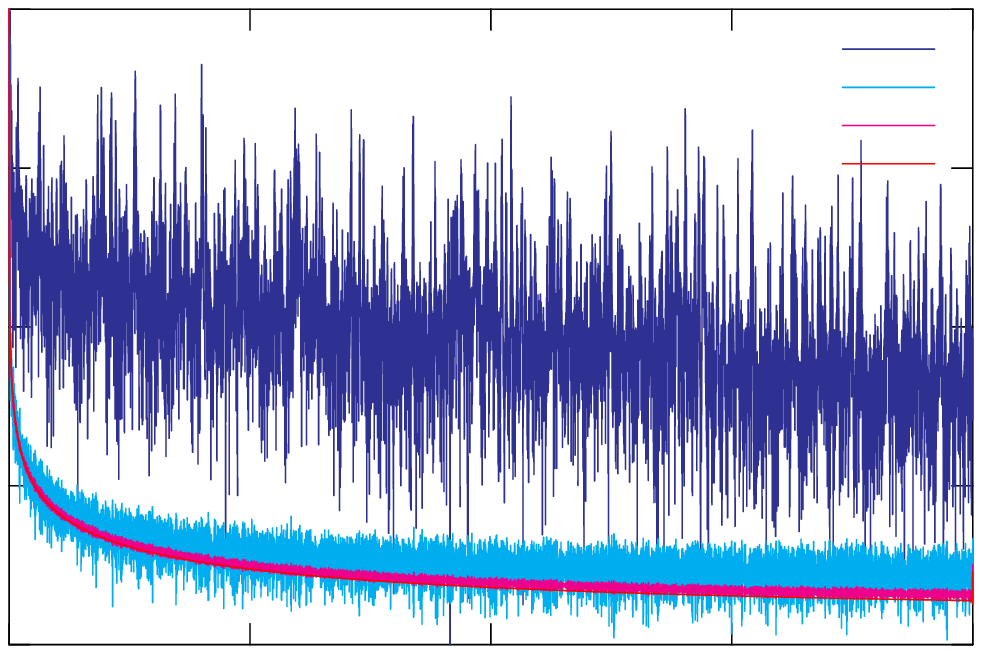}%
\end{picture}%
\setlength{\unitlength}{0.0200bp}%
\begin{picture}(18000,10800)(0,0)%
\put(3025,1100){\makebox(0,0)[r]{\strut{}$10^{-4}$}}%
\put(3025,3387){\makebox(0,0)[r]{\strut{}$10^{-3}$}}%
\put(3025,5675){\makebox(0,0)[r]{\strut{}$10^{-2}$}}%
\put(3025,7962){\makebox(0,0)[r]{\strut{}$10^{-1}$}}%
\put(3025,10250){\makebox(0,0)[r]{\strut{}$1$}}%
\put(3300,550){\makebox(0,0){\strut{}1}}%
\put(6768,550){\makebox(0,0){\strut{}2500}}%
\put(10237,550){\makebox(0,0){\strut{}5000}}%
\put(13706,550){\makebox(0,0){\strut{}7500}}%
\put(17175,550){\makebox(0,0){\strut{}10000}}%
\put(15094,593){\makebox(0,0)[l]{\strut{}$x/h$}}%
\put(1217,8873){\makebox(0,0)[l]{\strut{}$\vert {\mathcal F}_k(x) \vert$}}%
\put(15025,9675){\makebox(0,0)[r]{\strut{}$A=0.016$}}%
\put(15025,9125){\makebox(0,0)[r]{\strut{}$A=0.004$}}%
\put(15025,8575){\makebox(0,0)[r]{\strut{}$A=0.010$}}%
\put(15025,8025){\makebox(0,0)[r]{\strut{}$A=0$}}%
\end{picture}%
 

%% file: qspect_a3.tex
\begingroup%
  \makeatletter%
  \newcommand{\GNUPLOTspecial}{%
    \@sanitize\catcode`\%=14\relax\special}%
  \setlength{\unitlength}{0.1bp}%
{\GNUPLOTspecial{!
/gnudict 256 dict def
gnudict begin
/Color true def
/Solid true def
/gnulinewidth 5.000 def
/userlinewidth gnulinewidth def
/vshift -33 def
/dl {10 mul} def
/hpt_ 31.5 def
/vpt_ 31.5 def
/hpt hpt_ def
/vpt vpt_ def
/M {moveto} bind def
/L {lineto} bind def
/R {rmoveto} bind def
/V {rlineto} bind def
/vpt2 vpt 2 mul def
/hpt2 hpt 2 mul def
/Lshow { currentpoint stroke M
  0 vshift R show } def
/Rshow { currentpoint stroke M
  dup stringwidth pop neg vshift R show } def
/Cshow { currentpoint stroke M
  dup stringwidth pop -2 div vshift R show } def
/UP { dup vpt_ mul /vpt exch def hpt_ mul /hpt exch def
  /hpt2 hpt 2 mul def /vpt2 vpt 2 mul def } def
/DL { Color {setrgbcolor Solid {pop []} if 0 setdash }
 {pop pop pop Solid {pop []} if 0 setdash} ifelse } def
/BL { stroke userlinewidth 2 mul setlinewidth } def
/AL { stroke userlinewidth 2 div setlinewidth } def
/UL { dup gnulinewidth mul /userlinewidth exch def
      dup 1 lt {pop 1} if 10 mul /udl exch def } def
/PL { stroke userlinewidth setlinewidth } def
/LTb { BL [] 0 0 0 DL } def
/LTa { AL [1 udl mul 2 udl mul] 0 setdash 0 0 0 setrgbcolor } def
/LT0 { PL [] 1 0 0 DL } def
/LT1 { PL [4 dl 2 dl] 0 1 0 DL } def
/LT2 { PL [2 dl 3 dl] 0 0 1 DL } def
/LT3 { PL [1 dl 1.5 dl] 1 0 1 DL } def
/LT4 { PL [5 dl 2 dl 1 dl 2 dl] 0 1 1 DL } def
/LT5 { PL [4 dl 3 dl 1 dl 3 dl] 1 1 0 DL } def
/LT6 { PL [2 dl 2 dl 2 dl 4 dl] 0 0 0 DL } def
/LT7 { PL [2 dl 2 dl 2 dl 2 dl 2 dl 4 dl] 1 0.3 0 DL } def
/LT8 { PL [2 dl 2 dl 2 dl 2 dl 2 dl 2 dl 2 dl 4 dl] 0.5 0.5 0.5 DL } def
/Pnt { stroke [] 0 setdash
   gsave 1 setlinecap M 0 0 V stroke grestore } def
/Dia { stroke [] 0 setdash 2 copy vpt add M
  hpt neg vpt neg V hpt vpt neg V
  hpt vpt V hpt neg vpt V closepath stroke
  Pnt } def
/Pls { stroke [] 0 setdash vpt sub M 0 vpt2 V
  currentpoint stroke M
  hpt neg vpt neg R hpt2 0 V stroke
  } def
/Box { stroke [] 0 setdash 2 copy exch hpt sub exch vpt add M
  0 vpt2 neg V hpt2 0 V 0 vpt2 V
  hpt2 neg 0 V closepath stroke
  Pnt } def
/Crs { stroke [] 0 setdash exch hpt sub exch vpt add M
  hpt2 vpt2 neg V currentpoint stroke M
  hpt2 neg 0 R hpt2 vpt2 V stroke } def
/TriU { stroke [] 0 setdash 2 copy vpt 1.12 mul add M
  hpt neg vpt -1.62 mul V
  hpt 2 mul 0 V
  hpt neg vpt 1.62 mul V closepath stroke
  Pnt  } def
/Star { 2 copy Pls Crs } def
/BoxF { stroke [] 0 setdash exch hpt sub exch vpt add M
  0 vpt2 neg V  hpt2 0 V  0 vpt2 V
  hpt2 neg 0 V  closepath fill } def
/TriUF { stroke [] 0 setdash vpt 1.12 mul add M
  hpt neg vpt -1.62 mul V
  hpt 2 mul 0 V
  hpt neg vpt 1.62 mul V closepath fill } def
/TriD { stroke [] 0 setdash 2 copy vpt 1.12 mul sub M
  hpt neg vpt 1.62 mul V
  hpt 2 mul 0 V
  hpt neg vpt -1.62 mul V closepath stroke
  Pnt  } def
/TriDF { stroke [] 0 setdash vpt 1.12 mul sub M
  hpt neg vpt 1.62 mul V
  hpt 2 mul 0 V
  hpt neg vpt -1.62 mul V closepath fill} def
/DiaF { stroke [] 0 setdash vpt add M
  hpt neg vpt neg V hpt vpt neg V
  hpt vpt V hpt neg vpt V closepath fill } def
/Pent { stroke [] 0 setdash 2 copy gsave
  translate 0 hpt M 4 {72 rotate 0 hpt L} repeat
  closepath stroke grestore Pnt } def
/PentF { stroke [] 0 setdash gsave
  translate 0 hpt M 4 {72 rotate 0 hpt L} repeat
  closepath fill grestore } def
/Circle { stroke [] 0 setdash 2 copy
  hpt 0 360 arc stroke Pnt } def
/CircleF { stroke [] 0 setdash hpt 0 360 arc fill } def
/C0 { BL [] 0 setdash 2 copy moveto vpt 90 450  arc } bind def
/C1 { BL [] 0 setdash 2 copy        moveto
       2 copy  vpt 0 90 arc closepath fill
               vpt 0 360 arc closepath } bind def
/C2 { BL [] 0 setdash 2 copy moveto
       2 copy  vpt 90 180 arc closepath fill
               vpt 0 360 arc closepath } bind def
/C3 { BL [] 0 setdash 2 copy moveto
       2 copy  vpt 0 180 arc closepath fill
               vpt 0 360 arc closepath } bind def
/C4 { BL [] 0 setdash 2 copy moveto
       2 copy  vpt 180 270 arc closepath fill
               vpt 0 360 arc closepath } bind def
/C5 { BL [] 0 setdash 2 copy moveto
       2 copy  vpt 0 90 arc
       2 copy moveto
       2 copy  vpt 180 270 arc closepath fill
               vpt 0 360 arc } bind def
/C6 { BL [] 0 setdash 2 copy moveto
      2 copy  vpt 90 270 arc closepath fill
              vpt 0 360 arc closepath } bind def
/C7 { BL [] 0 setdash 2 copy moveto
      2 copy  vpt 0 270 arc closepath fill
              vpt 0 360 arc closepath } bind def
/C8 { BL [] 0 setdash 2 copy moveto
      2 copy vpt 270 360 arc closepath fill
              vpt 0 360 arc closepath } bind def
/C9 { BL [] 0 setdash 2 copy moveto
      2 copy  vpt 270 450 arc closepath fill
              vpt 0 360 arc closepath } bind def
/C10 { BL [] 0 setdash 2 copy 2 copy moveto vpt 270 360 arc closepath fill
       2 copy moveto
       2 copy vpt 90 180 arc closepath fill
               vpt 0 360 arc closepath } bind def
/C11 { BL [] 0 setdash 2 copy moveto
       2 copy  vpt 0 180 arc closepath fill
       2 copy moveto
       2 copy  vpt 270 360 arc closepath fill
               vpt 0 360 arc closepath } bind def
/C12 { BL [] 0 setdash 2 copy moveto
       2 copy  vpt 180 360 arc closepath fill
               vpt 0 360 arc closepath } bind def
/C13 { BL [] 0 setdash  2 copy moveto
       2 copy  vpt 0 90 arc closepath fill
       2 copy moveto
       2 copy  vpt 180 360 arc closepath fill
               vpt 0 360 arc closepath } bind def
/C14 { BL [] 0 setdash 2 copy moveto
       2 copy  vpt 90 360 arc closepath fill
               vpt 0 360 arc } bind def
/C15 { BL [] 0 setdash 2 copy vpt 0 360 arc closepath fill
               vpt 0 360 arc closepath } bind def
/Rec   { newpath 4 2 roll moveto 1 index 0 rlineto 0 exch rlineto
       neg 0 rlineto closepath } bind def
/Square { dup Rec } bind def
/Bsquare { vpt sub exch vpt sub exch vpt2 Square } bind def
/S0 { BL [] 0 setdash 2 copy moveto 0 vpt rlineto BL Bsquare } bind def
/S1 { BL [] 0 setdash 2 copy vpt Square fill Bsquare } bind def
/S2 { BL [] 0 setdash 2 copy exch vpt sub exch vpt Square fill Bsquare } bind def
/S3 { BL [] 0 setdash 2 copy exch vpt sub exch vpt2 vpt Rec fill Bsquare } bind def
/S4 { BL [] 0 setdash 2 copy exch vpt sub exch vpt sub vpt Square fill Bsquare } bind def
/S5 { BL [] 0 setdash 2 copy 2 copy vpt Square fill
       exch vpt sub exch vpt sub vpt Square fill Bsquare } bind def
/S6 { BL [] 0 setdash 2 copy exch vpt sub exch vpt sub vpt vpt2 Rec fill Bsquare } bind def
/S7 { BL [] 0 setdash 2 copy exch vpt sub exch vpt sub vpt vpt2 Rec fill
       2 copy vpt Square fill
       Bsquare } bind def
/S8 { BL [] 0 setdash 2 copy vpt sub vpt Square fill Bsquare } bind def
/S9 { BL [] 0 setdash 2 copy vpt sub vpt vpt2 Rec fill Bsquare } bind def
/S10 { BL [] 0 setdash 2 copy vpt sub vpt Square fill 2 copy exch vpt sub exch vpt Square fill
       Bsquare } bind def
/S11 { BL [] 0 setdash 2 copy vpt sub vpt Square fill 2 copy exch vpt sub exch vpt2 vpt Rec fill
       Bsquare } bind def
/S12 { BL [] 0 setdash 2 copy exch vpt sub exch vpt sub vpt2 vpt Rec fill Bsquare } bind def
/S13 { BL [] 0 setdash 2 copy exch vpt sub exch vpt sub vpt2 vpt Rec fill
       2 copy vpt Square fill Bsquare } bind def
/S14 { BL [] 0 setdash 2 copy exch vpt sub exch vpt sub vpt2 vpt Rec fill
       2 copy exch vpt sub exch vpt Square fill Bsquare } bind def
/S15 { BL [] 0 setdash 2 copy Bsquare fill Bsquare } bind def
/D0 { gsave translate 45 rotate 0 0 S0 stroke grestore } bind def
/D1 { gsave translate 45 rotate 0 0 S1 stroke grestore } bind def
/D2 { gsave translate 45 rotate 0 0 S2 stroke grestore } bind def
/D3 { gsave translate 45 rotate 0 0 S3 stroke grestore } bind def
/D4 { gsave translate 45 rotate 0 0 S4 stroke grestore } bind def
/D5 { gsave translate 45 rotate 0 0 S5 stroke grestore } bind def
/D6 { gsave translate 45 rotate 0 0 S6 stroke grestore } bind def
/D7 { gsave translate 45 rotate 0 0 S7 stroke grestore } bind def
/D8 { gsave translate 45 rotate 0 0 S8 stroke grestore } bind def
/D9 { gsave translate 45 rotate 0 0 S9 stroke grestore } bind def
/D10 { gsave translate 45 rotate 0 0 S10 stroke grestore } bind def
/D11 { gsave translate 45 rotate 0 0 S11 stroke grestore } bind def
/D12 { gsave translate 45 rotate 0 0 S12 stroke grestore } bind def
/D13 { gsave translate 45 rotate 0 0 S13 stroke grestore } bind def
/D14 { gsave translate 45 rotate 0 0 S14 stroke grestore } bind def
/D15 { gsave translate 45 rotate 0 0 S15 stroke grestore } bind def
/DiaE { stroke [] 0 setdash vpt add M
  hpt neg vpt neg V hpt vpt neg V
  hpt vpt V hpt neg vpt V closepath stroke } def
/BoxE { stroke [] 0 setdash exch hpt sub exch vpt add M
  0 vpt2 neg V hpt2 0 V 0 vpt2 V
  hpt2 neg 0 V closepath stroke } def
/TriUE { stroke [] 0 setdash vpt 1.12 mul add M
  hpt neg vpt -1.62 mul V
  hpt 2 mul 0 V
  hpt neg vpt 1.62 mul V closepath stroke } def
/TriDE { stroke [] 0 setdash vpt 1.12 mul sub M
  hpt neg vpt 1.62 mul V
  hpt 2 mul 0 V
  hpt neg vpt -1.62 mul V closepath stroke } def
/PentE { stroke [] 0 setdash gsave
  translate 0 hpt M 4 {72 rotate 0 hpt L} repeat
  closepath stroke grestore } def
/CircE { stroke [] 0 setdash 
  hpt 0 360 arc stroke } def
/Opaque { gsave closepath 1 setgray fill grestore 0 setgray closepath } def
/DiaW { stroke [] 0 setdash vpt add M
  hpt neg vpt neg V hpt vpt neg V
  hpt vpt V hpt neg vpt V Opaque stroke } def
/BoxW { stroke [] 0 setdash exch hpt sub exch vpt add M
  0 vpt2 neg V hpt2 0 V 0 vpt2 V
  hpt2 neg 0 V Opaque stroke } def
/TriUW { stroke [] 0 setdash vpt 1.12 mul add M
  hpt neg vpt -1.62 mul V
  hpt 2 mul 0 V
  hpt neg vpt 1.62 mul V Opaque stroke } def
/TriDW { stroke [] 0 setdash vpt 1.12 mul sub M
  hpt neg vpt 1.62 mul V
  hpt 2 mul 0 V
  hpt neg vpt -1.62 mul V Opaque stroke } def
/PentW { stroke [] 0 setdash gsave
  translate 0 hpt M 4 {72 rotate 0 hpt L} repeat
  Opaque stroke grestore } def
/CircW { stroke [] 0 setdash 
  hpt 0 360 arc Opaque stroke } def
/BoxFill { gsave Rec 1 setgray fill grestore } def
/Symbol-Oblique /Symbol findfont [1 0 .167 1 0 0] makefont
dup length dict begin {1 index /FID eq {pop pop} {def} ifelse} forall
currentdict end definefont pop
end
}}%
\begin{picture}(3600,2160)(0,0)%
{\GNUPLOTspecial{"
gnudict begin
gsave
0 0 translate
0.100 0.100 scale
0 setgray
newpath
1.000 UL
LTb
600 200 M
63 0 V
2787 0 R
-63 0 V
600 613 M
63 0 V
2787 0 R
-63 0 V
600 1027 M
63 0 V
2787 0 R
-63 0 V
600 1440 M
63 0 V
2787 0 R
-63 0 V
600 1853 M
63 0 V
2787 0 R
-63 0 V
600 200 M
0 63 V
0 1797 R
0 -63 V
1170 200 M
0 63 V
0 1797 R
0 -63 V
1740 200 M
0 63 V
0 1797 R
0 -63 V
2310 200 M
0 63 V
0 1797 R
0 -63 V
2880 200 M
0 63 V
0 1797 R
0 -63 V
3450 200 M
0 63 V
0 1797 R
0 -63 V
1.000 UL
LTb
600 200 M
2850 0 V
0 1860 V
-2850 0 V
600 200 L
1.000 UL
LT4
3103 1947 M
247 0 V
600 990 M
3 0 V
3 0 V
3 0 V
2 0 V
3 0 V
3 -1 V
3 0 V
3 0 V
3 0 V
2 0 V
3 0 V
3 0 V
3 0 V
3 0 V
3 0 V
3 0 V
2 1 V
3 0 V
3 0 V
3 0 V
3 1 V
3 0 V
3 0 V
2 1 V
3 0 V
3 0 V
3 0 V
3 0 V
3 0 V
2 0 V
3 0 V
3 0 V
3 0 V
3 0 V
3 1 V
3 0 V
2 0 V
3 0 V
3 0 V
3 1 V
3 0 V
3 0 V
3 0 V
2 1 V
3 0 V
3 0 V
3 0 V
3 1 V
3 0 V
3 0 V
2 1 V
3 0 V
3 0 V
3 1 V
3 0 V
3 1 V
2 0 V
3 0 V
3 1 V
3 0 V
3 1 V
3 0 V
3 0 V
2 1 V
3 0 V
3 0 V
3 1 V
3 0 V
3 0 V
3 1 V
2 0 V
3 0 V
3 1 V
3 0 V
3 1 V
3 0 V
2 0 V
3 1 V
3 0 V
3 0 V
3 1 V
3 0 V
3 0 V
2 1 V
3 0 V
3 1 V
3 0 V
3 0 V
3 1 V
3 0 V
2 1 V
3 1 V
3 0 V
3 1 V
3 1 V
3 0 V
2 1 V
3 1 V
3 0 V
3 1 V
3 1 V
3 0 V
3 1 V
2 0 V
3 1 V
3 1 V
3 0 V
3 1 V
3 1 V
3 1 V
2 1 V
3 0 V
3 1 V
3 1 V
3 0 V
3 1 V
2 1 V
3 0 V
3 1 V
3 1 V
3 0 V
3 1 V
3 0 V
2 1 V
3 1 V
3 0 V
3 1 V
3 1 V
3 0 V
3 1 V
2 1 V
3 0 V
3 1 V
3 0 V
3 1 V
3 0 V
2 1 V
3 1 V
3 1 V
3 0 V
3 1 V
3 1 V
3 1 V
2 1 V
3 0 V
3 1 V
3 1 V
3 0 V
3 1 V
3 1 V
2 1 V
3 0 V
3 1 V
3 1 V
3 1 V
3 1 V
2 1 V
3 1 V
3 1 V
3 0 V
3 1 V
3 1 V
3 1 V
2 1 V
3 1 V
3 0 V
3 1 V
3 1 V
3 1 V
3 1 V
2 0 V
3 1 V
3 1 V
3 1 V
3 0 V
3 1 V
2 1 V
3 0 V
3 1 V
3 0 V
3 1 V
3 0 V
3 0 V
2 1 V
3 0 V
3 0 V
3 0 V
3 1 V
3 0 V
3 0 V
2 -1 V
3 0 V
3 -1 V
3 -1 V
3 -2 V
3 -3 V
2 -5 V
3 -8 V
3 -15 V
3 -106 V
3 39 V
3 47 V
3 34 V
2 27 V
3 23 V
3 19 V
3 17 V
3 15 V
3 14 V
3 11 V
2 11 V
3 9 V
3 7 V
3 7 V
3 6 V
3 5 V
2 4 V
3 3 V
3 2 V
3 2 V
3 0 V
3 1 V
3 0 V
2 -1 V
3 0 V
3 0 V
3 -1 V
3 -1 V
3 -1 V
3 -1 V
2 -1 V
3 -1 V
3 -1 V
3 0 V
3 -1 V
3 0 V
2 0 V
3 0 V
3 0 V
3 0 V
3 0 V
3 0 V
3 0 V
2 0 V
3 0 V
3 0 V
3 0 V
3 -1 V
3 0 V
3 0 V
2 0 V
3 0 V
3 0 V
3 1 V
3 0 V
3 1 V
2 0 V
3 1 V
3 0 V
3 0 V
3 1 V
3 0 V
3 0 V
2 1 V
3 0 V
3 0 V
3 1 V
3 0 V
3 0 V
3 0 V
2 1 V
3 0 V
3 0 V
3 0 V
3 0 V
3 1 V
2 0 V
3 0 V
3 0 V
3 0 V
3 0 V
3 0 V
3 -1 V
2 0 V
3 -1 V
3 0 V
3 0 V
3 -1 V
3 0 V
3 -1 V
2 0 V
3 -1 V
3 0 V
3 -1 V
3 -1 V
3 0 V
2 -1 V
3 0 V
3 -1 V
3 -1 V
3 0 V
3 -2 V
3 0 V
2 -2 V
3 -1 V
3 -2 V
3 -1 V
3 -1 V
3 0 V
3 -1 V
2 -1 V
3 0 V
3 -1 V
3 0 V
3 -1 V
3 0 V
2 -1 V
3 -1 V
3 -1 V
3 -1 V
3 -1 V
3 -2 V
3 -1 V
2 -2 V
3 -1 V
3 -2 V
3 -1 V
3 -1 V
3 -2 V
3 -2 V
2 -1 V
3 -2 V
3 -1 V
3 -1 V
3 -2 V
3 -1 V
2 -1 V
3 -1 V
3 -1 V
3 -1 V
3 -1 V
3 0 V
3 -1 V
2 -1 V
3 -1 V
3 -1 V
3 -2 V
3 -2 V
3 -1 V
3 -2 V
2 -2 V
3 -1 V
3 -2 V
3 -2 V
3 -2 V
3 -1 V
2 -2 V
3 -2 V
3 -1 V
3 -2 V
3 -1 V
3 -2 V
3 -1 V
2 -2 V
3 -2 V
3 -1 V
3 -2 V
3 -1 V
3 -1 V
3 -2 V
2 -1 V
3 -1 V
3 -1 V
3 -1 V
3 -1 V
3 -2 V
2 -1 V
3 -1 V
3 -1 V
3 -1 V
3 -2 V
3 -1 V
3 -2 V
2 -2 V
3 -2 V
3 -1 V
3 -2 V
3 -1 V
3 -2 V
3 -1 V
2 -2 V
3 -1 V
3 -2 V
3 -1 V
3 -1 V
3 -2 V
2 -2 V
currentpoint stroke M
3 -1 V
3 -1 V
3 -1 V
3 -1 V
3 -2 V
3 -1 V
2 0 V
3 -2 V
3 -1 V
3 -2 V
3 -1 V
3 -1 V
3 -1 V
2 -2 V
3 -1 V
3 -1 V
3 -2 V
3 -1 V
3 -1 V
2 -1 V
3 0 V
3 -1 V
3 -1 V
3 -1 V
3 -1 V
3 -1 V
2 -1 V
3 -2 V
3 -1 V
3 -2 V
3 -2 V
3 -2 V
3 -2 V
2 -2 V
3 -2 V
3 -2 V
3 -2 V
3 -2 V
3 -2 V
2 -1 V
3 -2 V
3 -2 V
3 -1 V
3 -1 V
3 -1 V
3 -1 V
2 -2 V
3 -1 V
3 0 V
3 -1 V
3 -1 V
3 -1 V
3 0 V
2 -1 V
3 0 V
3 -1 V
3 -1 V
3 -1 V
3 -1 V
2 -1 V
3 0 V
3 -1 V
3 -1 V
3 -1 V
3 -1 V
3 -1 V
2 -1 V
3 -1 V
3 -1 V
3 -2 V
3 -1 V
3 0 V
3 -2 V
2 -1 V
3 -1 V
3 0 V
3 -1 V
3 -1 V
3 -1 V
2 0 V
3 0 V
3 0 V
3 0 V
3 -1 V
3 -1 V
3 -1 V
2 0 V
3 -1 V
3 -1 V
3 -1 V
3 -1 V
3 -1 V
3 -1 V
2 -1 V
3 -1 V
3 0 V
3 -1 V
3 0 V
3 -1 V
2 0 V
3 -1 V
3 -1 V
3 0 V
3 -1 V
3 0 V
3 0 V
2 -1 V
3 -1 V
3 -1 V
3 -1 V
3 -1 V
3 -1 V
3 -1 V
2 -1 V
3 0 V
3 -1 V
3 -1 V
3 -1 V
3 -1 V
2 -1 V
3 -1 V
3 -1 V
3 -1 V
3 -1 V
3 -1 V
3 -2 V
2 -1 V
3 -1 V
3 -1 V
3 -1 V
3 -1 V
3 0 V
3 -1 V
2 -1 V
3 0 V
3 -1 V
3 0 V
3 -1 V
3 0 V
2 0 V
3 0 V
3 -1 V
3 0 V
3 -1 V
3 -1 V
3 -1 V
2 0 V
3 -1 V
3 -1 V
3 -1 V
3 0 V
3 -1 V
3 -1 V
2 -1 V
3 0 V
3 0 V
3 0 V
3 0 V
3 0 V
2 -1 V
3 -1 V
3 0 V
3 -1 V
3 -1 V
3 0 V
3 -1 V
2 -1 V
3 -1 V
3 -1 V
3 0 V
3 0 V
3 0 V
3 -1 V
2 0 V
3 0 V
3 0 V
3 -1 V
3 0 V
3 0 V
2 0 V
3 0 V
3 0 V
3 0 V
3 0 V
3 0 V
3 -1 V
2 0 V
3 -1 V
3 -1 V
3 -1 V
3 0 V
3 -1 V
3 0 V
2 -1 V
3 0 V
3 0 V
3 -1 V
3 0 V
3 0 V
2 0 V
3 0 V
3 -1 V
3 0 V
3 0 V
3 0 V
3 0 V
2 0 V
3 1 V
3 0 V
3 0 V
3 1 V
3 0 V
3 0 V
2 0 V
3 0 V
3 0 V
3 0 V
3 0 V
3 -1 V
2 -1 V
3 -1 V
3 0 V
3 -1 V
3 -1 V
3 0 V
3 -1 V
2 -1 V
3 -1 V
3 -1 V
3 0 V
3 -1 V
3 0 V
3 -1 V
2 0 V
3 -1 V
3 0 V
3 -1 V
3 -1 V
3 0 V
2 -1 V
3 -1 V
3 0 V
3 -1 V
3 -1 V
3 -1 V
3 -1 V
2 -1 V
3 0 V
3 0 V
3 -1 V
3 0 V
3 0 V
3 -1 V
2 0 V
3 -1 V
3 0 V
3 -1 V
3 0 V
3 -1 V
2 -1 V
3 -1 V
3 0 V
3 -1 V
3 -1 V
3 0 V
3 -2 V
2 0 V
3 -1 V
3 0 V
3 -1 V
3 0 V
3 0 V
3 -1 V
2 0 V
3 -1 V
3 -1 V
3 -1 V
3 0 V
3 -2 V
2 0 V
3 -1 V
3 -1 V
3 0 V
3 -1 V
3 0 V
3 -1 V
2 0 V
3 0 V
3 -1 V
3 -1 V
3 -1 V
3 -1 V
3 -1 V
2 -2 V
3 -1 V
3 -2 V
3 -1 V
3 -2 V
3 -1 V
2 -1 V
3 -1 V
3 -1 V
3 0 V
3 -1 V
3 -1 V
3 0 V
2 -1 V
3 0 V
3 0 V
3 -1 V
3 0 V
3 -1 V
3 -1 V
2 -1 V
3 0 V
3 -1 V
3 -1 V
3 0 V
3 0 V
2 0 V
3 -1 V
3 1 V
3 0 V
3 0 V
3 0 V
3 0 V
2 0 V
3 0 V
3 0 V
3 0 V
3 0 V
3 -1 V
3 0 V
2 0 V
3 0 V
3 0 V
3 -1 V
3 0 V
3 0 V
2 0 V
3 0 V
3 0 V
3 0 V
3 0 V
3 -1 V
3 1 V
2 0 V
3 0 V
3 -1 V
3 0 V
3 0 V
3 0 V
3 0 V
2 0 V
3 0 V
3 0 V
3 0 V
3 -1 V
3 0 V
2 0 V
3 -1 V
3 -1 V
3 0 V
3 -1 V
3 0 V
3 -1 V
2 0 V
3 0 V
3 0 V
3 0 V
3 0 V
3 0 V
3 0 V
2 1 V
3 0 V
3 0 V
3 1 V
3 0 V
3 0 V
2 0 V
3 0 V
3 0 V
3 0 V
3 0 V
3 0 V
3 0 V
2 0 V
3 0 V
3 -1 V
3 0 V
3 0 V
3 -1 V
3 -1 V
2 -1 V
3 -1 V
3 -1 V
3 -1 V
3 -1 V
3 -1 V
2 -1 V
currentpoint stroke M
3 0 V
3 -1 V
3 0 V
3 0 V
3 -1 V
3 0 V
2 0 V
3 0 V
3 -1 V
3 0 V
3 -1 V
3 -1 V
3 0 V
2 0 V
3 -1 V
3 0 V
3 0 V
3 0 V
3 0 V
2 0 V
3 -1 V
3 0 V
3 0 V
3 0 V
3 -1 V
3 0 V
2 -1 V
3 0 V
3 0 V
3 -1 V
3 -1 V
3 0 V
3 0 V
2 1 V
3 0 V
3 0 V
3 1 V
3 0 V
3 0 V
2 0 V
3 0 V
3 0 V
3 -1 V
3 0 V
3 0 V
3 0 V
2 -1 V
3 0 V
3 0 V
3 -1 V
3 -1 V
3 0 V
3 0 V
2 -1 V
3 0 V
3 0 V
3 -1 V
3 0 V
3 0 V
2 -1 V
3 0 V
3 0 V
3 0 V
3 0 V
3 0 V
3 -1 V
2 0 V
3 0 V
3 0 V
3 0 V
3 0 V
3 0 V
3 0 V
2 1 V
3 0 V
3 0 V
3 0 V
3 0 V
3 0 V
2 0 V
3 0 V
3 0 V
3 0 V
3 1 V
3 1 V
3 1 V
2 0 V
3 1 V
3 0 V
3 0 V
3 0 V
3 0 V
3 0 V
2 -1 V
3 -1 V
3 -1 V
3 -1 V
3 -1 V
3 -1 V
2 -2 V
3 -1 V
3 -1 V
3 -1 V
3 0 V
3 0 V
3 -1 V
2 0 V
3 -1 V
3 0 V
3 -1 V
3 -1 V
3 0 V
3 -1 V
2 -1 V
3 0 V
3 -1 V
3 0 V
3 -1 V
3 0 V
2 -1 V
3 0 V
3 0 V
3 0 V
3 0 V
3 0 V
3 -1 V
2 0 V
3 0 V
3 0 V
3 0 V
3 0 V
3 0 V
3 -1 V
2 0 V
3 0 V
3 0 V
3 1 V
3 0 V
3 0 V
2 0 V
3 0 V
3 -1 V
3 0 V
3 0 V
3 -1 V
3 -1 V
2 0 V
3 -1 V
3 0 V
3 0 V
3 0 V
3 0 V
3 0 V
2 0 V
3 1 V
3 0 V
3 0 V
3 1 V
3 0 V
2 1 V
3 0 V
3 1 V
3 0 V
3 1 V
3 0 V
3 1 V
2 0 V
3 0 V
3 -1 V
3 0 V
3 -1 V
3 0 V
3 0 V
2 0 V
3 -1 V
3 0 V
3 -1 V
3 0 V
3 0 V
2 0 V
3 0 V
3 0 V
3 0 V
3 0 V
3 -1 V
3 0 V
2 0 V
3 0 V
3 0 V
3 0 V
3 1 V
3 -1 V
3 0 V
2 -1 V
3 0 V
3 0 V
3 0 V
3 0 V
3 0 V
2 0 V
3 0 V
3 -1 V
3 0 V
1.000 UL
LT1
3103 1847 M
247 0 V
600 948 M
3 0 V
3 0 V
3 1 V
2 0 V
3 0 V
3 0 V
3 0 V
3 0 V
3 -1 V
2 1 V
3 0 V
3 1 V
3 1 V
3 0 V
3 0 V
3 -1 V
2 1 V
3 -1 V
3 0 V
3 1 V
3 -1 V
3 0 V
3 0 V
2 1 V
3 0 V
3 0 V
3 0 V
3 0 V
3 1 V
2 1 V
3 0 V
3 1 V
3 0 V
3 0 V
3 0 V
3 0 V
2 0 V
3 0 V
3 0 V
3 0 V
3 1 V
3 0 V
3 1 V
2 0 V
3 1 V
3 0 V
3 0 V
3 1 V
3 0 V
3 1 V
2 0 V
3 0 V
3 0 V
3 1 V
3 1 V
3 0 V
2 1 V
3 0 V
3 1 V
3 0 V
3 0 V
3 1 V
3 0 V
2 1 V
3 0 V
3 1 V
3 0 V
3 0 V
3 0 V
3 1 V
2 1 V
3 0 V
3 0 V
3 1 V
3 0 V
3 0 V
2 0 V
3 1 V
3 1 V
3 0 V
3 0 V
3 1 V
3 1 V
2 0 V
3 1 V
3 1 V
3 0 V
3 1 V
3 1 V
3 1 V
2 0 V
3 0 V
3 1 V
3 0 V
3 2 V
3 1 V
2 0 V
3 2 V
3 1 V
3 0 V
3 0 V
3 1 V
3 1 V
2 1 V
3 0 V
3 1 V
3 0 V
3 0 V
3 1 V
3 2 V
2 1 V
3 0 V
3 1 V
3 1 V
3 1 V
3 1 V
2 1 V
3 1 V
3 1 V
3 1 V
3 1 V
3 1 V
3 0 V
2 0 V
3 1 V
3 1 V
3 0 V
3 1 V
3 1 V
3 1 V
2 1 V
3 1 V
3 1 V
3 1 V
3 1 V
3 1 V
2 1 V
3 1 V
3 1 V
3 1 V
3 1 V
3 1 V
3 1 V
2 2 V
3 1 V
3 1 V
3 1 V
3 2 V
3 1 V
3 0 V
2 1 V
3 1 V
3 1 V
3 1 V
3 1 V
3 1 V
2 1 V
3 1 V
3 1 V
3 2 V
3 2 V
3 2 V
3 1 V
2 1 V
3 1 V
3 1 V
3 1 V
3 1 V
3 2 V
3 1 V
2 2 V
3 1 V
3 1 V
3 2 V
3 1 V
3 2 V
2 2 V
3 1 V
3 2 V
3 1 V
3 2 V
3 1 V
3 2 V
2 1 V
3 1 V
3 1 V
3 2 V
3 2 V
3 2 V
3 1 V
2 1 V
3 2 V
3 1 V
3 1 V
3 -1 V
3 0 V
2 -2 V
3 -5 V
3 -14 V
3 -120 V
3 8 V
3 75 V
3 67 V
2 68 V
3 77 V
3 70 V
3 35 V
3 -22 V
3 -41 V
3 -39 V
2 -31 V
3 -27 V
3 -19 V
3 -14 V
3 -16 V
3 -9 V
2 -2 V
3 -7 V
3 -9 V
3 -6 V
3 -5 V
3 -4 V
3 -6 V
2 -8 V
3 -5 V
3 -3 V
3 -6 V
3 -8 V
3 -4 V
3 -1 V
2 -6 V
3 -2 V
3 2 V
3 -4 V
3 -2 V
3 3 V
2 3 V
3 0 V
3 -4 V
3 -2 V
3 -1 V
3 -2 V
3 -3 V
2 0 V
3 -1 V
3 -5 V
3 1 V
3 8 V
3 4 V
3 0 V
2 0 V
3 2 V
3 -1 V
3 -5 V
3 1 V
3 6 V
2 1 V
3 -6 V
3 -1 V
3 4 V
3 -3 V
3 1 V
3 2 V
2 4 V
3 0 V
3 6 V
3 8 V
3 7 V
3 -1 V
3 1 V
2 -2 V
3 -5 V
3 -5 V
3 0 V
3 2 V
3 -5 V
2 -2 V
3 5 V
3 6 V
3 3 V
3 -5 V
3 -2 V
3 -2 V
2 5 V
3 10 V
3 2 V
3 -4 V
3 -3 V
3 -5 V
3 -4 V
2 3 V
3 2 V
3 3 V
3 1 V
3 1 V
3 -1 V
2 -1 V
3 2 V
3 6 V
3 6 V
3 6 V
3 5 V
3 6 V
2 3 V
3 0 V
3 0 V
3 -2 V
3 1 V
3 5 V
3 5 V
2 -1 V
3 1 V
3 6 V
3 6 V
3 5 V
3 1 V
2 -1 V
3 7 V
3 4 V
3 7 V
3 5 V
3 0 V
3 3 V
2 4 V
3 1 V
3 1 V
3 12 V
3 15 V
3 3 V
3 -3 V
2 4 V
3 6 V
3 9 V
3 3 V
3 -1 V
3 4 V
2 1 V
3 6 V
3 5 V
3 7 V
3 8 V
3 8 V
3 6 V
2 2 V
3 11 V
3 12 V
3 9 V
3 2 V
3 1 V
3 6 V
2 5 V
3 5 V
3 2 V
3 4 V
3 3 V
3 2 V
2 2 V
3 3 V
3 6 V
3 9 V
3 3 V
3 4 V
3 8 V
2 5 V
3 1 V
3 -4 V
3 2 V
3 3 V
3 0 V
3 -5 V
2 -10 V
3 -1 V
3 1 V
3 -6 V
3 -9 V
3 -4 V
2 -3 V
3 -4 V
3 -7 V
3 -6 V
3 -6 V
3 -11 V
3 -8 V
2 -4 V
3 -6 V
3 -7 V
3 -7 V
3 -7 V
3 -5 V
3 3 V
2 -1 V
3 -6 V
3 -10 V
3 -7 V
3 -2 V
3 -3 V
2 -6 V
currentpoint stroke M
3 -8 V
3 -7 V
3 -2 V
3 1 V
3 -7 V
3 -8 V
2 -1 V
3 -2 V
3 -9 V
3 0 V
3 2 V
3 -4 V
3 -4 V
2 -5 V
3 -1 V
3 -1 V
3 -7 V
3 -7 V
3 -8 V
2 -3 V
3 -2 V
3 -7 V
3 -10 V
3 -10 V
3 -1 V
3 0 V
2 -3 V
3 4 V
3 4 V
3 -5 V
3 -9 V
3 -8 V
3 -4 V
2 1 V
3 0 V
3 1 V
3 -3 V
3 -2 V
3 1 V
2 -1 V
3 -5 V
3 -3 V
3 1 V
3 1 V
3 0 V
3 -6 V
2 -7 V
3 -4 V
3 0 V
3 2 V
3 1 V
3 1 V
3 0 V
2 -7 V
3 0 V
3 0 V
3 -2 V
3 -4 V
3 -3 V
2 2 V
3 -2 V
3 -7 V
3 -3 V
3 -2 V
3 3 V
3 -3 V
2 -3 V
3 2 V
3 1 V
3 0 V
3 -4 V
3 -9 V
3 -2 V
2 -3 V
3 -1 V
3 3 V
3 -3 V
3 -6 V
3 -5 V
2 -10 V
3 -2 V
3 1 V
3 0 V
3 3 V
3 1 V
3 -1 V
2 -5 V
3 -2 V
3 -5 V
3 -6 V
3 0 V
3 -4 V
3 -4 V
2 -3 V
3 5 V
3 5 V
3 -2 V
3 0 V
3 -2 V
2 -7 V
3 -8 V
3 -3 V
3 1 V
3 1 V
3 0 V
3 -4 V
2 -2 V
3 0 V
3 -2 V
3 -2 V
3 0 V
3 0 V
3 -2 V
2 -1 V
3 -1 V
3 -3 V
3 1 V
3 2 V
3 -1 V
2 -8 V
3 -7 V
3 2 V
3 -1 V
3 -3 V
3 2 V
3 4 V
2 4 V
3 -2 V
3 -2 V
3 0 V
3 3 V
3 -4 V
3 -2 V
2 -3 V
3 -5 V
3 3 V
3 4 V
3 1 V
3 1 V
2 -5 V
3 -5 V
3 4 V
3 3 V
3 0 V
3 -1 V
3 -2 V
2 -2 V
3 -11 V
3 -5 V
3 2 V
3 -2 V
3 -6 V
3 -4 V
2 -1 V
3 6 V
3 3 V
3 -3 V
3 -9 V
3 -4 V
2 -2 V
3 -1 V
3 0 V
3 -4 V
3 0 V
3 4 V
3 7 V
2 8 V
3 -3 V
3 -5 V
3 2 V
3 0 V
3 -1 V
3 -1 V
2 -2 V
3 -2 V
3 0 V
3 2 V
3 1 V
3 -4 V
2 -3 V
3 5 V
3 6 V
3 -1 V
3 -5 V
3 -6 V
3 -1 V
2 -6 V
3 -2 V
3 6 V
3 0 V
3 -2 V
3 1 V
3 -2 V
2 -1 V
3 0 V
3 2 V
3 -4 V
3 -4 V
3 0 V
2 3 V
3 4 V
3 3 V
3 -5 V
3 -6 V
3 2 V
3 3 V
2 3 V
3 1 V
3 -2 V
3 -10 V
3 -9 V
3 0 V
3 2 V
2 1 V
3 2 V
3 4 V
3 1 V
3 -3 V
3 -4 V
2 0 V
3 -1 V
3 -2 V
3 2 V
3 -2 V
3 -5 V
3 -5 V
2 -3 V
3 1 V
3 4 V
3 3 V
3 -4 V
3 -3 V
3 2 V
2 -5 V
3 -10 V
3 -5 V
3 1 V
3 5 V
3 3 V
2 -4 V
3 -3 V
3 -1 V
3 3 V
3 4 V
3 1 V
3 -3 V
2 -2 V
3 3 V
3 2 V
3 -2 V
3 1 V
3 1 V
3 -1 V
2 3 V
3 6 V
3 -2 V
3 -5 V
3 -7 V
3 -1 V
2 6 V
3 5 V
3 2 V
3 1 V
3 4 V
3 1 V
3 2 V
2 -2 V
3 -4 V
3 -2 V
3 0 V
3 -2 V
3 -2 V
3 -1 V
2 1 V
3 1 V
3 2 V
3 2 V
3 0 V
3 3 V
2 3 V
3 -3 V
3 -8 V
3 -7 V
3 -4 V
3 -3 V
3 1 V
2 3 V
3 -4 V
3 -7 V
3 -3 V
3 5 V
3 7 V
3 -1 V
2 -11 V
3 -9 V
3 -5 V
3 -1 V
3 2 V
3 2 V
2 5 V
3 1 V
3 1 V
3 -5 V
3 -6 V
3 -1 V
3 2 V
2 1 V
3 -5 V
3 3 V
3 8 V
3 1 V
3 -1 V
3 -4 V
2 -1 V
3 -4 V
3 -3 V
3 -5 V
3 -4 V
3 2 V
2 3 V
3 6 V
3 2 V
3 -6 V
3 -2 V
3 -2 V
3 -1 V
2 1 V
3 4 V
3 1 V
3 -1 V
3 -4 V
3 -4 V
3 -7 V
2 -1 V
3 5 V
3 3 V
3 3 V
3 1 V
3 2 V
2 10 V
3 8 V
3 -3 V
3 -6 V
3 -5 V
3 -9 V
3 2 V
2 2 V
3 5 V
3 5 V
3 -3 V
3 -4 V
3 5 V
3 2 V
2 2 V
3 1 V
3 -6 V
3 -5 V
3 -2 V
3 2 V
2 2 V
3 2 V
3 1 V
3 -1 V
3 -1 V
3 -7 V
3 -4 V
2 2 V
3 1 V
3 -1 V
3 4 V
3 -3 V
3 0 V
3 2 V
2 2 V
3 -3 V
3 -3 V
3 -7 V
3 -4 V
3 -1 V
2 -3 V
3 -3 V
3 5 V
3 9 V
3 9 V
3 0 V
3 -2 V
2 -5 V
3 -5 V
3 -3 V
3 -5 V
3 -2 V
3 2 V
3 0 V
2 1 V
3 -2 V
3 0 V
3 5 V
3 2 V
3 0 V
2 -3 V
currentpoint stroke M
3 -1 V
3 2 V
3 0 V
3 -3 V
3 1 V
3 10 V
2 6 V
3 1 V
3 -1 V
3 -6 V
3 -3 V
3 -4 V
3 -3 V
2 -2 V
3 -4 V
3 -7 V
3 -1 V
3 3 V
3 0 V
2 5 V
3 4 V
3 7 V
3 5 V
3 1 V
3 0 V
3 -5 V
2 -4 V
3 2 V
3 -1 V
3 -1 V
3 -1 V
3 -7 V
3 -1 V
2 2 V
3 -2 V
3 -7 V
3 -1 V
3 1 V
3 1 V
2 0 V
3 -1 V
3 2 V
3 1 V
3 0 V
3 0 V
3 2 V
2 0 V
3 0 V
3 -3 V
3 -5 V
3 2 V
3 4 V
3 -1 V
2 -4 V
3 -3 V
3 5 V
3 9 V
3 3 V
3 -5 V
2 -5 V
3 -2 V
3 -7 V
3 -2 V
3 1 V
3 4 V
3 1 V
2 2 V
3 0 V
3 -3 V
3 -5 V
3 -5 V
3 -4 V
3 -2 V
2 3 V
3 3 V
3 -1 V
3 1 V
3 -1 V
3 1 V
2 -1 V
3 0 V
3 4 V
3 5 V
3 6 V
3 0 V
3 -7 V
2 -9 V
3 -5 V
3 -3 V
3 5 V
3 2 V
3 -7 V
3 -1 V
2 2 V
3 -4 V
3 -3 V
3 0 V
3 5 V
3 3 V
2 -5 V
3 -2 V
3 4 V
3 -3 V
3 -8 V
3 -3 V
3 4 V
2 3 V
3 3 V
3 7 V
3 0 V
3 -11 V
3 -3 V
3 5 V
2 7 V
3 1 V
3 1 V
3 2 V
3 1 V
3 -1 V
2 -2 V
3 -3 V
3 4 V
3 -2 V
3 0 V
3 1 V
3 -3 V
2 -2 V
3 2 V
3 2 V
3 2 V
3 2 V
3 0 V
3 2 V
2 4 V
3 5 V
3 -2 V
3 0 V
3 2 V
3 0 V
2 1 V
3 3 V
3 0 V
3 -8 V
3 -4 V
3 -2 V
3 -2 V
2 -5 V
3 -5 V
3 2 V
3 4 V
3 -1 V
3 3 V
3 11 V
2 6 V
3 -1 V
3 -5 V
3 1 V
3 3 V
3 -3 V
2 -6 V
3 -5 V
3 -2 V
3 -1 V
3 1 V
3 -6 V
3 -9 V
2 0 V
3 3 V
3 3 V
3 5 V
3 3 V
3 -1 V
3 -6 V
2 -2 V
3 1 V
3 6 V
3 4 V
3 5 V
3 0 V
2 0 V
3 -1 V
3 1 V
3 2 V
3 3 V
3 5 V
3 0 V
2 2 V
3 4 V
3 -4 V
3 0 V
3 -3 V
3 -6 V
3 -7 V
2 0 V
3 3 V
3 -1 V
3 1 V
3 1 V
3 6 V
2 8 V
3 5 V
3 -1 V
3 -2 V
1.000 UL
LT2
3103 1747 M
247 0 V
600 923 M
3 0 V
3 0 V
3 0 V
2 0 V
3 0 V
3 0 V
3 1 V
3 -1 V
3 1 V
2 1 V
3 0 V
3 -1 V
3 1 V
3 0 V
3 0 V
3 1 V
2 0 V
3 0 V
3 -1 V
3 1 V
3 -1 V
3 0 V
3 0 V
2 0 V
3 0 V
3 0 V
3 0 V
3 1 V
3 0 V
2 0 V
3 0 V
3 1 V
3 0 V
3 0 V
3 1 V
3 0 V
2 0 V
3 1 V
3 -1 V
3 1 V
3 0 V
3 1 V
3 0 V
2 1 V
3 0 V
3 0 V
3 0 V
3 1 V
3 0 V
3 2 V
2 -1 V
3 0 V
3 1 V
3 0 V
3 1 V
3 -1 V
2 1 V
3 1 V
3 0 V
3 0 V
3 1 V
3 0 V
3 1 V
2 -1 V
3 1 V
3 1 V
3 0 V
3 1 V
3 1 V
3 0 V
2 0 V
3 1 V
3 1 V
3 1 V
3 0 V
3 1 V
2 1 V
3 1 V
3 0 V
3 0 V
3 1 V
3 1 V
3 0 V
2 0 V
3 1 V
3 0 V
3 1 V
3 1 V
3 0 V
3 0 V
2 1 V
3 2 V
3 0 V
3 0 V
3 1 V
3 1 V
2 1 V
3 1 V
3 0 V
3 1 V
3 0 V
3 2 V
3 0 V
2 1 V
3 1 V
3 1 V
3 1 V
3 0 V
3 0 V
3 1 V
2 1 V
3 1 V
3 1 V
3 0 V
3 2 V
3 1 V
2 1 V
3 1 V
3 0 V
3 1 V
3 1 V
3 0 V
3 1 V
2 1 V
3 1 V
3 0 V
3 1 V
3 2 V
3 1 V
3 1 V
2 1 V
3 0 V
3 1 V
3 1 V
3 1 V
3 2 V
2 1 V
3 1 V
3 1 V
3 1 V
3 1 V
3 1 V
3 1 V
2 1 V
3 2 V
3 1 V
3 1 V
3 1 V
3 1 V
3 2 V
2 1 V
3 1 V
3 1 V
3 2 V
3 1 V
3 1 V
2 2 V
3 0 V
3 2 V
3 1 V
3 2 V
3 2 V
3 1 V
2 1 V
3 1 V
3 2 V
3 2 V
3 1 V
3 2 V
3 1 V
2 2 V
3 1 V
3 2 V
3 2 V
3 2 V
3 1 V
2 2 V
3 2 V
3 2 V
3 2 V
3 2 V
3 2 V
3 2 V
2 2 V
3 2 V
3 3 V
3 3 V
3 2 V
3 2 V
3 3 V
2 3 V
3 1 V
3 4 V
3 2 V
3 2 V
3 3 V
2 1 V
3 -3 V
3 -8 V
3 -119 V
3 -13 V
3 82 V
3 241 V
2 42 V
3 -160 V
3 -120 V
3 -24 V
3 -47 V
3 34 V
3 -43 V
2 -17 V
3 -29 V
3 21 V
3 -60 V
3 21 V
3 19 V
2 -32 V
3 -22 V
3 12 V
3 3 V
3 1 V
3 11 V
3 0 V
2 -26 V
3 21 V
3 -8 V
3 -1 V
3 -16 V
3 10 V
3 -28 V
2 11 V
3 -12 V
3 0 V
3 -21 V
3 12 V
3 -10 V
2 -3 V
3 16 V
3 -4 V
3 1 V
3 -12 V
3 8 V
3 13 V
2 -35 V
3 46 V
3 8 V
3 -55 V
3 19 V
3 23 V
3 -21 V
2 -9 V
3 47 V
3 -32 V
3 18 V
3 -31 V
3 27 V
2 -42 V
3 11 V
3 14 V
3 2 V
3 8 V
3 0 V
3 -4 V
2 -11 V
3 -7 V
3 18 V
3 -15 V
3 6 V
3 -12 V
3 20 V
2 -3 V
3 8 V
3 5 V
3 -26 V
3 -28 V
3 11 V
2 4 V
3 12 V
3 10 V
3 -7 V
3 30 V
3 -10 V
3 14 V
2 3 V
3 -35 V
3 -1 V
3 -4 V
3 -6 V
3 4 V
3 -2 V
2 22 V
3 6 V
3 26 V
3 -7 V
3 21 V
3 -2 V
2 24 V
3 -6 V
3 2 V
3 -18 V
3 11 V
3 -3 V
3 -16 V
2 -25 V
3 13 V
3 -31 V
3 0 V
3 15 V
3 -6 V
3 0 V
2 27 V
3 -23 V
3 -1 V
3 23 V
3 21 V
3 -9 V
2 15 V
3 -20 V
3 -3 V
3 15 V
3 8 V
3 -20 V
3 13 V
2 -21 V
3 27 V
3 -8 V
3 35 V
3 -14 V
3 -9 V
3 7 V
2 -18 V
3 16 V
3 -18 V
3 17 V
3 0 V
3 12 V
2 10 V
3 -11 V
3 3 V
3 12 V
3 -6 V
3 11 V
3 16 V
2 19 V
3 -16 V
3 25 V
3 -12 V
3 -13 V
3 30 V
3 -33 V
2 7 V
3 -7 V
3 -15 V
3 6 V
3 -20 V
3 12 V
2 34 V
3 25 V
3 -7 V
3 -8 V
3 -9 V
3 -8 V
3 27 V
2 -9 V
3 25 V
3 -45 V
3 16 V
3 5 V
3 4 V
3 39 V
2 4 V
3 -9 V
3 -14 V
3 13 V
3 -11 V
3 22 V
2 2 V
3 12 V
3 15 V
3 35 V
3 -33 V
3 38 V
3 -16 V
2 3 V
3 -9 V
3 -11 V
3 27 V
3 -5 V
3 -2 V
3 -10 V
2 23 V
3 7 V
3 34 V
3 -23 V
3 32 V
3 0 V
2 -9 V
currentpoint stroke M
3 6 V
3 15 V
3 16 V
3 15 V
3 25 V
3 0 V
2 -6 V
3 8 V
3 6 V
3 26 V
3 18 V
3 22 V
3 3 V
2 27 V
3 43 V
3 -21 V
3 28 V
3 64 V
3 13 V
2 44 V
3 4 V
3 0 V
3 -32 V
3 -2 V
3 -38 V
3 -48 V
2 12 V
3 -44 V
3 0 V
3 -48 V
3 4 V
3 -28 V
3 -11 V
2 -22 V
3 -11 V
3 25 V
3 -37 V
3 -22 V
3 1 V
2 -29 V
3 5 V
3 -34 V
3 32 V
3 7 V
3 -36 V
3 -4 V
2 -12 V
3 0 V
3 28 V
3 -42 V
3 -21 V
3 -26 V
3 16 V
2 -18 V
3 -18 V
3 8 V
3 4 V
3 13 V
3 11 V
2 -26 V
3 9 V
3 -11 V
3 -11 V
3 7 V
3 -15 V
3 -22 V
2 1 V
3 1 V
3 -25 V
3 -7 V
3 16 V
3 4 V
3 -10 V
2 11 V
3 -6 V
3 -21 V
3 3 V
3 20 V
3 -4 V
2 -15 V
3 13 V
3 -30 V
3 -6 V
3 -6 V
3 -8 V
3 -33 V
2 10 V
3 9 V
3 25 V
3 -9 V
3 1 V
3 -8 V
3 -27 V
2 31 V
3 -7 V
3 28 V
3 -30 V
3 8 V
3 -26 V
2 40 V
3 -21 V
3 -18 V
3 9 V
3 -10 V
3 -9 V
3 1 V
2 -10 V
3 -2 V
3 6 V
3 -11 V
3 -2 V
3 1 V
3 12 V
2 -8 V
3 -13 V
3 2 V
3 17 V
3 -7 V
3 0 V
2 6 V
3 27 V
3 -6 V
3 11 V
3 -37 V
3 -9 V
3 7 V
2 -16 V
3 -19 V
3 -5 V
3 7 V
3 -23 V
3 22 V
3 2 V
2 14 V
3 -11 V
3 0 V
3 8 V
3 -50 V
3 18 V
2 30 V
3 0 V
3 -3 V
3 27 V
3 -19 V
3 19 V
3 1 V
2 -9 V
3 -8 V
3 -28 V
3 -26 V
3 -21 V
3 39 V
3 -37 V
2 5 V
3 0 V
3 11 V
3 13 V
3 -6 V
3 -23 V
2 37 V
3 -8 V
3 0 V
3 -2 V
3 12 V
3 -21 V
3 -15 V
2 -11 V
3 -13 V
3 0 V
3 -25 V
3 25 V
3 30 V
3 -6 V
2 -23 V
3 42 V
3 -14 V
3 5 V
3 9 V
3 1 V
2 -21 V
3 13 V
3 7 V
3 -4 V
3 -23 V
3 17 V
3 -3 V
2 14 V
3 -8 V
3 -20 V
3 21 V
3 -41 V
3 41 V
3 -13 V
2 -25 V
3 21 V
3 -3 V
3 17 V
3 -26 V
3 2 V
2 24 V
3 -8 V
3 -25 V
3 0 V
3 -29 V
3 32 V
3 -16 V
2 8 V
3 -30 V
3 6 V
3 4 V
3 28 V
3 -29 V
3 40 V
2 -9 V
3 12 V
3 -22 V
3 18 V
3 -3 V
3 10 V
2 3 V
3 -19 V
3 18 V
3 -6 V
3 10 V
3 -19 V
3 -1 V
2 -13 V
3 7 V
3 -32 V
3 -2 V
3 16 V
3 7 V
3 -2 V
2 -34 V
3 24 V
3 -35 V
3 -10 V
3 -2 V
3 60 V
2 -48 V
3 5 V
3 -3 V
3 2 V
3 9 V
3 -28 V
3 35 V
2 -23 V
3 5 V
3 8 V
3 -3 V
3 0 V
3 7 V
3 -11 V
2 38 V
3 17 V
3 -1 V
3 -31 V
3 -9 V
3 3 V
2 -9 V
3 34 V
3 -7 V
3 -17 V
3 4 V
3 -38 V
3 4 V
2 16 V
3 9 V
3 -35 V
3 -8 V
3 22 V
3 -14 V
3 10 V
2 1 V
3 13 V
3 20 V
3 -31 V
3 28 V
3 -24 V
2 -7 V
3 -4 V
3 3 V
3 0 V
3 9 V
3 -22 V
3 1 V
2 -8 V
3 17 V
3 -8 V
3 -13 V
3 27 V
3 11 V
3 -8 V
2 -1 V
3 -5 V
3 15 V
3 20 V
3 -16 V
3 -19 V
2 9 V
3 9 V
3 -3 V
3 -8 V
3 -4 V
3 5 V
3 -6 V
2 0 V
3 1 V
3 -10 V
3 11 V
3 5 V
3 30 V
3 -53 V
2 11 V
3 -22 V
3 -12 V
3 6 V
3 -2 V
3 -15 V
2 6 V
3 0 V
3 30 V
3 5 V
3 12 V
3 -19 V
3 -2 V
2 -20 V
3 -27 V
3 -11 V
3 33 V
3 19 V
3 10 V
3 -2 V
2 -3 V
3 -2 V
3 -1 V
3 10 V
3 -8 V
3 -35 V
2 14 V
3 0 V
3 5 V
3 -31 V
3 12 V
3 0 V
3 10 V
2 22 V
3 26 V
3 -22 V
3 -17 V
3 4 V
3 -4 V
3 10 V
2 -8 V
3 -5 V
3 7 V
3 16 V
3 5 V
3 -8 V
2 6 V
3 14 V
3 -5 V
3 22 V
3 -19 V
3 -16 V
3 -23 V
2 -14 V
3 5 V
3 18 V
3 11 V
3 -50 V
3 19 V
3 5 V
2 6 V
3 -10 V
3 12 V
3 -19 V
3 17 V
3 -19 V
2 9 V
3 -2 V
3 25 V
3 -47 V
3 8 V
3 29 V
3 -25 V
2 39 V
3 -17 V
3 30 V
3 -32 V
3 -2 V
3 -34 V
3 24 V
2 -15 V
3 36 V
3 -16 V
3 -9 V
3 33 V
3 -26 V
2 2 V
currentpoint stroke M
3 0 V
3 -17 V
3 10 V
3 9 V
3 4 V
3 -2 V
2 -23 V
3 18 V
3 -23 V
3 -4 V
3 57 V
3 -17 V
3 3 V
2 -20 V
3 -10 V
3 -4 V
3 15 V
3 8 V
3 -19 V
2 -4 V
3 32 V
3 8 V
3 -6 V
3 18 V
3 -7 V
3 -47 V
2 50 V
3 -30 V
3 -3 V
3 0 V
3 -6 V
3 16 V
3 -17 V
2 -3 V
3 -12 V
3 40 V
3 -11 V
3 14 V
3 -16 V
2 -18 V
3 -3 V
3 -8 V
3 29 V
3 -33 V
3 6 V
3 8 V
2 -5 V
3 6 V
3 -8 V
3 -19 V
3 8 V
3 4 V
3 16 V
2 -29 V
3 13 V
3 2 V
3 34 V
3 29 V
3 -16 V
2 -26 V
3 -5 V
3 1 V
3 -33 V
3 -8 V
3 27 V
3 11 V
2 5 V
3 20 V
3 -50 V
3 24 V
3 -24 V
3 14 V
3 22 V
2 -1 V
3 30 V
3 -40 V
3 -2 V
3 -22 V
3 5 V
2 3 V
3 8 V
3 24 V
3 -9 V
3 0 V
3 -7 V
3 -10 V
2 -3 V
3 -2 V
3 -1 V
3 9 V
3 -1 V
3 4 V
3 19 V
2 -4 V
3 -43 V
3 4 V
3 15 V
3 -7 V
3 15 V
2 8 V
3 -8 V
3 15 V
3 -16 V
3 -12 V
3 16 V
3 18 V
2 -24 V
3 5 V
3 -1 V
3 4 V
3 7 V
3 -11 V
3 4 V
2 -23 V
3 10 V
3 16 V
3 0 V
3 5 V
3 -27 V
2 3 V
3 18 V
3 24 V
3 -6 V
3 -6 V
3 16 V
3 -12 V
2 4 V
3 11 V
3 21 V
3 -21 V
3 -46 V
3 -1 V
3 20 V
2 -11 V
3 13 V
3 2 V
3 -6 V
3 32 V
3 -46 V
2 31 V
3 -9 V
3 0 V
3 1 V
3 -9 V
3 14 V
3 -15 V
2 -40 V
3 -8 V
3 21 V
3 -7 V
3 -9 V
3 27 V
3 1 V
2 -9 V
3 17 V
3 20 V
3 -15 V
3 -9 V
3 16 V
2 -42 V
3 20 V
3 -6 V
3 37 V
3 -37 V
3 9 V
3 -4 V
2 -3 V
3 34 V
3 -34 V
3 11 V
3 -27 V
3 6 V
3 15 V
2 12 V
3 -16 V
3 15 V
3 8 V
3 -21 V
3 -8 V
2 29 V
3 -11 V
3 -45 V
3 17 V
3 43 V
3 -37 V
3 6 V
2 -16 V
3 -2 V
3 9 V
3 10 V
3 -2 V
3 -11 V
3 2 V
2 -6 V
3 10 V
3 0 V
3 7 V
3 -10 V
3 -11 V
2 11 V
3 -7 V
3 12 V
3 21 V
1.000 UL
LT3
3103 1647 M
247 0 V
600 912 M
3 0 V
3 0 V
3 0 V
2 1 V
3 0 V
3 -1 V
3 1 V
3 0 V
3 0 V
2 0 V
3 0 V
3 0 V
3 0 V
3 0 V
3 1 V
3 0 V
2 0 V
3 0 V
3 0 V
3 0 V
3 0 V
3 0 V
3 0 V
2 1 V
3 0 V
3 0 V
3 0 V
3 0 V
3 1 V
2 0 V
3 0 V
3 0 V
3 1 V
3 0 V
3 0 V
3 0 V
2 0 V
3 2 V
3 -1 V
3 0 V
3 1 V
3 0 V
3 1 V
2 0 V
3 0 V
3 0 V
3 0 V
3 1 V
3 1 V
3 0 V
2 0 V
3 1 V
3 0 V
3 0 V
3 1 V
3 0 V
2 0 V
3 1 V
3 0 V
3 1 V
3 0 V
3 0 V
3 1 V
2 0 V
3 1 V
3 0 V
3 1 V
3 0 V
3 1 V
3 1 V
2 0 V
3 1 V
3 0 V
3 1 V
3 0 V
3 1 V
2 0 V
3 1 V
3 0 V
3 1 V
3 1 V
3 0 V
3 1 V
2 0 V
3 1 V
3 0 V
3 1 V
3 1 V
3 0 V
3 1 V
2 1 V
3 0 V
3 1 V
3 0 V
3 1 V
3 1 V
2 1 V
3 0 V
3 1 V
3 1 V
3 1 V
3 0 V
3 1 V
2 1 V
3 1 V
3 0 V
3 1 V
3 1 V
3 1 V
3 0 V
2 1 V
3 1 V
3 1 V
3 1 V
3 0 V
3 1 V
2 1 V
3 1 V
3 1 V
3 1 V
3 1 V
3 0 V
3 1 V
2 1 V
3 1 V
3 1 V
3 1 V
3 1 V
3 1 V
3 1 V
2 1 V
3 1 V
3 1 V
3 1 V
3 1 V
3 1 V
2 1 V
3 1 V
3 1 V
3 1 V
3 1 V
3 1 V
3 1 V
2 1 V
3 2 V
3 1 V
3 1 V
3 1 V
3 1 V
3 2 V
2 0 V
3 2 V
3 1 V
3 1 V
3 2 V
3 1 V
2 1 V
3 2 V
3 1 V
3 1 V
3 2 V
3 1 V
3 2 V
2 1 V
3 2 V
3 1 V
3 2 V
3 1 V
3 2 V
3 2 V
2 1 V
3 2 V
3 2 V
3 2 V
3 2 V
3 1 V
2 2 V
3 2 V
3 2 V
3 2 V
3 3 V
3 2 V
3 2 V
2 2 V
3 3 V
3 2 V
3 3 V
3 3 V
3 2 V
3 3 V
2 4 V
3 3 V
3 3 V
3 3 V
3 4 V
3 3 V
2 3 V
3 1 V
3 -5 V
3 -120 V
3 -22 V
3 185 V
3 6 V
2 -181 V
3 -108 V
3 -24 V
3 -59 V
3 -12 V
3 34 V
3 -95 V
2 45 V
3 -59 V
3 7 V
3 30 V
3 -44 V
3 -1 V
2 21 V
3 5 V
3 -29 V
3 -2 V
3 13 V
3 -57 V
3 20 V
2 46 V
3 -56 V
3 51 V
3 -21 V
3 -1 V
3 -17 V
3 -21 V
2 25 V
3 -22 V
3 3 V
3 24 V
3 67 V
3 -28 V
2 -21 V
3 -50 V
3 29 V
3 5 V
3 13 V
3 -24 V
3 -44 V
2 83 V
3 0 V
3 -46 V
3 13 V
3 -2 V
3 8 V
3 12 V
2 -64 V
3 -6 V
3 15 V
3 41 V
3 -12 V
3 7 V
2 47 V
3 -64 V
3 14 V
3 -94 V
3 23 V
3 9 V
3 4 V
2 0 V
3 76 V
3 28 V
3 -18 V
3 -52 V
3 -32 V
3 55 V
2 26 V
3 -29 V
3 3 V
3 -82 V
3 11 V
3 10 V
2 52 V
3 35 V
3 -19 V
3 -27 V
3 42 V
3 -41 V
3 53 V
2 -9 V
3 12 V
3 -53 V
3 -15 V
3 31 V
3 25 V
3 16 V
2 -46 V
3 7 V
3 -34 V
3 58 V
3 29 V
3 5 V
2 -47 V
3 -42 V
3 43 V
3 -2 V
3 1 V
3 19 V
3 -19 V
2 1 V
3 55 V
3 -16 V
3 -12 V
3 46 V
3 -71 V
3 12 V
2 -10 V
3 2 V
3 43 V
3 -65 V
3 20 V
3 -35 V
2 39 V
3 52 V
3 0 V
3 -59 V
3 -33 V
3 81 V
3 46 V
2 -33 V
3 -40 V
3 31 V
3 -13 V
3 58 V
3 -16 V
3 1 V
2 -116 V
3 73 V
3 -31 V
3 62 V
3 -25 V
3 -28 V
2 -10 V
3 -12 V
3 -22 V
3 48 V
3 -29 V
3 40 V
3 85 V
2 -70 V
3 54 V
3 -6 V
3 -84 V
3 -13 V
3 60 V
3 -21 V
2 50 V
3 -4 V
3 7 V
3 -72 V
3 16 V
3 71 V
2 55 V
3 -5 V
3 -70 V
3 -43 V
3 39 V
3 -8 V
3 70 V
2 -77 V
3 83 V
3 -14 V
3 -24 V
3 -43 V
3 23 V
3 -12 V
2 81 V
3 -79 V
3 26 V
3 72 V
3 -23 V
3 -42 V
2 12 V
3 -13 V
3 9 V
3 55 V
3 -6 V
3 26 V
3 -38 V
2 -32 V
3 47 V
3 -15 V
3 32 V
3 -65 V
3 20 V
3 6 V
2 4 V
3 -5 V
3 28 V
3 -14 V
3 53 V
3 14 V
2 -41 V
currentpoint stroke M
3 -24 V
3 112 V
3 -93 V
3 -45 V
3 35 V
3 49 V
2 -53 V
3 42 V
3 24 V
3 -16 V
3 46 V
3 15 V
3 -9 V
2 30 V
3 -85 V
3 26 V
3 59 V
3 24 V
3 -37 V
2 -18 V
3 -6 V
3 10 V
3 45 V
3 -28 V
3 96 V
3 -126 V
2 22 V
3 174 V
3 -85 V
3 -37 V
3 33 V
3 19 V
3 29 V
2 -2 V
3 20 V
3 2 V
3 99 V
3 -30 V
3 -22 V
2 20 V
3 105 V
3 -43 V
3 96 V
3 2 V
3 48 V
3 137 V
2 67 V
3 -119 V
3 -84 V
3 -59 V
3 -58 V
3 -40 V
3 -77 V
2 17 V
3 -31 V
3 30 V
3 -64 V
3 -5 V
3 8 V
2 -24 V
3 -15 V
3 -17 V
3 -62 V
3 112 V
3 -62 V
3 -34 V
2 -14 V
3 43 V
3 -14 V
3 25 V
3 -128 V
3 32 V
3 77 V
2 -54 V
3 -43 V
3 -24 V
3 46 V
3 -2 V
3 -88 V
2 56 V
3 10 V
3 -2 V
3 -50 V
3 -14 V
3 9 V
3 -10 V
2 50 V
3 18 V
3 -53 V
3 -50 V
3 67 V
3 -28 V
3 4 V
2 -43 V
3 8 V
3 -17 V
3 33 V
3 -9 V
3 5 V
2 -50 V
3 68 V
3 -28 V
3 -26 V
3 23 V
3 -5 V
3 35 V
2 -63 V
3 -35 V
3 49 V
3 -13 V
3 32 V
3 -14 V
3 -3 V
2 -61 V
3 64 V
3 -2 V
3 -82 V
3 62 V
3 -25 V
2 77 V
3 -91 V
3 35 V
3 34 V
3 -73 V
3 52 V
3 -49 V
2 61 V
3 -64 V
3 17 V
3 57 V
3 -23 V
3 -21 V
3 -40 V
2 -49 V
3 2 V
3 11 V
3 -30 V
3 21 V
3 -3 V
2 46 V
3 -17 V
3 -2 V
3 -13 V
3 23 V
3 22 V
3 -26 V
2 -8 V
3 59 V
3 -77 V
3 -19 V
3 -13 V
3 52 V
3 -55 V
2 3 V
3 46 V
3 -28 V
3 -1 V
3 32 V
3 -58 V
2 49 V
3 -20 V
3 25 V
3 26 V
3 -2 V
3 -35 V
3 -47 V
2 57 V
3 -36 V
3 18 V
3 43 V
3 -13 V
3 -21 V
3 -25 V
2 55 V
3 4 V
3 -84 V
3 -2 V
3 48 V
3 -10 V
2 -13 V
3 12 V
3 31 V
3 -9 V
3 -50 V
3 -19 V
3 10 V
2 30 V
3 8 V
3 -71 V
3 35 V
3 21 V
3 -3 V
3 -37 V
2 7 V
3 1 V
3 -71 V
3 29 V
3 64 V
3 -36 V
2 -20 V
3 -9 V
3 47 V
3 19 V
3 -27 V
3 -44 V
3 35 V
2 88 V
3 -64 V
3 8 V
3 -85 V
3 30 V
3 38 V
3 45 V
2 -77 V
3 22 V
3 34 V
3 -81 V
3 83 V
3 59 V
2 -54 V
3 -71 V
3 89 V
3 -56 V
3 -12 V
3 2 V
3 26 V
2 -88 V
3 55 V
3 16 V
3 -49 V
3 74 V
3 -78 V
3 31 V
2 9 V
3 14 V
3 23 V
3 -42 V
3 15 V
3 -13 V
2 4 V
3 -4 V
3 -44 V
3 33 V
3 47 V
3 -58 V
3 23 V
2 -24 V
3 34 V
3 49 V
3 -108 V
3 107 V
3 -71 V
3 7 V
2 -39 V
3 13 V
3 -6 V
3 56 V
3 -29 V
3 14 V
2 -69 V
3 -14 V
3 76 V
3 -9 V
3 -35 V
3 -40 V
3 73 V
2 0 V
3 6 V
3 -87 V
3 27 V
3 16 V
3 38 V
3 -60 V
2 -1 V
3 45 V
3 16 V
3 1 V
3 -21 V
3 -37 V
2 38 V
3 -8 V
3 67 V
3 -49 V
3 -23 V
3 -6 V
3 68 V
2 -45 V
3 9 V
3 21 V
3 -36 V
3 46 V
3 -25 V
3 19 V
2 -98 V
3 -38 V
3 117 V
3 -21 V
3 11 V
3 -19 V
2 -36 V
3 -10 V
3 54 V
3 -49 V
3 5 V
3 46 V
3 -5 V
2 35 V
3 46 V
3 -102 V
3 -3 V
3 12 V
3 32 V
3 -10 V
2 -73 V
3 3 V
3 42 V
3 -2 V
3 1 V
3 -49 V
2 -11 V
3 17 V
3 30 V
3 25 V
3 -8 V
3 -66 V
3 49 V
2 68 V
3 -74 V
3 59 V
3 -95 V
3 38 V
3 22 V
3 -24 V
2 95 V
3 -73 V
3 71 V
3 -119 V
3 126 V
3 -90 V
2 1 V
3 45 V
3 -50 V
3 38 V
3 -1 V
3 -50 V
3 -8 V
2 69 V
3 -32 V
3 6 V
3 72 V
3 -128 V
3 45 V
3 -35 V
2 33 V
3 28 V
3 -40 V
3 -4 V
3 47 V
3 -50 V
2 9 V
3 0 V
3 -42 V
3 25 V
3 -37 V
3 89 V
3 -92 V
2 26 V
3 45 V
3 -26 V
3 -49 V
3 69 V
3 -6 V
3 0 V
2 34 V
3 -21 V
3 -34 V
3 13 V
3 -14 V
3 -24 V
2 5 V
3 -2 V
3 -17 V
3 95 V
3 -64 V
3 -27 V
3 53 V
2 21 V
3 5 V
3 4 V
3 -67 V
3 43 V
3 -13 V
3 14 V
2 -70 V
3 1 V
3 68 V
3 -31 V
3 -5 V
3 -38 V
2 -11 V
currentpoint stroke M
3 56 V
3 -34 V
3 64 V
3 -80 V
3 91 V
3 -32 V
2 -28 V
3 51 V
3 19 V
3 -76 V
3 -9 V
3 7 V
3 27 V
2 49 V
3 -14 V
3 -92 V
3 83 V
3 -14 V
3 -16 V
2 -13 V
3 -18 V
3 8 V
3 -16 V
3 73 V
3 -45 V
3 2 V
2 -4 V
3 39 V
3 5 V
3 25 V
3 -67 V
3 -25 V
3 -42 V
2 10 V
3 54 V
3 -28 V
3 -28 V
3 -1 V
3 71 V
2 10 V
3 -2 V
3 2 V
3 -8 V
3 0 V
3 -22 V
3 54 V
2 -100 V
3 88 V
3 29 V
3 -4 V
3 -43 V
3 18 V
3 -23 V
2 -78 V
3 62 V
3 21 V
3 -7 V
3 -2 V
3 6 V
2 39 V
3 -18 V
3 -17 V
3 55 V
3 -63 V
3 -30 V
3 37 V
2 -57 V
3 107 V
3 -65 V
3 28 V
3 -42 V
3 -16 V
3 13 V
2 30 V
3 6 V
3 12 V
3 -35 V
3 -34 V
3 86 V
2 6 V
3 -94 V
3 90 V
3 -22 V
3 10 V
3 -43 V
3 11 V
2 1 V
3 -43 V
3 45 V
3 -66 V
3 45 V
3 66 V
3 -9 V
2 18 V
3 -61 V
3 -12 V
3 53 V
3 -47 V
3 -37 V
2 46 V
3 30 V
3 -47 V
3 2 V
3 -37 V
3 21 V
3 18 V
2 -30 V
3 4 V
3 53 V
3 -3 V
3 -13 V
3 -56 V
3 94 V
2 50 V
3 -68 V
3 14 V
3 -21 V
3 -6 V
3 -58 V
2 0 V
3 67 V
3 4 V
3 -86 V
3 33 V
3 -7 V
3 -28 V
2 77 V
3 28 V
3 22 V
3 -47 V
3 -39 V
3 8 V
3 27 V
2 -53 V
3 52 V
3 -8 V
3 7 V
3 -14 V
3 0 V
2 21 V
3 -21 V
3 -28 V
3 43 V
3 -47 V
3 46 V
3 -77 V
2 90 V
3 -25 V
3 -13 V
3 2 V
3 48 V
3 35 V
3 -133 V
2 51 V
3 55 V
3 -49 V
3 -32 V
3 53 V
3 -4 V
2 -38 V
3 92 V
3 -42 V
3 -58 V
3 109 V
3 -76 V
3 -45 V
2 62 V
3 -20 V
3 -32 V
3 -7 V
3 -20 V
3 20 V
3 3 V
2 44 V
3 -19 V
3 41 V
3 -8 V
3 1 V
3 -43 V
2 -1 V
3 10 V
3 10 V
3 -9 V
3 13 V
3 44 V
3 -76 V
2 7 V
3 -5 V
3 80 V
3 -54 V
3 9 V
3 -7 V
3 55 V
2 -32 V
3 14 V
3 9 V
3 -7 V
3 -71 V
3 -12 V
2 77 V
3 -8 V
3 -16 V
3 -105 V
1.000 UL
LT0
3103 1547 M
247 0 V
600 903 M
3 0 V
3 0 V
3 0 V
2 0 V
3 0 V
3 0 V
3 0 V
3 0 V
3 0 V
2 0 V
3 0 V
3 1 V
3 0 V
3 0 V
3 0 V
3 0 V
2 0 V
3 0 V
3 0 V
3 0 V
3 1 V
3 0 V
3 0 V
2 0 V
3 0 V
3 0 V
3 1 V
3 0 V
3 0 V
2 0 V
3 1 V
3 0 V
3 0 V
3 0 V
3 0 V
3 1 V
2 0 V
3 0 V
3 1 V
3 0 V
3 0 V
3 0 V
3 1 V
2 0 V
3 0 V
3 1 V
3 0 V
3 0 V
3 1 V
3 0 V
2 0 V
3 1 V
3 0 V
3 1 V
3 0 V
3 0 V
2 1 V
3 0 V
3 1 V
3 0 V
3 0 V
3 1 V
3 0 V
2 1 V
3 0 V
3 1 V
3 0 V
3 1 V
3 0 V
3 1 V
2 0 V
3 1 V
3 0 V
3 1 V
3 0 V
3 1 V
2 0 V
3 1 V
3 0 V
3 1 V
3 1 V
3 0 V
3 1 V
2 0 V
3 1 V
3 1 V
3 0 V
3 1 V
3 0 V
3 1 V
2 1 V
3 0 V
3 1 V
3 1 V
3 0 V
3 1 V
2 1 V
3 0 V
3 1 V
3 1 V
3 1 V
3 0 V
3 1 V
2 1 V
3 0 V
3 1 V
3 1 V
3 1 V
3 1 V
3 0 V
2 1 V
3 1 V
3 1 V
3 0 V
3 1 V
3 1 V
2 1 V
3 1 V
3 1 V
3 1 V
3 0 V
3 1 V
3 1 V
2 1 V
3 1 V
3 1 V
3 1 V
3 1 V
3 1 V
3 1 V
2 1 V
3 1 V
3 1 V
3 1 V
3 1 V
3 1 V
2 1 V
3 1 V
3 1 V
3 1 V
3 1 V
3 1 V
3 1 V
2 1 V
3 1 V
3 1 V
3 2 V
3 1 V
3 1 V
3 1 V
2 1 V
3 2 V
3 1 V
3 1 V
3 1 V
3 2 V
2 1 V
3 1 V
3 2 V
3 1 V
3 1 V
3 2 V
3 1 V
2 2 V
3 1 V
3 2 V
3 1 V
3 2 V
3 2 V
3 1 V
2 2 V
3 2 V
3 1 V
3 2 V
3 2 V
3 2 V
2 2 V
3 2 V
3 2 V
3 2 V
3 2 V
3 3 V
3 2 V
2 3 V
3 2 V
3 3 V
3 3 V
3 3 V
3 3 V
3 3 V
2 3 V
3 4 V
3 4 V
3 4 V
3 4 V
3 5 V
2 4 V
3 4 V
3 -1 V
3 -103 V
3 31 V
3 188 V
3 -283 V
2 -102 V
3 -62 V
3 -44 V
3 -35 V
3 -27 V
3 -23 V
3 -19 V
2 -16 V
3 -15 V
3 -12 V
3 -11 V
3 -10 V
3 -9 V
2 -8 V
3 -7 V
3 -6 V
3 -6 V
3 -5 V
3 -5 V
3 -4 V
2 -4 V
3 -3 V
3 -4 V
3 -2 V
3 -3 V
3 -2 V
3 -3 V
2 -1 V
3 -2 V
3 -2 V
3 -1 V
3 -1 V
3 -1 V
2 -1 V
3 -1 V
3 -1 V
3 0 V
3 -1 V
3 0 V
3 -1 V
2 0 V
3 0 V
3 0 V
3 0 V
3 0 V
3 0 V
3 0 V
2 0 V
3 1 V
3 0 V
3 0 V
3 1 V
3 0 V
2 1 V
3 0 V
3 1 V
3 0 V
3 1 V
3 1 V
3 0 V
2 1 V
3 1 V
3 1 V
3 0 V
3 1 V
3 1 V
3 1 V
2 1 V
3 1 V
3 0 V
3 1 V
3 1 V
3 1 V
2 1 V
3 1 V
3 1 V
3 1 V
3 1 V
3 1 V
3 2 V
2 1 V
3 1 V
3 1 V
3 1 V
3 1 V
3 1 V
3 1 V
2 2 V
3 1 V
3 1 V
3 1 V
3 1 V
3 2 V
2 1 V
3 1 V
3 1 V
3 2 V
3 1 V
3 1 V
3 2 V
2 1 V
3 1 V
3 1 V
3 2 V
3 1 V
3 2 V
3 1 V
2 1 V
3 2 V
3 1 V
3 1 V
3 2 V
3 1 V
2 2 V
3 1 V
3 2 V
3 1 V
3 2 V
3 1 V
3 2 V
2 1 V
3 2 V
3 1 V
3 2 V
3 1 V
3 2 V
3 1 V
2 2 V
3 1 V
3 2 V
3 2 V
3 1 V
3 2 V
2 1 V
3 2 V
3 2 V
3 1 V
3 2 V
3 2 V
3 2 V
2 1 V
3 2 V
3 2 V
3 1 V
3 2 V
3 2 V
3 2 V
2 2 V
3 1 V
3 2 V
3 2 V
3 2 V
3 2 V
2 2 V
3 2 V
3 1 V
3 2 V
3 2 V
3 2 V
3 2 V
2 2 V
3 2 V
3 2 V
3 2 V
3 2 V
3 2 V
3 2 V
2 3 V
3 2 V
3 2 V
3 2 V
3 2 V
3 2 V
2 3 V
3 2 V
3 2 V
3 2 V
3 3 V
3 2 V
3 2 V
2 3 V
3 2 V
3 3 V
3 2 V
3 3 V
3 2 V
3 3 V
2 2 V
3 3 V
3 2 V
3 3 V
3 3 V
3 2 V
2 3 V
currentpoint stroke M
3 3 V
3 3 V
3 3 V
3 3 V
3 2 V
3 3 V
2 3 V
3 3 V
3 4 V
3 3 V
3 3 V
3 3 V
3 3 V
2 4 V
3 3 V
3 4 V
3 3 V
3 4 V
3 3 V
2 4 V
3 4 V
3 3 V
3 4 V
3 4 V
3 4 V
3 4 V
2 5 V
3 4 V
3 4 V
3 5 V
3 4 V
3 5 V
3 5 V
2 5 V
3 5 V
3 5 V
3 5 V
3 6 V
3 5 V
2 6 V
3 6 V
3 6 V
3 6 V
3 7 V
3 6 V
3 7 V
2 8 V
3 7 V
3 8 V
3 8 V
3 8 V
3 9 V
3 10 V
2 9 V
3 11 V
3 11 V
3 11 V
3 13 V
3 13 V
2 14 V
3 16 V
3 16 V
3 19 V
3 20 V
3 24 V
3 26 V
2 31 V
3 38 V
3 47 V
3 64 V
3 100 V
3 243 V
3 -112 V
2 -167 V
3 -85 V
3 -57 V
3 -43 V
3 -35 V
3 -29 V
2 -25 V
3 -21 V
3 -20 V
3 -17 V
3 -16 V
3 -15 V
3 -13 V
2 -12 V
3 -12 V
3 -11 V
3 -10 V
3 -10 V
3 -9 V
3 -9 V
2 -8 V
3 -8 V
3 -7 V
3 -7 V
3 -7 V
3 -7 V
2 -6 V
3 -6 V
3 -6 V
3 -6 V
3 -5 V
3 -6 V
3 -5 V
2 -5 V
3 -5 V
3 -4 V
3 -5 V
3 -4 V
3 -5 V
3 -4 V
2 -4 V
3 -4 V
3 -4 V
3 -4 V
3 -3 V
3 -4 V
2 -4 V
3 -3 V
3 -4 V
3 -3 V
3 -3 V
3 -3 V
3 -3 V
2 -3 V
3 -3 V
3 -3 V
3 -3 V
3 -3 V
3 -3 V
3 -3 V
2 -2 V
3 -3 V
3 -3 V
3 -2 V
3 -3 V
3 -2 V
2 -3 V
3 -2 V
3 -2 V
3 -3 V
3 -2 V
3 -2 V
3 -2 V
2 -2 V
3 -3 V
3 -2 V
3 -2 V
3 -2 V
3 -2 V
3 -2 V
2 -2 V
3 -2 V
3 -2 V
3 -2 V
3 -1 V
3 -2 V
2 -2 V
3 -2 V
3 -2 V
3 -1 V
3 -2 V
3 -2 V
3 -1 V
2 -2 V
3 -2 V
3 -1 V
3 -2 V
3 -2 V
3 -1 V
3 -2 V
2 -1 V
3 -2 V
3 -1 V
3 -2 V
3 -1 V
3 -2 V
2 -1 V
3 -1 V
3 -2 V
3 -1 V
3 -2 V
3 -1 V
3 -1 V
2 -2 V
3 -1 V
3 -1 V
3 -1 V
3 -2 V
3 -1 V
3 -1 V
2 -1 V
3 -2 V
3 -1 V
3 -1 V
3 -1 V
3 -1 V
2 -2 V
3 -1 V
3 -1 V
3 -1 V
3 -1 V
3 -1 V
3 -1 V
2 -1 V
3 -2 V
3 -1 V
3 -1 V
3 -1 V
3 -1 V
3 -1 V
2 -1 V
3 -1 V
3 -1 V
3 -1 V
3 -1 V
3 -1 V
2 -1 V
3 -1 V
3 -1 V
3 -1 V
3 -1 V
3 -1 V
3 0 V
2 -1 V
3 -1 V
3 -1 V
3 -1 V
3 -1 V
3 -1 V
3 -1 V
2 -1 V
3 -1 V
3 0 V
3 -1 V
3 -1 V
3 -1 V
2 -1 V
3 -1 V
3 0 V
3 -1 V
3 -1 V
3 -1 V
3 -1 V
2 0 V
3 -1 V
3 -1 V
3 -1 V
3 0 V
3 -1 V
3 -1 V
2 -1 V
3 0 V
3 -1 V
3 -1 V
3 -1 V
3 0 V
2 -1 V
3 -1 V
3 0 V
3 -1 V
3 -1 V
3 -1 V
3 0 V
2 -1 V
3 -1 V
3 0 V
3 -1 V
3 -1 V
3 0 V
3 -1 V
2 0 V
3 -1 V
3 -1 V
3 0 V
3 -1 V
3 -1 V
2 0 V
3 -1 V
3 -1 V
3 0 V
3 -1 V
3 0 V
3 -1 V
2 0 V
3 -1 V
3 -1 V
3 0 V
3 -1 V
3 0 V
3 -1 V
2 -1 V
3 0 V
3 -1 V
3 0 V
3 -1 V
3 0 V
2 -1 V
3 0 V
3 -1 V
3 0 V
3 -1 V
3 0 V
3 -1 V
2 -1 V
3 0 V
3 -1 V
3 0 V
3 -1 V
3 0 V
3 -1 V
2 0 V
3 -1 V
3 0 V
3 -1 V
3 0 V
3 0 V
2 -1 V
3 0 V
3 -1 V
3 0 V
3 -1 V
3 0 V
3 -1 V
2 0 V
3 -1 V
3 0 V
3 -1 V
3 0 V
3 0 V
3 -1 V
2 0 V
3 -1 V
3 0 V
3 -1 V
3 0 V
3 0 V
2 -1 V
3 0 V
3 -1 V
3 0 V
3 -1 V
3 0 V
3 0 V
2 -1 V
3 0 V
3 -1 V
3 0 V
3 0 V
3 -1 V
3 0 V
2 0 V
3 -1 V
3 0 V
3 -1 V
3 0 V
3 0 V
2 -1 V
3 0 V
3 0 V
3 -1 V
3 0 V
3 -1 V
3 0 V
2 0 V
3 -1 V
3 0 V
3 0 V
3 -1 V
3 0 V
3 0 V
2 -1 V
3 0 V
3 0 V
3 -1 V
3 0 V
3 0 V
2 -1 V
3 0 V
3 0 V
3 -1 V
3 0 V
3 0 V
3 -1 V
2 0 V
3 0 V
3 0 V
3 -1 V
3 0 V
3 0 V
3 -1 V
2 0 V
3 0 V
3 -1 V
3 0 V
3 0 V
3 0 V
2 -1 V
currentpoint stroke M
3 0 V
3 0 V
3 -1 V
3 0 V
3 0 V
3 0 V
2 -1 V
3 0 V
3 0 V
3 0 V
3 -1 V
3 0 V
3 0 V
2 0 V
3 -1 V
3 0 V
3 0 V
3 0 V
3 -1 V
2 0 V
3 0 V
3 0 V
3 -1 V
3 0 V
3 0 V
3 0 V
2 -1 V
3 0 V
3 0 V
3 0 V
3 -1 V
3 0 V
3 0 V
2 0 V
3 -1 V
3 0 V
3 0 V
3 0 V
3 0 V
2 -1 V
3 0 V
3 0 V
3 0 V
3 0 V
3 -1 V
3 0 V
2 0 V
3 0 V
3 0 V
3 -1 V
3 0 V
3 0 V
3 0 V
2 0 V
3 -1 V
3 0 V
3 0 V
3 0 V
3 0 V
2 -1 V
3 0 V
3 0 V
3 0 V
3 0 V
3 0 V
3 -1 V
2 0 V
3 0 V
3 0 V
3 0 V
3 -1 V
3 0 V
3 0 V
2 0 V
3 0 V
3 0 V
3 0 V
3 -1 V
3 0 V
2 0 V
3 0 V
3 0 V
3 0 V
3 -1 V
3 0 V
3 0 V
2 0 V
3 0 V
3 0 V
3 0 V
3 -1 V
3 0 V
3 0 V
2 0 V
3 0 V
3 0 V
3 0 V
3 -1 V
3 0 V
2 0 V
3 0 V
3 0 V
3 0 V
3 0 V
3 0 V
3 -1 V
2 0 V
3 0 V
3 0 V
3 0 V
3 0 V
3 0 V
3 0 V
2 0 V
3 -1 V
3 0 V
3 0 V
3 0 V
3 0 V
2 0 V
3 0 V
3 0 V
3 0 V
3 0 V
3 -1 V
3 0 V
2 0 V
3 0 V
3 0 V
3 0 V
3 0 V
3 0 V
3 0 V
2 0 V
3 0 V
3 -1 V
3 0 V
3 0 V
3 0 V
2 0 V
3 0 V
3 0 V
3 0 V
3 0 V
3 0 V
3 0 V
2 0 V
3 0 V
3 0 V
3 -1 V
3 0 V
3 0 V
3 0 V
2 0 V
3 0 V
3 0 V
3 0 V
3 0 V
3 0 V
2 0 V
3 0 V
3 0 V
3 0 V
3 0 V
3 0 V
3 0 V
2 0 V
3 0 V
3 -1 V
3 0 V
3 0 V
3 0 V
3 0 V
2 0 V
3 0 V
3 0 V
3 0 V
3 0 V
3 0 V
2 0 V
3 0 V
3 0 V
3 0 V
3 0 V
3 0 V
3 0 V
2 0 V
3 0 V
3 0 V
3 0 V
3 0 V
3 0 V
3 0 V
2 0 V
3 0 V
3 0 V
3 0 V
3 0 V
3 0 V
2 0 V
3 0 V
3 0 V
3 0 V
stroke
grestore
end
showpage
}}%
\put(3053,1547){\makebox(0,0)[r]{$A=0.000$}}%
\put(3053,1647){\makebox(0,0)[r]{$A=0.002$}}%
\put(3053,1747){\makebox(0,0)[r]{$A=0.004$}}%
\put(3053,1847){\makebox(0,0)[r]{$A=0.008$}}%
\put(3053,1947){\makebox(0,0)[r]{$A=0.016$}}%
\put(657,1647){\makebox(0,0)[l]{$h^{-2}\sigma_e(q)$}}%
\put(1883,324){\makebox(0,0)[l]{$qh/\pi$}}%
\put(3450,100){\makebox(0,0){ 1}}%
\put(2880,100){\makebox(0,0){ 0.8}}%
\put(2310,100){\makebox(0,0){ 0.6}}%
\put(1740,100){\makebox(0,0){ 0.4}}%
\put(1170,100){\makebox(0,0){ 0.2}}%
\put(600,100){\makebox(0,0){ 0}}%
\put(550,1853){\makebox(0,0)[r]{$10^{4}$}}%
\put(550,1440){\makebox(0,0)[r]{$10^{2}$}}%
\put(550,1027){\makebox(0,0)[r]{$1$}}%
\put(550,613){\makebox(0,0)[r]{$10^{-2}$}}%
\put(550,200){\makebox(0,0)[r]{$10^{-4}$}}%
\end{picture}%
\endgroup
 

%% file: main.bbl
\begin{thebibliography}{34}
\bibitem[{fn2()}]{fn2}
 \bibinfo{note}{To the first order in $\varepsilon = \alpha/[1 - \alpha
  S(k,q_0)]$, where $q_0$ is the wavenumber such that $\vert 1 - \alpha
  S(k,q_0) \vert = {\rm min}$.}
\bibitem[{fn3()}]{fn3}
 \bibinfo{note}{In general, resonances take place at $q=\pm k + 2\pi
  l/h$, where $l$ is an arbitrary integer, with the restriction that $q$ must
  be in the first Brillouin zone of the lattice, $-\pi/h < q \leq \pi/h$. If
  $k<\pi/h$, as is the case in this paper, the only possible solutions are
  $q=\pm k$.}
\bibitem[{fn1()}]{fn1}
 \bibinfo{note}{Strictly speaking, there may be two solutions rather
  than one, since the curve can cross zero at two separate points. This happens
  in Fig.~\ref{fig:disp}(a) for $kh\le 0.2\pi$. However, since the peaks are
  exponentially narrow, the physical effect of exciting two SP with slightly
  different wave numbers is not expected to be physically observable.}
\bibitem[{fn4()}]{fn4}
 \bibinfo{note}{The singularity is integrable if we adopt the usual
  rule for bypassing the pole. Namely, the pole position at $q=q_1+i0$
  corresponds to $\gamma=+0$.}
\end{thebibliography}
